\documentclass{emulateapj}
\usepackage{graphicx}
\usepackage{lscape}

\slugcomment{Accepted for publication in the Astrophysical Journal Supplements}
\shortauthors{Tripp et al.}
\shorttitle{IGM \ion{O}{6} Statistics}

\begin{document}

\title{A High-Resolution Survey of Low-Redshift QSO Absorption Lines:
Statistics and Physical Conditions of O vi Absorbers\altaffilmark{1}}

\altaffiltext{1}{Based on observations with (1) the NASA/ESA {\it
Hubble Space Telescope}, obtained at the Space Telescope Science
Institute, which is operated by the Association of Universities for
Research in Astronomy, Inc., under NASA contract NAS 5-26555, and (2)
the NASA-CNES/ESA {\it Far Ultraoviolet Spectroscopic Explorer}
mission, operated by Johns Hopkins University, supported by NASA
contract NAS 5-32985.}

\author{Todd M. Tripp,\altaffilmark{2}, Kenneth
R. Sembach,\altaffilmark{3} David V. Bowen,\altaffilmark{4} Blair
D. Savage,\altaffilmark{5} Edward B. Jenkins,\altaffilmark{4}
Nicolas Lehner,\altaffilmark{6} Philipp Richter\altaffilmark{7}}

\altaffiltext{2}{Department of Astronomy, University of Massachusetts,
710 North Pleasant Street, Amherst, MA 01003-9305;
tripp@fcrao1.astro.umass.edu}

\altaffiltext{3}{Space Telescope Science Institute, 3700 San Martin
Drive, Baltimore, MD 21218}

\altaffiltext{4}{Princeton University Observatory, Peyton Hall, Ivy
Lane, Princeton, NJ 08544}

\altaffiltext{5}{Department of Astronomy, University of
Wisconsin-Madison, 475 North Charter Street, Madison, WI 53706}

\altaffiltext{6}{Department of Physics, University of Notre Dame, 225
Nieuwland Science Hall, Notre Dame, IN 46556}

\altaffiltext{7}{Institut f\"{u}r Physik, Universit\"{a}t Potsdam, Am Neuen Palais 10, 14469 Potsdam, Germany}

\begin{abstract}
Using high-resolution ultraviolet spectra of 16 low$-z$ QSOs obtained
with the E140M echelle mode of the Space Telescope Imaging
Spectrograph, we study the physical conditions and statistics of
\ion{O}{6} absorption in the intergalactic medium (IGM) at $z <
0.5$. We identify 51 intervening ($z_{\rm abs} \ll z_{\rm QSO}$)
\ion{O}{6} systems comprised of 77 individual components, and we find
14 ``proximate'' systems ($z_{\rm abs} \approx z_{\rm QSO}$)
containing 34 components. For intervening systems (components) with
rest-frame equivalent width $W_{\rm r} >$ 30 m\AA , the number of
\ion{O}{6} absorbers per unit redshift $dN/dz$ = 15.6$^{+2.9}_{-2.4}$
(21.0$^{+3.2}_{-2.8}$), and this decreases to $dN/dz$ =
0.9$^{+1.0}_{-0.5}$ (0.3$^{+0.7}_{-0.3}$) for $W_{\rm r} >$ 300 m\AA
. The number per redshift increases steeply as $z_{\rm abs}$
approaches $z_{\rm QSO}$; we find that $dN/dz$ is $\approx 3 - 10$
times higher within 2500 km s$^{-1}$ of $z_{\rm QSO}$. The most
striking difference between intervening and proximate systems is that
some proximate absorbers have substantially lower
\ion{H}{1}/\ion{O}{6} ratios. The lower ratios in proximate systems
could be partially due to ionization effects, but these proximate
absorbers must also have significantly higher metallicities.  We find
that 37\% of the intervening \ion{O}{6} absorbers have velocity
centroids that are well-aligned with corresponding \ion{H}{1}
absorption. If the \ion{O}{6} and the \ion{H}{1} trace the same gas,
the relatively small differences in line widths imply the absorbers
are cool with $T < 10^{5}$ K.  Most of these well-aligned absorbers
have the characteristics of metal-enriched photoionized gas.  However,
the \ion{O}{6} in the apparently simple and cold systems could be
associated with a hot phase with $T \approx 10^{5.5}$ K if the
metallicity is high enough to cause the associated broad Ly$\alpha$
absorption to be too weak to detect. We show that 53\% of the
intervening \ion{O}{6} systems are complex multiphase absorbers that
can accommodate both lower metallicity collisionally-ionized gas with
$T > 10^{5}$ K and cold photoionzed gas.
\end{abstract}


\section{Introduction\label{intro}}

Currently, several fundamental questions about the low-redshift
intergalactic medium (IGM) require better observational constraints.
What fraction of the ordinary baryonic matter in the universe is
located in the IGM at the present epoch?  What are the physical
conditions of the intergalactic baryons?  To what degree have
intergalactic gas clouds been enriched with metals, and what are the
physical processes that exchange matter and energy between galaxies
and the IGM?  Interest in the properties of the low-redshift IGM is
motivated by several broad issues in galaxy evolution and cosmology:

First, theoretical studies indicate that the IGM is the primary
reservoir of baryons throughout the history of the universe, but the
IGM is predicted to change from predominantly cool, photoionized gas
at high redshifts to a mixture of shock-heated gas, photoionized gas,
and condensed objects (stars) at low redshifts (Cen \& Ostriker 1999;
Dav\'{e} et al. 1999,2001; Cen \& Ostriker 2006).  The conversion of a
substantial fraction of the IGM from cool gas into moderately hot gas
could solve the long-standing ``missing baryons problem'', the fact
that the inventory of readily-observed low-z baryonic matter (e.g.,
Persic \& Salucci 1992; Fukugita et al. 1998) falls far short of the
quantity expected based on big bang nucleosynthesis and deuterium
measurements (e.g., O'Meara et al. 2006) and cosmic microwave
background observations (e.g., Spergel et al. 2006).  As time passes,
gas collects in deeper potential wells.  Some of that gas cools and
forms stars, but hydrodynamic simulations show that as gas accretes
into the potential wells of galaxies and groups/clusters,
gravitational shock heating drives much of it into the $10^{5} -
10^{7}$ K temperature range (Cen \& Ostriker 1999; Dav\'{e} et
al. 2001).  The models indicate that this shocked material is often
located in modest overdensity regions outside of galaxies, and it is
cooler and less dense than the X-ray emitting hot gas seen in galaxy
clusters, hence it has been dubbed the ``warm-hot intergalactic
medium'' (WHIM).  Dav\'{e} et al. (2001) have analyzed a set of six
hydrodynamic cosmological simulations with diverse computational
characteristics (e.g., different spatial resolutions and box sizes,
different numerical methods, and different assumptions about and
treatments of physical processes), and they find that the simulations
all robustly predict that a substantial fraction of the baryons ($30 -
50$\% ) should be found in the WHIM at $z \approx$ 0 (see also Cen \&
Fang 2006).  Hydrodynamic cosmological simulations are now being used
for a wide variety of purposes, so it is important to test the robust
predictions from these simulations with observations. WHIM
observations are also valuable for this purpose as well as for probing
fundamental questions about the physical conditions and distribution
of the baryons.  In addition, more recent simulations suggest that
WHIM models (and the very definition of the ``WHIM'') may require
refinements.  For example, while the original simulations placed the
low$-z$ missing baryons in the $10^{5} - 10^{7}$ K range, the
simulations of Kang et al. (2005) indicate that a significant portion
of the WHIM is heated and ionized by lower-velocity shocks and
photoionization and has $T < 10^{5}$ K.  Obtaining good constraints on
the properties of the WHIM is now a major observational goal of IGM
studies.

Second, it is becoming increasingly evident that the processes that
add or remove gas and energy from galaxies play an important role in
galaxy evolution (e.g., Keres et al. 2005; Veilleux et al. 2005; Voit
2005).  However, most observational constraints on processes such as
gas accretion and feedback via galactic winds are limited to regions
relatively close to galaxies.  Low-redshift absorption systems in the
spectra of quasi-stellar objects (QSOs) can provide detailed
information about the physical conditions and chemical enrichment of
intergalactic gas farther away from galaxies and thereby provide a
more complete view of how matter and energy are exchanged between
galaxies and the IGM.  For example, observations of nearby
star-forming galaxies show a strong relationship between stellar mass
and gas-phase metallicity (e.g., Tremonti et al. 2004) that has been
suggested to be a result of galactic winds: in the deeper potential
wells of more massive galaxies, supernova ejecta are retained, but
lower-mass objects lose metals into the IGM by the action of Galactic
winds (e.g., Mac Low \& Ferrara 1999).  But how much impact does this
``feedback'' from a lower-mass galaxy have on its environment?
Escaping outflows can, in principle, travel substantial distances from
the source galaxies. Direct evidence of winds has been found in the
immediate vicinity of starbursting and ultraluminous infrared galaxies
(Veilleux et al. 2005, and references therein), but how large is the
``sphere of influence'' of such galaxies?  Many commonly employed
techniques for the study of galactic outflows, e.g., optical and X-ray
emission observations, lose track of escaping material as it flows
away from the source, and absorption spectroscopy of star clusters
within the galaxies, while highly useful, mainly probes the outflows
within a few kpc of the galaxies.  QSO absorption lines provide a
unique opportunity to probe the lower density material at larger
distances away from interesting foreground galaxies.

Moreover, many recent papers have considered the possibility that
galaxy evolution is affected by feedback from black hole
accretion-driven processes in QSOs and active galactic nuclei (AGNs)
such as QSO/AGN winds that might develop when significant amounts of
matter are driven into the central regions of galaxies (e.g., Springel
et al. 2005a,b).  The dramatic ``broad absorption line'' (BAL) QSOs
(e.g., Turnshek 1988) provide evidence that such outflows exist, and
with outflow velocities that often exceed 10,000 km s$^{-1}$, BAL QSOs
could propel material to substantial distances and affect a large
region surrounding a QSO.  While it is not clear from observations how
much impact BAL QSOs really have on their surroundings, we do know
that BALs are relatively common: based on a large sample of QSOs from
the Sloan Digital Sky Survey (SDSS), Trump et al. (2006) report that
26\% of a sample of QSOs at $z \geq$ 1.7 show broad \ion{C}{4}
absorption features. Moreover, there is evidence that some {\it narrow}
absorption lines arise in QSO outflows (see \S \ref{absclass}), and
lower-luminosity AGNs are also known to drive high-velocity flows
(Crenshaw et al. 2003). These observations raise several questions:
How many of the absorbers with $z_{\rm abs} \ll z_{\rm QSO}$ arise in
high-velocity QSO/AGN outflows?  How do these outflows affect their
surroundings?  Are they important sources of feedback, and on what
scales?  When using QSOs to study the foreground ``ordinary'' IGM, we
can also simultaneously search for any evidence of narrow absorption
lines that arise in QSO outflows. 

For the last several years, we have used high-resolution ultraviolet
QSO spectra to study the properties of the intergalactic medium in the
nearby universe, galaxy-absorber connections, and the roles played by
the IGM in galaxy evolution.  We have provided complete surveys of all
absorption lines detected on individual QSO sight lines (Sembach et
al. 2001; Savage et al. 2002; Jenkins et al. 2003; Richter et
al. 2004; Sembach et al. 2004; Lehner et al. 2006, 2007) as well as
detailed studies of individual systems of interest (e.g., Tripp et
al. 2000,2001,2002,2005,2006a; Tripp \& Savage 2000; Savage et
al. 2005; Jenkins et al. 2005; Aracil et al. 2006).  The low$-z$ IGM
has been extensively studied with high-resolution spectra by other
groups as well (e.g., Penton et al. 2000a,b; 2002; 2004; Chen \&
Prochaska 2000; Chen et al. 2005; Prochaska et al. 2004, 2006; Shull
et al. 1998, 2003; Stocke et al. 2004,2006; Danforth \& Shull 2005;
Danforth et al. 2006; Tumlinson et al. 2005; Cooksey et al. 2007; Thom
\& Chen 2008).  These studies have provided clear observational
evidence that a substantial fraction of the baryons are in the IGM at
the present epoch, but many key questions remain open.

In this paper, we employ a larger sample in order to investigate the
statistical properties and physical conditions of the low$-z$ IGM.
For this purpose, we present a survey of low$-z$ \ion{O}{6} absorbers
based on high-resolution ultraviolet spectra of sixteen low-redshift
QSOs observed with the {\it Hubble Space Telescope (HST)} with the
Space Telescope Imaging Spectrograph (STIS).  Our observations were
made with the E140M echelle mode of STIS, which provides excellent
spectral resolution (7 km s$^{-1}$ FWHM).  Moreover, these spectra
have good sensitivity for detection of weak absorption lines; the data
are sufficient for detection of lines with rest-frame equivalent
widths $W_{\rm r} > 30$ m\AA\ over a substantial path length.
Consequently, these sight lines provide a unique opportunity to obtain
precise and deep measurements of low$-z$ QSO absorbers.  For seven of
our sight lines, we supplement the STIS observations with spectra
obtained with the {\it Far Ultraviolet Spectroscopic Explorer (FUSE)},
which extend the spectral coverage farther into the far ultraviolet.
In this paper, we present the first results from this survey: we
report the measurements of all intervening and proximate \ion{O}{6}
absorption-line systems in these spectra with a focus on the
statistics and physical conditions of the absorbers.  While our
primary interest is in the IGM, to assess the issues discussed above,
we also compare and contrast the intervening and proximate (i.e.,
$z_{\rm abs} \approx z_{\rm QSO}$) absorbers.

The paper is organized as follows: In \S 2, we present the ultraviolet
spectra that we use for the survey, and we discuss the data reduction
(\S \ref{uvspecsec}), our procedures for line identifications and
measurements (\S \ref{idsec} and \ref{secmeas}, respectively), and our
classification of the absorbers (\S \ref{classsec}); we describe in \S
\ref{classmulti} how we distinguish between apparently
simple/single-phase absorbers and complex multiphase systems, and in
\S \ref{absclass} we also separate the systems into intervening
absorbers or ``proximate'' ($z_{\rm abs} \approx z_{\rm QSO}$) cases.
A variety of statistics of the intervening and proximate absorbers are
reviewed in \S 3 including the number of \ion{O}{6} absorbers detected
per unit redshift (\S \ref{dndzsec}), the column density and Doppler
parameter distributions of the \ion{O}{6} lines (\S \ref{basicstats}),
the correlation (or lack thereof) between the \ion{O}{6} column
densities and Doppler parameters (\S \ref{nbcor}), a detailed
comparison of the {\it shapes} of the \ion{O}{6} and \ion{H}{1} lines
and the fractions of the absorbers that have well-aligned \ion{O}{6}
and \ion{H}{1} profiles (\S \ref{shapesection}), and the correlations
of $N$(\ion{O}{6}) and $N$(\ion{H}{1})/$N$(\ion{O}{6}) with
$N$(\ion{H}{1}) (\S \ref{ratiosec}).  We analyze the physical
conditions of the intervening absorbers in \S 4.  In this section, we
first derive constraints on the absorber plasma temperatures for
components with well-aligned \ion{O}{6} and \ion{H}{1} profiles under
the assumption that the good correspondence of the \ion{O}{6} and
\ion{H}{1} profile centroids indicates that the \ion{O}{6} and
\ion{H}{1} lines arise in the same gas (\S \ref{tempsec}). We then
show that while purely collisionally ionized models work poorly given
the implied cool temperatures (\S \ref{collionsec}), models including
photoionization are highly consistent with the properties of the
well-matched \ion{O}{6} and \ion{H}{1} lines (\S \ref{photoionsec},\S
\ref{hybridsec}).  In \S \ref{complexmulti} we demonstrate that the
complex multicomponent/multiphase \ion{O}{6} absorbers can accommodate
warm-hot gas, but additional information is needed to constrain these
complex systems. We also examine in \S \ref{hiddenblasec} whether the
apparently simple and cold \ion{O}{6} absorbers could actually be
multiphase systems, including hot collisionally ionized gas, in which
the broad \ion{H}{1} absorption expected to go with the hot \ion{O}{6}
phase is hidden in the noise.  We find that this is possible if the
\ion{O}{6} absorbers are essentially always multiphase systems with
the \ion{O}{6} frequently located in a quiescent interface layer on
the surface of a cooler phase. We close with discussion of the
implications of our measurements, including comments on the baryonic
content of the intervening absorbers and some suggestions for future
observations (\S 5) before we summarize the paper in \S 6.  The
Appendix provides comments about individual absorption systems
including line identification and problems caused by blending, hot
pixels, and saturation.  In this work, we assume the solar oxygen and
carbon abundances reported by Allende Prieto et al. (2001,2002):
(O/H)$_{\odot} = 4.9 \times 10^{-4}$, (C/H)$_{\odot} = 2.5 \times
10^{-4}$. We note that these abundances are currently a topic of
debate (e.g., Basu \& Antia 2004; Bahcall et al. 2005a,b), and the
solar C and O abundances could be $\approx 0.2$ dex higher.  While
important for some questions about low$-z$ \ion{O}{6} absorbers, this
level of uncertainty does not significantly impact the results
presented in this paper.  In this paper we also assume $H_{0}$ = 75 km
s$^{-1}$ Mpc$^{-1}$, $\Omega_{\rm m}$ = 0.3, and $\Omega_{\Lambda}$ =
0.7.

\section{Data\label{datasec}}

\subsection{Ultraviolet Spectroscopy\label{uvspecsec}}

Our survey of low-redshift \ion{O}{6} absorbers is based primarily on
high-resolution ultraviolet spectroscopy of QSOs acquired with STIS.
We only use data obtained with the E140M echelle mode of STIS for two
reasons: (1) this setup provides the best combination of spectral
resolution and signal-to-noise (S/N) ratio that can be obtained in
practical amounts of {\it HST} time, and (2) single exposures with
this STIS mode cover a relatively large wavelength range.  STIS E140M
observations typically cover the 1150$-$1730 \AA\ range with 7 km
s$^{-1}$ resolution (FWHM) with 2 pixels per resolution
element.\footnote{The E140M spectral resolution is a factor of 3 (or
more) higher than any of the STIS first-order grating modes, and the
wavelength range is $\sim$6 times larger.}  The E140M wavelength scale
calibration is excellent: the STIS Handbook (Kim Quijano et al. 2003)
reports that the relative wavelength scale is accurate to $0.25 - 0.5$
pixels ($\sim 1-2$ km s$^{-1}$) across the full spectral range, and
the absolute wavelength calibration is accurate to $0.5 - 1.0$ pixels
($\sim 2-4$ km s$^{-1}$).  We have occasionally identified slightly
larger relative wavelength scale errors (Tripp et al. 2005), but we
generally find agreement with the STIS Handbook wavelength
accuracies. We reduced the STIS data using the STIS ID Team version of
CALSTIS at the Goddard Space Flight Center. Our procedure for
reduction of the STIS E140M data is described in Tripp et
al. (2001,2005) and includes the two-dimensional echelle scattered
light correction (Valenti et al. 2002) and the algorithm for automatic
repair of hot pixels (Lindler 2003). 

Warm/hot pixels are a difficult problem suffered by the STIS MAMA
detector used in the E140M mode, and the possible effects of warm/hot
pixels must always be borne in mind.  The hot pixel problem often
affects several adjacent pixels (see examples in \S \ref{classmulti}
and the Appendix), and thus hot pixel ``features'' can have a more
severe impact than a single occasional bad pixel.  In some cases, the
automatic hot pixel repair algorithm successfully corrects hot pixels
by interpolation, but many warm/hot pixels are not identified by the
algorithm and are evident in the final spectra.  Up until 2002 August,
the position of the spectrum on the E140M MAMA detector was moved
roughly monthly.\footnote{After 2002 August, this procedure was halted
to enable a sensitivity correction that better accounts for the
echelle blaze.}  This offsetting procedure provides another means to
identify and suppress hot pixels: when the STIS observations were
obtained on multiple dates separated by more than a month (see
Table~\ref{obslog}), hot pixels will only be present in the extracted
spectrum in observations obtained on a particular date (because the
spectrum was shifted to a different place on the detector on the other
dates).  By masking and rejecting the pixels in the affected
individual exposures obtained at the ``bad'' position, the hot pixels
can be suppressed.  Of course, this reduces the S/N ratio in the
affected region because only a subset of the exposures are coadded in
that region, but by judiciously masking a minimum number of pixels in
the vicinity of the hot pixel feature, the impact of S/N loss is
minimized.  Hot pixels are often quite obvious, even in individual
exposures. However, low-level ``warm'' pixels can be difficult to
recognize, and many of the QSOs were observed on essentially a single
date (i.e., with the spectrum located at the same place on the
detector for all exposures) or were observed after 2002 August, so
unfortunately, in some cases we were unable to mitigate the effect of
warm/hot pixel features.  An example of hot pixel supression and
comments on hot pixel problems are provided in the Appendix.  Further
information on the design and performance of STIS can be found in
Kimble et al. (1998) and Woodgate et al. (1998).

For this survey, we used two simple criteria to select low-redshift
QSOs from the STIS E140M archive: (1) we required the QSO redshift
($z_{\rm QSO}$) to be greater than 0.15 in order to provide a
sufficiently large redshift range that could be searched for the
\ion{O}{6} $\lambda \lambda$ 1031.93, 1037.62 doublet, and (2) the S/N
had to be sufficient to detect lines with rest-frame equivalent width
$W_{\rm r} > 30$ m\AA\ over a substantial portion of the \ion{O}{6}
redshift range.  These criteria resulted in a sample of 16
low-redshift QSOs.  We summarize in Table~\ref{obslog} some basic
properties of the 16 sight lines including $z_{\rm QSO}$, Galactic
coordinates, and column densities of the Galactic \ion{H}{1} and
H$_{2}$ due to the foreground ISM in each direction.\footnote{Galactic
ISM lines block portions of the spectrum and reduce the effective path
that can be searched for extragalactic lines of interest; the H~I
and H$_{2}$ column densities in Table~\ref{obslog} provide
an indication of the degree of blocking and confusion caused by the
ISM.}  Table~\ref{obslog} also provides a log of the STIS E140M
observations.  Ten of the targets are from our own {\it HST} programs
that were explicitly designed to study low$-z$ \ion{O}{6} absorbers;
the other 6 targets were not observed specifically to investigate the
\ion{O}{6} systems but were generally selected to study various types
of low$-z$ QSO absorbers.  Because the observations were obtained for
different programs, some used the $0\farcs 06 \times 0\farcs 2$
aperture while others used the $0\farcs 2 \times 0\farcs 2$ slit as
noted in Table~\ref{obslog}.  The spectroscopic line-spread function
has more prominent broad wings when the $0\farcs 2 \times 0\farcs 2$
is used (Kim Quijano et al. 2003), and we take this into account when
fitting Voigt profiles to the absorption lines (see below). Most of
the targets in our sample were observed to similar S/N ratios, but
there is some S/N variation among the sight lines, and a few of the
QSO spectra have better S/N (e.g., 3C 273.0 and H1821+643).  For
purposes of comparison, the final column in Table~\ref{obslog} lists
the mean S/N ratio per resolution element in a 2 \AA\ continuum
region\footnote{We used a 2 \AA\ region centered on 1300 \AA\ unless a
line was found in that region, in which case a small shift was applied
so that the S/N was calculated in a line-free continuum region.}  at
$\lambda _{\rm ob} \approx$ 1300 \AA .

Many of the QSOs in our sample have also been observed with {\it FUSE}
for various purposes (e.g., Savage et al. 2003; Sembach et al. 2003).
The STIS E140M wavelength range covers \ion{O}{6} $\lambda
\lambda$1031.93,1037.62 doublets with redshifts ranging from $z_{\rm
abs} \gtrsim 0.12$ out to the redshift of the QSO for all 16
QSOs.\footnote{The STIS E140M sensitivity drops precipitously at
$\lambda \lesssim 1175$ \AA ; the S/N of the data and the lower
redshift limit for which the STIS data provide useful coverage of
O~VI vary from sightline to sightline depending on the flux of
the QSO at $\lambda < 1200$ \AA\ and the total integration time.}  The
{\it FUSE} spectrographs cover the $912 - 1187$ \AA\ range, and
therefore {\it FUSE} data can be used to extend the survey to cover
all \ion{O}{6} absorbers with $0 \leq z_{\rm abs} \leq z_{\rm QSO}$.
However, the {\it FUSE} bandpass has a high density of lines arising
from the Milky Way ISM, particularly molecular hydrogen lines from
various rotational levels (see, e.g., Wakker 2006; Gillmon et
al. 2006), as well as extragalactic Lyman series \ion{H}{1} lines and
a rich array of (sometimes unfamiliar) metal lines (Verner, Barthel,
\& Tytler 1994) that are redshifted into the {\it FUSE} bandpass for
extragalactic absorbers.  Consequently, one of the lines of the
\ion{O}{6} doublet is often blocked by some unrelated line and is
unmeasureable. For this reason, extra caution is warranted with the
{\it FUSE} spectra because ISM and extragalactic lines can be
incorrectly identified as redshifted \ion{O}{6}.  We have supplemented
our STIS survey with {\it FUSE} observations, but only for the seven
sight lines in the sample for which complete identifications of all
lines in the {\it FUSE} bandpass have been published, including 3C
273.0 (Sembach et al. 2001), H1821+643 (Sembach et al. 2008),
HE0226-4110 (Lehner et al. 2006), PG0953+415 (Savage et al. 2002),
PG1116+215 (Sembach et al. 2004), PG1259+593 (Richter et al. 2004),
and PHL1811 (Jenkins et al. 2003,2005).  In addition, we have used
{\it FUSE} spectra to measure higher Lyman series \ion{H}{1} lines and
\ion{C}{3} $\lambda$977.020 lines, when they can be securely
identified, for other sightlines.

We acquired the {\it FUSE} spectra from the archive and reduced them
following the method described in Tripp et al. (2005).  The {\it FUSE}
spectral resolution is lower than the STIS E140M resolution: the {\it
FUSE} resolution is $\sim 20-25$ km s$^{-1}$ (FWHM).  The uncertainty
in the {\it FUSE} wavelength-scale zero point is also larger.  Moos et
al. (2002) report that the $\lambda$ zero point uncertainty is
typically $\pm$30 km s$^{-1}$ but can be as large as $\pm$100 km
s$^{-1}$.\footnote{Improvements of the CALFUSE data reduction software
implemented after the analysis of Moos et al. (2002) have improved the
{\it FUSE} wavelength-scale uncertainty; see Dixon et al. (2007) and
Bowen et al. (2007) for details.}  However, we have bootstrap
calibrated the {\it FUSE} data by aligning well-detected lines in the
{\it FUSE} spectra with appropriate, comparable-strength lines in the
corresponding STIS spectrum.\footnote{For example, we always aligned
the Milky Way Fe~II $\lambda$1144.94 line ({\it FUSE}) with the
Galactic Fe~II $\lambda$1608.45 line (STIS), and we aligned
Galactic C~II $\lambda$1036.34 ({\it FUSE}) with C~II
$\lambda$1334.53 (STIS).  These strong ISM lines are always detected,
and because they have similar $f\lambda$ values, the {\it FUSE} and
STIS profiles have similar shapes and structure that can be used to
accurately align the {\it FUSE} data with the STIS data. The STIS data
are binned to the resolution of the {\it FUSE} spectra before the
lines are compared and aligned.}  The bootstrap calibration reduces
the uncertainty in the {\it FUSE} wavelength-scale zero point to
$5-10$ km s$^{-1}$.  Additional information about the design and
on-orbit performance of {\it FUSE} can be found in Moos et
al. (2000,2002) and Sahnow et al. (2000).

\subsection{Absorption-Line Identification\label{idsec}}

Several groups have studied the properties of low$-z$ QSO absorption
lines (see \S 1), and the various groups have employed different
methods and criteria for identifying lines and selecting samples.  The
differing methods affect the statistics and analysis outcomes and must
be borne in mind when comparing results from different papers.  We
molded our line identification procedure based on several issues that
affect the identification of extragalactic \ion{O}{6} lines: (1) The
first high-resolution studies of low$-z$ \ion{O}{6} absorbers showed
that they have a wide range of properties.  For example, Tripp et
al. (2000) found that the $N$(\ion{O}{6})/$N$(\ion{H}{1}) ratio varies
by a factor of $\geq$37 in the \ion{O}{6} systems observed toward
H1821+643, and subsequent studies have confirmed that the
\ion{O}{6}/\ion{H}{1} column density ratio is highly variable in these
systems (e.g., Danforth \& Shull 2005).  Because the \ion{H}{1}
Ly$\alpha$ line can be quite weak in \ion{O}{6} systems, we cannot
compile an unbiased sample by first selecting \ion{H}{1} Ly$\alpha$
lines and then searching for the corresponding \ion{O}{6} doublet;
this approach would miss \ion{O}{6} systems that have little or no
detectable \ion{H}{1} absorption (such systems do exist, as we show
below).  (2) Extragalactic \ion{O}{6} and affiliated lines can be
partially or fully blocked (hidden) due to blending with unrelated
absorption lines, either from the foreground ISM or from absorption
systems at other redshifts.  (3) Emission features can partially or
fully contaminate absorption lines.  In the STIS spectra, the
``emission'' features that cause contamination are mainly warm/hot
pixels, which can extend across several adjacent pixels and thereby
fill in real absorption lines (see \S \ref{classmulti} and
Appendix). Some of the hot pixels can be assuaged by the procedures
discussed above, but many of the warm/hot pixels are not adequately
removed in the fully reduced data.  In the {\it FUSE} data, airglow
emission lines from the Earth's atmosphere are present at various
wavelengths (Feldman et al. 2001).  Some of the the airglow lines are
excited by sunlight and can be significantly suppressed by using {\it
FUSE} data recorded on the night side of the orbit only, but this
reduces the total integration time and S/N of the final spectrum, and
some residual airglow emission features remain in the night-only
spectra.  In this paper, we use all of the available {\it FUSE} data
(day and night) in order to maximize the data S/N.

To identify extragalactic \ion{O}{6} absorbers (and the ancillary
lines in their absorption systems) in a way that considers these
issues, we employed a two-pass search procedure:

In the first pass, we searched each spectrum for lines that have the
relative wavelength separation and the relative line strengths of the
redshifted \ion{O}{6} $\lambda \lambda$1031.93, 1037.62 doublet.  In
order to be included in our samples that are used for statistical
measurements and physical conditions analyses in the rest of the
paper, we required that lines have measured rest-frame equivalent
widths ($W_{\rm r}$) that are recorded at $3\sigma$ significance or
better.  However, for identification of doublets or \ion{H}{1} Lyman
series lines, it is useful to examine marginally detected lines as
well.  For example, in several cases we detected the \ion{O}{6}
$\lambda$1031.93 line at the $4-5 \sigma$ level, and the weaker
corresponding \ion{O}{6} $\lambda$1037.62 line was found at $\approx 2
\sigma$ significance.  In these cases, the marginally detected
$\lambda$1037.62 line bolsters the \ion{O}{6} identification, so we
report the marginal $\lambda 1037.62$ $W_{\rm r}$ measurement but
mainly rely on the well-detected line for subsequent analyses.
Marginal (and undetected) lines such as higher \ion{H}{1} Lyman series
lines are also useful for establishing that the well-detected line
measurements are not badly affected by unresolved saturation.  For
example, if a moderately strong \ion{H}{1} Ly$\alpha$ line is detected
but the corresponding Ly$\beta$ line is not present, this provides
some assurance that the measurements based on Ly$\alpha$ are not badly
corrupted by saturation.  When we identified an \ion{O}{6} doublet in
this first pass, we then searched for and identified all affiliated
\ion{H}{1} Lyman series and metal lines (e.g., \ion{C}{3}
$\lambda$977.020, \ion{C}{4} $\lambda \lambda$1548.20, 1550.78, and
\ion{Si}{3} $\lambda$1206.50) at the same redshift.  For a large
fraction of our \ion{O}{6} absorbers, we only detect \ion{O}{6} and
\ion{H}{1} lines, but we do detect \ion{C}{3} and \ion{Si}{3} in a
useful number of absorbers, and we occasionally find additional metal
lines such as the \ion{C}{4} doublet, \ion{C}{2} $\lambda$1334.53, or
the \ion{N}{5} doublet.  In a few cases, a rich array of
low-ionization lines are detected as well as \ion{O}{6} (e.g., Chen \&
Prochaska 2000; Sembach et al. 2004; Savage et al. 2005; Ganguly et
al. 2006).  In this paper, we will concentrate on the implications of
the \ion{H}{1}, \ion{O}{6}, and \ion{C}{3} measurements and limits.
Analyses of other detected extragalactic metals in the 16 sight lines
in this paper will be presented in subsequent papers or have already
been published (Tripp et al. 2000,2001,2002,2005,2006a; Tripp \&
Savage 2000; Oegerle et al. 2000; Chen \& Prochaska 2000; Sembach et
al. 2001,2004; Savage et al. 2002,2005; Yuan et al. 2002; Jenkins et
al. 2003,2005; Prochaska et al. 2004,2006; Richter et al. 2004; Stocke
et al. 2004; Bregman et al. 2004; Narayanan et al. 2005; Aracil et
al. 2006; Lehner et al. 2006; Ganguly et al. 2006; Cooksey et
al. 2007).

In the second pass, we identified absorption systems based on other
lines such as multiple \ion{H}{1} Lyman series lines or \ion{H}{1} +
metals (but not \ion{O}{6}).  With these identifications in hand, we
then searched for either the \ion{O}{6} $\lambda$1031.93 or the
\ion{O}{6} $\lambda$1037.62 line at the same redshift determined from
\ion{H}{1} and metal lines.  This second pass identified \ion{O}{6}
lines not found in the first pass for cases in which one of the
members of the \ion{O}{6} doublet is blocked by blending with an
interloping line and cases in which the \ion{O}{6} $\lambda$1031.93
line is detected but $\lambda$1037.62 is lost in the noise. We have
consistently applied this line-identification procedure to all of our
sight lines.  Other papers have employed different line-identification
and sample-selection procedures, and the different methods can affect
statistics such as $dN/dz$, the number of absorbers per unit redshift.
For example, Thom \& Chen (2008) note some disagreements with our line
identifications.  These discrepancies partly arise because in cases
where the targets were observed with both STIS and {\it FUSE}, Thom \&
Chen use only the STIS observations, while we use both the STIS and
the {\it FUSE} data.  In the \ion{O}{6} redshift range where STIS and
{\it FUSE} data overlap, the {\it FUSE} spectra often have
substantially better S/N ratios and hence reveal lines that are
difficult to detect in the corresponding STIS spectra.  We also more
comprehensively employ the information provided by the spectra in
order to find \ion{O}{6} absorbers. Detailed comments about the system
identifications, including discrepancies in comparisons with other
papers, are provided in the Appendix.  Of course, lines that are
marginally detected can be challenging to identify, and it would be
highly informative to obtain higher S/N spectra.  Hopefully, the
Cosmic Origins Spectrograph (COS, Green et al. 1999) will be installed
in {\it HST} in late 2008 or early 2009.  With this instrument, it
will be possible to obtain significantly higher S/N UV spectra of
low$-z$ QSOs.

\subsection{Absorption-Line Measurements\label{secmeas}}

\subsubsection{Equivalent Widths, Column Densities, Centroids, and $b-$Values\label{measbasics}}

The complete sample of \ion{O}{6} systems that we have identified in
the 16 sight lines is presented in Tables~\ref{intprop} and
\ref{compprop}.  The median redshift of the intervening \ion{O}{6}
systems in this sample is 0.213, and the proximate absorber median
redshift is 0.267. Many of the \ion{O}{6} absorbers show multiple
components (see examples below), and the \ion{O}{6} lines can be
offset in velocity from lower-ionization transitions in the same
absorber.  In this paper, the ``system'' redshift is defined to be the
redshift of the centroid of the strongest {\it individual} component
detected in the \ion{O}{6} profiles.  We use two complementary methods
to measure the absorption-line properties: (1) direct apparent
optical depth integration, and (2) Voigt-profile fitting.  We employ
both methods because they each have advantages and disadvantages, and
comparison of measurements from each technique provides an indication
of the uncertainty introduced by various systematics. Detailed notes
on line identification issues as well as comments and warnings
regarding hot pixels, line blending, and line saturation in individual
absorbers are provided in the Appendix of this paper.  For
convenience, the comments in the Appendix are cross-referenced to the
entries in Table~\ref{compprop}: the number listed in the final column
of Table~\ref{compprop} corresponds to the comment number in the
Appendix.

Table~\ref{intprop} lists the rest-frame equivalent widths and column
densities obtained from direct integration of the apparent optical depth of the 
\ion{O}{6} doublet
lines and the \ion{H}{1} Ly$\alpha$ line along with the spectrograph
that was used to make the measurement.  Note that for \ion{H}{1}, we
mainly list the directly integrated measurements of the Ly$\alpha$
line in Table~\ref{intprop}. Because it is the most frequently
detected transition in quasar spectra, the Ly$\alpha$ equivalent width
is often used as a fiducial measurement when comparing
absorbers. However, the Ly$\alpha$ line is often strong and saturated,
and in that case, better measurements of the \ion{H}{1} column density
are provided by the Voigt-profile fits to multiple \ion{H}{1} Lyman
series lines presented below. To measure the directly integrated
equivalent widths and column densities, we use the methods of Sembach
\& Savage (1992) including their formalism for evaluating the
contribution from continuum placement uncertainty and flux zero level
uncertainty in the overall error budget of the measurements.
Advantages of direct integration are that the measurements are
straightforward, easily implemented, and relatively objective; profile
fitting, in contrast, can yield different results depending on the
number of components that are fitted to the profiles of interest.
Disadvantages of direct line integration are that it does not
disentangle significantly blended components, and if a line is
affected by saturation (which is sometimes unresolved), direct
integration underestimates the line column density.

\begin{figure*}
\centering
    \includegraphics[width=18.0cm, angle=0]{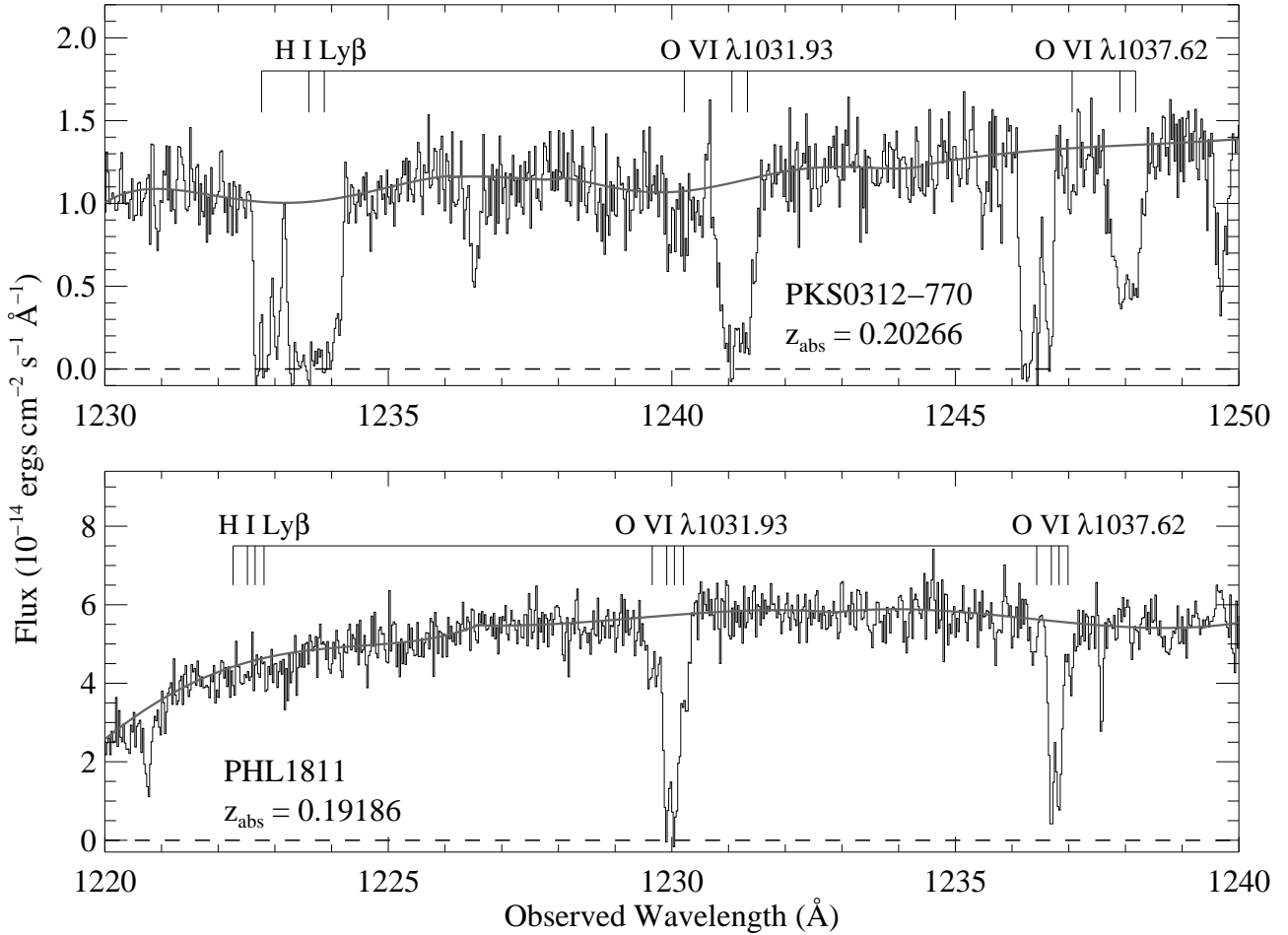}
\caption{Examples of strong O~VI absorbers detected in STIS
E140M echelle spectra of low$-$redshift QSOs.  Small portions of the
flux-calibrated echelle spectra of PKS0312-770 (upper panel) and
PHL1811 (lower panel) are plotted versus heliocentric observed
wavelength, and in both panels, the bar with tick marks indicates the
centroids of components detected in the O~VI $\lambda
\lambda$1031.92, 1037.62 doublet as labeled.  The thick gray lines
indicate our adopted continuum placement in the vicinity of lines of
interest.  Toward PKS0312-770, a strong O~VI system is detected
at $z_{\rm abs}$ = 0.20266, and toward PHL1811 a multicomponent
O~VI absorber is found at $z_{\rm abs}$ = 0.19186.  The tick
marks also show the positions of the corresponding H~I
Ly$\beta$ absorption regardless of whether it is
detected.\label{specsample}}
\end{figure*}

\begin{figure*}
\centering
    \includegraphics[width=8.0cm, angle=0]{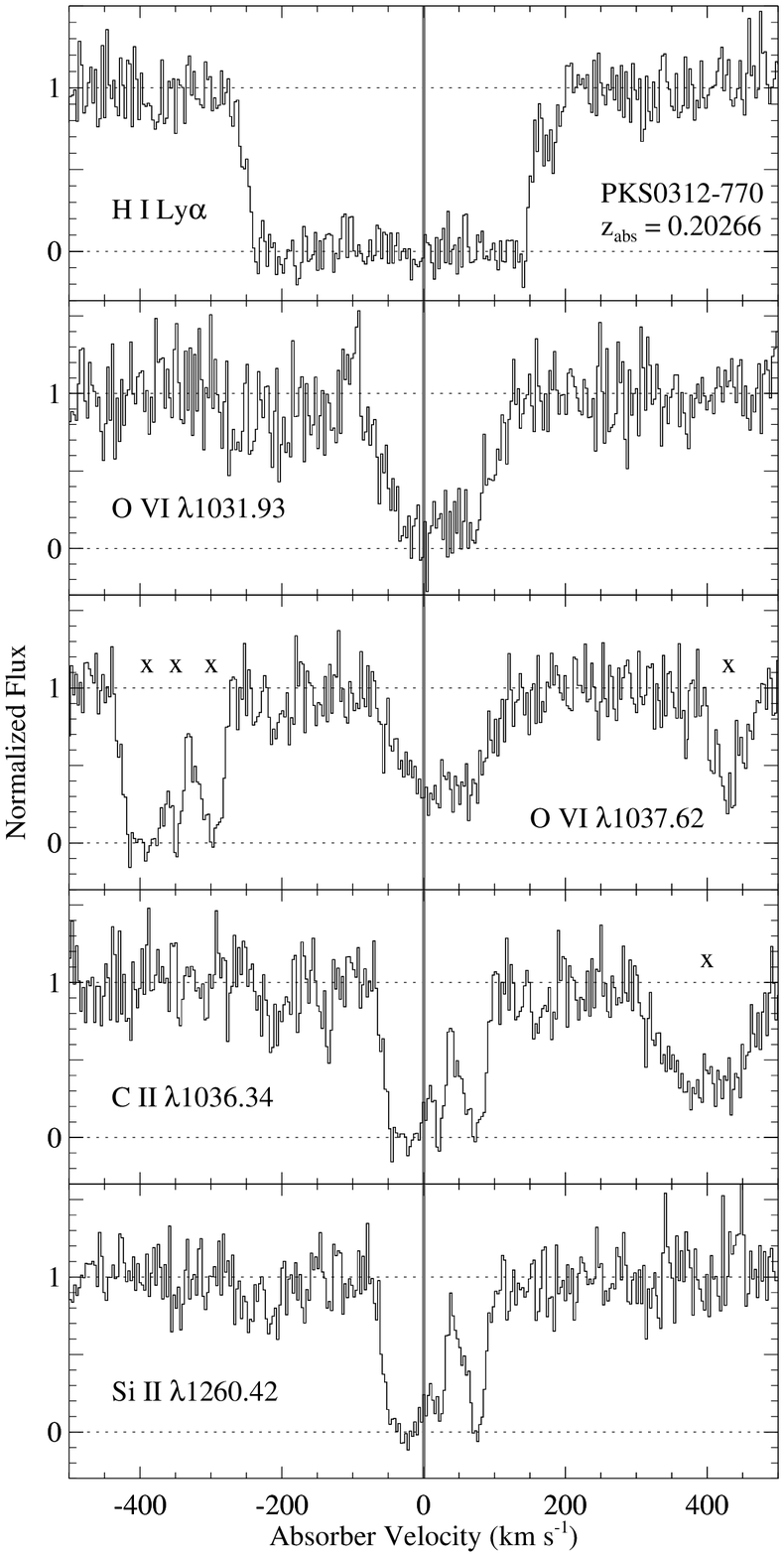}
    \includegraphics[width=8.0cm, angle=0]{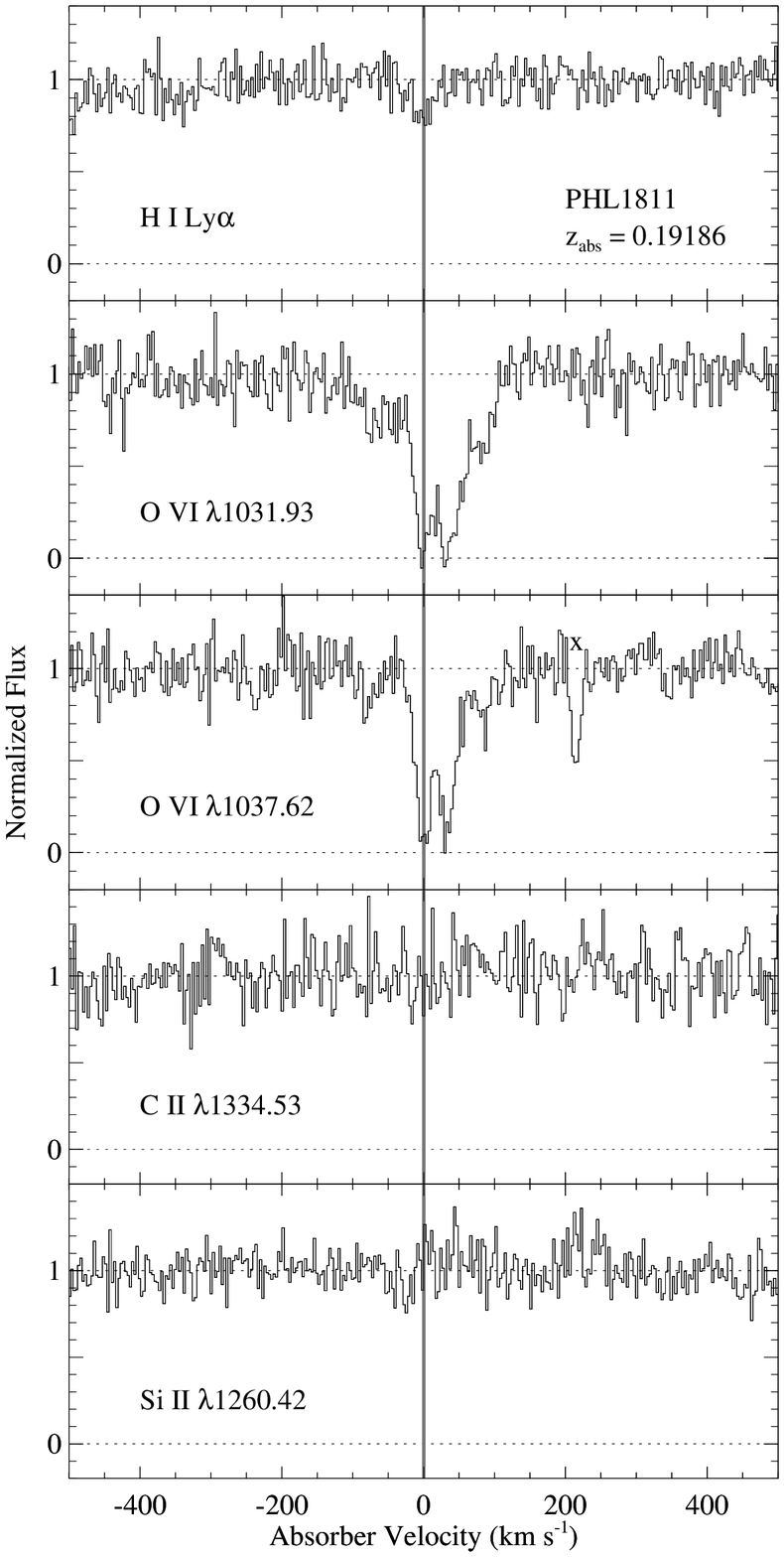}
\caption{{\it Left:} Examples of continuum-normalized absorption
profiles detected in the intervening O~VI absorber at $z_{\rm
abs}$ = 0.20266 in the spectrum of PKS0312-770, including the
H~I Ly$\alpha$, O~VI $\lambda \lambda$1031.93, 1037.62,
C~II $\lambda$1036.34, and Si~II $\lambda$1260.42
transitions, as labeled. Unrelated lines are marked with crosses. {\it
Right:} Examples of analogous absorption profiles detected in the
proximate O~VI absorption system at $z_{\rm abs}$ = 0.19186
toward PHL1811.\label{speccontrast}}
\end{figure*}

\begin{figure}
\centering
    \includegraphics[width=8.0cm, angle=0]{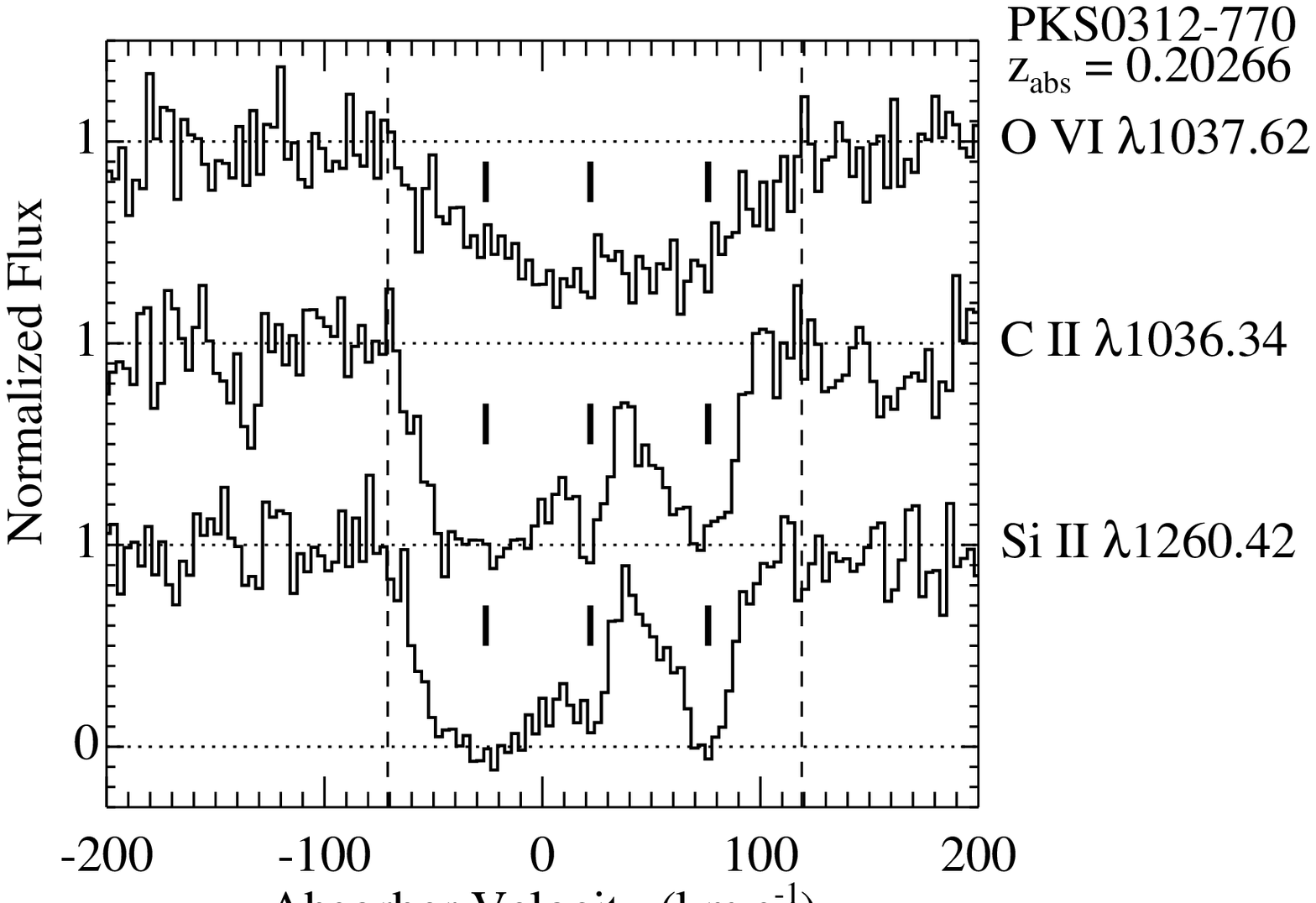}
\caption{Comparison of the detailed absorption profiles of the
O~VI $\lambda$1037.62 ({\it top}), C~II $\lambda$1036.34
({\it middle}), and Si~II $\lambda$1260.42 ({\it bottom}) lines
detected at $z_{\rm abs}$ = 0.20266 toward PKS0312-770. For purposes
of comparison, the vertical dashed lines indicate the approximate
velocity extent of the O~VI absorption profile, and the heavy
tickmarks indicate the velocity centroids of the three primary
components evident in the low-ion profiles.\label{pks0312multi}}
\end{figure}

However, we can check for unresolved saturation by comparing the
``apparent'' column density profiles (Savage \& Sembach 1991; Jenkins
1996) of two or more transitions of a species with significantly
different $f\lambda$ values.  To construct an apparent column density
profile, we first calculate the apparent optical depth $\tau _{\rm
a}(v)$ in each pixel of an absorption profile as a function of velocity,
\begin{equation}
\tau_{\rm a}(v) = {\rm ln} \left[ \frac{I_{\rm c}(v)}{I_{\rm ob}(v)}\right] ,
\end{equation}
where $I_{\rm c}(v)$ is the estimated continuum intensity in the pixel
at velocity $v$ and $I_{\rm ob}(v)$ is the observed intensity in that
same pixel.  To estimate the continuum shape and intensity, we fit a
low-order Legendre polynomial fitted to line-free regions of the
spectrum in the $\pm 1000$ km s$^{-1}$ interval centered on the
absorption line of interest.  The apparent column density is then
determined from the apparent optical depth,
\begin{equation}
N_{\rm a}(v) = \frac{m_{e}c}{\pi e^{2}} \ \frac{\tau _{\rm
a}(v)}{f\lambda} = 3.768 \times 10^{14} \ \frac{\tau _{\rm
a}(v)}{f\lambda} ,\label{naeqn}
\end{equation}
where $f$ is the oscillator strength and $\lambda$ is the wavelength
of the transition, and the other symbols have their usual
meanings. The numerical coefficient in eqn.~\ref{naeqn} requires a
wavelength measured in \AA\ and produces an $N_{\rm a}(v)$ profile in
atoms cm$^{-2}$ (km s$^{-1}$)$^{-1}$.  $N_{\rm a}(v)$ is a
representation of the true column density profile broadened by the
instrument line spread function.  If the line is fully resolved by the
instrument, then $N_{\rm a}(v)$ can be integrated to determine the
total column density.  Even if the line is not well-resolved, the
total column can still be measured reliably provided that the profile
is not significantly affected by unresolved saturation.  If two or
more lines with log $f\lambda$ differing by 0.3 dex or more are
compared and the $N_{\rm a}(v)$ profiles are found to be in agreement,
then the profiles are not badly affected by saturation and can be
integrated to measure the total column density, $N_{\rm total} = \int
N_{\rm a}(v) dv$. If, conversely, the $N_{\rm a}(v)$ profiles are
affected by saturation, then weaker lines will indicate higher
apparent columns than their corresponding stronger lines in the
velocity range affected by saturation.  Apparent column density
profiles have additional virtues as we will discuss in \S
\ref{shapesection}.

Since direct integration does not separate blended components, we list
the {\it total} column densities, integrated across all components
within an absorption feature, in Table~\ref{intprop}.  For the
measurements of {\it individual} component properties, we employ the
Voigt profile fitting code developed by Fitzpatrick \& Spitzer (1997)
with the STIS E140M line-spread functions from the STIS Handbook (Kim
Quijano et al. 2003).  We fit all transitions from a particular
species (e.g., the \ion{O}{6} $\lambda$1031.93 and $\lambda$1037.62
transitions) simultaneously and obtain a single set of velocity
centroids, column densities, and Doppler parameters for the components
in the profiles of that species.  Different species (e.g., \ion{H}{1}
vs. \ion{O}{6} or \ion{C}{3} vs. \ion{O}{6}) can exist in separate gas
phases and therefore are fitted independently. Table~\ref{compprop}
lists the velocity centroids, Doppler parameters ($b-$values), and
column densities of {\it individual} components detected in each
\ion{O}{6} absorption system (some \ion{O}{6} absorbers show only one
component, but many of the systems have multiple components).  For
cross referencing with the profile-fitting results, column 6 in
Table~\ref{intprop} lists the number of components that are identified
within each \ion{H}{1} and \ion{O}{6} profile and their velocity
centroids.  Voigt profile fitting is a valuable measurement technique,
but it is important to recognize the limitations and systematic
uncertainties inherent in the method.  Components that are free from
blending, or components that are within blends but have a distinctive
and well-constrained Gaussian shape in optical depth, are generally
well-constrained by profile fitting.  However, many components are
revealed by well-detected but strongly blended inflections and profile
asymmetries, and extra components that fit profile
inflections/asymmetries are sometimes required to obtain an acceptable
fit.  Profile-fitting results for such blended components can be
sensitive to the number of components selected to fit the profile. In
these cases, the main (strongest) components are usually
well-constrained, but the parameters of the ``inflection'' components
can be sensitive to the number of components chosen for the
fit. Components that are poorly constrained due to problems such as
severe blending with adjacent components, significant saturation of
all available lines, or low S/N are marked with a colon in
Table~\ref{compprop}.  However, it is important to bear in mind that
fitting results for complex multicomponent profiles can change
considerably if more (or fewer) components are used in the fit. As
emphasized by Spitzer \& Fitzpatrick (1995) and Fitzpatrick \& Spitzer
(1997), the parameter uncertainties estimated by this Voigt profile
fitting code increase appropriately when components are strongly
blended, but the code cannot fully account for typical sources of
systematic error.  In our analyses below, we will consider how
additional systematic uncertainties could affect our results.

\subsubsection{Alignment of \ion{O}{6} and \ion{H}{1} Components\label{aligndefsec}}

In our previous papers on low$-z$ \ion{O}{6} absorbers, we have found
that some \ion{O}{6} and \ion{H}{1} components are remarkably
well-aligned, and in some cases the narrow width of the aligned
\ion{H}{1} indicates a surprisingly low temperature (e.g., Tripp \&
Savage 2000; Lehner et al. 2006; Tripp et al. 2006a).  We will revisit
this aspect of the \ion{O}{6} systems with the larger sample of this
paper in \S \ref{tempsec}.  We quantitatively identify aligned
components based on the velocity difference between the \ion{O}{6} and
\ion{H}{1} component centroids, $\Delta v$(\ion{H}{1}-\ion{O}{6}), and
the uncertainty in that velocity difference, $\sigma (\Delta v)$.  We
calculate $\sigma (\Delta v)$ by combining the uncertainties in the
\ion{O}{6} and \ion{H}{1} component velocity centroids (from
Table~\ref{compprop}) in quadrature along with a term ($\sigma _{\rm
wavecal}$) to account for the uncertainy in the wavelength scale
calibration: $\sigma (\Delta v) = \sqrt{\sigma _{\rm O~VI}^{2} +
\sigma _{\rm H~I}^{2} + \sigma _{\rm wavecal}^{2}}$.  As noted above,
the STIS and {\it FUSE} data have uncertainties in the
wavelength-scale zero points as well as the relative wavelength
calibration across the wavelength range of the observations.  We
bootstrap calibrated the {\it FUSE} data by comparing similar lines in
the {\it FUSE} and STIS spectra, and this procedure reduces the {\it
FUSE} zero-point uncertainty to $\approx 5$ km s$^{-1}$. When
comparing \ion{O}{6} and \ion{H}{1} components from STIS data only, we
adopt $\sigma _{\rm wavecal}$ = 2 km s$^{-1}$ (based on the relative
wavelength accuracies from the STIS Handbook), and when comparing
components measured from STIS and {\it FUSE} data, we use $\sigma
_{\rm wavecal}$ = 5 km s$^{-1}$.  If an \ion{O}{6} and \ion{H}{1}
component pair have $\Delta v$(\ion{H}{1}-\ion{O}{6}) $< 2\sigma
(\Delta v)$, we consider the components to be aligned.  In the
majority of the absorbers, the component matching is unambiguous: only
one \ion{H}{1} component is aligned with an \ion{O}{6} component
within the 2$\sigma$ velocity difference uncertainty, or the
\ion{H}{1} and \ion{O}{6} components are simply not aligned.  However,
in a very small number of cases, an \ion{O}{6} component is aligned
with two \ion{H}{1} components to within the 2$\sigma$ uncertainties.
In these cases, we use the components with the smallest $\Delta
v$(\ion{H}{1}-\ion{O}{6}) for our analysis, but we discuss how such
ambiguities could affect the results.  In our analysis of aligned
\ion{O}{6} and \ion{H}{1} components, we only use the measurements
that are robust; we exclude the components that are marked with a
colon in Table~\ref{compprop} because those measurements suffer from
large uncertainties for various reasons (see Appendix).

For convenience, we number the components in each absorption system in
column 8 of Table~\ref{compprop}, and we indicate whether the
\ion{O}{6} and \ion{H}{1} components are aligned or offset from each
other.  If the components are aligned, we give the \ion{H}{1} and
\ion{O}{6} components the same number, and if a component is offset,
it is given a unique number.  For example, in the 3C 249.1 system at
$z_{\rm abs}$ = 0.24676, we identify two \ion{H}{1} components at $v =
-79\pm 3$ km s$^{-1}$ (component 1) and $v = 0\pm 1$ km s$^{-1}$
(component 2), and we find one \ion{O}{6} component at $v = 0\pm 3$ km
s$^{-1}$.  The \ion{O}{6} component is aligned with \ion{H}{1}
component 2, and thus the \ion{O}{6} component is also identified as
component 2.  By using the same component numbers for aligned
\ion{H}{1} and \ion{O}{6} cases, the reader can easily identify which
components are matched together from Table~\ref{compprop}.

\subsection{Absorber Classification\label{classsec}}

\begin{figure*}
\centering
    \includegraphics[width=8.0cm, angle=0]{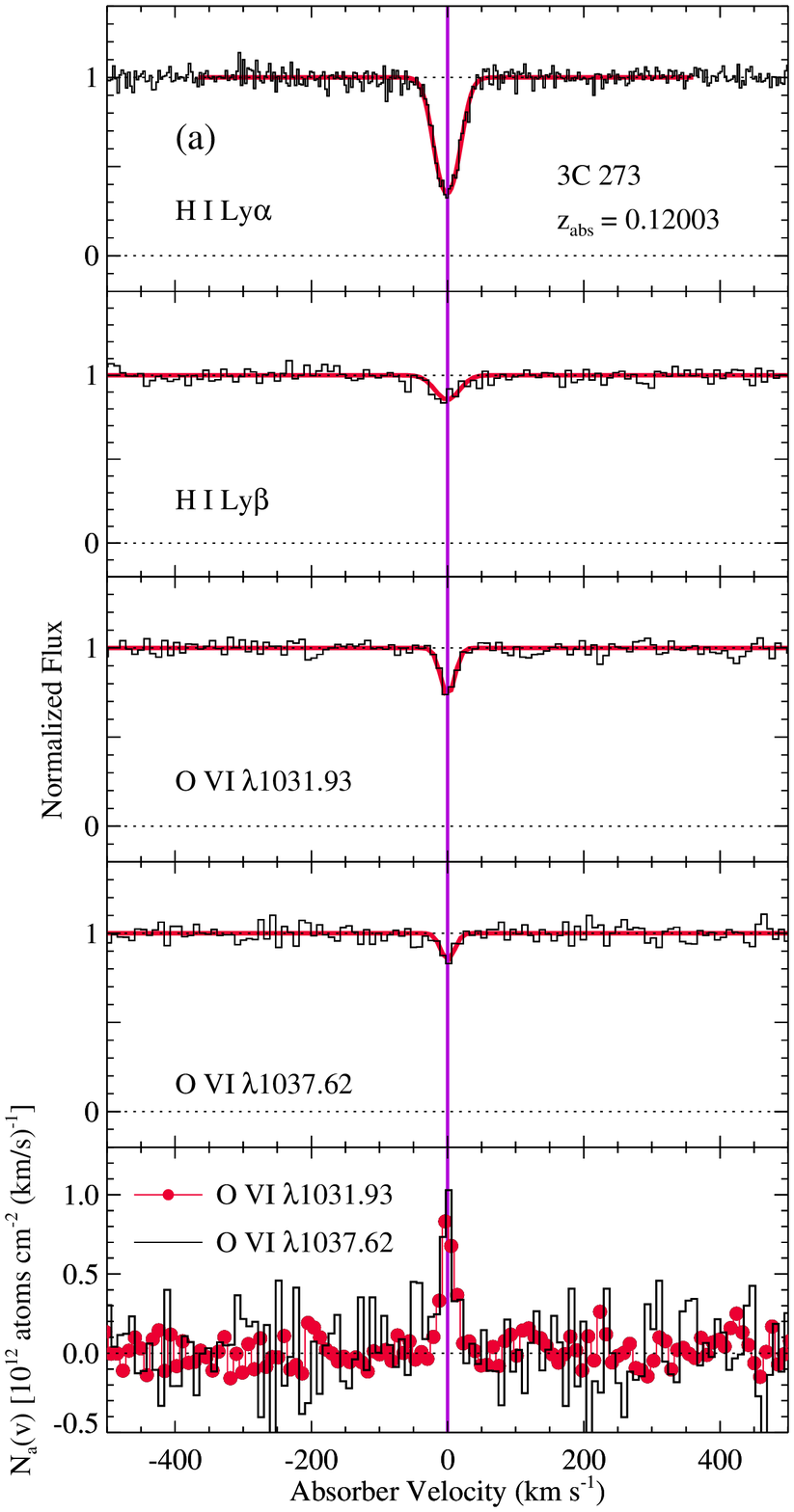}
    \includegraphics[width=8.0cm, angle=0]{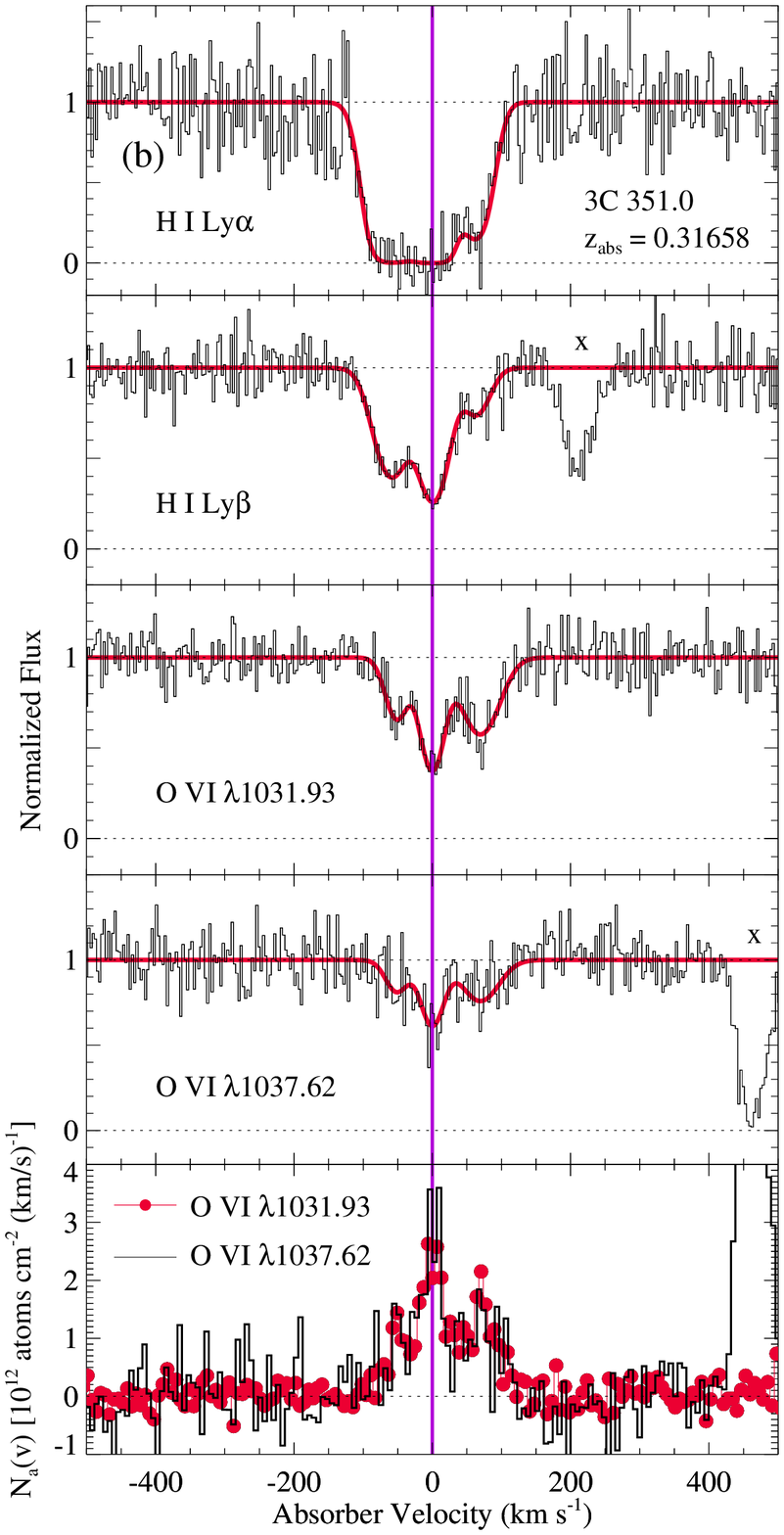}
\caption{Examples of continuum-normalized absorption profiles (black
lines) of H~I and O~VI absorption lines detected in
intervening absorbers.  Voigt-profile fits are overplotted with a red
line.  The bottom panel in each stack shows the apparent column
density profiles (see \S \ref{secmeas}) of the O~VI $\lambda
1031.93$ transition (red histogram + gray dots) and O~VI $\lambda
1037.62$ line (black histogram). The panels show the following
systems: (a) the 3C 273.0 absorber at $z_{\rm abs}$ = 0.12003, (b) the
3C 351.0 absorber at $z_{\rm abs}$ = 0.31658, (c) the H1821+643
absorber at $z_{\rm abs}$ = 0.26656, and (d) the PG1216+069 absorber
at $z_{\rm abs}$ = 0.12360. \label{intsample1}}
\end{figure*}

\begin{figure*}
\figurenum{4}
\centering
    \includegraphics[width=8.0cm, angle=0]{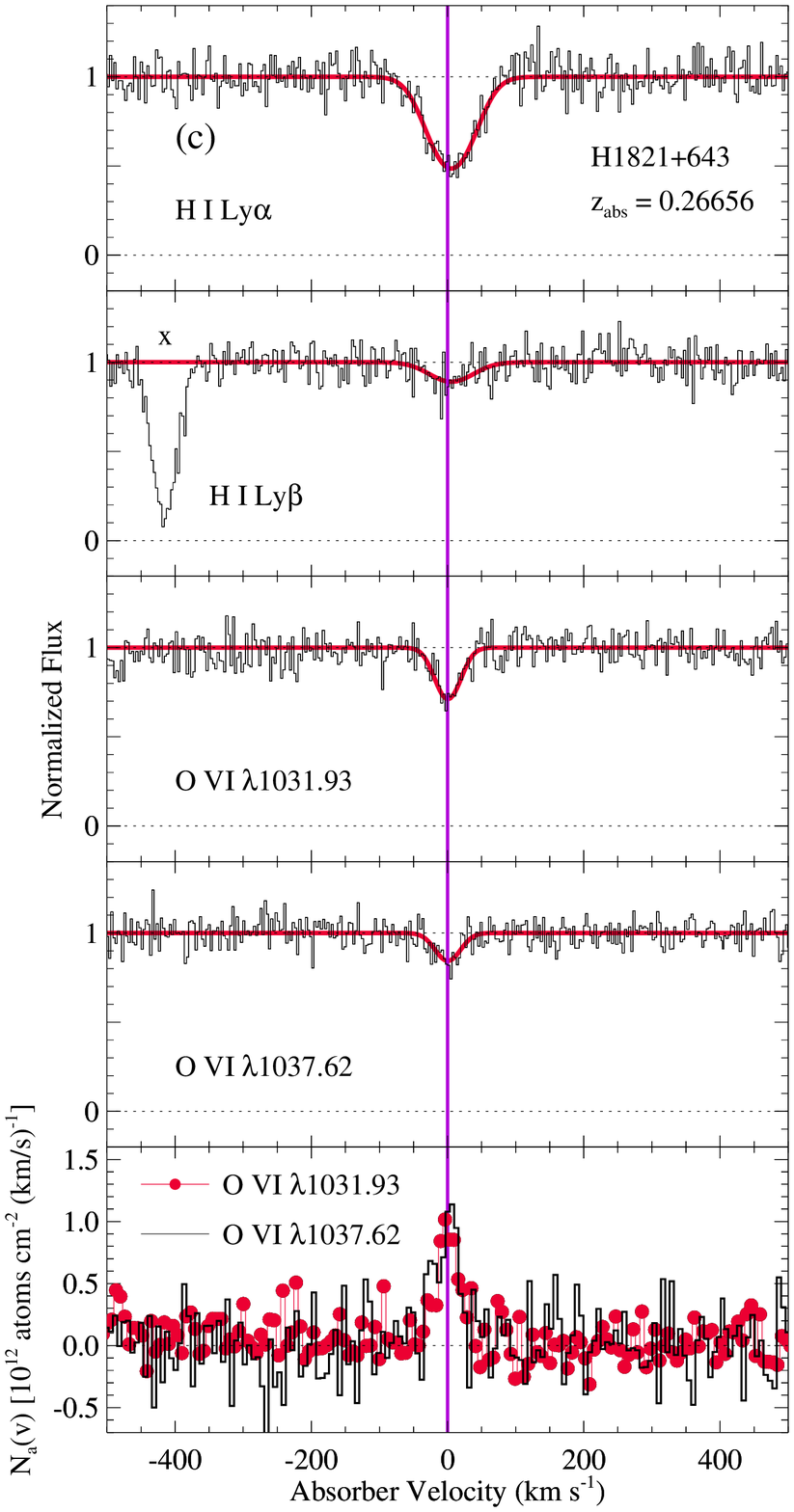}
    \includegraphics[width=8.0cm, angle=0]{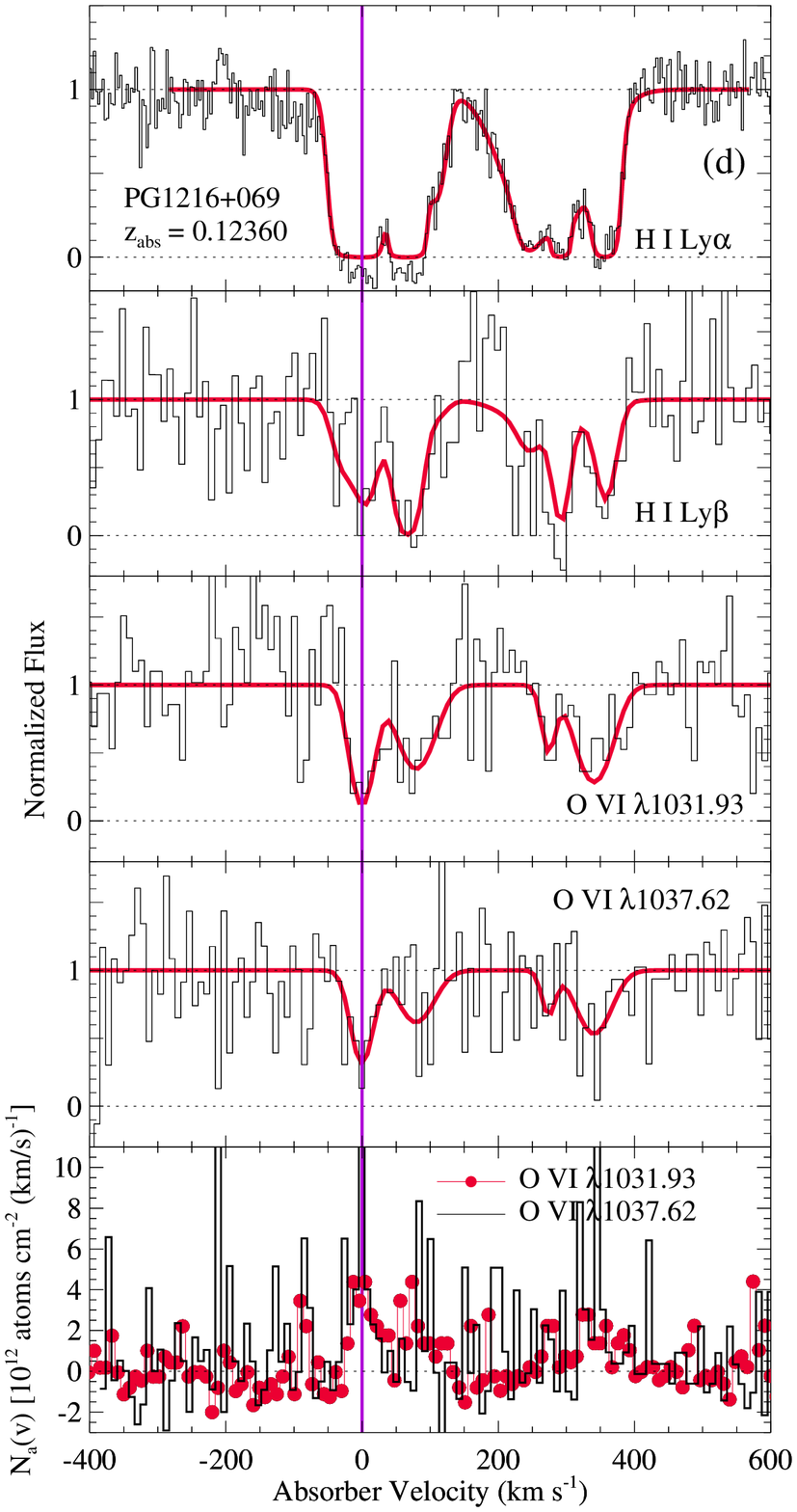}
\caption{continued.\label{intsample2}}
\end{figure*}

\begin{figure}
\centering
    \includegraphics[width=8.0cm, angle=0]{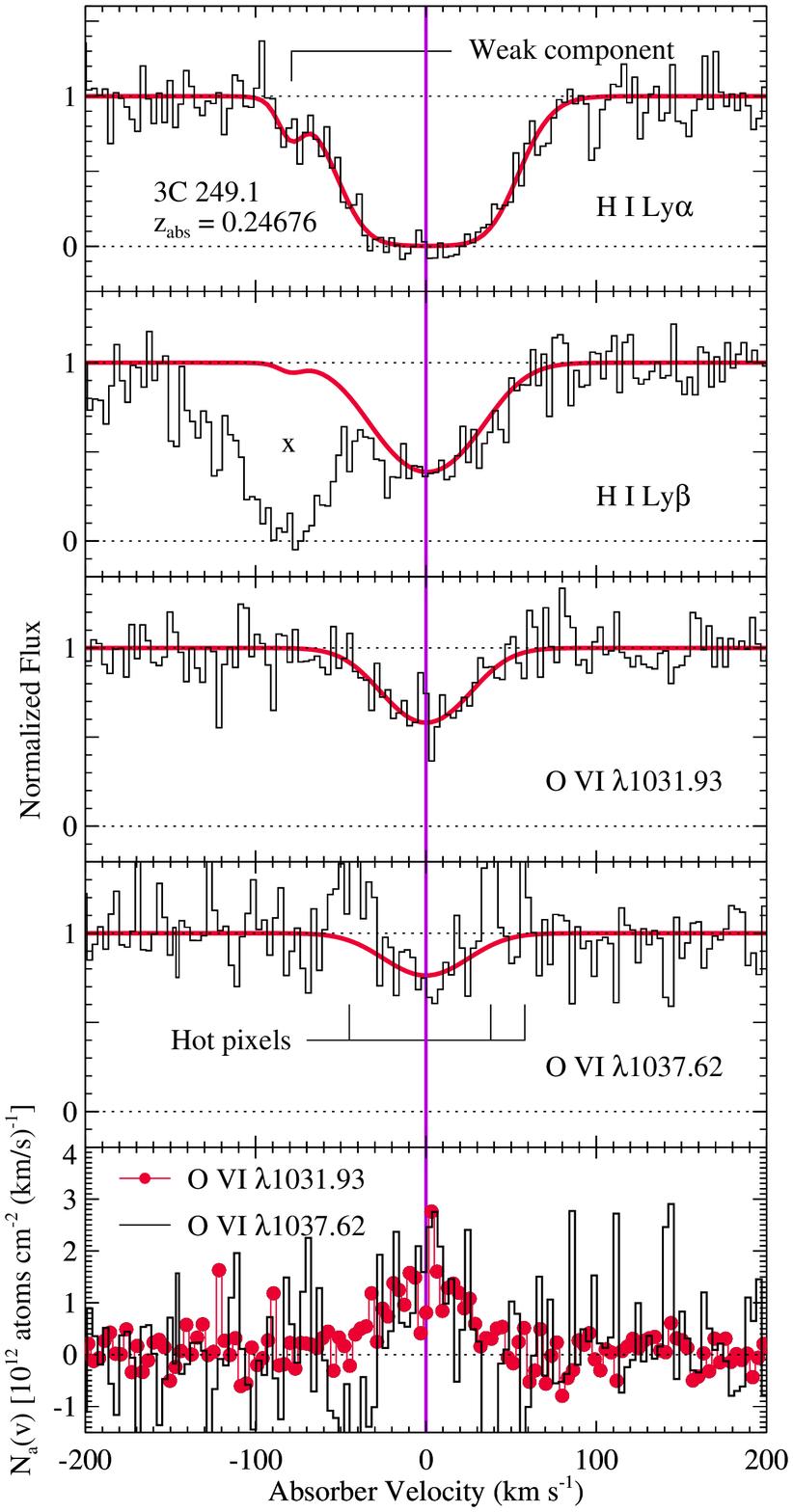}
\caption{Example of an O~VI absorber in which the O~VI and
H~I are well-aligned with the exception of a weak, outlying
component detected only in the Ly$\alpha$ profile.  Apart from the
labeled weak H~I component at $v = -79$ km s$^{-1}$, the
O~VI and H~I profiles are quite similar.  Given the range
of H~I/O~VI ratios found in O~VI systems, any
O~VI associated with the weak component is likely to be
undetectable at this signal-to-noise level.  We classify this absorber
as ``simple*'' (\S \ref{classmulti}) to indicate that the majority of
the O~VI and H~I profiles are well-aligned and can be
analyzed as a single-phase system. As in Figure~\ref{intsample1}, the
lowest panel compares the $N_{\rm a}(v)$ profiles of the O~VI
$\lambda \lambda$1031.93,1037.62 lines, and unrelated lines from other
redshifts are marked with an $\times$.  Note that several hot-pixel
features that are not readily removed are located near the O~VI
$\lambda$1037.62 line.\label{exsimstar}}
\end{figure}

\begin{figure}
\centering
\includegraphics[width=8.0cm, angle=0]{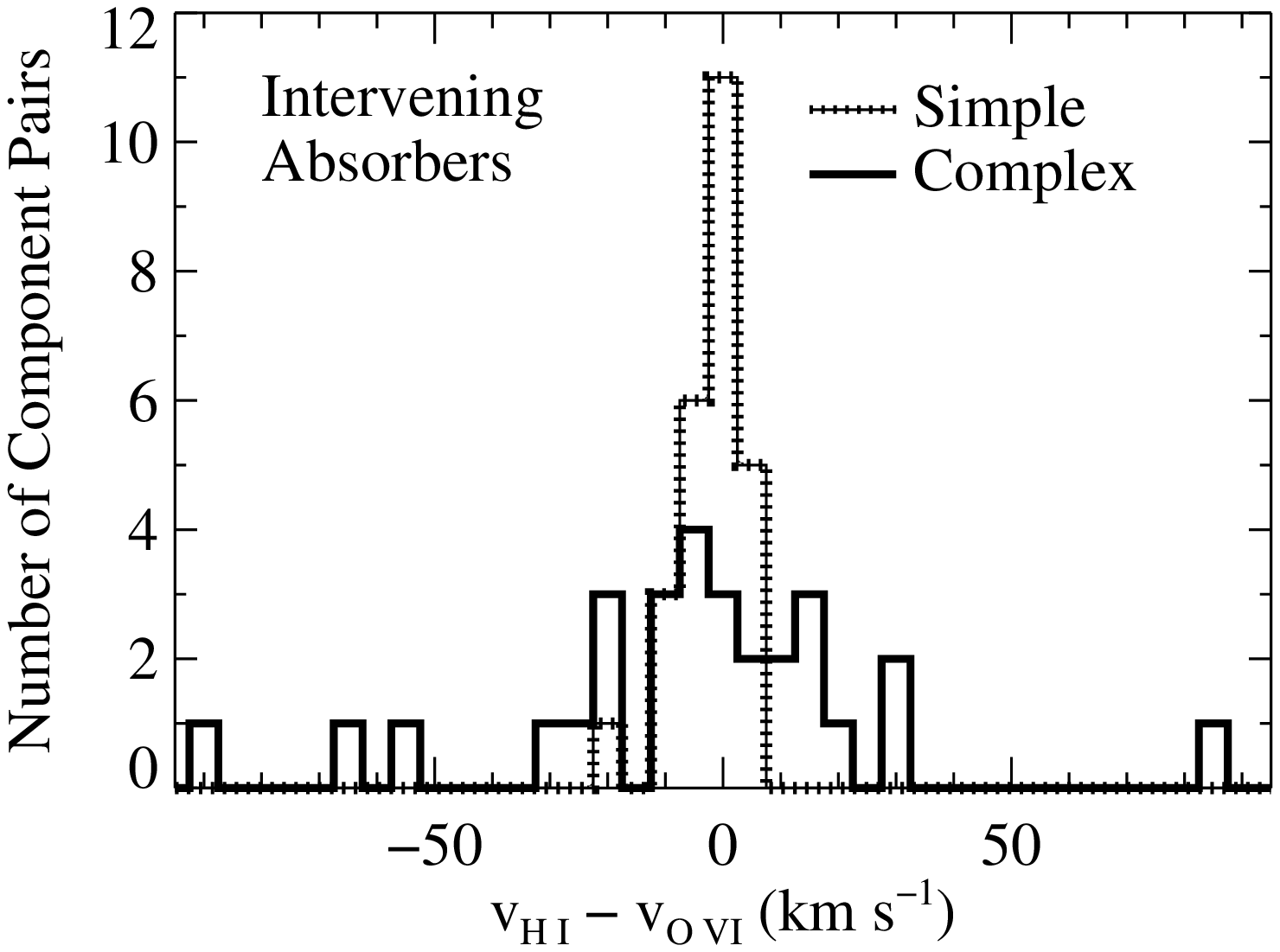}
\caption{Velocity offsets between H~I and O~VI components
in intervening absorbers measured in simple (single-phase) systems
(hatched thin line) and complex (multiphase) absorbers (thick line).
Each O~VI component is matched with the H~I component
that is closest in velocity in order to calculate the velocity
difference.\label{veloffsets}}
\end{figure}

\begin{figure*}
\centering
    \includegraphics[width=8.0cm, angle=0]{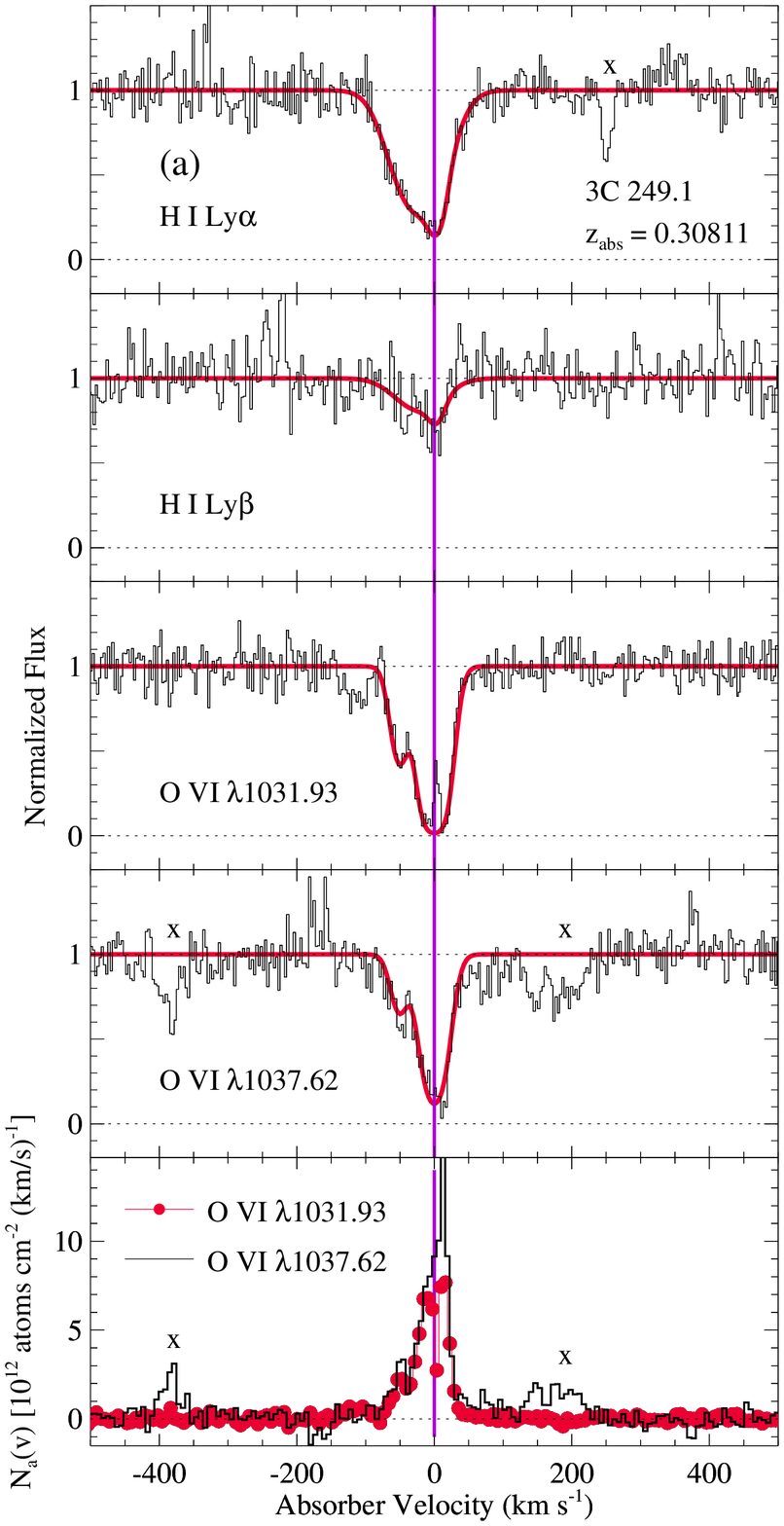}
    \includegraphics[width=8.0cm, angle=0]{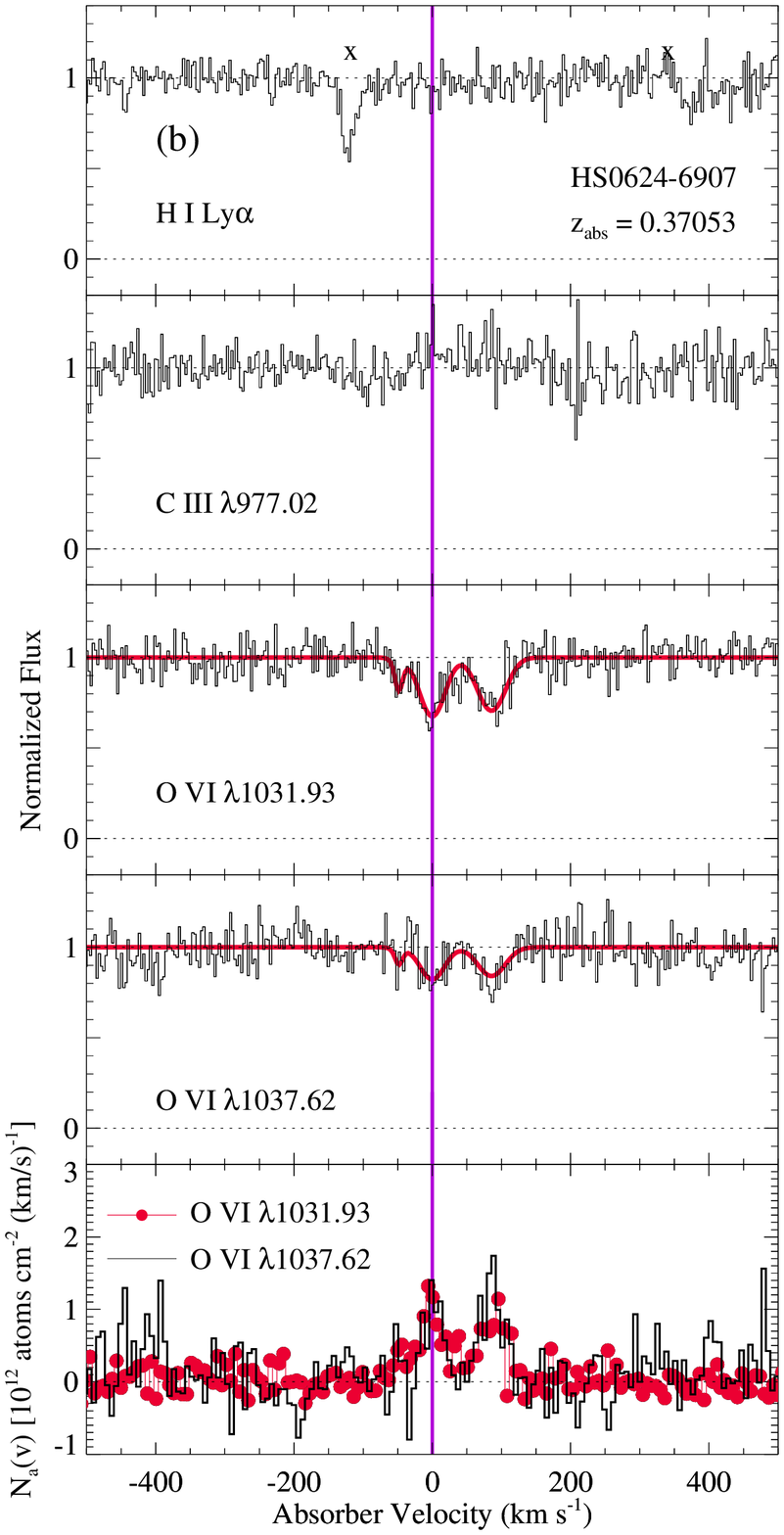}
\caption{Examples of continuum-normalized absorption profiles (black
lines) of H~I and O~VI absorption lines detected in {\it
proximate} absorbers (i.e., absorption systems located within 5000 km
s$^{-1}$ of the QSO redshift); the symbols and line colors are
analogous to those in Figure~\ref{intsample1}. The panels show the
following systems: (a) the 3C 249.1 absorber at $z_{\rm abs}$ =
0.30811, (b) the HS0624+6907 absorber at $z_{\rm abs}$ = 0.37053, (c)
the PG0953+415 absorber at $z_{\rm abs}$ = 0.23351, and (d) the TON28
absorber at $z_{\rm abs}$ = 0.33021. \label{ascsample1}}
\end{figure*}

\begin{figure*}
\figurenum{7}
\centering
    \includegraphics[width=8.0cm, angle=0]{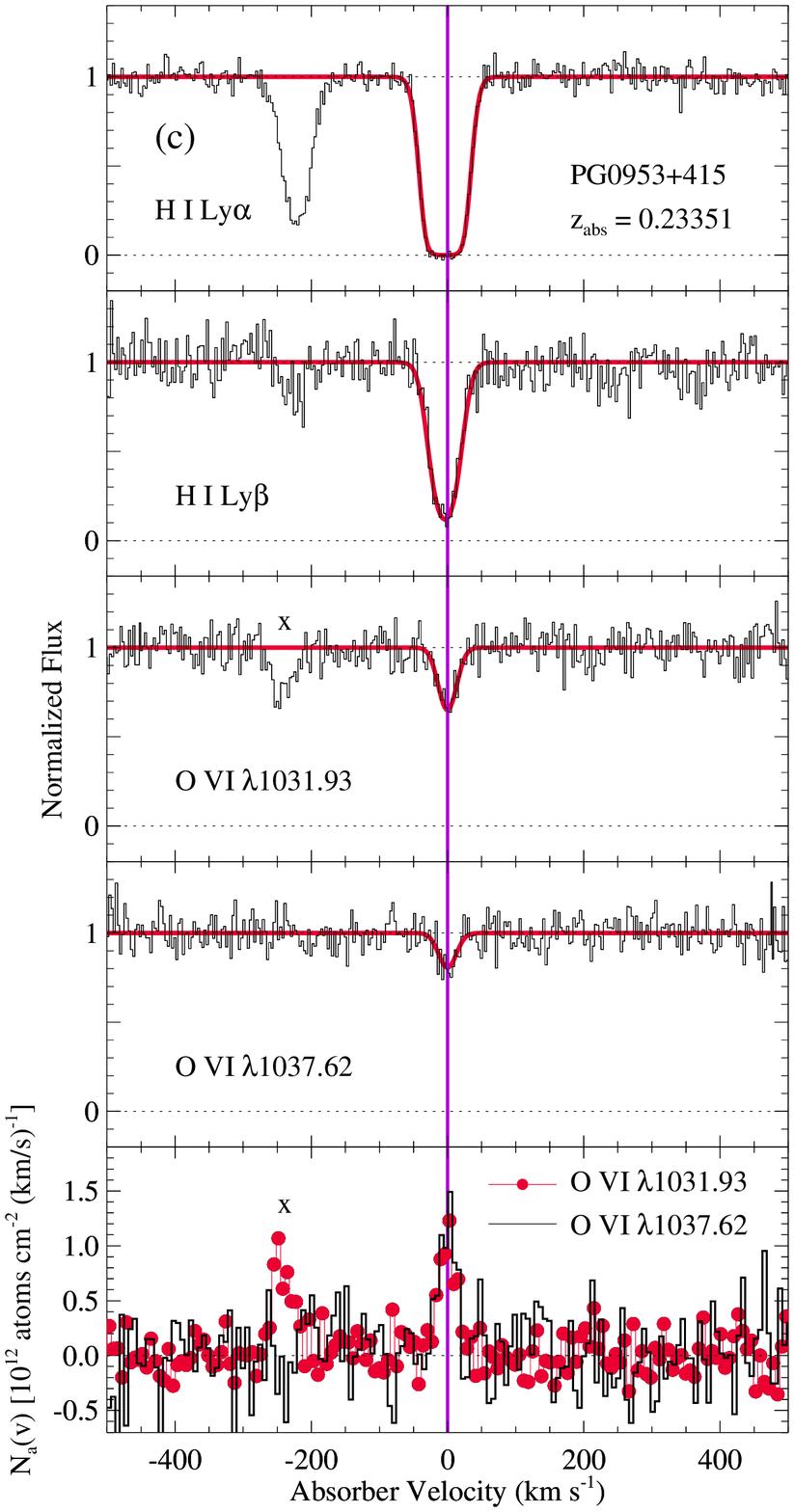}
    \includegraphics[width=8.0cm, angle=0]{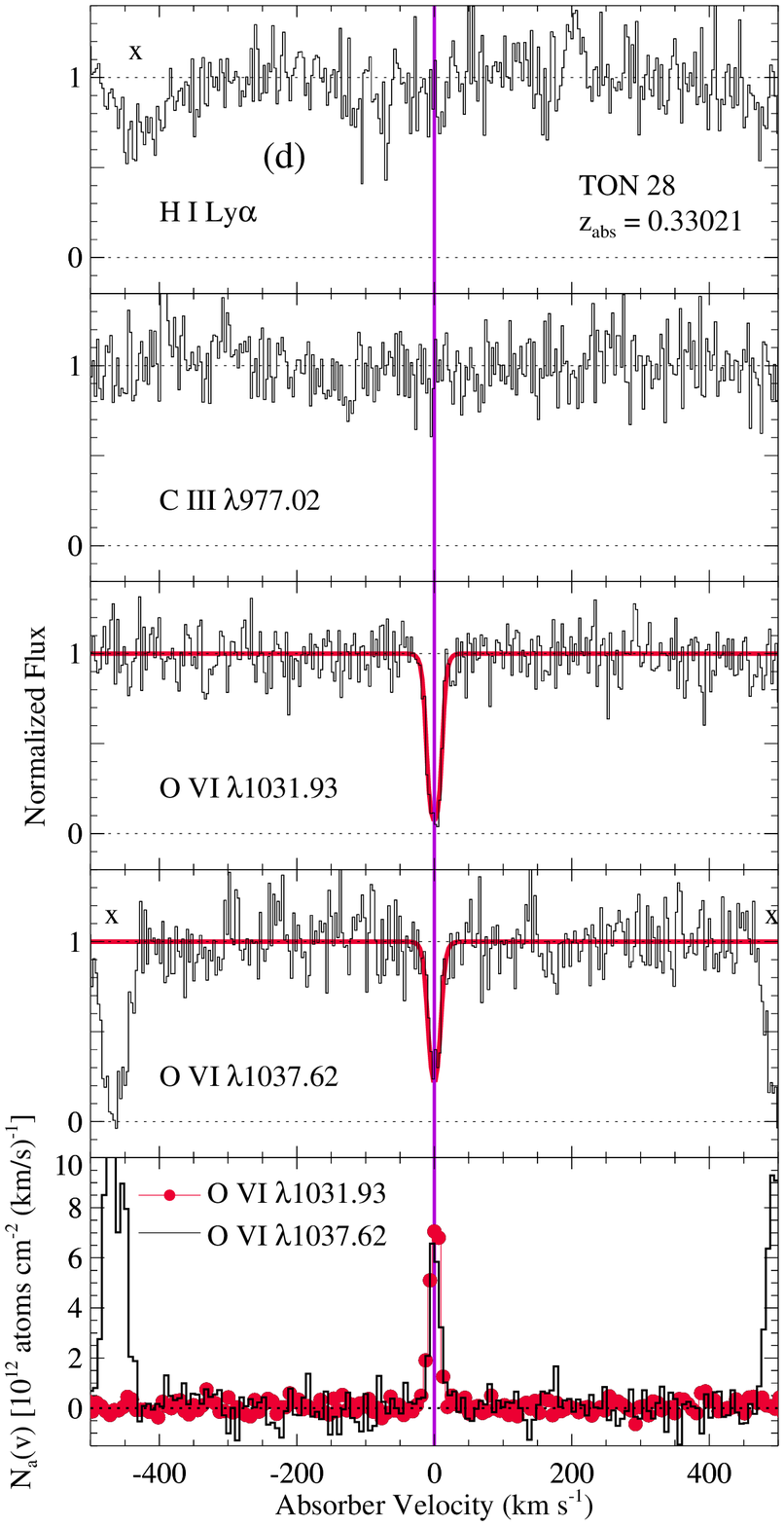}
\caption{continued.\label{ascsample2}}
\end{figure*}

\subsubsection{Single-Phase/Simple vs. Multiphase/Complex Absorbers\label{classmulti}}

It is useful to classify the low$-z$ \ion{O}{6} absorbers in various
ways.  To motivate the classifications used in this paper, we begin
with some examples of the absorption-line systems that demonstrate
their diverse characteristics and categories. Two examples of strong
\ion{O}{6} absorbers identified in our STIS spectra are shown in
Figure~\ref{specsample}. This figure shows small portions of the STIS
spectra of PKS0312-770 and PHL1811 covering \ion{O}{6} systems at
$z_{\rm abs}$ = 0.20266 and 0.19186, respectively.  The examples in
Figure~\ref{specsample} demonstrate the tremendous diversity of
low$-z$ \ion{O}{6} absorbers: from Tables~\ref{intprop} and
\ref{compprop}, we see that the \ion{O}{6} column densities are
similar in these PKS0312-770 and PHL1811 systems, but the
$N$(\ion{H}{1})/$N$(\ion{O}{6}) ratio is ostensibly $\gtrsim 10^{4}$
times larger in the PKS0312-770 example. This can partially be seen by
comparing the paucity of \ion{H}{1} Ly$\beta$ absorption in the
PHL1811 case to the strongly saturated, multicomponent Ly$\beta$
profile in the PKS0312-770 system.  There are several reasons that
\ion{O}{6} absorbers can show such a large range in the
\ion{H}{1}/\ion{O}{6} ratio.  First, when good constraints are
available, it has been shown that the metallicities\footnote{In this
paper, we use the usual notation for logarithmic metallicity, i.e.,
[X/H] = log (X/Y) - log (X/Y)$_{\odot}$, and we express linear
metallicities with the variable $Z$.} of apparently intergalactic,
low$-z$ QSO absorbers span a large range from $Z = 0.02 - 0.06
Z_{\odot}$ (e.g., Tripp et al. 2002, 2005) up to $Z = 0.5 - 1.0
Z_{\odot}$ (e.g., Savage et al. 2002; Prochaska et al. 2004; Jenkins
et al. 2005; Aracil et al. 2006). If the metallicity varies by a
factor of 50, the \ion{H}{1}/\ion{O}{6} ratio will vary by a similar
ratio. Second, the \ion{H}{1}/\ion{O}{6} ratio is very sensitive to
the ionization conditions.  As we will discuss in detail in \S 4, in
collisional ionization equilibrium, for example, relatively small
changes in the gas temperature can lead to large changes in the
\ion{H}{1}/\ion{O}{6} ratio.  Similarly, in photoionized gas, small
changes in the gas density, ionizing flux, or ionizing radiation field
shape can lead to substantial changes in the \ion{H}{1}/\ion{O}{6}
ratio.  Third, many of the \ion{O}{6} absorbers are {\it multiphase}
absorbers.  Lower-ionization phases can substantially increase the
\ion{H}{1} absorption strength without increasing the \ion{O}{6}
strength at all.

Indeed, looking more closely at the examples in
Figure~\ref{specsample}, we can find strong evidence that multiphase
effects contribute to the huge difference in the \ion{H}{1}/\ion{O}{6}
ratios in these particular cases. Figure~\ref{speccontrast} shows the
continuum-normalized absorption profiles of selected \ion{H}{1},
\ion{O}{6}, \ion{C}{2}, and \ion{Si}{2} transitions in the PKS0312-770
and PHL1811 absorbers.  Strong, multicomponent \ion{C}{2} and
\ion{Si}{2} absorption is clearly detected in the PKS0312-770 system,
but these low ions are not evident in the PHL1811 absorber.  Moreover,
the profiles of the \ion{C}{2} and \ion{Si}{2} lines in the
PKS0312-770 case are quite similar to each other with at least three
distinct components, two of which are relatively
narrow. Figure~\ref{pks0312multi} shows a closer look at some of the
PKS0312-770 metal lines (over a smaller velocity range).  From this
figure, we immediately see that the \ion{O}{6} component structure
differs from the low-ion component structure.  The \ion{O}{6}
absorption extends over a similar velocity range but does not show the
three distinct components seen in \ion{C}{2} and \ion{Si}{2}.  This
indicates either that the \ion{O}{6} lines originate in a different
phase or that additional components are present in the \ion{O}{6}
profile that blend together and smear out the component structure;
either explanation requires at least some of the \ion{O}{6} absorption
to arise in separate phases from the low ions.  The former explanation
is more natural: if the \ion{O}{6} absorption arises in hotter gas (as
expected since it peaks in abundance at $T \approx 10^{5.5}$ K in
collisionally ionized gas), its Doppler parameter $b$ will be broader
($b = \sqrt{2kT/m}$, where $T$ is the temperature and $m$ is the
atomic mass) and the \ion{O}{6} components would hence be more blended
as seen in Figure~\ref{pks0312multi}.  It is worthwhile to note that
some high-velocity clouds (HVCs) in the vicinity of the Milky Way have
similar characteristics to the profiles shown in
Figure~\ref{pks0312multi}, i.e., \ion{O}{6} and low-ion absorption
spread over the same velocity range but with somewhat different
centroids and/or line widths (see, e.g., Sembach et al. 2003; Tripp et
al. 2003; Fox et al. 2004,2006; Ganguly et al. 2005; Collins et
al. 2007).

It is not surprising to conclude that \ion{O}{6} and low-ion lines
originate in different phases; this is expected intuitively because
these species exist mainly in very different temperature ranges in
collisionally ionized gas in equilibrium (e.g., Bryans et
al. 2006).\footnote{However, O~VI and low-ion lines can arise in
a single phase in non-equilibrium collisional ionization (\S
\ref{collionsec} ) or photoionized gas (\S \ref{photoionsec}) under
the right conditions.} A more surprising result is that we also find
that a substantial fraction of our \ion{O}{6} absorbers show similar
\ion{O}{6} and \ion{H}{1} profiles suggestive of single-phase clouds.
We show four examples of intervening \ion{O}{6} absorbers in
Figure~\ref{intsample1}.  In these four examples, we detect both
\ion{O}{6} and \ion{H}{1} with good significance, and the \ion{O}{6}
and \ion{H}{1} lines are well-aligned and have similar shapes.  In
these aligned systems, the \ion{H}{1}/\ion{O}{6} ratio varies
significantly from one system to the next (and the ratio even varies
among components within a single absorber as, e.g., in the $z_{\rm
abs}$ = 0.31658 system toward 3C 351.0), but it is critical to notice
how the \ion{O}{6} and \ion{H}{1} centroids are well-aligned in
velocity, and the profile shapes are quite similar (but are not
identical). This suggests that the \ion{O}{6} and \ion{H}{1} originate
in a single gas phase in these cases and provides an important
constraint on the physical conditions of the gas (as we will discuss
further in \S 4).

Motivated by the mixture of evidence regarding the number of phases in
the \ion{O}{6} absorbers, we have classified the \ion{O}{6} systems as
either (1) ``simple'' absorbers that are entirely composed of aligned
\ion{H}{1} and \ion{O}{6} componenents (as defined in \S
\ref{aligndefsec}), which suggests that the components could be
single-phase absorbing entities, and (2) ``complex'' systems that show
significant velocity offsets between low-ionization and
high-ionization lines, which indicates that these absorbers are
multiphase systems.  These classifications are listed in column 2 of
Table~\ref{compprop}.  In a few absorbers, the primary components are
aligned in the \ion{O}{6} and \ion{H}{1} profiles, but weak components
are detected in \ion{H}{1} that do not show corresponding \ion{O}{6}.
An example of such a system is shown in Figure~\ref{exsimstar}.  In
this absorber in the spectrum of 3C 249.1, we see that the main
component at $v = 0$ km s$^{-1}$ is well-aligned in \ion{H}{1} and
\ion{O}{6}, but an additional, rather weak \ion{H}{1} component that
shows no \ion{O}{6} is evident at $v = -79$ km s$^{-1}$. The
\ion{H}{1} Ly$\alpha$ transition is very sensitive to low-density,
low-$N$(\ion{H}{1}) gas, and when $N$(\ion{H}{1}) is very low, it is
likely that any associated \ion{O}{6} absorption is simply too weak to
be detected at the S/N level of our data.  In cases where the bulk of
the \ion{H}{1} and \ion{O}{6} absorption is aligned but weak
\ion{H}{1} components are present with no corresponding \ion{O}{6} (as
in Figure~\ref{exsimstar}), we classify the absorber as a simple case,
but we mark the system with an asterisk in Table~\ref{compprop} to
note the presence of weak \ion{H}{1} without corresponding \ion{O}{6}.
In some instances, the \ion{H}{1} profiles and component centroids are
poorly constrained, e.g., if the absorber is a black Lyman-limit
system presenting only badly saturated Lyman series lines or if
\ion{H}{1} is not detected at all (as found in several systems near
the QSO redshifts); in these cases, we list the classification as
uncertain.  In the intervening absorber sample, we classify 37\% of
the absorbers as simple (single-phase) systems and 53\% as complex
(multiphase) systems (10\% have uncertain classifications).

To show the quantitative distinction between the simple (single-phase)
and complex (multiphase) intervening absorbers, we show in
Figure~\ref{veloffsets} the velocity offsets between \ion{H}{1} and
\ion{O}{6} components, $\Delta v$(\ion{H}{1} - \ion{O}{6}) $= v_{{\rm
H~I}} - v_{{\rm O~VI}}$.  For this figure, we match each \ion{O}{6}
component with the \ion{H}{1} component that is closest in velocity.
For this figure only, all \ion{O}{6} components are matched with an
\ion{H}{1} component, including cases where there are more \ion{O}{6}
components than \ion{H}{1} components (in these cases, multiple
\ion{O}{6} components are paired with the same \ion{H}{1}
line).\footnote{As we show in \S \ref{absclass}, some of the proximate
absorbers with $z_{\rm abs} \approx z_{\rm QSO}$ are detected in the
O~VI doublet without any affiliated H~I absorption.  In
contrast, in the intervening systems there are often clear velocity
offsets between the O~VI and H~I absorption components,
but we always find at least some significant H~I absorption
within $\pm$100 km s$^{-1}$ of the O~VI.} Systems that have
uncertain classification are excluded from Figure~\ref{veloffsets},
and components that are significantly uncertain due to problems such
as blending and saturation (components marked with a colon in
Table~\ref{compprop}) are also excluded.  From this figure, we see
that there is a clear kinematical distinction between
simple/single-phase and complex/multiphase absorbers: some of our
\ion{O}{6} absorbers are characterized by highly-aligned \ion{O}{6}
and \ion{H}{1} components, but an equally large fraction of the
intervening systems have more complex relationships between \ion{H}{1}
and \ion{O}{6} with significantly different kinematics.  We will
return to this distinction below.

\subsubsection{Intervening vs. Proximate/Intrinsic Absorbers\label{absclass}}

We classify the absorbers a second way for an orthogonal purpose: we
list in column 2 of Table~\ref{compprop} whether the system is
classified as an {\it intervening} absorber (Int) or a {\it proximate}
absorber (Prox).  It has long been recognized that some of the
so-called ``associated'' absorption systems (often called
``proximate'' absorbers in the recent literature) close to the
redshift of the background QSO ($z_{\rm abs} \approx z_{\rm QSO}$)
have a fundamentally different nature from intervening absorbers that
arise in the IGM or ISM of foreground objects (e.g., Lynds 1967;
Burbidge 1970; Foltz et al. 1986).  The ``broad-absorption line''
(BAL) QSOs are easily recognized by their dramatic P Cygni-like
absorption troughs with outflow velocities approaching a substantial
fraction of the speed of light (see, e.g., Turnshek 1988), but apart
from BAL QSOs, there is a statistical excess of metal-bearing
absorbers within $\approx$ 5000 km s$^{-1}$ of $z_{\rm QSO}$ (Foltz et
al. 1986) that in many regards look like the ordinary, narrow
absorption lines that arise in foreground galaxies and the IGM.
However, evidence such as temporal variability (on timescales of a few
years) and/or incomplete covering of the background flux source
indicates that at least some ``narrow'' absorption lines originate in
material located quite close to the QSO itself (see, e.g., Hamann
1997; Yuan et al. 2002; Ganguly et al. 2003, and references therein).

Following the seminal work of Weymann et al. (1979) and Foltz et
al. (1986), narrow absorption lines found within 5000 km s$^{-1}$ of
the QSO redshift are usually classified as associated (proximate)
absorbers, and we adhere to this standard definition in this paper
(however, in some cases we show how the measurements change if we
change the velocity cutoff for the proximate absorber
classification). The velocity of displacement $v_{\rm displ}$ must
account for relativistic effects but can be easily determined from the
QSO and absorber redshifts using the usual $\beta$ formula,
\begin{equation}
\beta = \frac{v_{\rm displ}}{c} = \frac{(1 + z_{\rm QSO})^2 - (1 +
z_{\rm abs})^2}{(1 + z_{\rm QSO})^2 + (1 + z_{\rm abs})^2}.
\end{equation}
We also distinguish between associated absorbers and ``mini-BAL''
systems.  Mini-BALs show many of the characteristics of BALs but are
spread over a smaller velocity range (e.g., Barlow et al. 1997).  One
of our sight lines (3C 351.0) shows a clear example of a mini-BAL with
strong evidence that the absorption lines arise in intrinsic material
close to the QSO. This is clearly a different type of absorption
system, and we do not include it in our samples.  A full analysis of
the 3C 351.0 mini-BAL has been presented by Yuan et al. (2002).

However, the 5000 km s$^{-1}$ velocity interval used to define
proximate absorbers corresponds to a relatively large Hubble-flow
distance, and it is inevitable that some truly intervening absorbers
will be classified as associated just because they are located in the
5000 km s$^{-1}$ region of space near the QSO; indeed, Sembach et
al. (2004) show examples of this problem (see their \S 10). On the
other hand, compelling direct evidence of narrow absorption systems
associated with the QSO with $v_{\rm displ} \gg$ 5000 km s$^{-1}$ has
been found at high redshifts; Hamann, Barlow, \& Junkkarinen (1997)
reported clear temporal variability and partial coverage of the flux
source in an absorber displaced by $\approx$24,000 km s$^{-1}$ from
the emission redshift of Q2343+125 ($z_{\rm QSO}$ = 2.24), for
example. Based on statistical excesses of absorbers observed toward
different types of QSOs, Richards et al. (1999,2001) have argued that
up to 36\% of \ion{C}{4} absorbers at $z_{\rm abs} \sim$ 2.5 with 5000
$\leq v_{\rm displ} \leq$ 75,000 km s$^{-1}$ arise in ejected material
intrinsic to the background QSO.  More recently, Misawa et al. (2007)
have searched for partial coverage in narrow metal-line absorbers
detected in high-resolution Keck spectra of 37 high$-z$ QSOs, and they
conclude that $10-17$\% of narrow \ion{C}{4} absorbers are intrinsic
(associated with the QSO) in the 5000 $\leq v_{\rm displ} \leq$ 75,000
km s$^{-1}$ interval.

Evidently, displacement velocity might not provide a straightforward
means of distinguishing between intervening and {\it intrinsic}
absorbers.  The absorbers that are usually referred to as associated
systems are a mixture of intervening and intrinsic absorbers.  To
avoid the confusing connotations of the term ``associated'', in this
paper we will follow recent papers (e.g., Ellison et al. 2002; Russell
et al. 2006; Hennawi \& Prochaska 2007) and refer to any absorber with
$v_{\rm displ} \leq$ 5000 km s$^{-1}$ as a ``proximate'' absorber
regardless of its nature/origin, but we use ``intrinsic'' to connote
that the system is physically close to the QSO.  It is important to
bear in mind the possible confusion that intrinsic systems could cause
in samples of intervening systems defined based on $v_{\rm displ}$.
Because nearby galaxy redshifts are more straightforwardly measured,
an advantage of low$-z$ samples is that if $10-30$\% of the systems in
the $5000 - 75,000$ km s$^{-1}$ interval are intrinsic/ejected, we
should find that this fraction of the metal systems are randomly
distributed with respect to the foreground galaxies. Studies of
correlations between metal systems and foreground galaxies so far have
not found any evidence of randomly distributed metal systems; instead,
the metal absorbers appear to be clearly correlated with galaxies
(e.g., Savage, Tripp, \& Lu 1998; Tripp \& Savage 2000; Savage et
al. 2002; Sembach et al. 2004; Tumlinson et al. 2005; Prochaska et
al. 2006; Stocke et al. 2006; Williger et al. 2006; Aracil et
al. 2006; Tripp et al. 2006a). We are currently expanding the sample
of sight lines for low$-z$ galaxy-absorber correlation analyses, but
this investigation will be presented in a separate paper.  In this
paper, however, we will search for any evidence that some of the
\ion{O}{6} absorbers at 5000 $\leq v_{\rm displ} \leq$ 75,000 km
s$^{-1}$ have characteristics that indicate an ejected/intrinsic
origin.  We will find that in our sample, ejected/intrinsic systems
appear to be predominantly confined to $v_{\rm displ} \lesssim$ 2500
km s$^{-1}$ (see \S 3).

\begin{figure*}
\centering
    \includegraphics[width=17.0cm, angle=0]{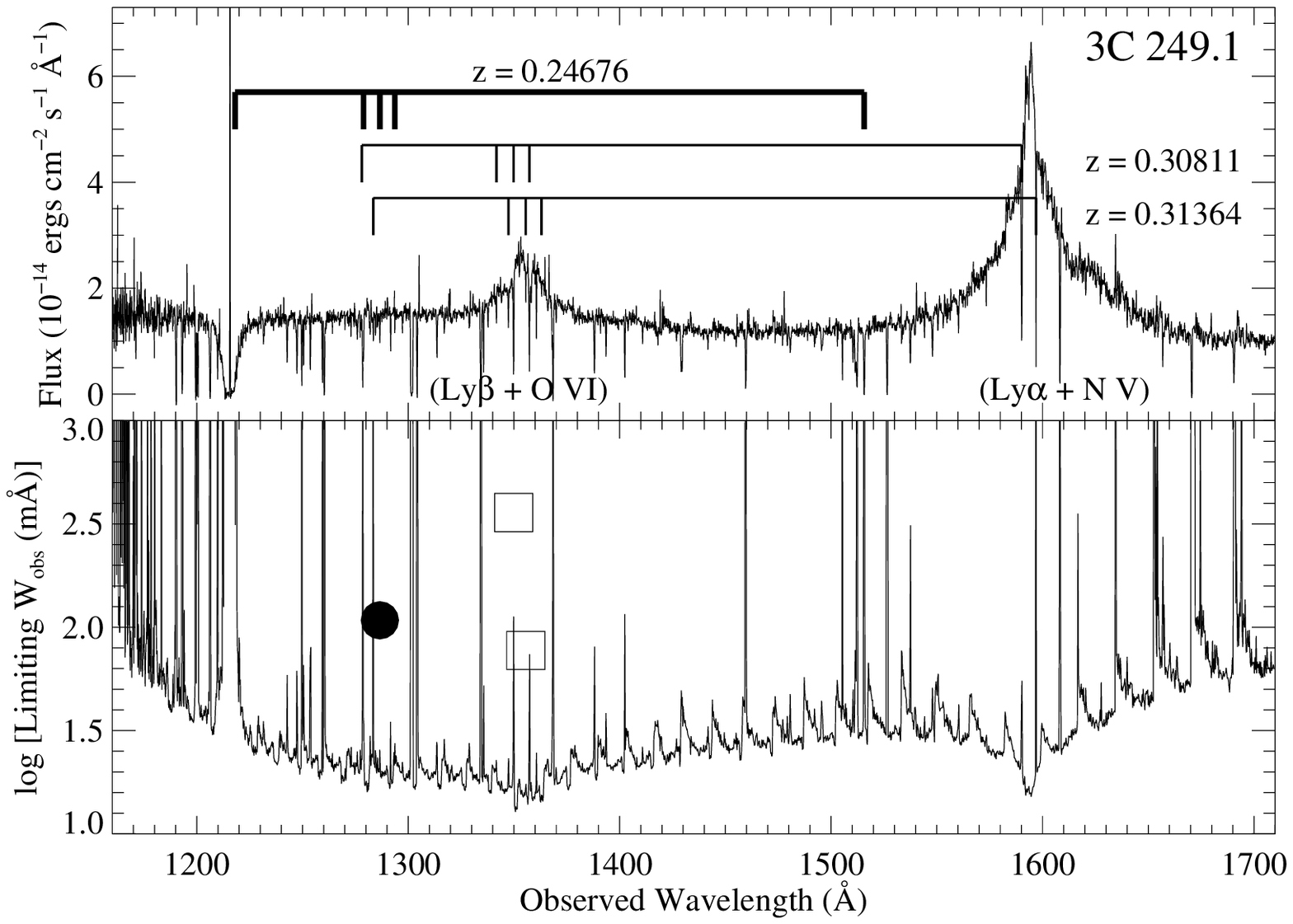}
\caption{Example of the sensitivity of our STIS E140M data.  The top
panel shows the STIS E140M spectrum of 3C 249.1, binned to $\approx$35
km s$^{-1}$ pixels to show the continuum shape and emission lines
clearly.  QSO broad emission lines are labeled in parentheses
underneath the spectrum. The bars and tickmarks show the locations of
the intervening O~VI absorbers at $z_{\rm abs}$ = 0.23641 and
0.24676 as well as the proximate O~VI systems at $z_{\rm abs}$ =
0.30811 and 0.31364.  For each system, the ticks indicate the
redshifted wavelengths of the C~III $\lambda$977.02, H~I
Ly$\beta$, O~VI $\lambda \lambda$1031.93, 1037.62, and
H~I Ly$\alpha$ lines from left to right, respectively.  The
thick bars are the intervening cases; the thin bars mark the proximate
systems. The lower panel shows the 3$\sigma$ limiting equivalent width
obtained by integrating the full-resolution data over 15-pixel
windows.  The sawtooth pattern in the limiting equivalent width is due
to the blaze function of the echelle grating, which causes the S/N of
the spectrum to vary significantly across a given echelle order.  This
is mitigated in some regions because adjacent echelle orders overlap,
and coaddition of the overlapping regions offsets the lower S/N at the
edges of the echelle orders.  The large positive spikes are due to
strong interstellar or intergalactic absorption lines, which block
portions of the spectrum and prevent the detection of weak
extragalactic lines within the wavelength range of the strong
absorption profile. The measured equivalent widths of the detected
O~VI $\lambda$1031.93 lines in the intervening and proximate
O~VI systems are shown with filled circles and open squares,
respectively, in the lower panel.\label{pathcalc}}
\end{figure*}

Figure~\ref{ascsample1} shows several examples of the proximate
\ion{O}{6} absorbers that we have identified in the 16 sight lines
studied in this paper.  This figure shows that some of the proximate
systems have characteristics that are quite similar to those of the
intervening systems.  For example, the proximate absorbers of 3C 249.1
(at $z_{\rm abs}$ = 0.30811) and PG0953+415 ($z_{\rm abs}$ = 0.23351)
have strong \ion{H}{1} lines as well as \ion{O}{6}, and the \ion{H}{1}
and \ion{O}{6} lines are well-aligned.  However, some of the proximate
systems have very weak affiliated \ion{H}{1} absorption.  The
HS0624+6907 and TON28 proximate systems in Figure~\ref{ascsample1} are
two examples of proximate systems with very weak \ion{H}{1}. In
several cases, \ion{H}{1} is not detected at all despite good S/N.  As
noted in the previous section, some intervening systems show
substantial velocity offsets between the \ion{H}{1} and \ion{O}{6}
absorption, but intervening systems almost always show some detectable
\ion{H}{1} absorption in the vicinity of the \ion{O}{6}.  As we will
discuss further below, one unique characteristic of the proximate
systems is that some show no affiliated \ion{H}{1} absorption
whatsoever.  Using the same classification criteria discussed above,
we categorize 21\% of the proximate absorbers as ``simple'' systems
composed of \ion{O}{6} and \ion{H}{1} components that are aligned, and
50\% of the proximate systems are complex.\footnote{The proximate
absorber sample is smaller and a larger fraction of the proximate
absorbers have uncertain classifications, hence a substantial portion
of these absorbers (29\% ) are not classified.}
 
\section{Statistics\label{statsec}}

\subsection{Number of \ion{O}{6} Absorbers per Unit Redshift\label{dndzsec}}

We now turn to the statistical properties of the intervening and
proximate absorbers presented in Tables~\ref{intprop} and
\ref{compprop}.  We begin with the number of \ion{O}{6} absorbers per
unit redshift, $dN/dz$.  This is a fundamental quantity that
characterizes the various types of QSO absorbers.  Moreover, it is
straightforward to use theoretical models and simulations to predict
$dN/dz$ for \ion{O}{6} absorbers (e.g., Cen et al. 2001; Fang \& Bryan
2001; Chen et al. 2003; Furlanetto et al. 2005; Tumlinson \& Fang
2005; Cen \& Fang 2006), and measurement of $dN/dz$ provides an
important test of the validity of these large-scale cosmological
models.

\subsubsection{$dN/dz$ Measurement Method and Uncertainties\label{dndzmeas}}

To estimate $dN/dz$ for a sample of absorbers with equivalent widths
greater than a specified limiting equivalent width $W_{\rm lim}$, we
simply tabulate the number of \ion{O}{6} systems detected with $W_{\rm
r} \geq W_{\rm lim}$ and divide by the total path $\Delta z$ over
which we were able to search for the absorbers.  However, there are a
few minor complications in this simple calculation.  First, the S/N
ratios of our spectra are not completely uniform; some sight lines
have higher over S/N ratios than others, and even along a single sight
line, some wavelength regions have higher S/N.  Second, some redshifts
are blocked by strong Galactic ISM lines or unrelated extragalactic
lines.  These blocked regions reduce the total $\Delta z$ that is
effectively searched for absorption systems. The total $\Delta z$ and
number of absorbers in a given sample must carefully account for the
variable S/N and blocking of the spectra.  For example, suppose that a
line with $W_{\rm r} = $ 150 m\AA\ is detected in a spectral region
with S/N that is only sufficient to detect lines with $W_{\rm r} \geq$
100 m\AA . That region would not be included in the total $\Delta z$
for a sample with $W_{\rm r} \geq$ 30 m\AA , and thus the 150 m\AA\
line cannot be counted in a $W_{\rm r} \geq$ 30 m\AA\ sample.
However, this absorber would be included in a sample with $W_{\rm r}
\geq$ 100 m\AA or larger.

To account for the variable S/N of the data and blocking by unrelated
lines, we calculate for each sight line the rest-frame $3\sigma$
limiting equivalent width at each observed wavelength $\lambda$ where
\ion{O}{6} absorbers can be detected,
\begin{equation}
W_{\rm lim}(\lambda) = \frac{3 \sigma _{W(\lambda)}}{1 + z_{\rm abs}},
\end{equation}
where $\sigma _{W(\lambda)}$ is the total uncertainty in an equivalent
width integrated over some number of pixels.  For a single pixel $i$,
the equivalent width uncertainty $\sigma _{W(i)}$ is 
\begin{equation}
\sigma _{W(i)} = \Delta \lambda(i) \left[ \frac{\sigma _{I(\lambda
_{i})}}{I(\lambda _{i})}\right],
\end{equation}
where $\Delta \lambda(i)$ is the pixel width, $I(\lambda _{i})$ is the
flux in pixel $i$, and $\sigma _{I(\lambda _{i})}$ is the $1\sigma$
flux uncertainty in pixel $i$. However, both the STIS and {\it FUSE}
data have at least two pixels per resolution element, and in fact, all
of the clearly detected \ion{O}{6} lines are spread over at least
several 2-pixel resolution elements.  Therefore, to evaluate $W_{\rm
lim}(\lambda)$, we must integrate over $n$ pixels,
\begin{equation}
\sigma _{W(\lambda)}^2 = \sum_{i = 1}^{n} \sigma _{W(i)}^2. \label{limeqwint}
\end{equation}
For a given limiting equivalent width, we have empirically determined
the number of pixels to integrate over based on the {\it detected}
\ion{O}{6} lines with measured equivalent widths close to the limiting
equivalent width of interest (see further details below). After
$W_{\rm lim}(\lambda)$ has been evaluated over the full range of
\ion{O}{6} coverage for all 16 sight lines, we evaluate $\Delta z$ for
various samples of lines with strengths greater than $W_{\rm r}$ by
summing the total redshift path over which $W_{\rm lim}(\lambda) \leq
W_{\rm r}$. We first check for detectability of the \ion{O}{6}
$\lambda$1031.93 line when evaluating the total $\Delta z$.  When
\ion{O}{6} $\lambda$1031.93 is redshifted into a region blocked by a
strong ISM or extragalactic line, our algorithm checks whether the
corresponding \ion{O}{6} $\lambda$1037.62 can be detected
instead. \ion{O}{6} $\lambda$1037.62 is redshifted to a different
observed wavelength, and the different observed $\lambda$ usually
places $\lambda$1037.62 outside of the blocked region. However,
because of its lower $f\lambda$ value, the \ion{O}{6} $\lambda$1037.62
is expected to be a factor of two weaker than $\lambda$1031.93, so
when checking the corresponding $\lambda$1037.62 region, we require
$W_{\rm lim}(\lambda) \leq 0.5 W_{\rm r}$ in order to add to the total
$\Delta z$ (e.g., if we cannot detect a 30 m\AA\ \ion{O}{6}
$\lambda$1031.93 line, we have to be able to detect a 15 m\AA\
\ion{O}{6} $\lambda$1037.62 line instead).  In this way, we estimate
the total path over which we can detect {\it either} \ion{O}{6}
$\lambda$1031.93 or \ion{O}{6} $\lambda$1037.63.  This approach also
accounts for regions with steep S/N gradients (e.g., in the vicinity
of the broad QSO emission lines) that might make one line of the
doublet detectable while the other lines falls below the detection
threshold.

Figure~\ref{pathcalc} shows an example of the resulting limiting
equivalent width calculated for equivalent widths greater than 30
m\AA\ for the STIS observations of 3C 249.1. The top panel of
Figure~\ref{pathcalc} shows the 3C 249.1 STIS spectrum (binned to
$\approx$ 35 km s$^{-1}$ pixels to more clearly show the
characteristics of the spectrum), and the lower panel shows the
$3\sigma$ limiting equivalent width obtained by integrating over
15-pixel detection windows (15 full-resolution, unbinned pixels).
Figure~\ref{pathcalc} shows several important characteristics of the
data that deserve comment. First, the proper calculation of $W_{\rm
lim}(\lambda)$ automatically corrects for portions of the spectra that
are blocked by strong interstellar or extragalactic lines.  In these
regions, the S/N drops precipitously and $W_{\rm lim}(\lambda)$
increases accordingly; if neither the $\lambda 1031.93$ line nor the
$\lambda 1037.62$ line falls in a region with sufficient sensitivity,
then that redshift window does not contribute to the total $\Delta z$.
In addition, when moderate strength lines are present, $W_{\rm
lim}(\lambda)$ increases, but these locations {\it can} add to $\Delta
z$ for samples of stronger lines.  For example, a contaminating 150
m\AA\ ISM line might effectively hide a 30 m\AA\ extragalactic
\ion{O}{6} line, but it would not hide a 300 m\AA\ \ion{O}{6} line,
and this is properly evaluated using this approach.  Second, the S/N
and $W_{\rm lim}(\lambda)$ are not uniform across a STIS E140M
spectrum for several reasons. The STIS E140M sensitivity peaks at
$\lambda _{\rm obs} \approx$1340 \AA\ and decreases at shorter and
longer wavelengths.  However, superimposed on this broad and
slowly-varying sensitivity is a sawtooth pattern due to the blaze
function of the echelle grating: within {\it each} echelle order, the
S/N decreases significantly when moving away from the blaze peak.
This is partly mitigated in regions where adjacent orders overlap; in
these overlapping regions, coaddition of the two orders recovers some
of the lost signal-to-noise.  Given these various effects, it is clear
that the limiting equivalent width is a complicated array that is not
easily approximated by a simple function.  However, our method fully
accomodates the various sources of sensitivity variations.

As can be seen from Figures~\ref{specsample}-\ref{ascsample2} and
Tables~\ref{intprop}-\ref{compprop}, many of the \ion{O}{6} systems
show multiple (often blended) components spread over 200$-$500 km
s$^{-1}$ intervals. The way that the {\it individual} components
within these blends are counted introduces another source of
uncertainty in $dN/dz$ measurements.  For example, the \ion{O}{6}
absorber at $z_{\rm abs}$ = 0.31658 toward 3C 351.0 shows three
blended (but clearly detected and distinct) components at $v_{\rm
abs}$ = $-51, 0,$ and 70 km s$^{-1}$ (see Figure~\ref{intsample1}).
Does this case count as one or three absorption systems?  These
velocity differences correspond to a substantial distance in a pure
Hubble flow, and it is conceivable that the three components in this
example arise in gas associated with three discrete galaxies.  But,
the three components could alternatively be due to three clouds within
the surroundings of a single galaxy. Published theoretical $dN/dz$
predictions often do not specify how the absorbers how counted, and
indeed, the appropriate method likely depends on the purpose of the
$dN/dz$ measurement.  This issue has greatest impact on our $dN/dz$
measurements for \ion{O}{6} absorbers with the largest equivalent
widths; these cases are rare and almost always show evidence multiple
components, so spliting the high-$W_{\rm r}$ cases into several
lower-$W_{\rm r}$ cases reduces $dN/dz$ for the highest $W_{\rm r}$
bins.  To show how this issue affects our results, we report two
separate $dN/dz$ measurements.  We estimate $dN/dz$ for {\it systems},
in which we count multiple components that are contiguously connected
with each other as a {\it single} \ion{O}{6} case, and we also
determine $dN/dz$ for {\it components} where every identified
individual component is counted as a separate case.

\subsubsection{$dN/dz$ Results: Cumulative Distribution\label{dndzresults_sec}}

\begin{figure}
\centering
\includegraphics[width=9.0cm, angle=0]{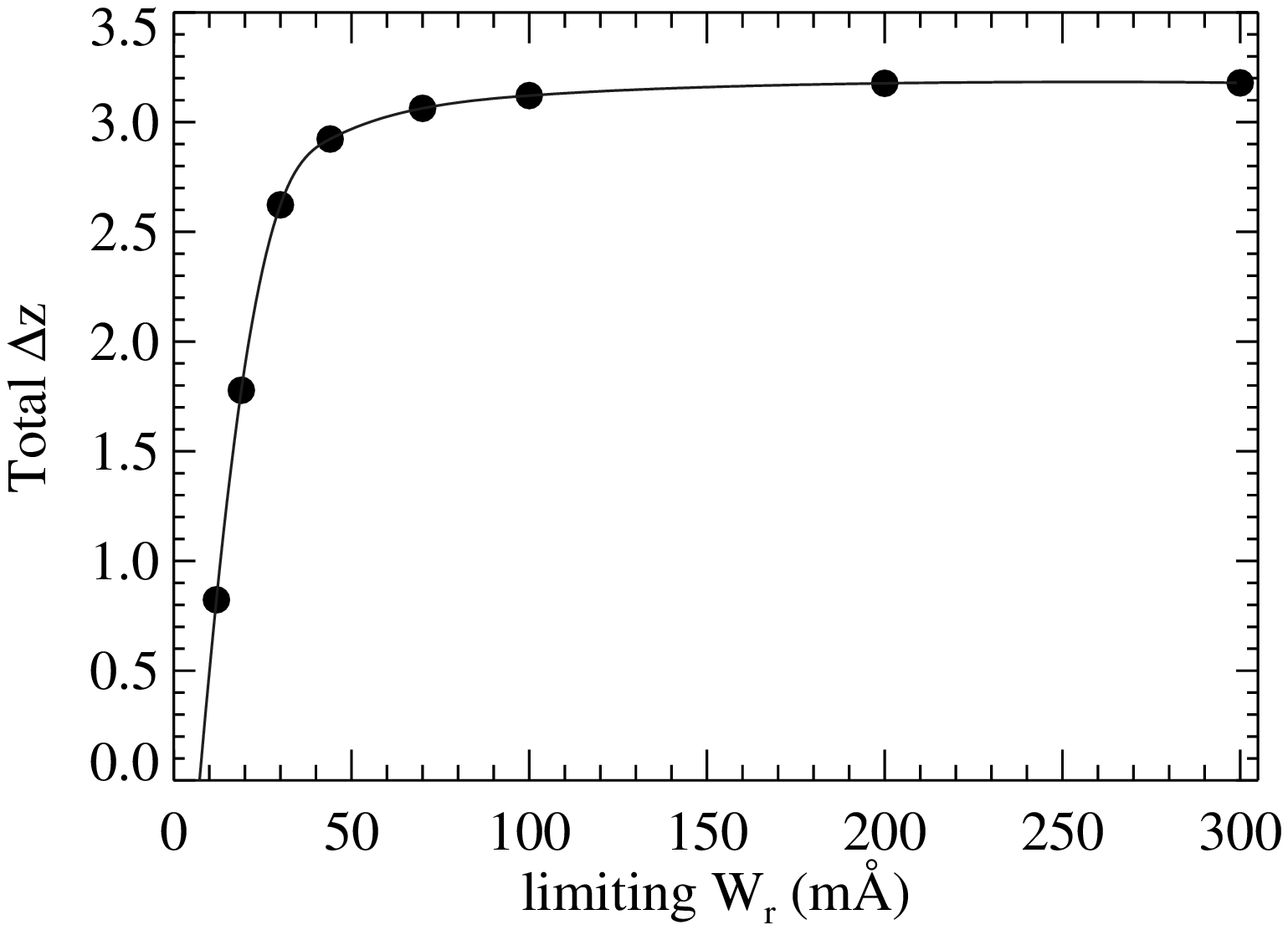}
\caption{The total path ($\Delta z$) over which the 16 QSO sight lines
have sufficient signal-to-noise ratio to detect intervening O~VI
lines with rest-frame equivalent width $W_{\rm r} \geq$ limiting
$W_{\rm r}$, calculated as described in \S \ref{dndzsec}.  The total
$\Delta z$ was explicitly calculated at the points indicated by filled
circles, and the solid line shows a spline fit that is used to
interpolate to find $\Delta z$ at values between the filled
circles.\label{pathplot}}
\end{figure}

In Table~\ref{dndztab}, we report our $dN/dz$ measurements for
intervening O~VI absorbers for samples with $W_{\rm r} >$ 30,
70, 100, 200, and 300 m\AA . To determine the limiting equivalent
equivalent width and total $\Delta z$ for these bins (as described in
\S \ref{dndzmeas}), we integrate over 15, 22, 30, 50, and 70 STIS
pixels, respectively.\footnote{The number of pixels that we integrate
over is based on the number of pixels spanned by {\it detected}
O~VI features with equivalent widths comparable to the limiting
equivalent width of interest.}  For convenience, after explicitly
calculating $\Delta z$ for these specific limiting $W_{\rm r}$ values,
we fitted cubic spline functions to the $\Delta z$ vs. limiting
$W_{\rm r}$ data.  The total $\Delta z$ values and the cubic spline
fit for the intervening absorbers are shown in Figure~\ref{pathplot}.
For each intervening sample, Table~\ref{dndztab} lists the number of
\ion{O}{6} systems identified, the total path $\Delta z$ over which we
could detect an \ion{O}{6} $\lambda$1031.93 line with $W_{\rm r}$
greater than the limiting value of that sample, and the implied
$dN/dz$ found by counting either systems (column 4) or components
(column 5) as explained above.

Table~\ref{asscdndztab} lists the same quantities but for the {\it
proximate} \ion{O}{6} absorbers.  As discussed above, typically
associated/proximate absorbers are defined in the literature to be any
system with $v_{\rm displ} < 5000$ km s$^{-1}$, so we derive $dN/dz$
assuming this standard definition, and the results are shown in the
upper half of Table~\ref{asscdndztab}.  However, we will show below
that the excess of proximate \ion{O}{6} systems for our sight lines
appears to be confined to the $-1000 \lesssim v_{\rm displ} \lesssim
2500$ km s$^{-1}$ interval, so we also list in Table~\ref{asscdndztab}
the results that we obtain by alternatively defining proximate
systems to be those found with $v_{\rm displ} \leq 2500$ km
s$^{-1}$. In Tables~\ref{dndztab} and \ref{asscdndztab}, we use the
Gehrels (1986) small-sample tables to estimate the statistical
uncertainties in $dN/dz$.

We now offer several comments on our $dN/dz$ measurements; we compare
these results to theoretical predictions in \S \ref{dndztheory}.  We
begin by comparing our intervening absorber measurements to the
\ion{O}{6} $dN/dz$ results reported by Danforth \& Shull (2005; see
also Danforth et al. 2006).  Danforth \& Shull (2005) have conducted a
search for \ion{O}{6} lines at $z_{\rm abs} < 0.15$ in the spectra of
31 active galactic nuclei observed with {\it FUSE}.  For their $dN/dz$
calculation, Danforth \& Shull (2005) count each component as a
separate case (see Table 3 in Danforth et al. 2006), so we should
compare our {\it component} $dN/dz$ results to the Danforth \& Shull
(2005) measurements. Comparing the numbers in column 5 of
Table~\ref{dndztab} to the analogous quantities from Danforth \&
Shull's Table 1, we see that the $dN/dz$ measurements from the two
papers agree within the reported $1-2\sigma$ uncertainties.  However,
our $dN/dz$ measurements are systematically higher than those of
Danforth \& Shull (2005) by roughly $0.1 - 0.3$ dex.

We can identify several reasons why we find systematically higher
$dN/dz$ values than Danforth \& Shull (2005) for intervening
\ion{O}{6} systems including the following: (1) Danforth \& Shull
(2005) only searched for \ion{O}{6} affiliated with absorption systems
that have \ion{H}{1} Ly$\alpha$ lines with equivalent widths greater
than 80 m\AA . In our sample, on the other hand, we include all
\ion{O}{6} absorbers regardless of the strength of the corresponding
\ion{H}{1} Ly$\alpha$ line.  From Table~\ref{intprop}, we see that we
do find \ion{O}{6} absorbers with weak corresponding Ly$\alpha$ lines
with $W_{\rm r}$(Ly$\alpha$) $<$80 m\AA .  Our inclusion of these
weak-Ly$\alpha$ \ion{O}{6} absorbers increases $dN/dz$ compared to the
results of Danforth \& Shull (2005). (2) We do not classify any
absorbers within 5000 km s$^{-1}$ of the QSO redshift as an
intervening absorber, while Danforth et al. (2006) use 1800 km
s$^{-1}$ as the cutoff for intervening vs. proximate absorbers.  The
Danforth \& Shull proximate absorber definition results in a larger
$\Delta z$ searched for intervening \ion{O}{6} per sight line, and
since we do not find many \ion{O}{6} lines in the $1800 \leq v_{\rm
displ} \leq 5000$ km s$^{-1}$ interval (see below), this has the net
effect of increasing $\Delta z$ palpably without increasing the total
number of systems much. Using the 1800 km s$^{-1}$ cutoff therefore
decreases $dN/dz$. If we were to adopt the Danforth \& Shull proximate
absorber cutoff, our $dN/dz$ measurements would decrease by 0.03
dex. (3) Danforth \& Shull integrate over a {\it single} resolution
element in order to determine their limiting equivalent width.  Since
very few (if any) of our detected \ion{O}{6} systems are completely
unresolved and spread over a single resolution element, we choose to
integrate over a larger number of pixels to determine $W_{\rm lim}$ as
discussed above. This results in higher $W_{\rm lim}$ in our sample
and a {\it lower} total $\Delta z$ over which our data have sufficient
S/N to reveal an \ion{O}{6} system of a given strength.  If we were to
also integrate over a single resolution element, our $\Delta z$ values
would increase and our $dN/dz$ numbers would decrease.  This issue is
most important for the samples that include the weakest lines (e.g.,
the $W_{\rm r} \geq$ 30 m\AA\ sample).  Individually, these three
factors have small effects on the measured $dN/dz$, but they all
change $dN/dz$ in the same direction when comparing our results to
those of Danforth \& Shull (2005).  The combined effect of these three
issues can easily account for the systematic difference between our
findings and those of Danforth \& Shull (2005).

\begin{figure}
\centering
\includegraphics[width=9.0cm, angle=0]{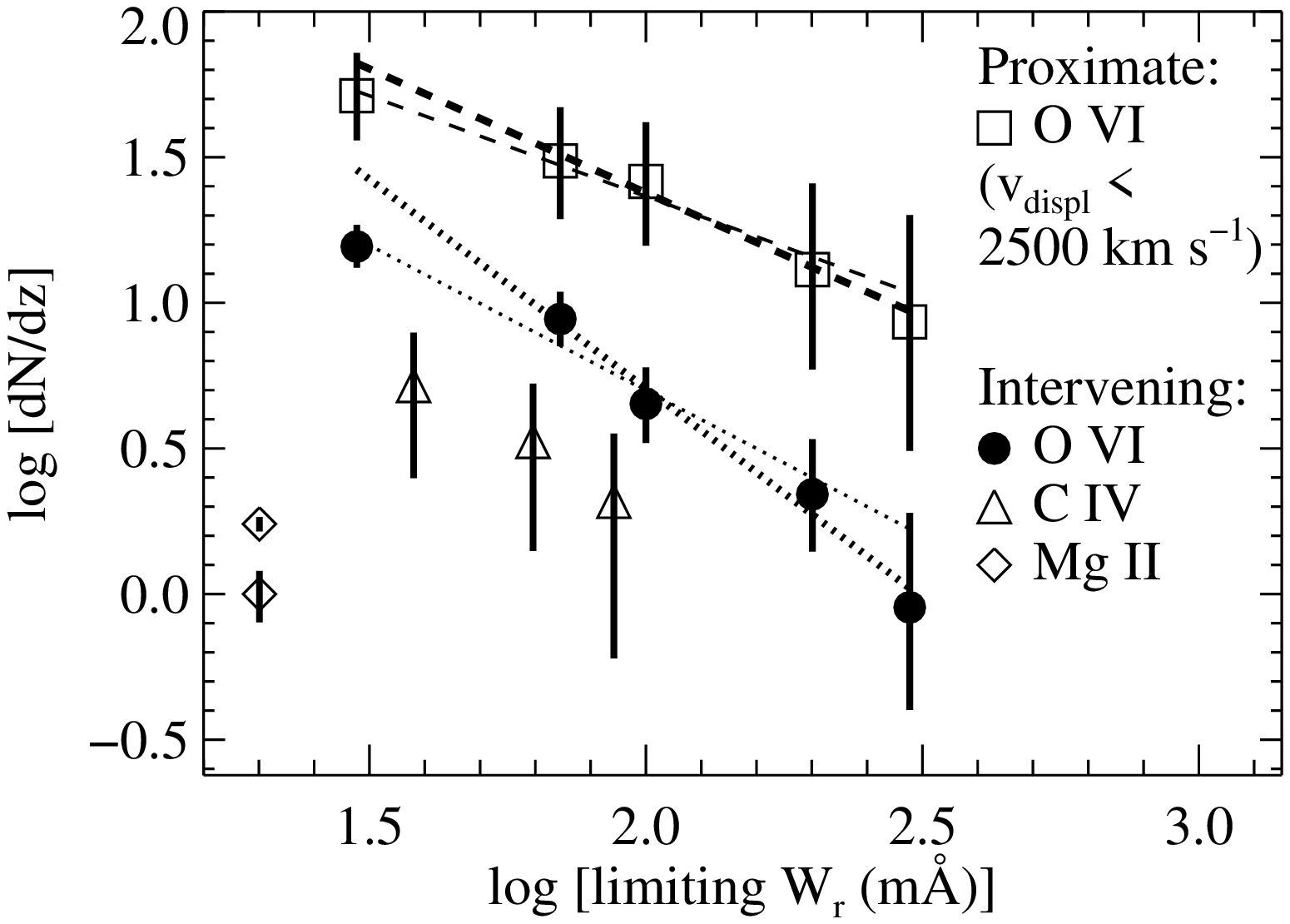}
\caption{Measurements of the cumulative number of QSO absorbers per
unit redshift ($dN/dz$) for O~VI, C~IV, and Mg~II
absorbers with $1\sigma$ statistical uncertainties. The O~VI
measurements are from this paper (see Tables~\ref{dndztab} and
\ref{asscdndztab}), the C~IV results are from Frye et
al. (2003), and the Mg~II numbers are from Narayanan et
al. (2005) and Churchill et al. (1999). For the O~VI systems,
$dN/dz$ is plotted for both the intervening systems (filled circles)
and the proximate absorbers found within 2500 km s$^{-1}$ of the QSO
redshift (open squares; as shown in Table~\ref{asscdndztab}, the
proximate system $dN/dz$ decreases if we use $v_{\rm displ} < 5000$ km
s$^{-1}$ as the cutoff).  In this figure, we show $dN/dz$ for
O~VI {\it systems} (see text \S \ref{dndzmeas}).  All of these
measurements are for systems at $z_{\rm abs} < 0.6$ except for the
Churchill et al. point (the higher Mg~II $dN/dz$), which was
derived from a somewhat higher redshift sample ($0.4 \leq z_{\rm abs}
\leq 1.4$). The dotted lines show power-law fits, weighted by inverse
variance, to the intervening O~VI absorbers using all of the
measurements (thin dotted line) and using all of the measurements
except the 30 m\AA\ point (thick dotted line). The dashed lines show
analogous power-law fits for the proximate absorbers. \label{obsdndz}}
\end{figure}

Figure~\ref{obsdndz} compares our \ion{O}{6} $dN/dz$ measurements to
analogous measurements for low$-z$ \ion{C}{4} absorbers from Frye et
al. (2003) and low$-z$ \ion{Mg}{2} systems from Churchill et
al. (1999) and Narayanan et al. (2005).  All of the measurements in
Figure~\ref{obsdndz} were derived from low$-z$ samples ($z_{\rm abs} <
0.6$) except for the \ion{Mg}{2} measurement from Churchill et
al. (1999), which was derived from a sample extending to moderately
higher redshifts ($0.4 \leq z_{\rm abs} \leq 1.4$).  \ion{O}{6},
\ion{C}{4}, and \ion{Mg}{2} are the most commonly studied metals in
optically-thin QSO absorption systems, and we see from
Figure~\ref{obsdndz} that \ion{O}{6} lines have a substantially higher
number per unit $z$ at low redshifts than either the \ion{Mg}{2}
absorbers or the \ion{C}{4} systems. This is not surprising in the
case of \ion{Mg}{2}; this species is easily photoionized by the UV
background from QSOs, and it is likely to be present only when an
absorber has a relatively high \ion{H}{1} column density.  However, it
is interesting to note that $dN/dz$ of \ion{O}{6} is substantially
higher than $dN/dz$ of \ion{C}{4}.  Because \ion{C}{4} is often
redshifted outside of the Ly$\alpha$ forest, it is by far the most
studied metal in the high-redshift regime, and it would be valuable to
compare the low$-z$ \ion{C}{4} absorbers to their high$-z$ analogs. It
is unfortunate that for our low$-z$ \ion{O}{6} absorber sample, the
\ion{C}{4} doublet is usually redshifted beyond the long-wavelength
cutoff of our observations.  However, IGM ionization models can be
constrained by the different $dN/dz$ trends for \ion{O}{6} and
\ion{C}{4} shown in Figure~\ref{obsdndz}.

From Tables~\ref{dndztab}-\ref{asscdndztab} and Figure~\ref{obsdndz},
we also see that statistically, the \ion{O}{6} $dN/dz$ increases
substantially when the absorber redshift range is close to the
background QSO redshift (compare the open squares and filled circles
in Figure~\ref{obsdndz}).  Using the standard $v_{\rm disp} < 5000$ km
s$^{-1}$ definition of proximate absorbers, we find that there are
2$-$6 times more proximate \ion{O}{6} systems per unit $z$ than
intervening \ion{O}{6} systems (we find similar results comparing
$dN/dz$ of individual components instead of $dN/dz$ of \ion{O}{6}
systems).  As we have already noted, the excess of \ion{O}{6} lines
appears to be largely confined to a smaller velocity interval ($v_{\rm
displ} < 2500 $ km s$^{-1}$) closer to the QSO redshift than the
standard 5000 km s$^{-1}$ definition (see further discussion below).
If we use the 2500 km s$^{-1}$ velocity cutoff to define the proximate
absorbers, we find an even more dramatic excess: in this case, there
are $3-10$ times more proximate \ion{O}{6} systems per unit redshift
than we find in the intervening sample.  The magnitude of the
proximate absorber excess depends on the sample limiting equivalent
width, and it is worth noting that the excess is greatest when
comparing the stronger \ion{O}{6} lines.

It is of interest to assess whether a convenient function, such as a
power law, can be fitted to the \ion{O}{6} $dN/dz$ data shown in
Figure~\ref{obsdndz}. As shown in Figure~\ref{obsdndz}, for the system
$dN/dz$ results, an inverse variance-weighted power law of the form
log $dN/dz$ = $\alpha \ {\rm log} \ W_{\rm lim} + constant$ provides a
satisfactory fit for both the intervening and proximate \ion{O}{6}
measurements with $\alpha$(intervening) = $-1.00$ and
$\alpha$(proximate) = $-0.69$. However, we can also see from
Figure~\ref{obsdndz} that for the intervening systems, the slope of
the observed log $dN/dz$ vs. log $W_{\rm lim}$ data appears to change
at $W_{\rm lim} <$ 70 m\AA . To show this, we also plot in
Figure~\ref{obsdndz} the power law obtained by fitting only the bins
with $W_{\rm lim} \geq$ 70 m\AA . Exclusion of the 30 m\AA\ bin
results in a steeper slope and clearly improves the fit to the higher
equivalent width bins. This type of turnover is sometimes interpreted
as evidence of incompleteness, i.e., the measured $dN/dz$ is lower
than the true $dN/dz$ because the data were inadequate to detect all
of the absorbers in the weakest $W_{\rm r}$ bin.  We argue that this
explanation is unlikely to explain the change in slope apparent in our
\ion{O}{6} measurements.  We have corrected for incompleteness by
carefully calculating the total $\Delta z$ path over which 30 m\AA\
lines can be detected.  In principle, we could have overestimated
$\Delta z$, e.g., by integrating equation~\ref{limeqwint} over an
insufficient number of pixels, which in turn would underestimate
$dN/dz$.  However, this is an unlikely explanation because we would
have to overestimate $\Delta z$ by an unreasonable amount to reconcile
the intervening \ion{O}{6} data with the steeper power law shown in
Figure~\ref{obsdndz}.  Figure~\ref{obsdndz} shows the results and fits
for \ion{O}{6} {\it systems}; as shown in Table~\ref{dndztab}, if we
count components instead of systems, the $dN/dz$ vs. $W_{\rm lim}$
slope steepens significantly (single strong systems split up into
multiple weaker components).  Power-law fits to the {\it component}
$dN/dz$ measurements are acceptable for the proximate absorbers, but
we find that a single power law provides a rather poor description of
the $dN/dz$ measurements for the intervening \ion{O}{6} components.

\subsubsection{$dN/dz$ Results: Differential Distribution}

\begin{figure}
\centering
    \includegraphics[width=9.0cm, angle=0]{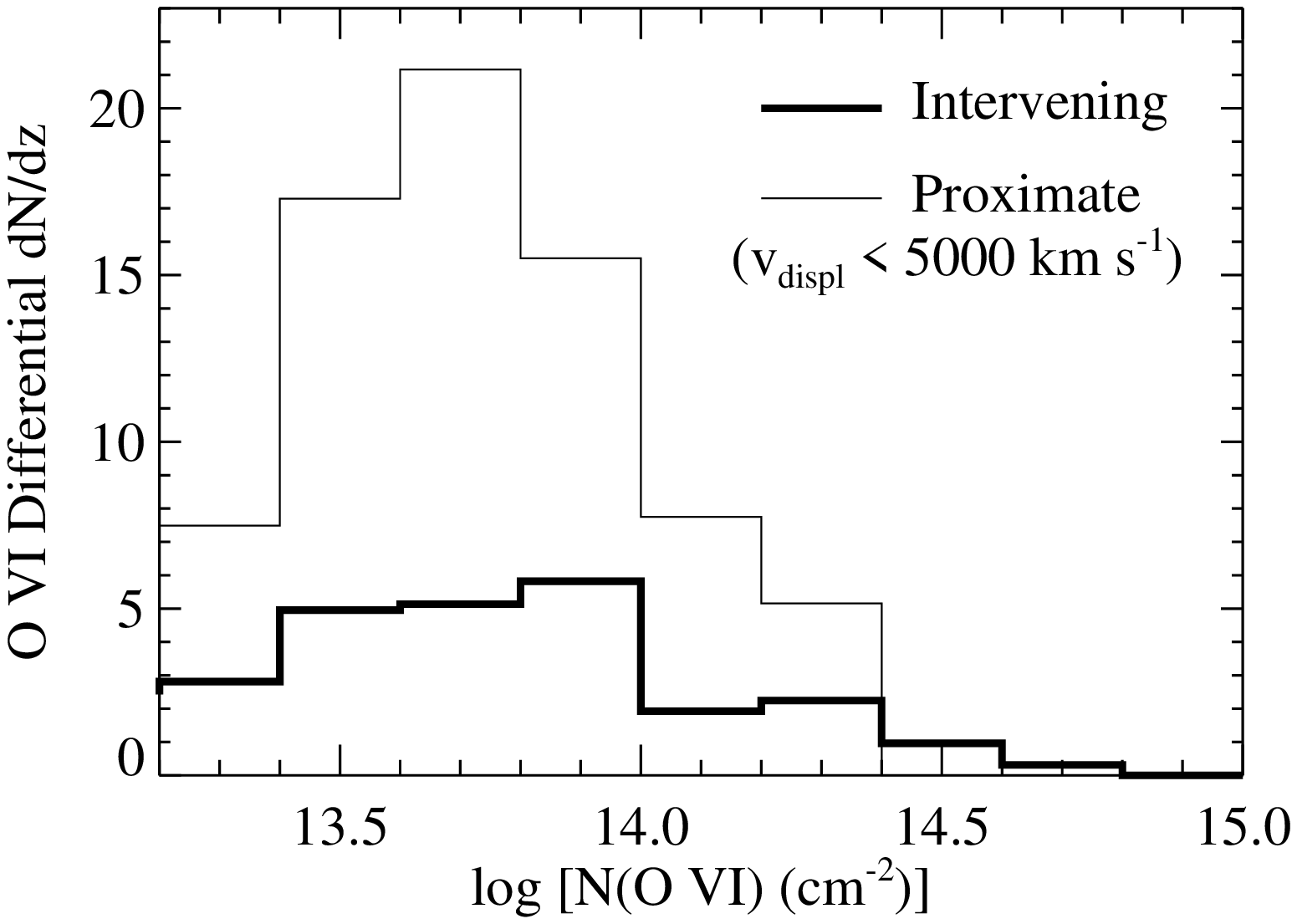}
\caption{The differential O~VI $dN/dz$ distribution for
intervening absorber components (thick line) and proximate absorber
components with $v_{\rm displ} < 5000$ km s$^{-1}$ (thin line). The
differential $dN/dz$ is shown in 0.2 dex bins for log $N$(O~VI)
$\geq$ 13.2.\label{diffdndz}}
\end{figure}

In the previous section we discussed the cumulative $dN/dz$
distribution of \ion{O}{6} absorbers.  We show the {\it differential}
\ion{O}{6} $dN/dz$ distribution for intervening and proximate
absorbers in Figure~\ref{diffdndz} as a function of $N$(\ion{O}{6})
using 0.2 dex bins extending from log $N$(\ion{O}{6}) = 13.2 to 14.8.
In Figure~\ref{diffdndz}, we show the $dN/dz$ that results from
counting individual components, and we exclude the marginal
measurements (those marked with a colon in Table~\ref{compprop}).  The
total path ($\Delta z$) for each column density bin is from
Figure~\ref{pathplot}.  As discussed in \S \ref{dndzmeas}, we
calculate $\Delta z$ as a function of limiting equivalent width.  To
estimate $\Delta z$ for a limiting column density, we converted
$W_{\rm lim}$ to $N_{\rm lim}$ assuming the median $b-$value found for
the intervening and proximate absorbers (see \S \ref{basicstats}): $b$
= 24 km s$^{-1}$ for intervening absorbers, and $b$ = 16 km s$^{-1}$
for proximate systems.

\subsubsection{Comparison with Theoretical $dN/dz$ Predictions for Intervening Absorbers\label{dndztheory}}

Most theoretical studies of the low$-z$ \ion{O}{6} absorbers have made
predictions regarding the cumulative $dN/dz$ distribution, so we focus
on the cumulative distribution in our comparison to theoretical
work. It is not clear if there is a physical motivation for fitting
the $dN/dz$ data in Figure~\ref{obsdndz} with a power law.  Some
theoretical papers have argued that $dN/dz$ vs. $W_{\rm lim}$ should
turn over as $W_{\rm lim}$ decreases.  For example, Tumlinson \& Fang
(2005) suggest that this turnover could be a result of the processes
that transport metals out of galaxies. They argue that if galaxies can
only pollute metals into limited volumes in their immediate
vicinities, then $dN/dz$ should turnover as observed.  In their
recent survey of low$-z$ \ion{O}{6} absorber environments, Stocke et
al. (2006) find no \ion{O}{6} systems in galaxy voids, which is
consistent with the Tumlinson \& Fang (2005) hypothesis.

\begin{figure}
\centering
    \includegraphics[width=9.5cm, angle=0]{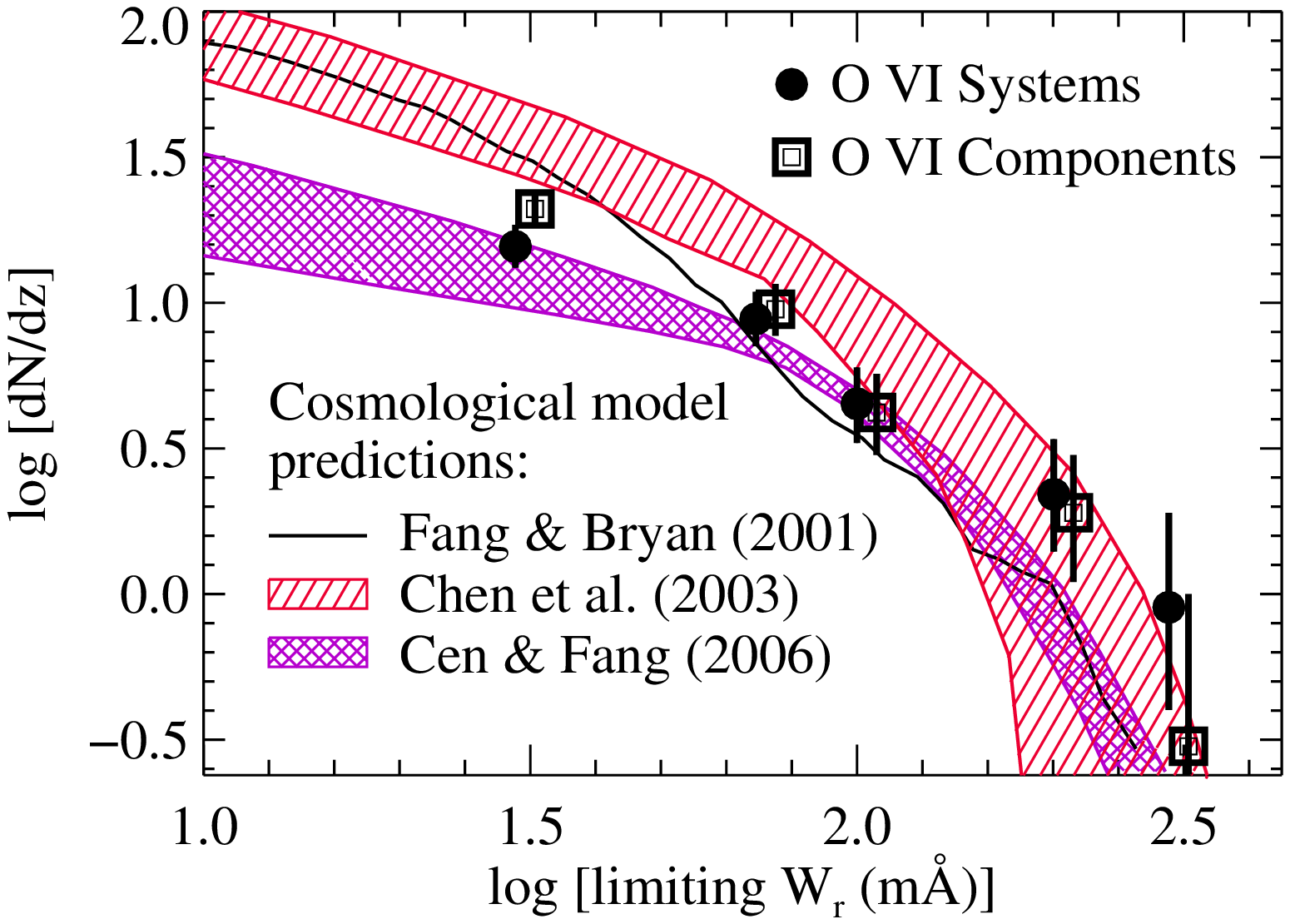}
\caption{Comparison of the observed number of intervening O~VI
absorbers per unit redshift ($dN/dz$) from this paper to the
predictions of hydrodynamic cosmological simulations from Fang \&
Bryan (2001, solid black line), Chen et al. (2003, red hatched
region), and Cen \& Fang (2006, purple hatched region).  These
theoretical predictions apply to $z \approx 0$.  The observations are
indicated by points with $1\sigma$ statistical error bars; filled
circles show the results obtained by counting blended components as a
single ``system'', and double squares show the $dN/dz$ measurements
found by counting each component (including components that are close
in velocity and blended) as a separate case. Chen et al. (2003) and
Cen \& Fang (2006) present a variety of models that vary the physics
implementations and assumptions.  The hatched regions indicate the
{\it range} of $dN/dz$ predicted by their models.\label{dndzplot}}
\end{figure}

Cosmological simulations also predict a $dN/dz$ turnover. Several
papers have employed hydrodynamic simulations of large-scale structure
growth to predict the number of \ion{O}{6} absorbers per unit redshift
for various samples of intervening absorbers defined by the limiting
equivalent width.  We compare these theoretical predictions to our
measurements in Figure~\ref{dndzplot}.  The earlier predictions of Cen
et al. (2001) are superseded by the more recent predictions of Cen \&
Fang (2006), so we do not show the Cen et al. (2001) predictions in
Figure~\ref{dndzplot}.  Also, Chen et al. (2003) and Cen \& Fang
(2006) have presented predictions for multiple models that make
different assumptions about, e.g., the processes that transport metals
out of galaxies, nonequilibrium ionization effects, and the overall
level and scatter of the IGM metallicity.  Current cosmological
simulations have limited ability to fully model complicated processes
such as metal transport, and it should be borne in mind that
significantly uncertain assumptions are sometimes required to predict
statistics of metal absorption line systems from hydrodynamic
simulations.  For this reason, we do not show the predictions of
individual models in Figure~\ref{dndzplot} but instead use hatched
regions to show the {\it range} of $dN/dz$ predictions for intervening
\ion{O}{6} systems from several theoretical papers, i.e., the lower
envelope of the hatched region shows the lowest prediction from the
indicated paper, and the upper envelope indicates the highest
prediction.\footnote{However, one of the models reported by Chen et
al. (2003) predicts $dN/dz$ for O~VI assuming that the IGM is
uniformly enriched to $Z = 1.0 Z_{\odot}$.  This particular model
predicts vastly more O~VI absorbers than observed, and we do not
include this model in the range of Chen et al. models shown in
Figure~\ref{dndzplot}.} 

As we have already noted, it is not always clear how the \ion{O}{6}
absorbers are counted in the cosmological simulation papers. Chen et
al. (2003) note that they {\it do not} attempt to deblend blended
lines, so for comparison to their predictions, the system $dN/dz$
measurements from column (4) of Table~\ref{dndztab} should be
employed.  Other simulation papers employ Voigt-profile fitting (e.g.,
Fang \& Bryan 2001), so for those predictions, the component $dN/dz$
should be used.  In Figure~\ref{dndzplot}, we show the measurements of
the \ion{O}{6} redshift density counting systems (filled circles) and
components (double squares).  Overall, we see broad but imperfect
agreement between the simulation predictions and the observations.
The simulations predict that the $dN/dz$ distribution should turnover
with decreasing $W_{\rm lim}$ as observed.  The Chen et al. (2003)
models predict too many \ion{O}{6} systems at $W_{\rm lim} \lesssim
70$ m\AA\ but roughly agree with the observations at higher equivalent
width limits.  The Cen \& Fang (2006) model that results in the
highest \ion{O}{6} redshift density (i.e., their model that includes
galactic superwinds but neglects nonequilibrium ionization effects)
agrees with the measurements for weak \ion{O}{6} systems but
underpredicts the observed number of relatively strong \ion{O}{6}
absorbers.  It is not surprising that the models and observations do
not agree in detail because the physical processes that are relevant
(e.g., metal transport) are very challenging to model, and approximate
assumptions are required.  Bearing this in mind, the rough agreement
between the models and the observations is encouraging.

\subsection{Doppler Parameter and Column Density Distributions\label{basicstats}}

\begin{figure}
\centering
    \includegraphics[width=9.0cm, angle=0]{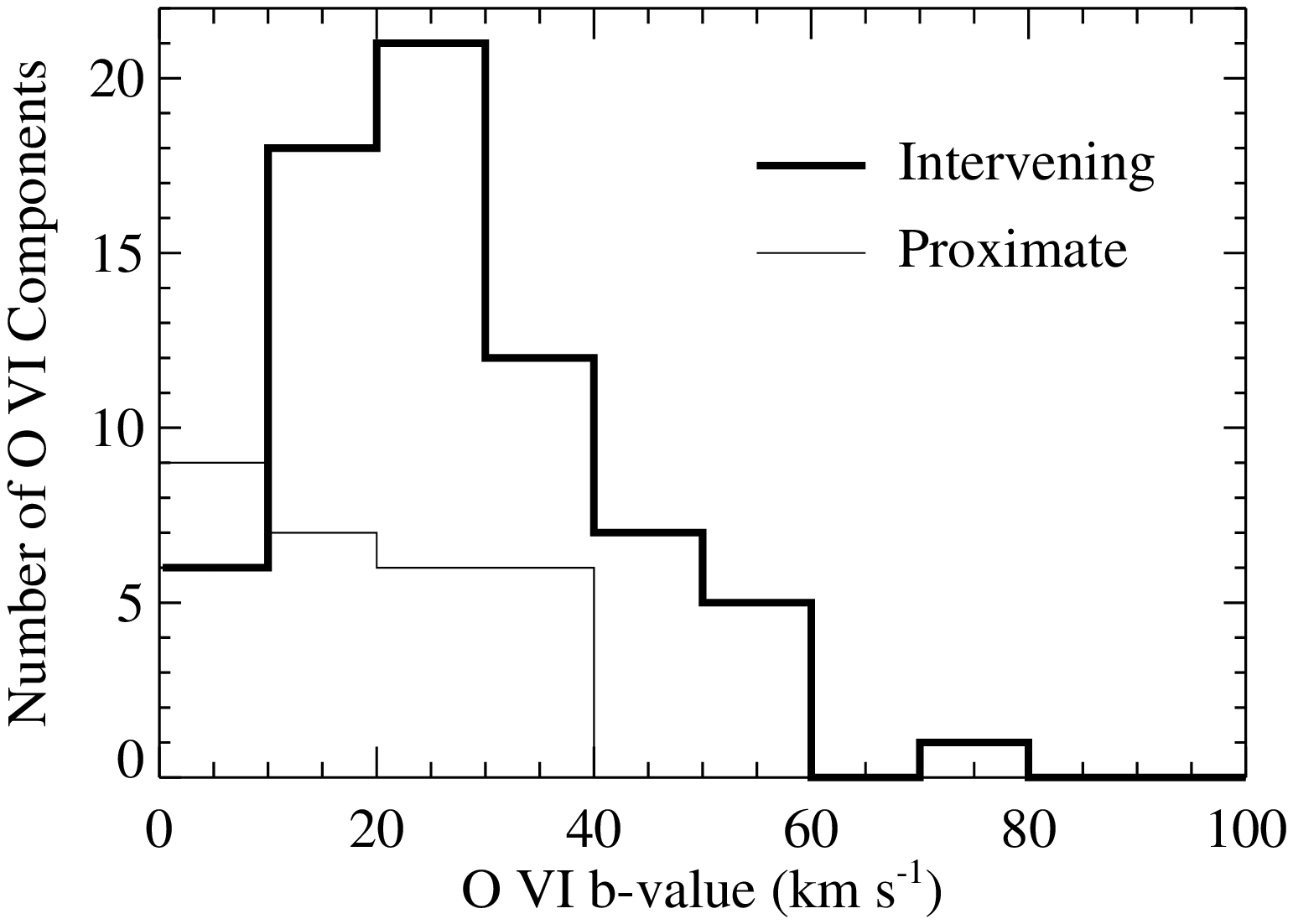}
\caption{The distribution of O~VI $b-$values measured in
individual components from Voigt-profile fitting (see
Table~\ref{compprop}). The thick lines show the measurements in
intervening systems, and the thin lines show the proximate absorber
measurements (with $v_{\rm displ} <$ 5000 km s$^{-1}$). Marginal
measurements (those flagged with a colon in Table~\ref{compprop}) are
excluded from this plot. The $b-$value for O~VI at $T = 3 \times
10^{5}$ K (where O~VI peaks in abundance in CIE) is 18 km
s$^{-1}$.\label{bvaldist}}
\end{figure}

Are the physical conditions implied by the \ion{O}{6} observations
consistent with the expectations from the cosmological simulations?
To address this question, we now examine the implications of the
detailed \ion{O}{6} line measurements.  We begin by comparing the
Doppler parameter and column density distributions of the intervening
and proximate absorber components. Our motivation for this comparison
is twofold: First, considering the findings and issues noted in \S
\ref{absclass}, we would like to search for differences in the
properties of intervening and proximate absorbers that might be useful
for recognizing ejected/intrinsic absorbers at high displacement
velocities. Second, the proximate absorbers are highly likely to be
photoionized. Some proximate systems show evidence that they are
located close to the QSO central engine (see \S \ref{absclass}) where
photoionization by the intense UV emission from the QSO is likely to
be dominant. Even if the proximate absorbers arise in an entirely
different galaxy from the QSO host, they are still likely to be
predominantly photoionized by the QSO light because the Stromgren
spheres of luminous QSOs can extend over large distances/redshift
intervals (e.g., Zheng \& Davidsen 1995; Smette et
al. 2002). Therefore proximate absorbers provide a valuable
photoionized control sample for understanding the ionization of the
intervening systems.

We have noted in Table~\ref{compprop} that some of the individual
component measurements are more uncertain due to systematic problems
such as strong blending and line saturation.  We evaluate the effects
of including or excluding these uncertain cases by defining two
samples. Our ``robust'' sample excludes the uncertain cases that are
flagged with a colon in Table~\ref{compprop}, and the ``full'' sample
uses all of the measurements (including the highly uncertain cases).
Table~\ref{mediantab} summarizes some basic measurements for these
samples including the median and mean values of the \ion{O}{6} column
densities and $b-$values and the number of components in each sample.
Inspecting Table~\ref{mediantab}, we see that the mean and median
\ion{O}{6} column densities of the intervening and proximate absorbers
are similar, with the intervening absorbers having somewhat lower
columns on average.  Comparing the $b-$values, we notice that the
proximate systems tend to be somewhat narrower than then intervening
cases.  To statistically test whether the intervening and proximate
absorbers have different $N$(\ion{O}{6}) and $b$ distributions, we
have applied Kolmogorov-Smirnov (KS) tests. Table~\ref{mediantab}
lists the results of our KS tests comparing the column density and
$b-$value distributions of the proximate and intervening
samples. Comparing the column-density distributions of the robust
intervening and proximate absorbers with a KS test, we find the KS
statistic $D = 0.171$ and the probability of the null hypothesis
(i.e., that the intervening and proximate absorbers are drawn from the
same parent distribution) is 55.8\% . Therefore we find no compelling
evidence that the intervening and proximate absorbers have different
column density distributions. If we use the full sample instead of the
robust sample, we reach the same conclusion.

We have also used a KS test to compare the differential $dN/dz$
distribution of the intervening and proximate \ion{O}{6} components
shown in Figure~\ref{diffdndz}.  By comparing the differential $dN/dz$
distributions instead of the straight column-density distributions, we
account for possible differences in the sensitivity of the spectra
used to search for intervening vs. proximate absorbers.  By
definition, proximate absorbers are located near the peak of the
Ly$\beta$ + \ion{O}{6} emission line of the background QSO, and
consequently, the proximate absorbers are found in regions where the
flux is elevated and the S/N ratio is higher.  Thus, the proximate
absorber regions can reveal weaker lines than the regions probed for
intervening systems.  However, a KS test applied to the differential
$dN/dz$ distributions shown in Figure~\ref{diffdndz} indicates a
49.7\% probability that the samples are drawn from the same parent
distribution, so there is no statistically significant difference
between the intervening and proximate absorbers.

The mean and median $b-$values of the proximate absorbers are lower
than those of the intervening lines. The proximate and intervening
absorber $b-$value distributions are compared in
Figure~\ref{bvaldist}. Applying the KS test to the robust intervening
and proximate $b-$value distributions, we find that the probability
that the two samples are drawn from the same parent distribution is
3.6\% .  This weakly suggests that the intervening and proximate
absorbers have different $b-$value distributions, but the result is
only marginally significant. The full sample yields a KS probability
of 5.3 \%, i.e., a similar result.

However, as we have discussed, the $v_{\rm displ} <$ 5000 km s$^{-1}$
definition of the proximate absorbers may be somewhat arbitrary, and
the truly intrinsic absorbers might be mostly confined to a smaller
velocity interval closer to $z_{\rm QSO}$.  In Table~\ref{mediantab},
we show the mean and median statistics obtained by defining proximate
absorbers to be those cases with $v_{\rm displ} <$ 2500 km s$^{-1}$.
We see that this alternative definition only results in a slightly
lower median and mean column densities and $b-$values.  However, the
probability that the intervening and proximate $b-$values have the
same distribution decreases if we use the $v_{\rm displ} <$ 2500 km
s$^{-1}$ definition: the probability of the null hypothesis decreases
to $1.6 - 2.7$\%.  Thus, we find a somewhat stronger indication that
the proximate absorbers are statistically more narrow than the
intervening lines with the hypothesis that proximate/intrinsic
systems are predominantly confined to $v_{\rm displ} <$ 2500 km
s$^{-1}$.

\begin{figure}
\centering
\includegraphics[width=9.0cm, angle=0]{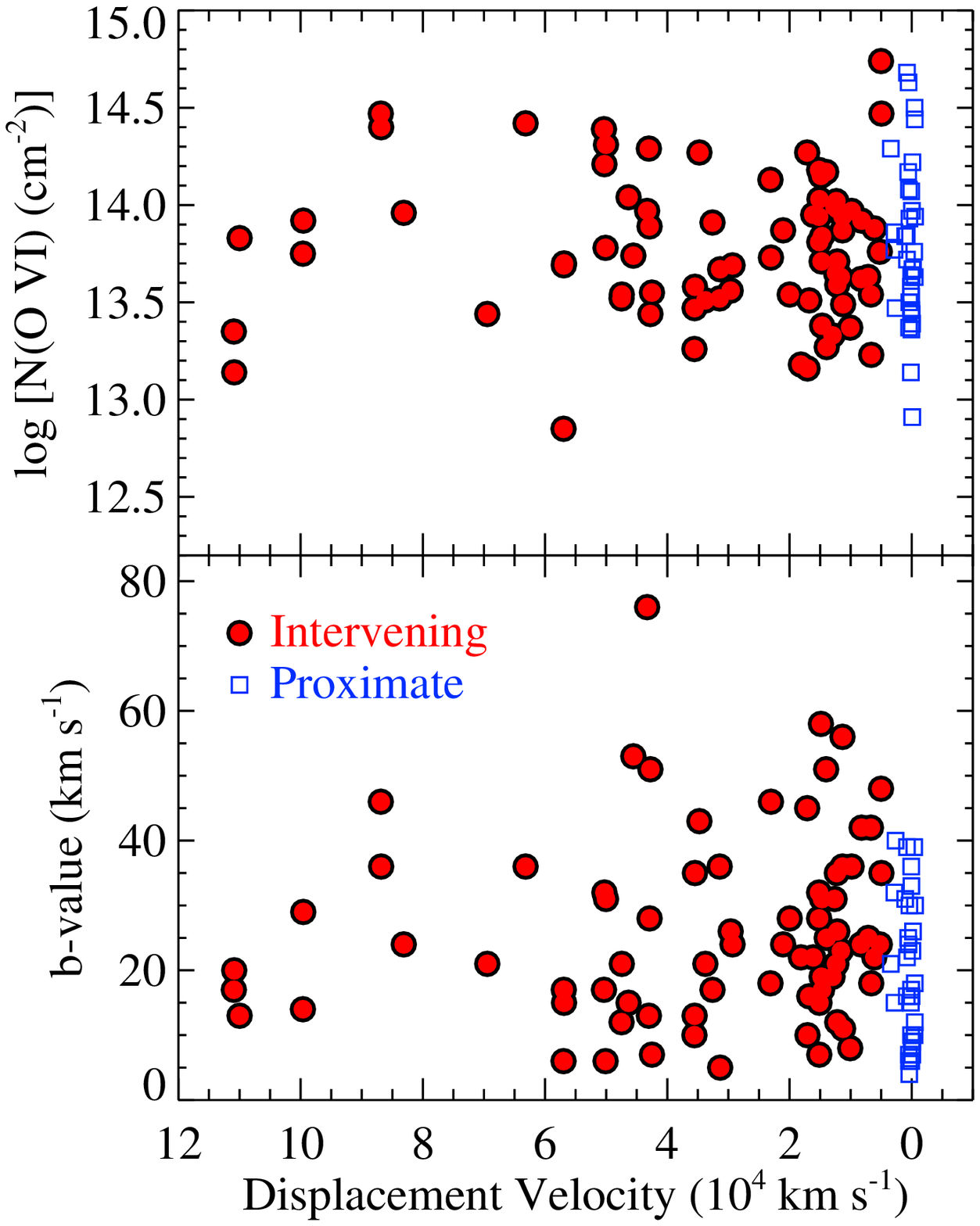}
\caption{{\it Upper panel:} Individual-component O~VI column
densities as a function of velocity of displacement from the QSO
redshift, $v_{\rm displ} = c[(1 + z_{\rm QSO})^2 - (1 + z_{\rm
abs})^2]/[(1 + z_{\rm QSO})^2 + (1 + z_{\rm abs})^2]$, for intervening
components (red filled circles) and proximate systems within 5000 km
s$^{-1}$ of the QSO redshift (blue open squares). {\it Lower panel:}
O~VI $b-$values vs. displacement velocity for intervening and
proximate absorbers.\label{ejectplot}}
\end{figure}

Is there any {\it direct} evidence that the absorber distributions
and properties change at some value of $v_{\rm displ}$?  Or, is there
evidence that a substantial fraction of the proximate/intrinsic
absorbers are dispersed to large displacement velocities and are
intermingled with the intervening systems as suggested by some of the
papers noted in \S \ref{absclass}?  In Figure~\ref{ejectplot}, we show
the \ion{O}{6} column densities and $b-$values of the intervening
absorbers (filled circles) and the proximate systems (open squares)
as a function of $v_{\rm displ}$. While some intrinsic systems with
high $v_{\rm displ}$ have been reported in the literature (\S
\ref{absclass}), the majority of the known intrinsic systems are found
relatively close to redshift of the QSO.  Therefore, we might expect
the absorber properties to change as $v_{\rm displ}$ approaches zero
(i.e., as $z_{\rm abs}$ approaches $z_{\rm QSO}$) if there are a
significant number of intrinsic systems at $v_{\rm displ} < 5000$ km
s$^{-1}$. This would be manifested as a correlation between
$N$(\ion{O}{6}) or $b$(\ion{O}{6}) and $v_{\rm displ}$.  However, we
have used Spearman tests to evaluate this hypothesis, and we find that
there are no statistically significant correlations of any combination
of the data shown in Figure~\ref{ejectplot}.  Considering only the
intervening absorbers, we find that $N$(\ion{O}{6}) and
$b$(\ion{O}{6}) are not correlated with $v_{\rm displ}$.  Likewise,
the proximate absorbers show no significant correlations.  Finally,
combining the intervening and proximate systems into a single sample,
we find no compelling correlations.

\begin{figure}
\centering
\includegraphics[width=9.0cm, angle=0]{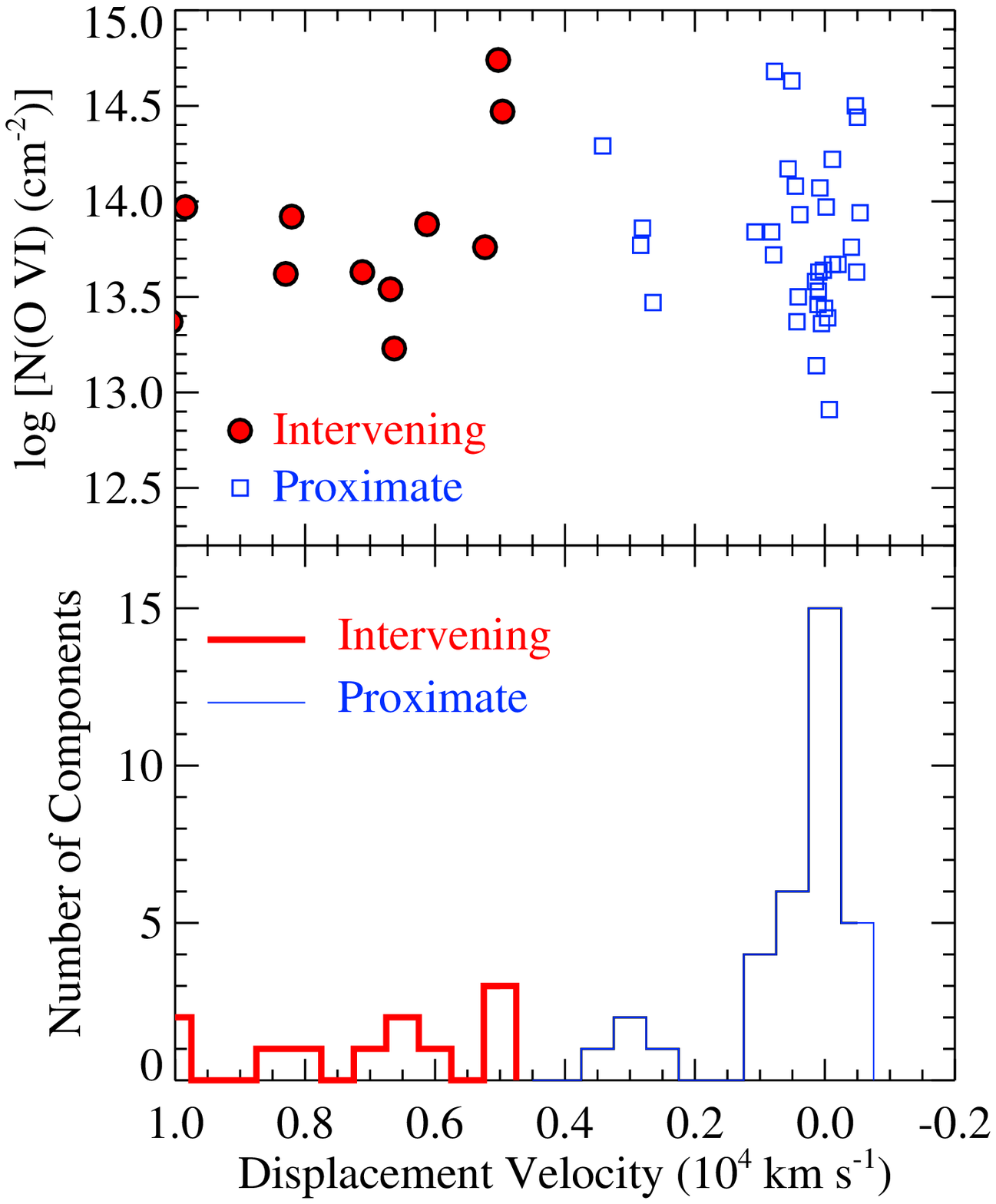}
\caption{{\it Upper panel:} Invidual-component O~VI column
densities vs. displacement velocity (as in Figure~\ref{ejectplot} but
expanded to more clearly show the absorbers within 10,000 km s$^{-1}$
of the QSO redshift).  {\it Lower panel:} Number of O~VI
components detected in 500 km s$^{-1}$ displacement-velocity bins
extending from $v_{\rm displ} = -2000$ km s$^{-1}$ to $+10,000$ km
s$^{-1}$.  The thick line shows the systems formally classified as
intervening, and the thin line shows the proximate systems.  A
substantial excess of absorption lines is clearly evident within
$\approx$1000 km s$^{-1}$ of the QSO redshifts.\label{ejectblowup}}
\end{figure}

Ostensibly, Figure~\ref{ejectplot} does show an apparent increase in
the simple number of intervening \ion{O}{6} lines at $v_{\rm displ}
\lesssim 6 \times 10^{4}$ km s$^{-1}$.  However, this is not a real
effect.  Because of the combination of QSO redshifts and the observed
wavelength range of the various spectra, some of the sight lines probe
larger ranges of $v_{\rm displ}$ than others.  All of the sight lines
cover the $v_{\rm displ} \approx$ 0 km s$^{-1}$ region, but the upper
bound on the observed $v_{\rm displ}$ range varies significantly among
the sight lines.  Consequently, the number of lines detected thins out
as $v_{\rm displ}$ increases because fewer sight lines cover the
larger $v_{\rm displ}$ values.

We do notice, however, that the distribution of proximate absorber
velocities is sharply peaked around $v_{\rm displ} \approx 0$ km
s$^{-1}$.  This is a real effect since all of the sight lines fully
cover the $-2000 \leq v_{\rm displ} \leq 5000$ km s$^{-1}$ range. To
show this more clearly, we show in the lower panel of
Figure~\ref{ejectblowup} the number of \ion{O}{6} lines found in 500
km s$^{-1}$ bins in the $-2000 \leq v_{\rm displ} \leq 10000$ km
s$^{-1}$ range; the thick-line histogram shows the intervening
systems, and the thin-line histogram represents the proximate cases
(assuming the $v_{\rm displ} \leq 5000$ km s$^{-1}$ definition).  For
reference, the upper panel of Figure~\ref{ejectblowup} shows the
column densities of the individual components that contribute to each
bin in the lower panel. We see that for $v_{\rm displ} \gtrsim 1500$
km s$^{-1}$, all of the bins, intervening or proximate, have $0 - 3$
counts. At $v_{\rm displ} < 1500$ km s$^{-1}$, however, the number of
counts in each bin increases dramatically with 15 components in the
bin centered at $v_{\rm displ} = 0$ km s$^{-1}$ and $4 - 6$ counts in
the other three bins.  There is clearly a substantial excess of
\ion{O}{6} lines at $v_{\rm displ} < 1500$ km s$^{-1}$.  The excess is
also evident in the contrasting $dN/dz$ measurements listed in
Tables~\ref{dndztab} and \ref{asscdndztab}. This excess is similar to
the excess of \ion{C}{4} absorption lines at $v_{\rm displ} \approx 0$
km s$^{-1}$ observed at higher redshifts (e.g., Foltz et al. 1986).
Apparently, the standard $v_{\rm displ}$ value used to classify
proximate absorbers (i.e., $v_{\rm displ} < 5000$ km s$^{-1}$) could
be too large; the majority of the truly intrinsic absorbers appear to
be located at smaller displacement velocities.  There are known
proximate systems at $z_{\rm abs} < 0.4$ with intrinsic absorption
lines identified at $v_{\rm displ} \gg 1500$ km s$^{-1}$ (e.g., Yuan
et al. 2002; Ganguly et al. 2003), but those systems have the
characteristics of ``mini-BAL'' systems.

\subsection{Correlation of $N$(\ion{O}{6}) with $b$(\ion{O}{6})\label{nbcor}}

Some theoretical studies have predicted that $N$(\ion{O}{6}) and
$b$(\ion{O}{6}) should be highly correlated.  Using the basic ideas
and formalism of Edgar \& Chevalier (1986), Heckman et al. (2002) have
argued that a correlation of $N$(\ion{O}{6}) and $b$(\ion{O}{6}) is a
natural consequence of radiative cooling in a wide variety of
contexts, and they present a heterogeneous sample of \ion{O}{6}
measurements that are consistent with the expected $N-b$ correlation.
Heckman et al. (2002) include in their sample a small number of
extragalactic \ion{O}{6} systems as well as measurements in the Milky
Way ISM, the Magellanic Clouds, and starburst galaxies.  Subsequently,
Lehner et al. (2006) and Danforth et al. (2006) revisited the Heckman
et al. model with more homogeneous samples that only included
extragalactic \ion{O}{6} systems, and they report mixed results.
Danforth et al. (2006) report that they find no correlation between
$N$(\ion{O}{6}) and $b$(\ion{O}{6}).  Conversely, Lehner et al. (2006)
present a sample of measurements that are generally consistent with
the Heckman et al. models, but they note that the models predict
\ion{Ne}{8} columns that are too large compared to their observations.

\begin{figure}
\centering
    \includegraphics[width=9.0cm, angle=0]{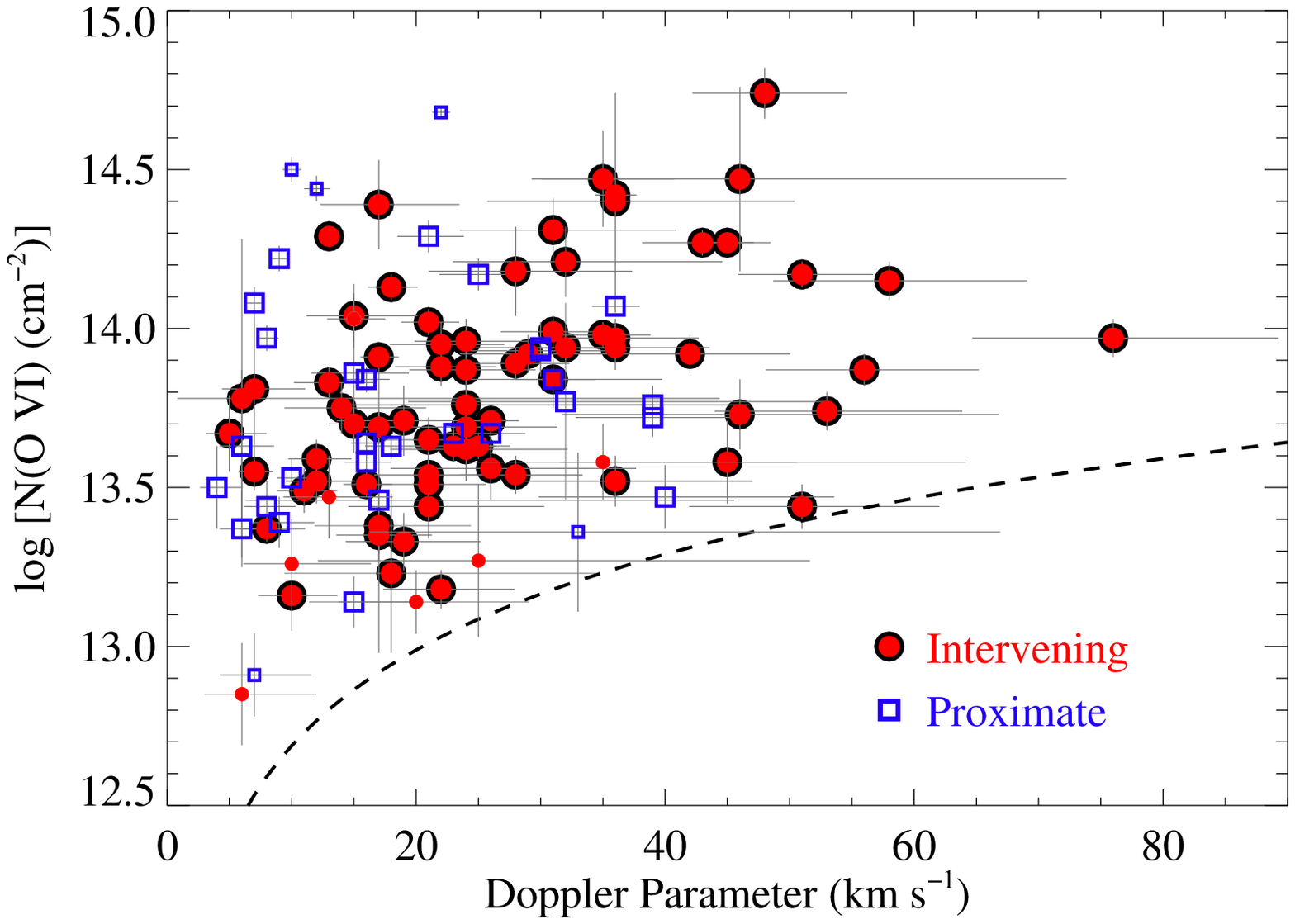}
\caption{The O~VI column density and $b-$value combinations
measured in individual components using Voigt-profile fitting (see
Table~\ref{compprop}).  Filled red circles show measurements of the
column densities and $b-$values of components in intervening systems,
and open blue squares show the proximate absorber measurements.  This
figure shows all measurements reported in Table~\ref{compprop}, but
the measurements that are marginal (due to problems such as line
saturation) are plotted with smaller symbols. The dashed line shows
the locus of $b-$value - $N$(O~VI) combinations for a Gaussian
line with a central optical depth of 10\% ; this is an approximate
detection threshold for the data used in this paper. \label{bvsn}}
\end{figure}

Using our sample of component measurements from Table~\ref{compprop},
Figure~\ref{bvsn} plots log $N$(\ion{O}{6}) vs. $b$(\ion{O}{6}).  As
in previous sections, we show both the intervening and proximate
systems (with $v_{\rm displ} <$ 5000 km s$^{-1}$) with filled circles
and open squares, respectively.  The dashed line in Figure~\ref{bvsn}
indicates the locus of $N$(\ion{O}{6}) and $b$(\ion{O}{6}) values for
a single Gaussian line with a central optical depth of 10\% .  Lines
that are broad and have a low column density are difficult or
impossible to detect because the optical depth in individual pixels in
their absorption profiles are comparable to (or less than) the noise
in the continuum. Moreover, such lines are difficult to distinguish
from undulations in the intrinsic continuum of the QSO.  Consequently,
this $N - b$ locus is effectively a detection threshold; lines below
the dashed line in Figure~\ref{bvsn} are usually undetectable. Neglect
of this factor can produce an artificial correlation, as can be seen
from Figure~\ref{bvsn}.  We have used Spearman tests to evaluate
whether $N$(\ion{O}{6}) and $b$(\ion{O}{6}) measurements from this
paper are correlated.  To account for the detection threshold, we only
use lines with log $N$(\ion{O}{6}) $\geq 13.6$ in the Spearman test;
from Figure~\ref{bvsn}, we can see that we can detect lines with log
$N$(\ion{O}{6}) $\geq 13.6$ over the full range of measured
$b-$values.  From Spearman tests, we find that the proximate
absorbers are not significantly correlated using either the robust
sample or the full sample.  On the other hand, the Spearman test
indicates that in the case of the intervening absorbers,
$N$(\ion{O}{6}) and $b$(\ion{O}{6}) are correlated at the $2.7\sigma$
level.\footnote{Most of the problematic measurements in the full
intervening sample are below the log $N$(O~VI) = 13.6 cutoff, so
we obtain nearly identical results for the intervening systems using
either the robust or the full sample.} 

While the correlation in the intervening sample is suggestive, its
significance is insufficient to provide compelling support for the
Heckman et al. (2002) model.  Moreover, comparing our measurements to
the predictions from Heckman et al. (see their Figure 1), we see that
the scatter in the intervening absorbers is substantially larger than
the scatter expected in the radiative cooling model.  This may not be
surprising because the intervening sample contains absorbers that are
quite possibly photoionized (see \S 4) and would not be
expected to have correlated values of $N$ and $b$.  However, if we
omit the absorbers that appear to be photoionized, the significance of
the correlation remains marginal ($< 3\sigma$).  The intervening
absorber sample considered here has a narrower range of $b$ and $N$
than the sample studied by Heckman et al. (2002).  Their sample
contained much larger columns and velocity widths since it included
measurements of \ion{O}{6} absorption in starburst outflows.  The
correlation between $N$ and $b$ is strongest in these types of
environments, and recent results seem to confirm that the trend at the
higher end of the $N-b$ distribution persists (T. M. Heckman 2008,
private communication).  In such environments, collisional processes
dominate, and radiative cooling is an important, perhaps dominant,
process in establishing line profile shapes. While the simple Heckman
et al. (2002) model may not reflect the \ion{O}{6} intervening
absorber population characteristics in all cases, it is likely that
radiative cooling contributes in part to the observed distribution of
$N$ and $b$ values.

\subsection{Profile Shapes: \ion{O}{6} vs. \ion{H}{1}\label{shapesection}}

The detailed correspondence (or lack thereof) between \ion{O}{6} and
\ion{H}{1} provides useful insight on the physical conditions and
nature of the absorbers, so we now present comparisons of the
\ion{O}{6} and \ion{H}{1} profile shapes.  In \S \ref{classmulti}, we
noted that some of the \ion{O}{6} absorbers are characterized by
kinematical simplicity with all the \ion{O}{6} and \ion{H}{1}
component velocity centroids aligned to within their 2$\sigma$
uncertainties (\S \ref{aligndefsec}), but other systems are more
complex with velocity offsets between the nearest \ion{O}{6} and
\ion{H}{1} components that can exceed 50 km s$^{-1}$ (see
Figure~\ref{veloffsets}).  When the \ion{H}{1} and \ion{O}{6}
centroids in a component are well-aligned, the component can be
modeled as a single-phase gas cloud (conversely, significant velocity
offsets preclude single-phase models). Additional information such as
the low-ion profiles presented in \S \ref{classmulti} and
Figure~\ref{pks0312multi} (for the PKS0312-770 system at $z_{\rm abs}
= 0.20266$) can also establish that an absorber is a multiphase
entity, and when available, we have used information on low ions to
evaluate whether a system is truly a multiphase absorber.

As noted in \S \ref{classsec}, we classify 37\% of the intervening
absorbers as simple systems with all of their \ion{O}{6} and
\ion{H}{1} components aligned to within the measurement uncertainties.
We can go beyond rough classifications and gain better insight by
directly examining the \ion{O}{6} and \ion{H}{1} profiles.  To do
this, we employ apparent column density $[N_{\rm a}(v)]$ profiles as
described in \S \ref{secmeas}.  $N_{\rm a}(v)$ profiles have a
significant advantage for comparing different species: $N_{\rm a}(v)$
profiles are linear functions, and they can be simply scaled to
compare the detailed correspondence of two species. When good
detections of multiple transitions are available for a given species
(e.g., the 1031.93 and 1037.62 \AA\ lines or multiple \ion{H}{1} Lyman
series lines), we use the method of Jenkins \& Peimbert (1997) to
construct composite $N_{\rm a}(v)$ profiles.  These composite profiles
have better S/N ratios. More importantly, weak lines can be used to
constrain the $N_{\rm a}(v)$ profile in high-column density regions
where the strong lines are saturated, and conversely, strong lines can
be employed to flesh out the profile in the low-density regions where
the weak lines are undetectable.

\begin{figure}
\centering
    \includegraphics[width=8.0cm, angle=0]{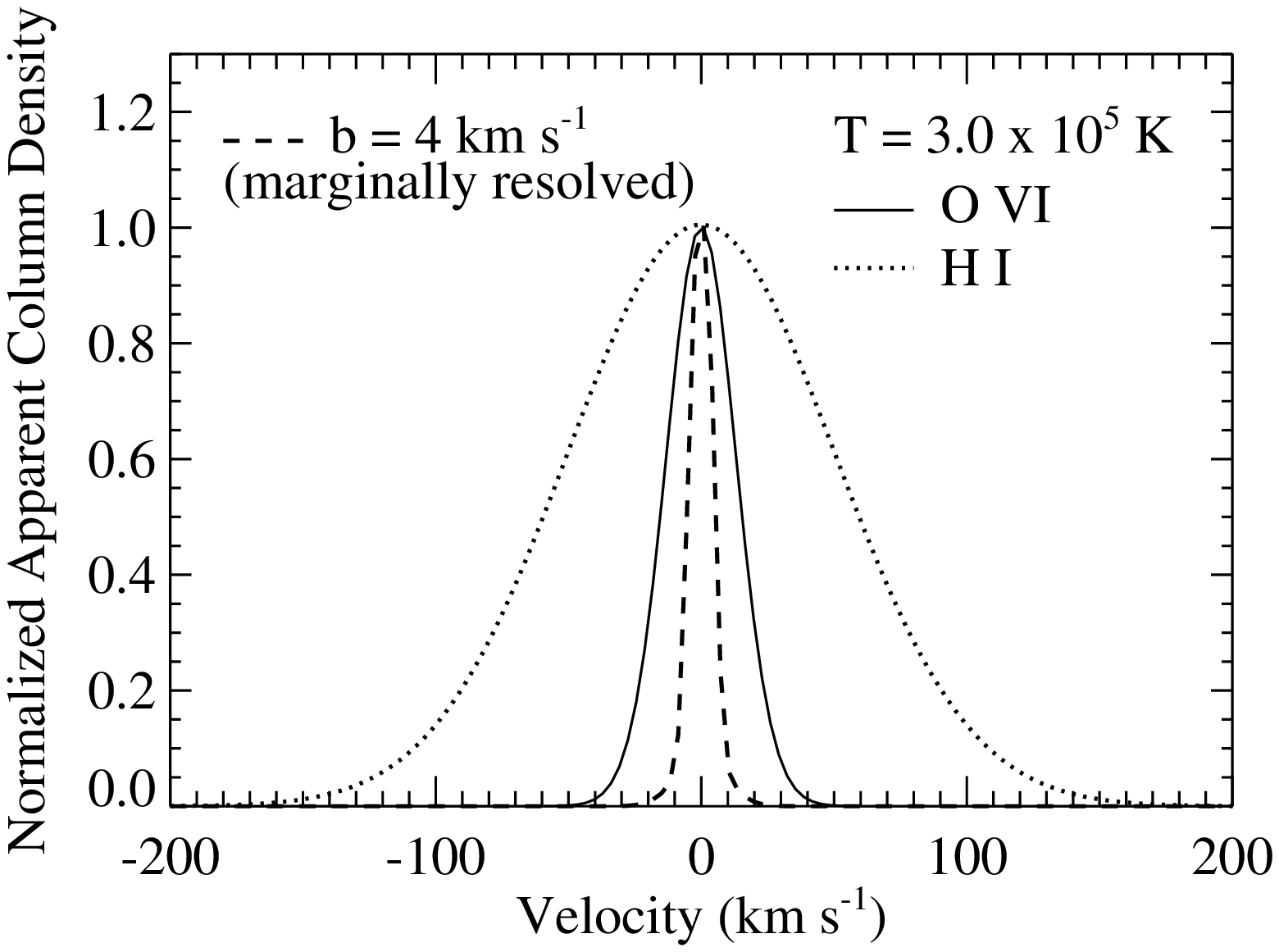}
\caption{Theoretical apparent column density profiles observed at the
resolution of the STIS E140M echelle spectrograph (based on the E140M
line-spread functions from the STIS Handbook for the $0\farcs 06
0\farcs 2$ slit). All of the profiles are normalized so that the peak
of the profile = 1.0.  The dashed line shows the profile of a
marginally resolved line with $b$ = 4 km s$^{-1}$.  The solid line
shows the predicted profile for an O~VI line broadened only by
thermal motions and arising in gas at $T = 3.0 \times 10^{5}$ K, and
the dotted line shows the corresponding H~I profile, i.e., an
H~I line that is only broadened by thermal motions at $T = 3.0
\times 10^{5}$ K. At this temperature, O~VI and H~I lines
are well-resolved with the E140M spectrograph and are expected to have
dramatically different $N_{\rm a}(v)$ profiles if they are
predominantly thermally broadened.\label{navdemo}}
\end{figure}

\begin{figure*}
\centering
    \includegraphics[width=8.2cm, angle=0]{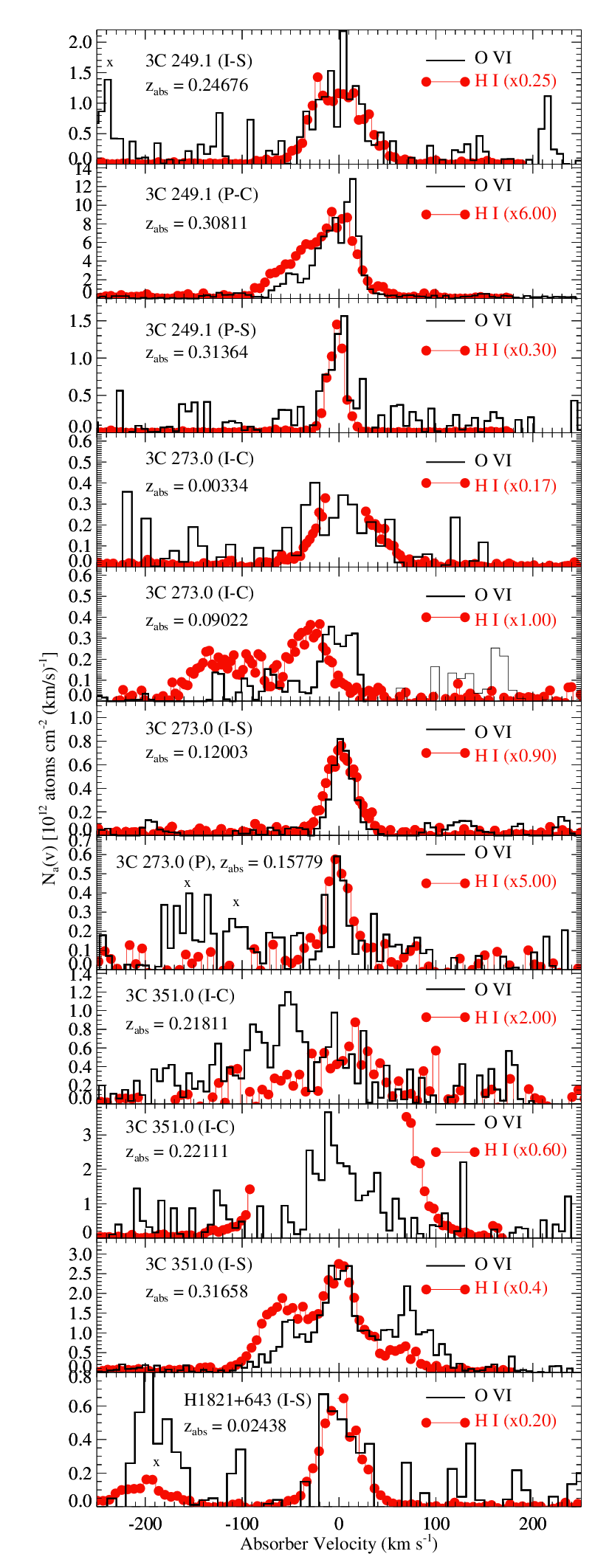}
    \includegraphics[width=8.2cm, angle=0]{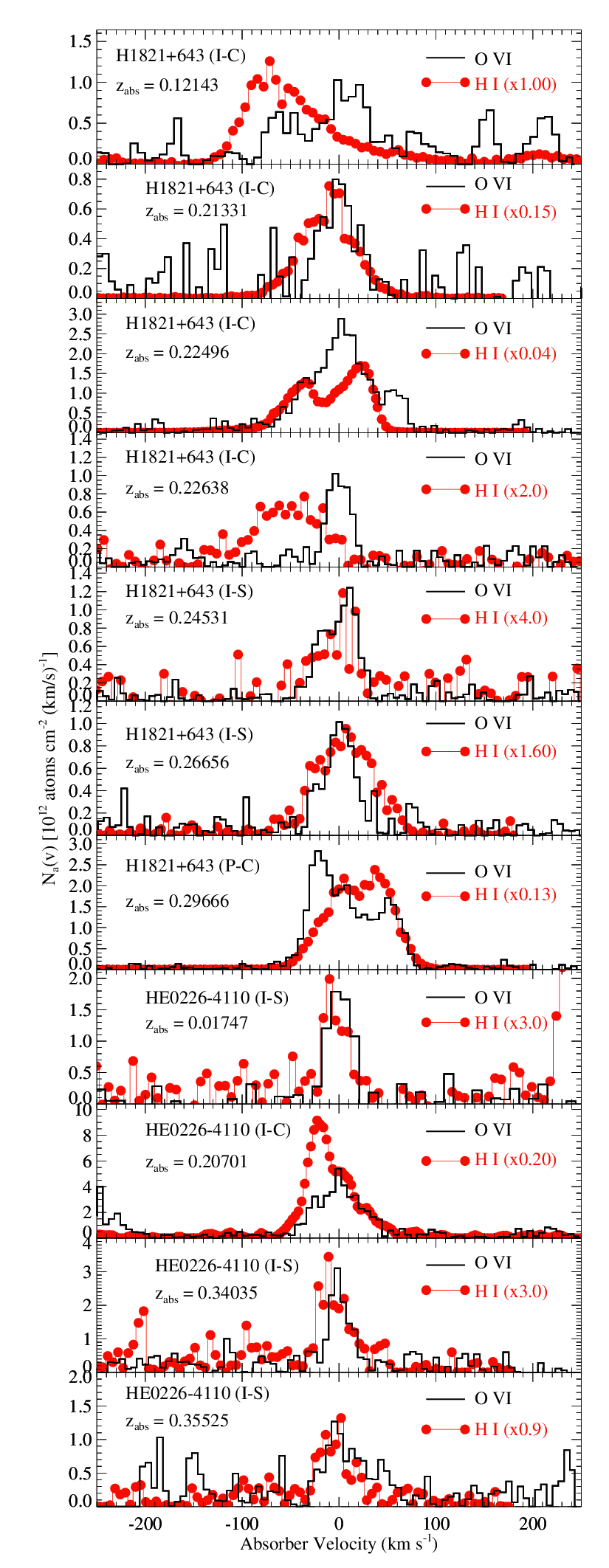}

\caption{Comparison of the O~VI apparent column density profiles
(black histograms) to the affiliated H~I $N_{\rm a}(v)$
profiles (red-dot histograms) for all O~VI absorbers reported in
this paper. Each panel is labeled with the QSO sight line and the
redshift of the system shown in that panel. Codes in parentheses
denote whether the system is intervening (I) or proximate (P) and
whether it is classified as simple (S) or complex (C).  The H~I 
profiles have been scaled as indicated in each panel for comparison
with the O~VI profile (the H~I scale factors are not
based on fits; these are just convenient scalings for purposes of
comparison); in some cases a different scaling is shown with a small
black-dot histogram. Whenever possible, the profiles are weighted
composite profiles derived from multiple transitions using the method
of Jenkins \& Peimbert (1997).  By constructing composite profiles,
weaker lines can be used to determine $N_{\rm a}(v)$ in the line cores
where the strong lines are saturated. However, in some cases, all
available data are strongly saturated in some portion of the
profile. In these cases, there are gaps in velocity ranges where all
available data are saturated. In this figure, the STIS data are binned
to 7 km s$^{-1}$ pixels and the {\it FUSE} data are shown with
$\approx$10 km s$^{-1}$ pixels. Unrelated features are marked with
``x''. \label{navplots}}

\end{figure*}

\begin{figure*}
\figurenum{18}
\centering
    \includegraphics[width=8.9cm, angle=0]{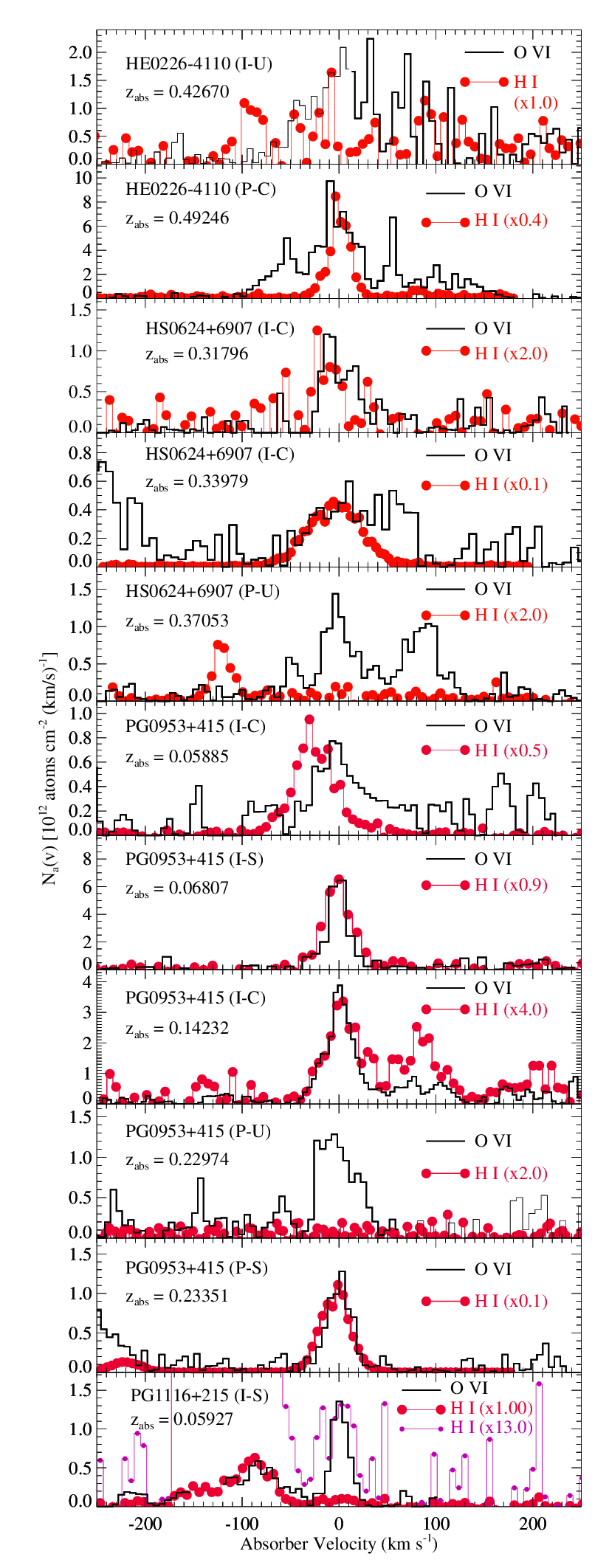}
    \includegraphics[width=8.9cm, angle=0]{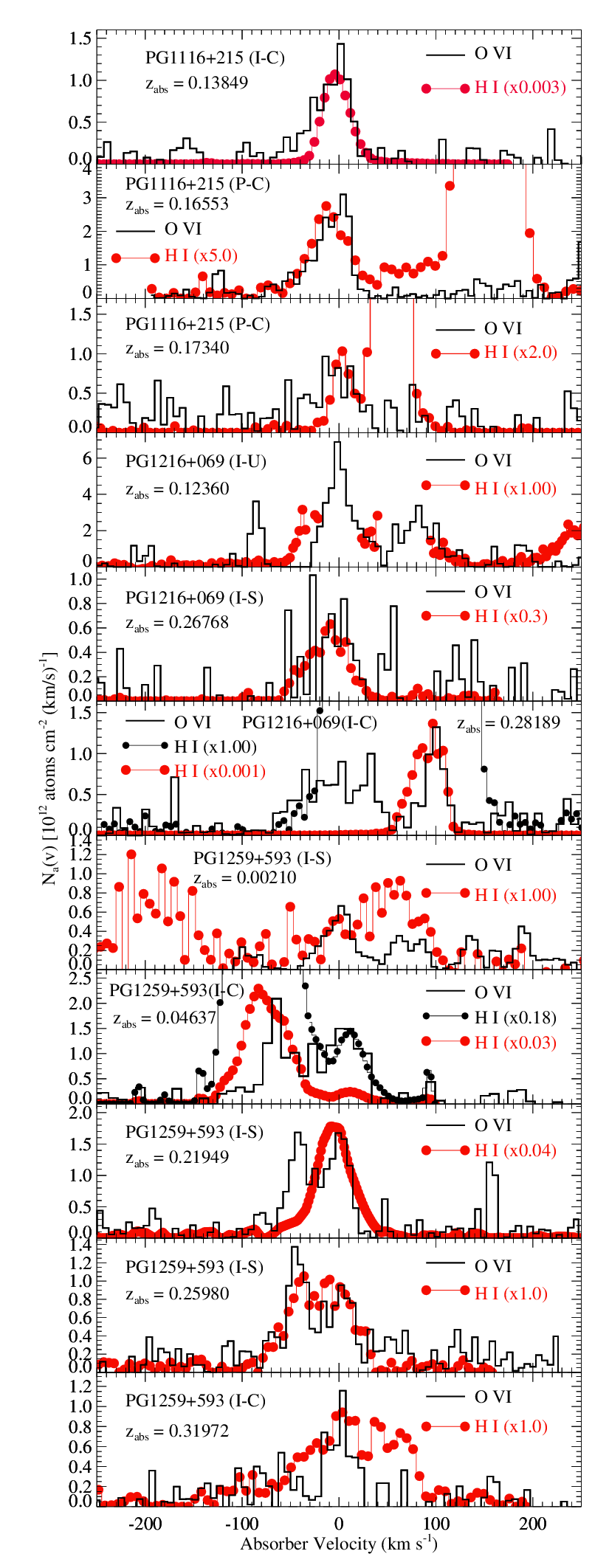}
\caption{continued}
\end{figure*}

\begin{figure*}
\figurenum{18}
\centering
    \includegraphics[width=8.9cm, angle=0]{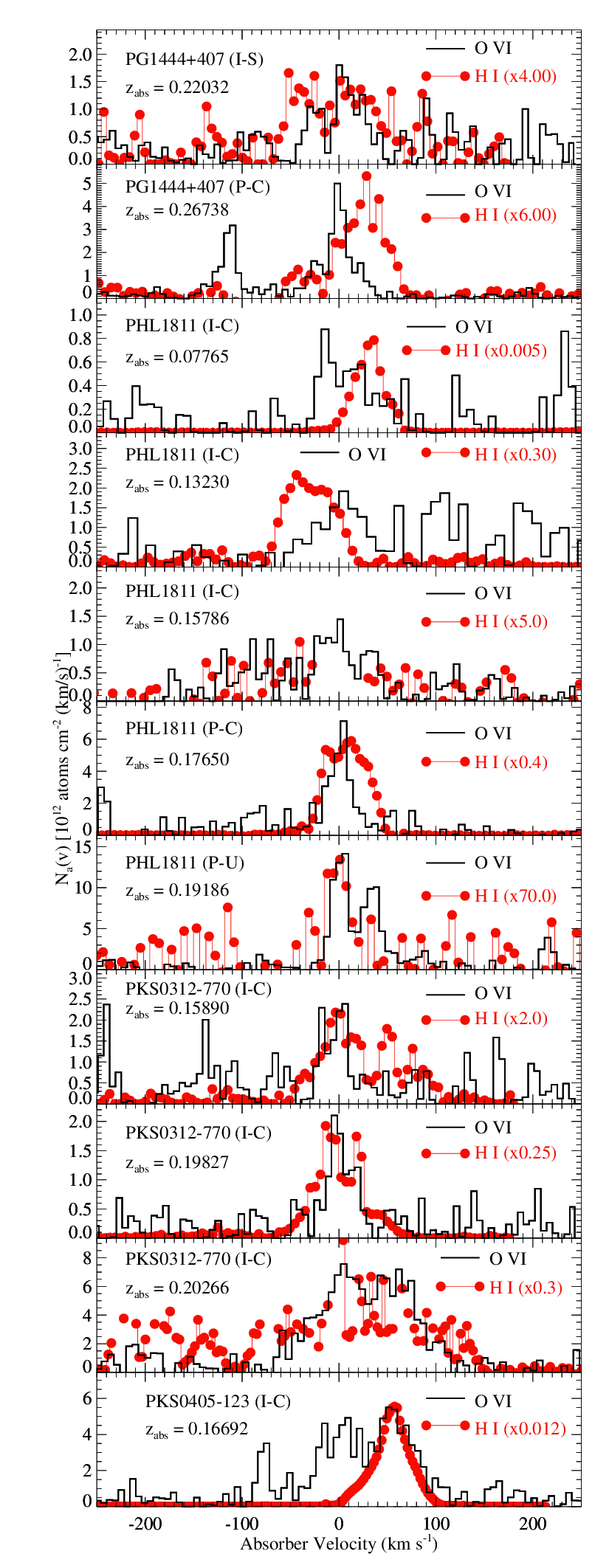}
    \includegraphics[width=8.9cm, angle=0]{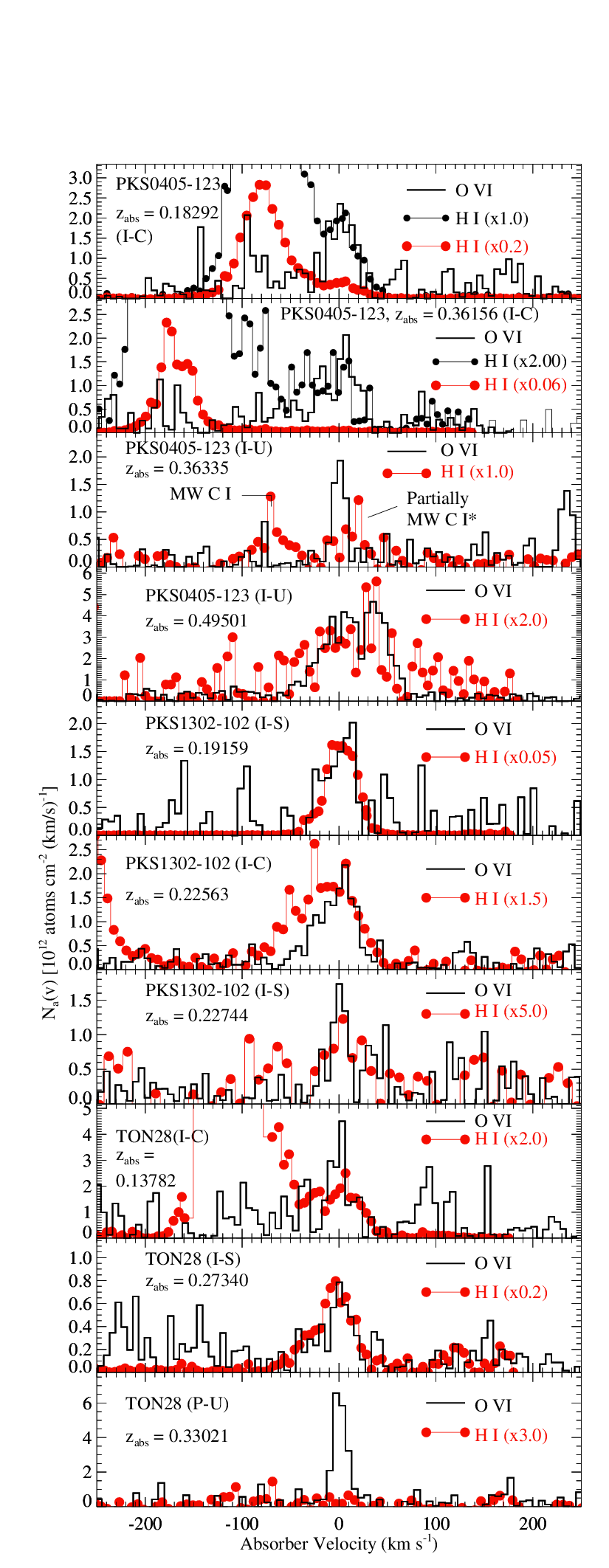}
\caption{continued}
\end{figure*}

To guide the interpretation of the $N_{\rm a}(v)$ profiles, we show in
Figure~\ref{navdemo} several theoretical $N_{\rm a}(v)$ profiles
constructed using the STIS E140M line-spread function (using the LSF
for the $0\farcs 06 \times 0\farcs 2$ slit).  The E140M mode has
excellent spectral resolution, and consequently, relatively narrow
lines are resolved with this spectrograph. The dashed line in
Figure~\ref{navdemo} shows a marginally resolved, hypothetical line
with $b = 4$ km s$^{-1}$; the solid line shows a thermally broadened
\ion{O}{6} profile at $T = 3.0 \times 10^{5}$ K [i.e., close to the
temperature where \ion{O}{6} peaks in abundance in collisional
ionization equilibrium (CIE)]. We can see that at this gas
temperature, an \ion{O}{6} line is well-resolved, and the {\it
apparent} column density profile is therefore a close approximation of
the the {\it true} column density profile. The dotted line in
Figure~\ref{navdemo} shows the $N_{\rm a}(v)$ profile of a thermally
broadened \ion{H}{1} line, also at $T = 3.0 \times 10^{5}$ K,
normalized to have the same peak value as the \ion{O}{6}
line. Comparison of the \ion{O}{6} and \ion{H}{1} profiles in
Figure~\ref{navdemo} makes an important point: because its mass is 16
times smaller, \ion{H}{1} profiles should be four times broader than
\ion{O}{6} lines arising in the same gas if the lines are
predominantly thermally broadened.  At the resolution provided by our
STIS data, detection of this effect is, in principle, trivial at the
temperatures expected for the WHIM gas.  As we shall see, \ion{H}{1}
profiles that are significantly broader than the corresponding
\ion{O}{6} profiles (like those shown in Figure~\ref{navdemo}) are
relatively rare in the real data of our sample.  We note that the
sensitivity of our STIS data is sufficient to detect \ion{H}{1} with
good significance at the temperatures where \ion{O}{6} peaks in CIE
(see, e.g., Richter et al. 2006a,b).  However, there are some
important caveats.  As we will show in \S \ref{hiddenblasec}, if the
temperature range is tuned to maximize the \ion{O}{6}/\ion{H}{1} ratio
or the absorber metallicity is intrinsically high, then a broad
\ion{H}{1} component corresponding with hot \ion{O}{6} could be quite
difficult to detect.  Also, in complex multicomponent systems, broad
Ly$\alpha$ components are often easily hidden (\S \ref{complexmulti}).

The observed $N_{\rm a}(v)$ profiles derived from the \ion{O}{6} and
\ion{H}{1} lines for all of the absorbers (intervening and proximate)
are shown in Figure~\ref{navplots}.  The profiles are ordered
alphabetically by sight line name and presented sequentially by
redshift for a given sight line. Following the sight line name, the
codes ``I'' and ``P'' indicate whether the absorber is an intervening
or proximate system, and ``S'' and ``C'' represent the simple and
complex classifications.  The black histograms show the \ion{O}{6}
profiles and the large-dot histograms represent the \ion{H}{1}
profiles.  The \ion{H}{1} profiles have been scaled to show their
correspondence with the \ion{O}{6} data, and the \ion{H}{1} scale
factor is indicated in each panel of Figure~\ref{navplots}.  In some
cases, a second scaling of the \ion{H}{1} profile is shown with
small-dot histograms.

Inspection of Figure~\ref{navplots} reveals an interesting result:
very few of these absorption systems clearly have \ion{H}{1} lines
that are four times broader than the corresponding \ion{O}{6}
lines. The \ion{H}{1} lines are broader than the \ion{O}{6} lines, as
expected if the lines arise in the same gas, but they are not four
times broader.  This indicates that the absorbers are not
predominantly thermally broadened; nonthermally broadening is
apparently important. Moreover, many of the \ion{H}{1} lines in these
cases are narrower than expected in WHIM plasma. As we will show in \S
\ref{tempsec}, these well-aligned \ion{H}{1} and \ion{O}{6} components
frequently indicate that $T \ll 10^{5}$ K if the lines originate in
the same plasma.  As noted above, 37\% of the intervening \ion{O}{6}
absorbers have \ion{O}{6} and \ion{H}{1} profiles that are
well-aligned in velocity and are classified as single-phase systems.
However, from Figure~\ref{navplots} we see that even in the complex
absorbers there are velocity ranges where the \ion{O}{6} and
\ion{H}{1} profiles are well-aligned.  The good velocity alignment
of the \ion{O}{6} and \ion{H}{1} profiles does not preclude warm-hot
gas; this only indicates that non-thermal broadening is important.
However, as we will show in \S \ref{tempsec}, the implied temperatures
tend to be less than $10^{5}$ K.  Warm-hot gas is still possible in
some of these cases (\S \ref{hiddenblasec}), but this requires that
the warm-hot \ion{O}{6} is usually found in close proximity to cooler,
low-ionization gas.  We shall consider these issues in detail in \S 4.

\subsection{\ion{H}{1}/\ion{O}{6} Column Density Ratios\label{ratiosec}}

\begin{figure}
\centering
    \includegraphics[width=9.0cm, angle=0]{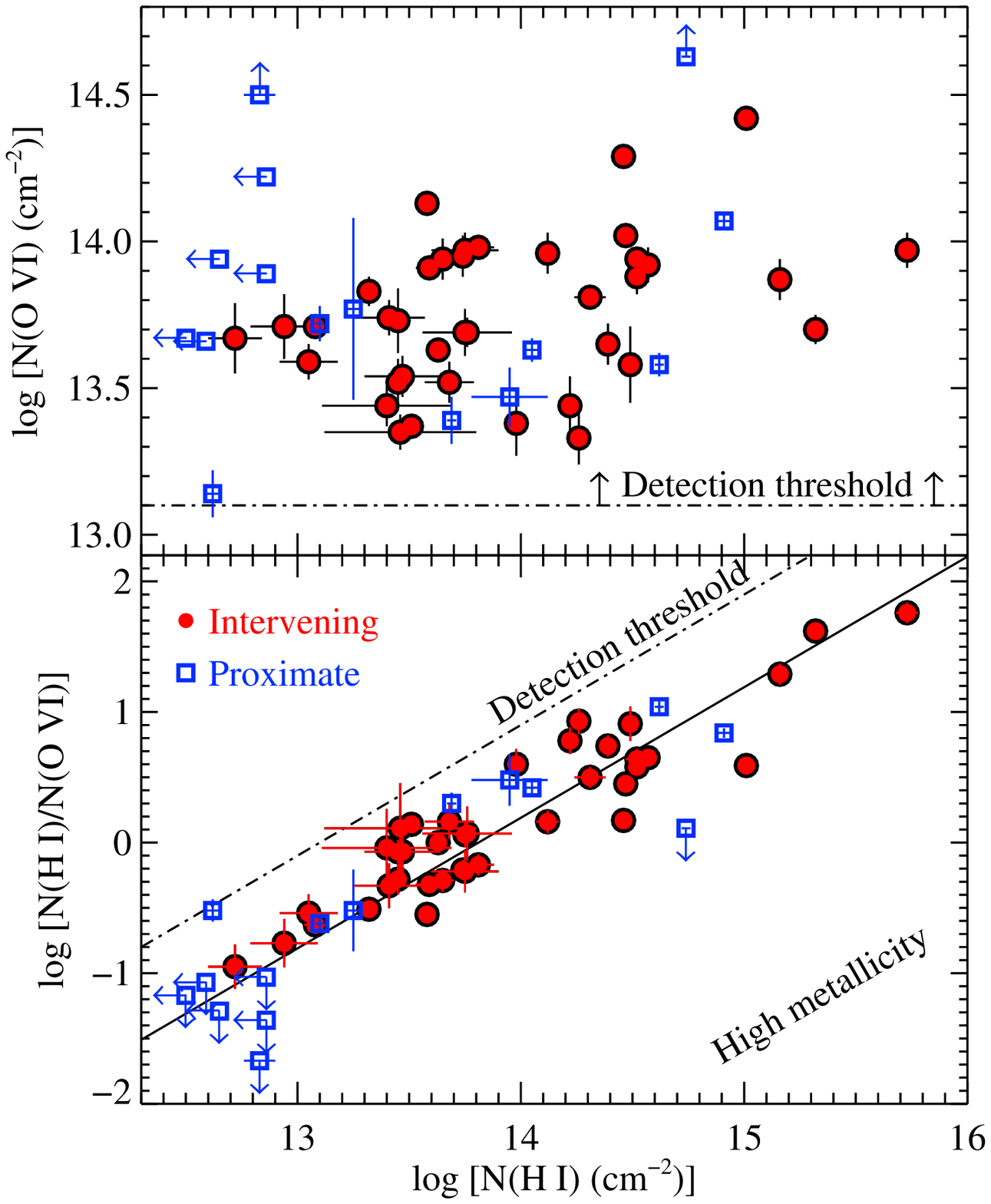}
\caption{{\it Upper panel:} The O~VI vs. H~I column
densities for the O~VI and H~I components that have
velocity centroids that agree within their $2\sigma$ uncertainties
(see \S \ref{aligndefsec}). Intervening systems are shown with filled
circles and proximate absorbers are plotted with open squares with
$1\sigma$ error bars.  For several proximate absorbers, H~I is
not detected, and the plotted point marks the $3\sigma$ upper limit on
$N$(H~I).  In two cases, the O~VI is saturated, and
$N$(O~VI) is indicated as a lower limit. {\it Lower panel:} The
log [H~I/O~VI] column density ratio vs. log
$N$(H~I) for the aligned components shown in the upper
panel. The solid line shows the relation between
$N$(H~I)/$N$(O~VI) and $N$(O~VI) if $N$(O~VI)
is a constant equal to the median of the robust intervening sample
(see Table~\ref{mediantab}). The solid line is plotted for
illustration (see text) and is {\it not} a fit to the data.  The
dash-dot line shows the O~VI detection threshold.  In the lower
panel, points in the upper-left region would have O~VI column
densities below our detection threshold, and points in the lower right
would require high metallicities.  Thus, the correlation of
H~I/O~VI with $N$(H~I) can be largely due to
selection effects.\label{plainh1o6}}
\end{figure}

Finally, we consider the statistics of $N$(\ion{O}{6})
vs. $N$(\ion{H}{1}) and the \ion{H}{1}/\ion{O}{6} column density
ratios in our \ion{O}{6} samples.  These statistics can, in principle,
provide insight about the physical conditions of the absorbers.
However, because these absorbers can have a multiphase nature with
contributions to $N$(\ion{H}{1}) from low-ionization gas as well as
high-ionization phases (see, e.g., \S \ref{absclass}; Chen \&
Prochaska 2000; Tripp et al. 2000; Richter et al. 2004; Sembach et
al. 2004; Tumlinson et al. 2005; Savage et al. 2005), the
\ion{H}{1}/\ion{O}{6} ratio should be interpreted carefully.  The
\ion{H}{1} in the PKS0312-770 absorber shown in
Figures~\ref{speccontrast} and \ref{pks0312multi}, for example, must
include a substantial contribution from the low-ionization gas, and it
is not at all clear how much of the \ion{H}{1} is affiliated with the
\ion{O}{6}-bearing gas.  Comparison of the {\it total} \ion{H}{1} and
\ion{O}{6} column densities can introduce systematic errors because
this can mix in some \ion{H}{1} gas that is known to be unrelated to
the \ion{O}{6} into the measurements.  To overcome this problem, we
strive to identify cases where the data indicate that the \ion{H}{1}
and \ion{O}{6} absorption lines are more likely to arise in the same
gas. To achieve this, we focus in this section on the components which
have well-aligned \ion{O}{6} and \ion{H}{1} absorption components.  As
discussed in the previous section, Figure~\ref{navplots} shows that
many of the \ion{O}{6} and \ion{H}{1} profiles are well-aligned.  This
similarity is not necessarily expected in multiphase entities;
different temperatures, turbulence, and/or bulk kinematics in the
different phases would likely lead to detailed differences in the
profiles in the low-ionization and high-ionization phases.  The
remarkable alignment of some of the components strongly suggests that
in these cases, the \ion{O}{6} and \ion{H}{1} are strongly mixed in a
single gas phase, hence we concentrate on these aligned cases in this
section.  We note that there are other components with clear velocity
offsets between the \ion{O}{6} and \ion{H}{1} (see
Figure~\ref{veloffsets}); we discuss these more complicated cases in
\S \ref{complexmulti}.

\begin{figure}
\centering
    \includegraphics[width=9.0cm, angle=0]{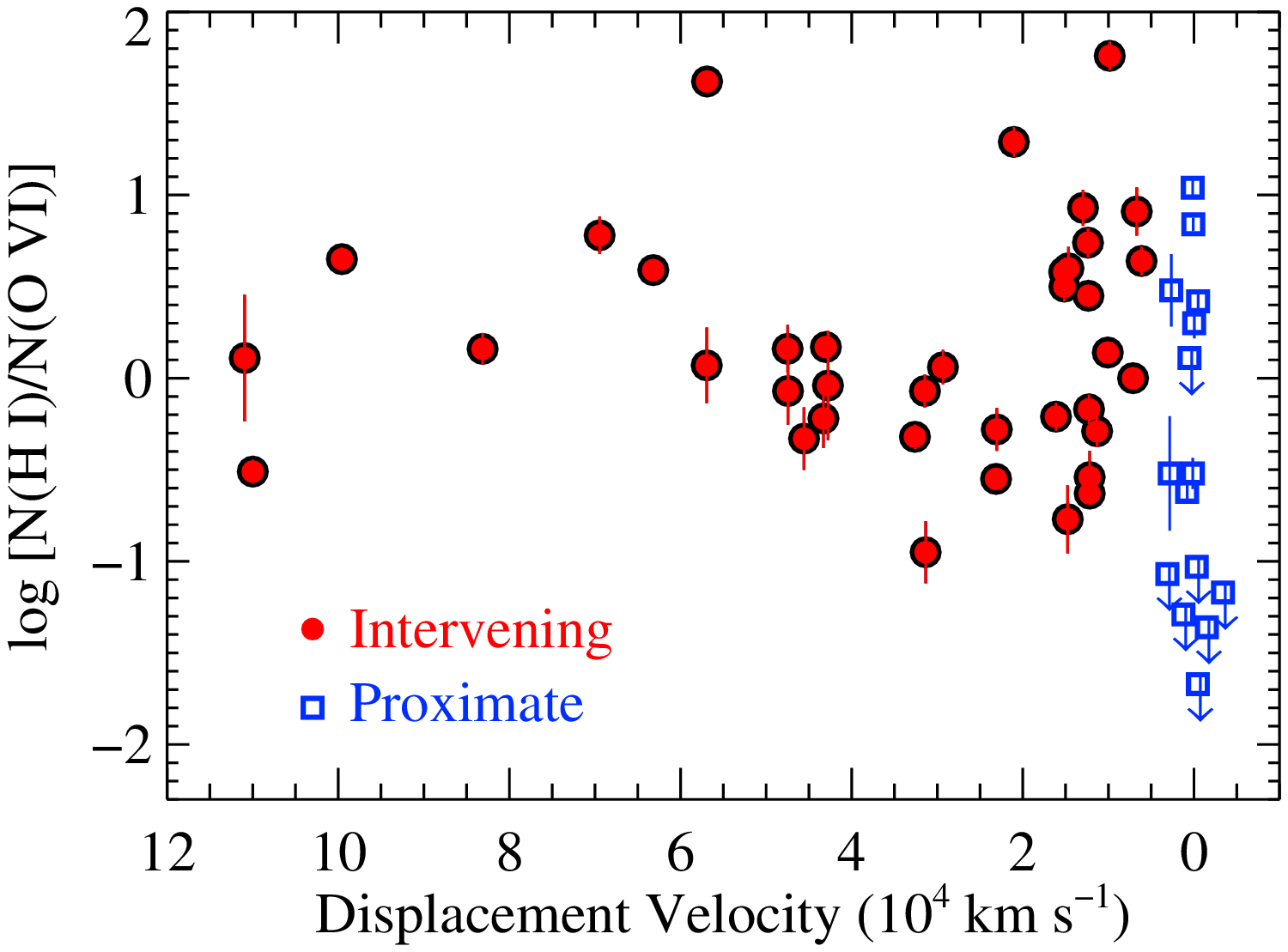}
\caption{The H~I/O~VI ratio vs. velocity of displacement
from the QSO redshift.  This figure uses the well-aligned component
sample that is shown in Figure~\ref{plainh1o6} with the same symbols.
For clarity, some small artificial velocity offsets were applied to
the clump of upper limits on the H~I/O~VI ratio for the
proximate absorbers in the lower-right corner of the
plot.\label{assco6h1}}
\end{figure}

From our robust samples of intervening and proximate absorbers, we
have selected all components whose \ion{O}{6} and \ion{H}{1} velocity
centroids are aligned to within their $2\sigma$ velocity-difference
uncertainties.  We employ these ``well-matched'' \ion{O}{6} +
\ion{H}{1} components for several purposes in the rest of this paper,
and various measurements for the well-matched cases are plotted in
Figures~\ref{plainh1o6} - \ref{hybridmod}. Several useful properties
of the well-aligned intervening components are listed in
Table~\ref{alignedtab} including constraints on the plasma temperature
and nonthermal broadening of the lines (see \S \ref{tempsec}),
measured \ion{H}{1}/\ion{O}{6} and \ion{C}{3}/\ion{O}{6}
column-density ratios (including upper limits), and the classification
(simple vs. complex) of the absorption system in which the particular
component is found.\footnote{As can be seen from Table~\ref{compprop},
there are well-aligned components located within complex,
multicomponent absorbers that also show clear offsets in other
components; we include these aligned components in this section and in
Table~\ref{alignedtab}, but we note that they are drawn from complex
absorbers in column 11 of Table~\ref{alignedtab}.}

We show $N$(\ion{O}{6}) vs. $N$(\ion{H}{1}) for the well-matched,
robust sample in the upper frame of Figure~\ref{plainh1o6}.  Applying
a Spearman test to the intervening sample in Figure~\ref{plainh1o6},
we find that the probability of the null hypothesis (i.e., no
correlation) is 2.7\% .  Thus, we find a weak indication that
$N$(\ion{O}{6}) and $N$(\ion{H}{1}) are correlated in the intervening
absorbers. Interpretation of the proximate system data is complicated
by the fact that \ion{H}{1} is not detected at all in several
proximate absorbers (see Figure~\ref{ascsample1}), but Spearman tests
nevertheless indicate that there is no significant evidence for any
correlation of any combination of the proximate system \ion{H}{1} and
\ion{O}{6} column densities shown in the upper panel of
Figure~\ref{plainh1o6}.

In the lower panel of Figure~\ref{plainh1o6} we show the correlation
between log [$N$(\ion{H}{1})/$N$(\ion{O}{6})] and log $N$(\ion{H}{1})
for the robust well-matched sample.  Danforth \& Shull (2005) have
argued that this correlation is an important indicator of the
multiphase nature of low$-z$ \ion{O}{6} systems and that the strong
correlation of $N$(\ion{H}{1})/$N$(\ion{O}{6}) with $N$(\ion{H}{1})
indicates that the \ion{O}{6} arises in collisionally ionized gas.  We
find from a Spearman test applied to the well-aligned sample in
Figure~\ref{plainh1o6} that the \ion{H}{1}/\ion{O}{6} ratio is indeed
highly correlated with \ion{H}{1}, and we find that the slope in
Figure~\ref{plainh1o6} is fully consistent with the slope reported by
Danforth \& Shull (2005).  However, this correlation is automatic at
some level since \ion{H}{1}/\ion{O}{6} is being correlated with
$N$(\ion{H}{1}). The solid line in Figure~\ref{plainh1o6} shows the
correlation that would result from a sample with a single (constant)
\ion{O}{6} column density equal to the median $N$(\ion{O}{6}) that we
measure from our robust intervening sample.  The correlation in the
lower panel of Figure~\ref{plainh1o6} simply shows that
$N$(\ion{O}{6}) is {\it not} strongly correlated with
$N$(\ion{H}{1}). Danforth \& Shull (2005) argue that $N$(\ion{O}{6})
should be directly proportional to $N$(\ion{H}{1}) if the gas is
photoionized.  While $N$(\ion{O}{6}) and $N$(\ion{H}{1}) might be
weakly correlated, Figure~\ref{plainh1o6} clearly shows that
$N$(\ion{O}{6}) is not linearly proportional to $N$(\ion{H}{1}); there
is substantial scatter in the data.  However, the amount of \ion{O}{6}
in an \ion{H}{1} cloud depends on both the metallicity and the
physical conditions of the gas, and it is not clear that
$N$(\ion{O}{6}) should be strictly proportional to $N$(\ion{H}{1}) in
photoionized gas.  We will return to this issue in \S
\ref{photoionsec}, and we will show that the correlation in the lower
panel of Figure~\ref{plainh1o6} can arise naturally in photoionized
gas.

Also, there are some important selection effects that should be borne
in mind when analyzing the lower panel of Figure~\ref{plainh1o6}.
First, if the \ion{O}{6} column density is too low, the line will be
lost in the noise.  In our sample, we detect very few lines with log
$N$(\ion{O}{6}) $<$13.1 (see Figure~\ref{bvsn}), and we take this as
our \ion{O}{6} detection threshold.  The dash-dot line in
Figure~\ref{plainh1o6} shows this detection threshold.  We can see
that this threshold explains the absence of points in the upper-left
region of the lower panel of Figure~\ref{plainh1o6}; absorbers in that
region would have undetectable \ion{O}{6} lines.\footnote{As we have
shown in previous papers (e.g., Savage et al. 2002; Richter et
al. 2004; Sembach et al. 2004; Tripp et al. 2005; Aracil et al. 2006;
Lehner et al. 2006, 2007), there are many Ly$\alpha$
absorbers,including relatively high$-N$(\ion{H}{1}) systems, that show
no corresponding O~VI absorption.  Therefore, it is highly
possible that systems that populate the upper-left portion of the
lower panel of Figure~\ref{plainh1o6} exist but have not yet been
identified due to the limited S/N of the currently available data.} We
can detect absorbers that would populate the lower-right region of
that plot, but any ionization mechanism would require a high
metallicity to place points in that region.  The lack of points in the
lower right likely indicates that the \ion{O}{6} absorbers do not have
such high metallicities.

We noted in \S \ref{absclass} that there is a notable difference
between the intervening and proximate absorbers: some of the
proximate systems show no detectable, affiliated
\ion{H}{1}. Figure~\ref{plainh1o6} shows that the majority of these
\ion{H}{1}-deficient proximate systems are located in a part of the
$N$(\ion{O}{6}) - $N$(\ion{H}{1}) parameter space where no intervening
absorbers are found.  More measurements of proximate absorbers are
needed to determine if low \ion{H}{1}/\ion{O}{6} ratios are a
distinguishing signature of intrinsic absorbers, but the lack of
analogous low \ion{H}{1}/\ion{O}{6} components in the intervening
absorbers suggests that there is very little contamination of the
intervening sample with intrinsic absorbers that have been accelerated
to high ejection velocities.  Figure~\ref{assco6h1} shows the
\ion{H}{1}/\ion{O}{6} ratio of the matched intervening and proximate
absorbers as a function of displacement velocity.  We see that roughly
one third of the of proximate systems have log
[$N$(\ion{H}{1})/$N$(\ion{O}{6})] $< -1$.  If, for example, up to a
third of the intervening systems are high-velocity ejected systems (as
suggested by some papers, see \S \ref{absclass}), and one third of
those ejected cases have log [$N$(\ion{H}{1})/$N$(\ion{O}{6})] $< -1$,
then we would expect to find $4 - 5$ absorbers in the intervening
sample with similarly low \ion{H}{1}/\ion{O}{6} ratios.  This number
of low \ion{H}{1}/\ion{O}{6} systems is clearly not present in the
matched intervening sample (see Figure~\ref{assco6h1}).  It remains
possible that some of the intervening systems that are classified as
multiphase absorbers because of substantial differences between the
\ion{H}{1} and \ion{O}{6} profile shapes are the high-velocity ejected
systems, but at this juncture, we do not find any clear evidence that
ejected, intrinsic absorbers are a significant source of contamination
of the intervening \ion{O}{6} sample defined by $v_{\rm displ} > 5000$
km s$^{-1}$.

\section{Physical Conditions\label{secphyscon}}

Next we consider the physical conditions implied by our \ion{O}{6} and
\ion{H}{1} measurements.  We first derive quantitative constraints on
the gas temperatures in the components with well-matched \ion{O}{6}
and \ion{H}{1} components (\S \ref{tempsec}), and then we discuss the
implications of the well-aligned absorber properties regarding the
process by which the gas is ionized (\S \S \ref{collionsec},
\ref{photoionsec}, \ref{hybridsec}).  We will find that the simplest
models favor an origin in cool, photoionized gas for a substantial
fraction of the absorbers.  However, in \S \ref{complexmulti} we point
out some of the complexities encountered in multiphase systems and the
implications regarding hot gas in such absorbers, and we examine
whether the broad Ly$\alpha$ components associated with hot \ion{O}{6}
could be hidden in the noise (\S \ref{hiddenblasec}). In this section,
we will focus primarily on the intervening absorbers, which are more
pertinent to the primary goals of our survey, and we exclude the more
uncertain measurements that are flagged with a colon in
Table~\ref{compprop}.

\subsection{\ion{O}{6} Systems with Well-Aligned \ion{H}{1} Components}

\subsubsection{Temperature Constraints\label{tempsec}}

The line width of an individual absorption component provides a
fundamental constraint on the temperature of the absorbing gas.  If
the line width is dominated by thermal motions, then the $b-$value is
directly related to the gas temperature $T$ and the mass $m$ of the
species,
\begin{equation}
T = \frac{mb^{2}}{2k} = A\left( \frac{b}{0.129} \right) ^{2} ,\label{tvsb}
\end{equation}
where $A$ is the atomic mass number and the numerical coefficient
assumes that $b$ is in km s$^{-1}$ and $T$ is in K.  In many
situations, however, other factors will contribute to the line
broadening such as turbulence or multiple unresolved components. If
other factors are likely to contribute to the line broadening, then
equation \ref{tvsb} can still provide a useful upper limit on $T$.
However, if absorption lines from two or more species with
significantly different masses are available (and are believed to
arise in the same gas), then $b$ can be expresssed as a combination of
a thermal broadening term and a nonthermal broadening term $b_{\rm nt}$,
\begin{equation}
b^{2} = (0.129)^{2} \frac{T}{A} + b_{\rm nt}^{2} , \label{fulltvsb}
\end{equation}
which can be solved for $T$ and $b_{\rm nt}$, assuming the nonthermal
motions are adequately characterized by a Gaussian profile.  This
equation can provide valuable insights about the nature of the
absorbers.

\begin{figure*}
\centering
\includegraphics[width=17.0cm, angle=0]{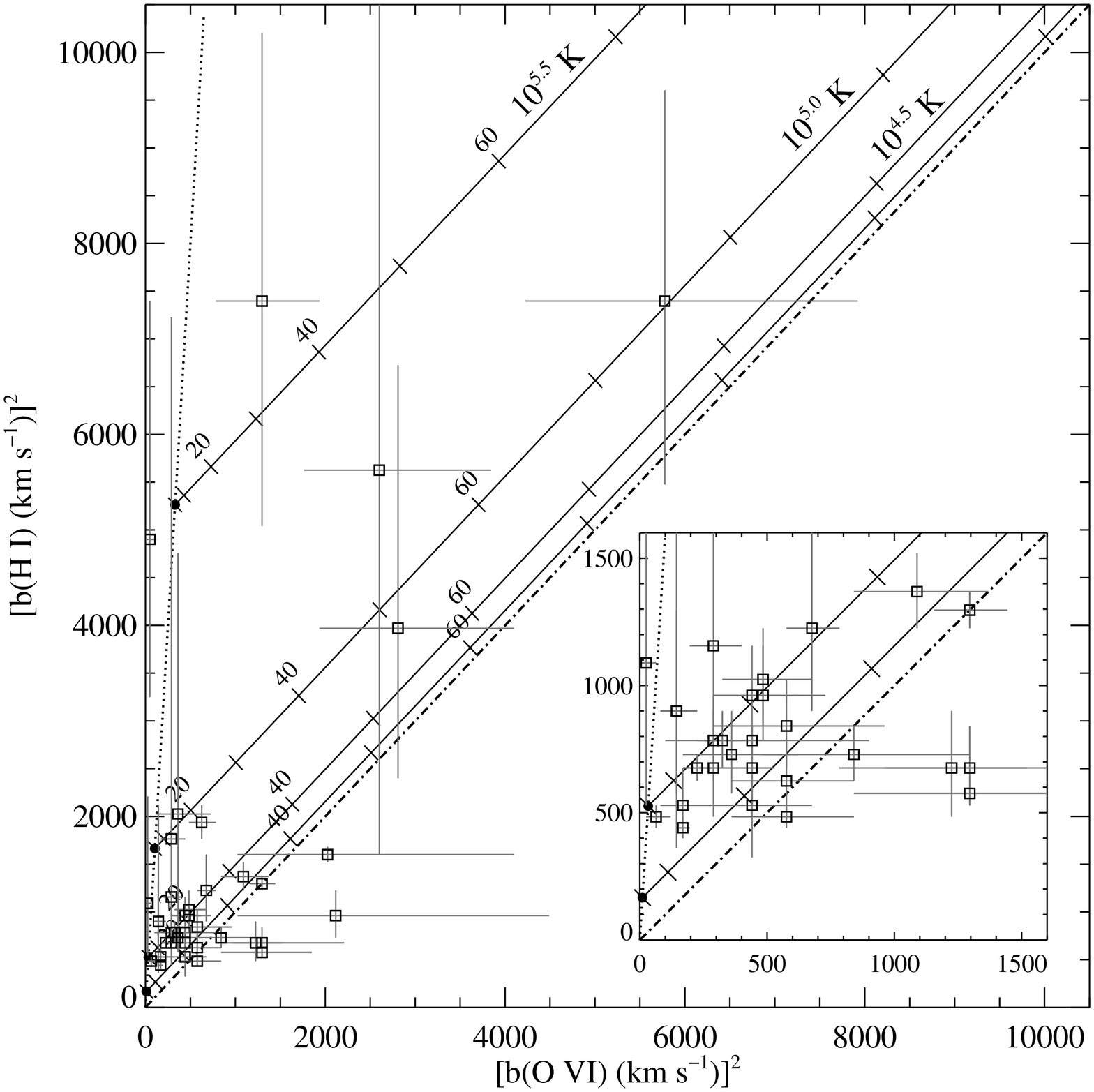}
\caption{Squared $b-$values of H~I vs. O~VI for the
well-aligned components (\S \ref{aligndefsec}). The measured points
are shown by open squares with 1$\sigma$ error bars.  For most of
these aligned components, $b$(H~I) $> \ b$(O~VI), as
expected if the H~I and O~VI are mixed together and arise
in the same gas phase [the dash-dot line indicates the $b$(H~I)
= $b$(O~VI) line of equality].  The dotted line shows the
$b$(H~I) = $4b$(O~VI) locus; this is the relation for
purely thermally broadened lines and represents the other boundary of
physically possible measurements.  Some of the cases that are outside
of the physically-allowed region can be attributed to measurement
errors, but we show in \S \ref{complexmulti} that some of these cases
could be due to multiphase media in which the O~VI is close in
velocity to narrow and broad H~I components, and our algorithm
has incorrectly assigned the O~VI to the narrow H~I
line. The solid lines show loci of constant-temperature for log $T$ =
4.0, 4.5, 5.0, and 5.5 (from bottom to top, respectively), and the
tick marks on each solid line indicate the values of $b_{\rm nt}$ that
produce the $b$(H~I), $b$(O~VI) combination at the given
temperature.  The $b_{\rm nt}$ tick marks increase in increments of 10
km s$^{-1}$ from left to right. The $b_{\rm nt} = 0$ km s$^{-2}$
points are indicated with a filled circle, and the $b_{\rm nt}$ = 20,
40, and 60 km s$^{-1}$ ticks are labeled on each constant-$T$
line. The dense cluster of measurements in the lower left indicates
that the aligned components are relatively cold with $b_{\rm nt}
\lesssim 30$ km s$^{-1}$. To show this cluster more clearly, the inset
shows the same plot over a smaller $b^{2}$ range.\label{bsqplot}.}
\end{figure*}

One of the main goals of this survey is to search for evidence of the
shock-heated warm-hot intergalactic gas in order to understand the
inventory of baryons in the universe (\S 1).  In most
theoretical studies, the WHIM is defined to have $T \geq 10^{5}$ K
(but see Kang et al. 2005 for evidence that a substantial portion of
the WHIM could be cooler).  For \ion{O}{6}, $T \geq 10^{5}$ K implies
that $b$(\ion{O}{6}) $\geq 10$ km s$^{-1}$. From Table~\ref{compprop}
and Figure~\ref{bvsn}, we see that the majority of the intervening
\ion{O}{6} absorption lines have sufficiently large $b-$values to be
consistent with a WHIM origin.  This is a necessary but insufficient
requirement to establish that these absorbers originate in the WHIM,
however, because if $b_{\rm nt}$ is substantial, $T \ll 10^{5}$
remains possible.  Much more stringent constraints can be derived from
\ion{H}{1}, which has $b$(\ion{H}{1}) $\geq$ 41 km s$^{-1}$ if $T \geq
10^{5}$ K.

Therefore, we next consider the temperatures implied by the
well-matched sample of \ion{H}{1} and \ion{O}{6} components that are
well-aligned in velocity.\footnote{In principle, this
technique can be applied to other combinations such as C~III and
O~VI.  However, these combinations usually are not particularly
useful because of the similar masses of elements like C and O combined
with the measurement uncertainties in the $b-$values.}  The good
correspondence of many of the aligned \ion{O}{6} and \ion{H}{1}
profiles (see Figs.~\ref{intsample1} and \ref{navplots}) suggests that
these \ion{O}{6} and \ion{H}{1} lines arise in the same gas, and
therefore we can employ the two versions of equation~\ref{fulltvsb}
for \ion{H}{1} and \ion{O}{6} to solve for $T$ and $b_{\rm nt}$.
Using the \ion{O}{6} and \ion{H}{1} $b-$values for the aligned
components and equations \ref{tvsb} and \ref{fulltvsb}, we obtain the
gas temperature constraints listed in Table~\ref{alignedtab}.  Columns
4 and 5 in Table~\ref{alignedtab} list the constraints implied from
the \ion{O}{6} $b-$values only (eqn.\ref{tvsb}); column 4 gives the
temperature implied by the best value of $b$(\ion{O}{6}) and column 5
lists the $3\sigma$ upper limit implied by the uncertainties in
$b$(\ion{O}{6}).  We treat these temperature constraints as upper
limits because nonthermal broadening could be a significant source of
line broadening.  Columns 6 and 7 list the temperatures found by
solving equation~\ref{fulltvsb} with $b$(\ion{H}{1}) and
$b$(\ion{O}{6}); column 6 reports the best temperature and column 7
provides the $3\sigma$ upper limit found by propagating the $b-$value
uncertainties through equation~\ref{fulltvsb}.

\begin{figure}
\centering \includegraphics[width=8.0cm, angle=0]{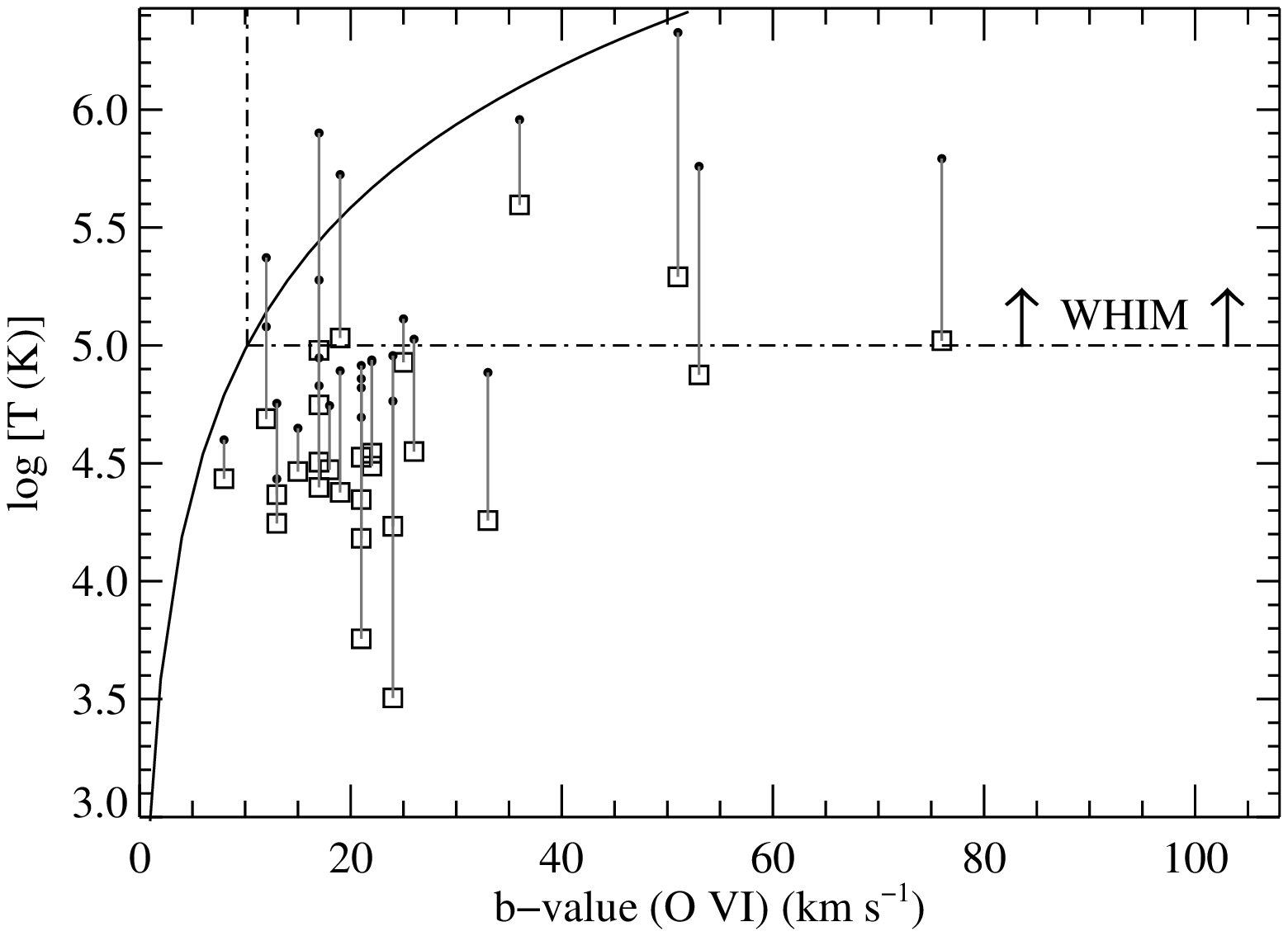}
\caption{Plasma temperatures implied by the H~I and O~VI
$b-$values measured for well-aligned intervening components.  Squares
indicate the temperature implied by the best-fit O~VI and
H~I $b-$values (solving eqn.~\ref{fulltvsb}), and the small
dots connected to the squares by a vertical line show the $3\sigma$
upper limits on the temperature from propagation of the $b-$value
uncertainties through eqn.~\ref{fulltvsb}.  The solid line indicates
the temperature implied by the O~VI $b-$value if the line is
predominantly thermally broadened (eqn.~\ref{tvsb}). The region above
and to the right of the dash-dot line is consistent with the canonical
WHIM temperature ($T \geq 10^{5}$ K). \label{temperatureplot}}
\end{figure}

\begin{figure}
\centering
    \includegraphics[width=8.0cm, angle=0]{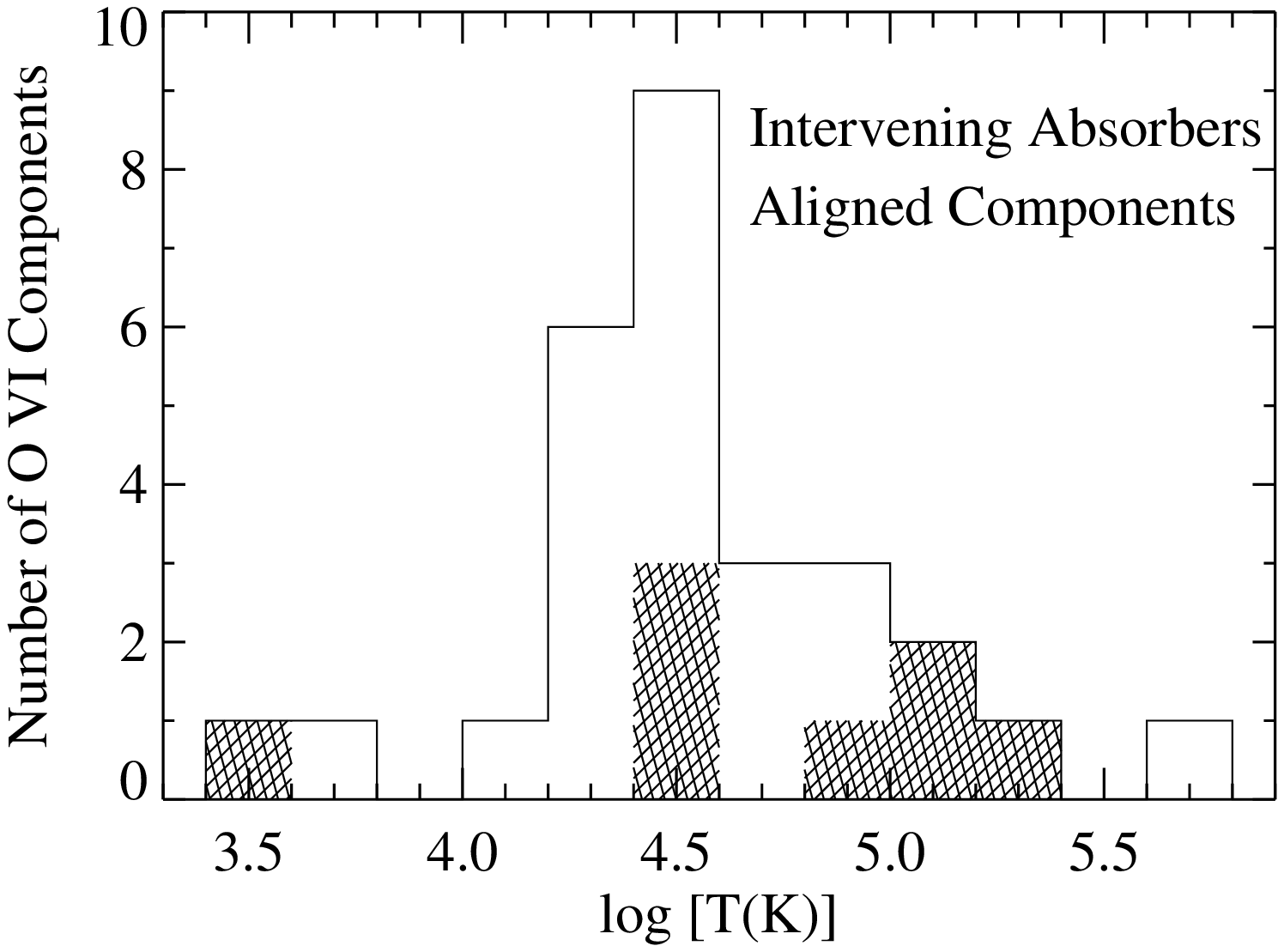}
\caption{Distribution of temperatures determined for the aligned
intervening O~VI components in 0.2 dex bins. The temperatures
were found by solving equation~\ref{fulltvsb} with the measured
H~I and O~VI $b-$values. The temperatures and amount of
nonthermal broadening for individual components are summarized in
Table~\ref{alignedtab}. The open histogram indicates the total number
of O~VI components in a bin regardless of its classification;
the hatched histogram shows the components in complex/multiphase
systems only.\label{tempdist}}
\end{figure}

From equation~\ref{fulltvsb}, we see that at a constant temperature,
the relationship between $b$(\ion{H}{1})$^{2}$ and
$b$(\ion{O}{6})$^{2}$ is a straight line with unity slope.  Therefore,
by plotting $b$(\ion{H}{1})$^{2}$ vs. $b$(\ion{O}{6})$^{2}$, we can
visually show where an \ion{H}{1} + \ion{O}{6} absorber is located in
($T,b_{\rm nt}$) space. We show this visual representation of the
$b-$values of the well-aligned absorbers in Figure~\ref{bsqplot}.  The
advantage of this presentation is that the combinations of $T$ and
$b_{\rm nt}$ can be quickly assessed by inspection.  In addition,
Figure~\ref{bsqplot} shows whether the lines are predominantly
thermally broadened (or not) and whether the assertion that the
\ion{H}{1} and \ion{O}{6} are mixed together is physically
reasonable. There are two physical limits on allowed $b$(\ion{H}{1}),
$b$(\ion{O}{6}) combinations if the \ion{H}{1} and \ion{O}{6} are
mixed together.  The dash-dot line shows the physical limit of
$b$(\ion{H}{1}) = $b$(\ion{O}{6}); $b$(\ion{H}{1}) $< b$(\ion{O}{6})
is unphysical but can arise from noise in the $b-$value measurements
if the lines are mainly broadened by nonthermal motions and/or have
large $b-$value uncertainties.  The dotted line indicates the other
physical limit, $b$(\ion{H}{1}) = $4b$(\ion{O}{6}), which occurs if
the lines are entirely thermally broadened.  The solid lines are loci
of constant temperature with log $T$ = 4.0, 4.5, 5.0, and 5.5 (from
bottom to top, respectively).  On each constant$-T$ locus, the tick
marks indicate values of $b_{\rm nt}$ starting from $b_{\rm nt}$ = 0
km s$^{-1}$ and increasing in increments of 10 km s$^{-1}$ from left
to right (the $b_{\rm nt}$ = 20, 40, and 60 km s$^{-1}$ ticks are
labeled).

Several aspects of Figure~\ref{bsqplot} deserve comment.  First, most
of the well-aligned intervening components are {\it not} predominantly
thermally broadened.  Relatively few of the measurements are close to
the $b$(\ion{H}{1}) = $4b$(\ion{O}{6}) dotted-line locus expected for
predominantly thermally broadened absorbers.  Second, many of the
points indicate temperatures well below the canonical 10$^{5}$ K lower
limit for WHIM plasma (see further comments below).  Third, some
points fall outside of the physically allowed regions.  While some of
these apparently unphysical combinations could be artifacts due to
noise, we note that many of these cases have complex multicomponent
structure, and this can lead to ambiguity in the assignment of
\ion{O}{6} components with \ion{H}{1} components.  It remains possible
that broad \ion{H}{1} components are present in these absorbers, and
the \ion{O}{6} has been inadvertently aligned with a narrow \ion{H}{1}
line that happens to have a similar velocity. We discuss this
possibility in \S \ref{complexmulti}.  In most of the aligned
absorption lines, the \ion{H}{1} $-$ \ion{O}{6} matching is clear and
unambiguous (see examples shown in Figure~\ref{intsample1}), but a
broad \ion{H}{1} component could still be hidden in the noise (see \S
\ref{hiddenblasec}).

\begin{figure}
\centering
    \includegraphics[width=8.0cm, angle=0]{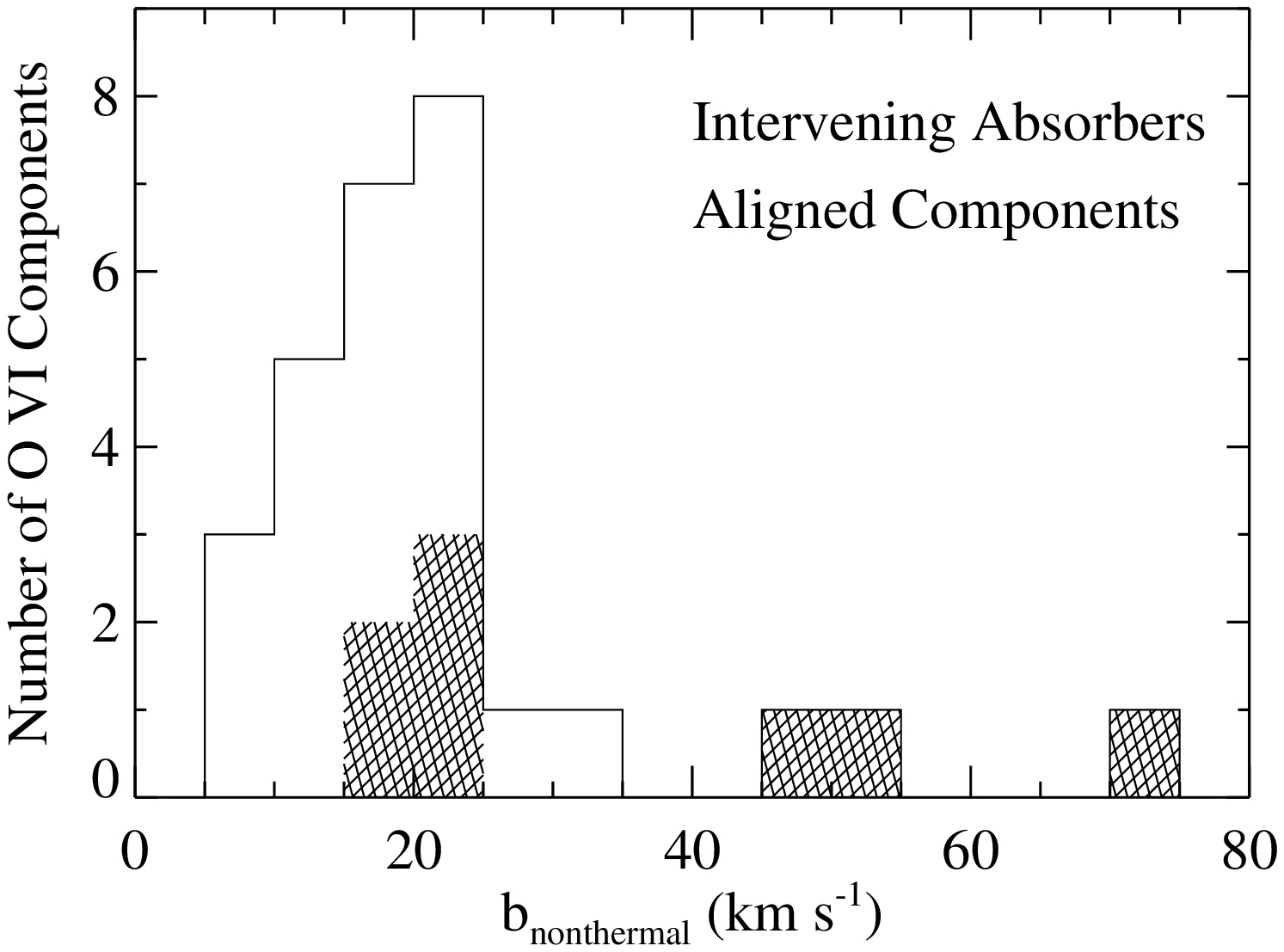}
\caption{Distribution of nonthermal broadening parameters (see
equation~\ref{fulltvsb}) for the well-aligned intervening components
in 5 km s$^{-1}$ bins (see Table~\ref{alignedtab}).  As in
Figure~\ref{tempdist}, all of the aligned components from intervening
systems are shown with an open histogram.  The hatched histogram
represents components from complex/multiphase systems
only.\label{bntdist}}
\end{figure}

Nevertheless, the simplest model is to assume that if the \ion{H}{1}
and \ion{O}{6} line centroids are well-aligned and the profiles have
similar shapes, then the \ion{H}{1} and \ion{O}{6} absorption lines
arise in the same (single-phase) gas cloud.  We adopt this assumption
for the rest of this section. We plot the best temperatures and upper
limits derived from $b$(\ion{H}{1}) and $b$(\ion{O}{6}) (columns 6 and
7 from Table~\ref{alignedtab}) as a function of $b$(\ion{O}{6}) in
Figure~\ref{temperatureplot}, and we show the differential
distributions of these estimations of $T_{\rm best}$ and $b_{\rm nt}$
in Figures~\ref{tempdist} and \ref{bntdist}, respectively.  From these
figures, we see that the quantitative results derived in this section
corroborate the qualitative conclusion reached from comparison of the
\ion{O}{6} and \ion{H}{1} profiles in \S \ref{shapesection}: {\it a
large fraction of the well-matched components appear to be too cold to
arise in WHIM gas} (with the standard $T > 10^{5}$ K WHIM definition).
Sixty-two percent of the $T_{\rm best}$ values in column 6 of
Table~\ref{alignedtab} are below $10^{5}$ K.  The $3\sigma$ upper
limits provide more conservative constraints, but even the
$T_{3\sigma}$ upper limits are inconsistent with WHIM temperatures for
44\% of the aligned components.  Evidently, many of the well-matched
components are not consistent with collisional ionization at the
standard temperatures where \ion{O}{6} is found in CIE, and most of
these aligned components are not consistent with the standard
temperatures expected for WHIM gas.  It is possible that most of the
WHIM gas is located in the multiphase absorbers (as we discuss below),
but in the simple absorbers we find strong evidence that a substantial
fraction of these systems are either photoionized or arise in
collisionally ionized gas that is out of equilibrium.

It is interesting to examine the temperatures in
Figures~\ref{temperatureplot} and \ref{tempdist} more closely.  While
most of aligned components imply temperatures lower than log $T$ =
5.5, ten of the aligned components (26\% ) in Table~\ref{alignedtab}
imply log $T \geq$ 4.6, which is warmer than the temperatures expected
in standard photoionization models (see, e.g., the photoionization
models reported in Sembach et al. 2004 and Lehner et al. 2006).  These
moderately warm systems could arise in shock-heated gas that has
cooled more rapidly than it can recombine and thus is in an
overionized, moderately warm condition.  It is also interesting to
note that aligned components that are found within multiphase systems
(indicated with hatching in Figures~\ref{tempdist} and \ref{bntdist})
tend to indicate higher temperatures and greater amounts of nonthermal
broadening.  Multiphase absorption could be more likely to occur in
gas accreting into deeper potential wells because such locations are
likely sites for shock heating but could also be places where more gas
cools quickly due to processes such as thermal instability (e.g.,
Maller \& Bullock 2004). Motivated by this possibility, we consider
whether available non-equilibrium collisional ionization models agree
with these observations in the next section (\S \ref{collionsec})

We have commented above on the fractions of the {\it aligned}
components that are relatively cool. While we have not placed
constraints on the temperatures of the non-aligned \ion{O}{6}
components (i.e., those with significant velocity offsets between
\ion{O}{6} and \ion{H}{1}), we can nevertheless place a lower limit on
the fraction of the \ion{O}{6} components that are relatively
cold. There are 70 components in the robust intervening sample.  From
Table~\ref{alignedtab}, we see that there are 24 components with log
$T_{\rm best} < 5.0$ in the simplest model.  This indicates that at
least 34\% of the intervening components are cooler than the standard
$10^{5}$ K WHIM cutoff.  Conversely, at least 14\% of the robust
sample of intervening components has log $T_{\rm best} > 4.6$ (hotter
than expected in typical photoionization models).

\subsubsection{Collisional Ionization Models\label{collionsec}}

We have presented strong evidence that many of the low$-z$ \ion{O}{6}
lines arise in cool gas.  For many of the \ion{O}{6} components, the
temperature constraints in Table~\ref{alignedtab} preclude a WHIM
origin at the collisional ionization equilibrium temperatures where
\ion{O}{6} is expected to peak in abundance (e.g., Sutherland \&
Dopita 1993). But could the absorption lines arise in collisionally
ionized gas that is out of equilibrium?  

In the context of the Milky Way ISM, the possibility that high ions
could be detectable in non-equilibrium, overionized gas has long been
considered (e.g., Kafatos 1973; Shapiro \& Moore 1976; Edgar \&
Chevalier 1986; Indebetouw \& Shull 2004; Gnat \& Sternberg 2007).  If
shock-heated, highly ionized gas can cool more rapidly than it
recombines, then species like \ion{O}{6} might be found in
surprisingly cool material. ISM densities are substantially higher
than IGM densities though, and in the context of the IGM, it is
sometimes argued that this scenario is unlikely because the cooling
timescale, $\tau _{\rm cool} = 3kT/2n\Lambda(T)$, is quite long when
the cooling function $\Lambda(T)$ for low-metallicity material (e.g.,
Sutherland \& Dopita 1993; Gnat \& Sternberg 2007) is combined with
the theoretically expected low densities of IGM gas.  If we use the
parameters typically adopted for $\tau _{\rm cool}$ calculations
(e.g., the gas is initially heated to $T > 10^{6}$ K, $n < 10^{-5}$
cm$^{-3}$, $Z < 0.1 Z_{\odot}$), we do find that it will not cool in a
Hubble time (e.g., Dopita \& Sutherland 2003; Shull et al. 2003).
However, there are several factors that could make $\tau _{\rm cool}$
shorter: First, in low$-z$ IGM clouds where several metals are
detected, studies have indicated metallicities significantly greater
than 0.1 $Z_{\odot}$ (e.g., Savage et al. 2002; Jenkins et al. 2005;
Tumlinson et al. 2005; Aracil et al. 2006; Lehner et al. 2006).
Higher metallicity increases $\Lambda(T)$ (see Figure 4 in Gnat \&
Sternberg 2007) and thus decreases $\tau _{\rm cool}$. On the other
hand, similar studies of low$-z$ systems with well-constrained
metallicities have indicated $Z < 0.1 Z_{\odot}$ (e.g., Tripp et
al. 2002,2005).  There appears to be a very large range in the
metallicity of the low$-$ IGM, and it is not clear what metallicity
should be assumed for a random IGM cloud, but it has been shown that
some clouds are significantly enriched and therefore could cool more
quickly.  Second, if the WHIM is heated by lower-velocity shocks, the
initial temperature might be lower than usually assumed, which
increases $\Lambda(T)$.  If the initial temperature is only $10^{5.5}$
K, for example, the gas will cool more rapidly (see Figure 4 in Gnat
\& Sternberg).  Thirdly, the gas densities could be somewhat higher
than $10^{-5}$ cm$^{-3}$.  For the well-aligned sample, log
$N$(\ion{H}{1}) ranges from $\approx$ 13.3 to 14.5 (see
Figure~\ref{plainh1o6}). The \ion{H}{1} number density $n$(\ion{H}{1})
= $3.2 \times 10^{-25} N$(\ion{H}{1})/$L$, where $L$ is the size of
the cloud in Mpc.  At log $T$ = 5.5, the \ion{H}{1} ion fraction is
$10^{-5.9}$ (Sutherland \& Dopita 1993; Gnat \& Sternberg 2007), so if
$L$ = 100 kpc and $T = 10^{5.5}$ K, the total H density in the
well-aligned absorbers ranges from $\approx 5 \times 10^{-5}$
cm$^{-3}$ to $\approx 8 \times 10^{-5}$ cm$^{-3}$. The combination of
higher densities, higher metallicities, and lower initial temperatures
than usually assumed can lead to cooling times that are significantly
shorter than the Hubble time.

The timescale to approach ionization equilibrium is approximately
$\tau _{\rm ioneq} = [\tau _{\rm ion}^{-1} + \tau _{\rm
rec}^{-1}]^{-1}$, where $\tau _{\rm rec}$ is the recombination time
scale and the $\tau _{\rm ion}$ is the ionization
timescale. Considering estimates of $\tau_{\rm ion}$ and $\tau _{\rm
rec}$ for \ion{O}{6}, \ion{O}{7}, and \ion{O}{8} at WHIM temperatures
and densities (see, e.g., Figure 3 in Yoshikawa \& Sasaki 2006), we
see that $\tau _{\rm ioneq} \ll \tau _{\rm Hubble}$, and for many
combinations of IGM physical conditions, $\tau _{\rm ioneq} < \tau
_{\rm cool}$. Therefore CIE could be a good approximation for
\ion{O}{6}, \ion{O}{7}, and \ion{O}{8} in WHIM absorbers. 

\begin{figure}
\centering
    \includegraphics[width=9.0cm, angle=0]{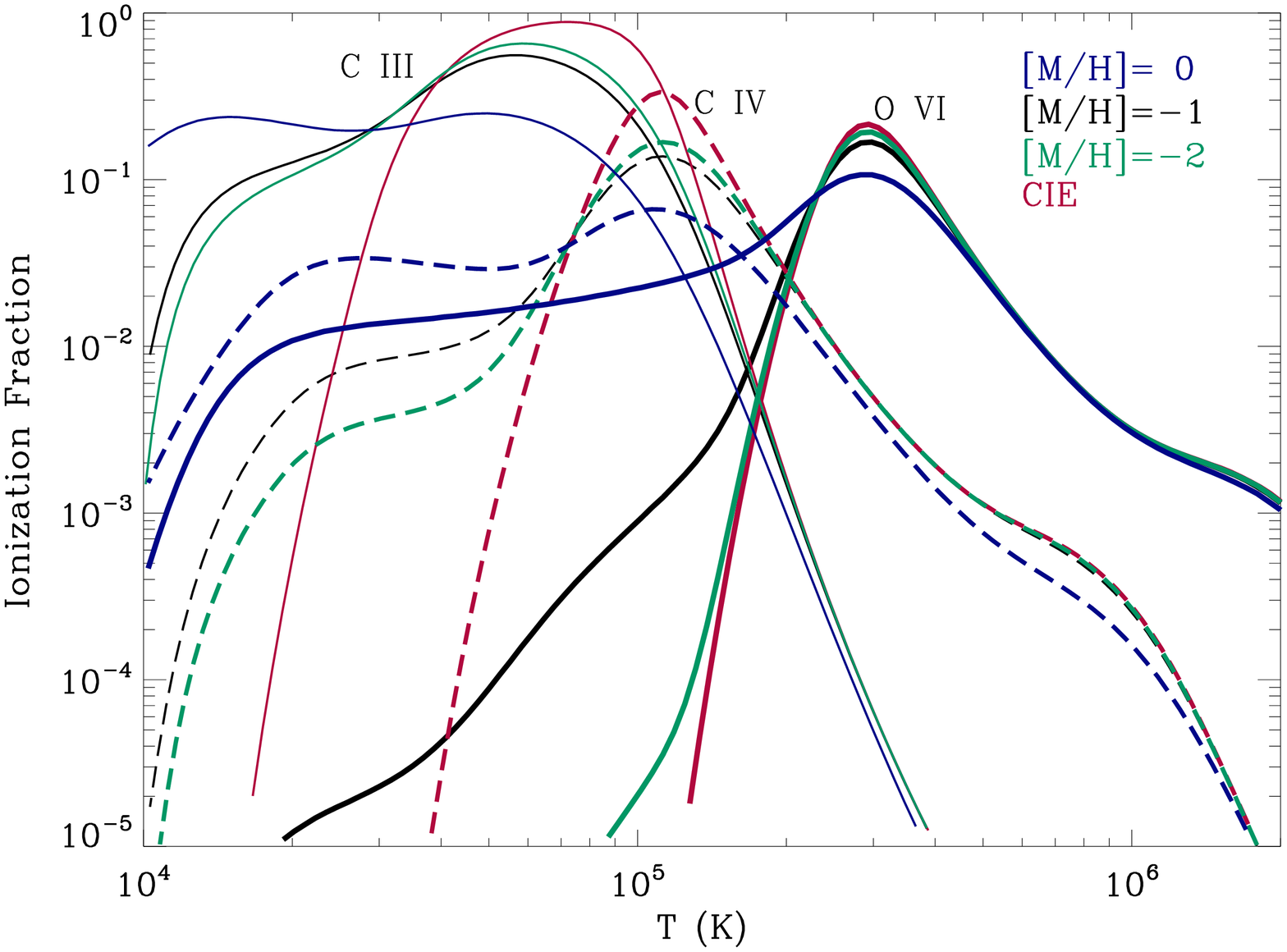}
\caption{Comparison of nonequilibrium ion fractions in isochoric
radiatively cooling gas for O~VI (thick solid lines), C~IV
(dashed lines), and C~III (thin solid lines) calculated by Gnat
\& Sternberg (2007) for logarithmic metallicities of [M/H] = 0 (blue
curves), [M/H] = $-1$ (black curves), and [M/H] = $-2$ (green curves).
The collisional ionization equilibrium (CIE) ion fractions are also
shown with red lines.\label{gnatfracs}}
\end{figure}

\begin{figure}
\centering
    \includegraphics[width=9.0cm, angle=0]{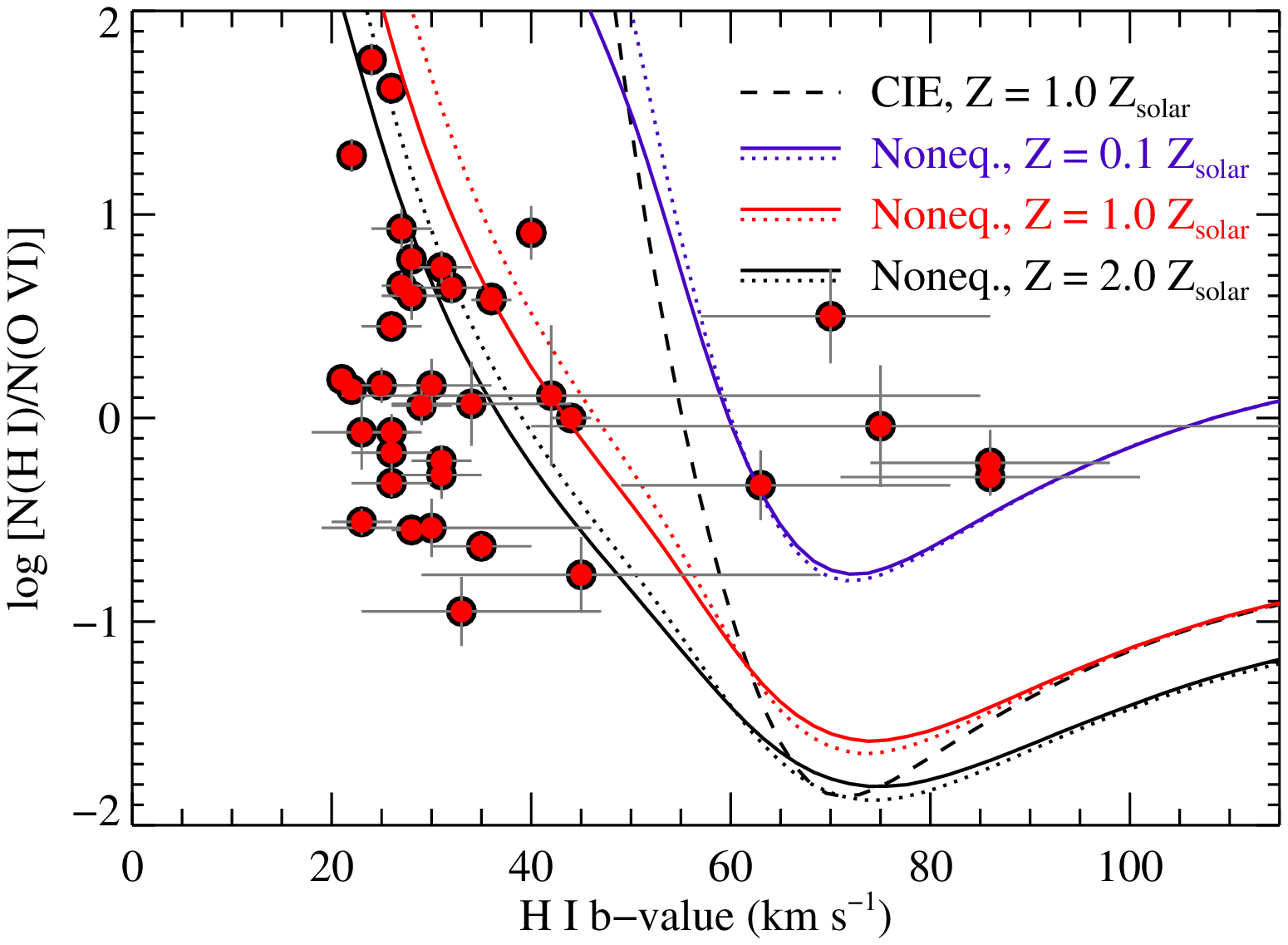}
\caption{Observed H~I/O~VI ratios vs. H~I
$b-$values measured in the intervening absorber components that are
aligned to within their 2$\sigma$ component centroid uncertainties
(filled circles; these are the same data presented in
Figure~\ref{plainh1o6} but only for the intervening systems). The
observations are compared to predictions from the ionization models of
Gnat \& Sternberg (2007) for collisional ionization equilibrium (CIE,
dashed line) and for nonequilibrium, radiatively cooling models for
gas with $Z = 0.1 Z_{\odot}$ (purple lines), $Z = 1.0 Z_{\odot}$ (red
lines), and $Z = 2.0 Z_{\odot}$ (black lines). For each metallicity,
the solid lines show the results for gas cooling isochorically
(constant density), and the dotted lines show the isobaric (constant
pressure) models. The models fail to explain the majority of the
observed measurements.  The models predict ion fractions as a function
of gas temperature; we have converted the model temperatures to
$b-$values assuming pure thermal broadening of the lines.  If
non-thermal motions contribute to the line broadening, the discrepancy
between the models and the observations is even
worse.\label{noneqmod}}
\end{figure}

Since these absorbers could be cooling rapidly, we will now consider
whether the observed properties of the \ion{O}{6} systems could be
consistent with predictions from radiatively cooling, overionized gas
models.  For this purpose, we list in columns 9 and 10 of
Table~\ref{alignedtab} two observables that can be readily compared to
nonequilibrium model predictions: the $N$(\ion{H}{1})/$N$(\ion{O}{6})
and $N$(\ion{C}{3})/$N$(\ion{O}{6}) column density ratios measured in
the well-matched components of intervening absorbers.  In many of our
systems, \ion{C}{3} is not detected, so we provide $3\sigma$ limits
based on the measured upper limit on the \ion{C}{3} rest-frame
equivalent width assuming the linear portion of the curve of growth
applies.  In some cases, \ion{C}{3} is detected but is strong enough
so that it could be underestimated due to saturation (see, e.g.,
Savage et al. 2002); in these cases we treat the measured
$N$(\ion{C}{3}) as a lower limit.

Gnat \& Sternberg (2007) have recently modeled nonequilibrium
radiatively cooling gas, and they provide predictions for these column
density ratios as a function of gas temperature.  They have modeled
both isochoric and isobaric cooling gas, and they provide predictions
for models with a wide range of metallicities. In
Figure~\ref{gnatfracs}, we compare the ion fractions of \ion{C}{3},
\ion{C}{4}, and \ion{O}{6} predicted by Gnat \& Sternberg (2007), for
isochoric radiatively cooling gas, to the ion fractions for these
species from CIE.  This figure demonstrates several important aspects
of nonequilibrium radiatively cooling models when applied to
extragalactic \ion{O}{6} absorbers.  First, nonequilibrium ionization
effects depend on the metallicity of the gas and become increasingly
important as the metallicity increases.  This is expected from the
timescale arguments: the cooling rate depends on metallicity, so when
the metallicity is low, $\tau _{\rm ioneq} < \tau _{\rm cool}$ and CIE
is a good approximation.  However, the metallicity at which
nonequilibrium effects set in depends on the species.  For example, if
the overall metallcity [M/H] = $-2$, then the nonequilibrium
\ion{O}{6} ion fraction is almost identical to the equilibrium ion
fraction (compare the thick green and red lines in
Figure~\ref{gnatfracs}).  In comparison, nonequilibrium effects are
already significant for \ion{C}{3} and \ion{C}{4} at [M/H] = $-2$.
Even at [M/H] = $-1$, the nonequilibrium \ion{O}{6} ion fraction is
quite low at log $T <$ 5.  From Figure~\ref{gnatfracs}, it appears the
nonequilibrium effects will lead to detectable \ion{O}{6} in cool gas
only if the metallicity is relatively high.  Nevertheless, we
inevitably must compare \ion{O}{6} with other species to discriminate
bewteen various ionization models, so nonequilbrium effects on other
species can affect conclusions about \ion{O}{6} systems.

Figure~\ref{noneqmod} shows the observed \ion{H}{1}/\ion{O}{6} ratios
vs. the measured \ion{H}{1} $b-$values for the well-matched components
(filled circles).  These measurements are compared to the predicted
ratios from the Gnat \& Sternberg (2007) isochoric models (solid
lines) and isobaric models (dotted lines) for the three metallicities.
For reference, the dashed line plots the CIE ratio from Gnat \&
Sternberg (2007).  Gnat \& Sternberg predict these ratios as a
function of temperature $T$; we have converted their temperatures to
the corresponding $b-$value that \ion{H}{1} would have at that $T$ for
a purely thermally broadened line.  We see from Figure~\ref{noneqmod}
that none of these models agree with the bulk of the matched
components.  In general, at the temperature implied by the narrowness
of the observed \ion{H}{1} lines, the models predict
\ion{H}{1}/\ion{O}{6} ratios that are too high.  A fraction of the
matched absorbers are consistent with the model predictions, but only
if the metallicity is relatively high.

\begin{figure}
\centering
    \includegraphics[width=9.0cm, angle=0]{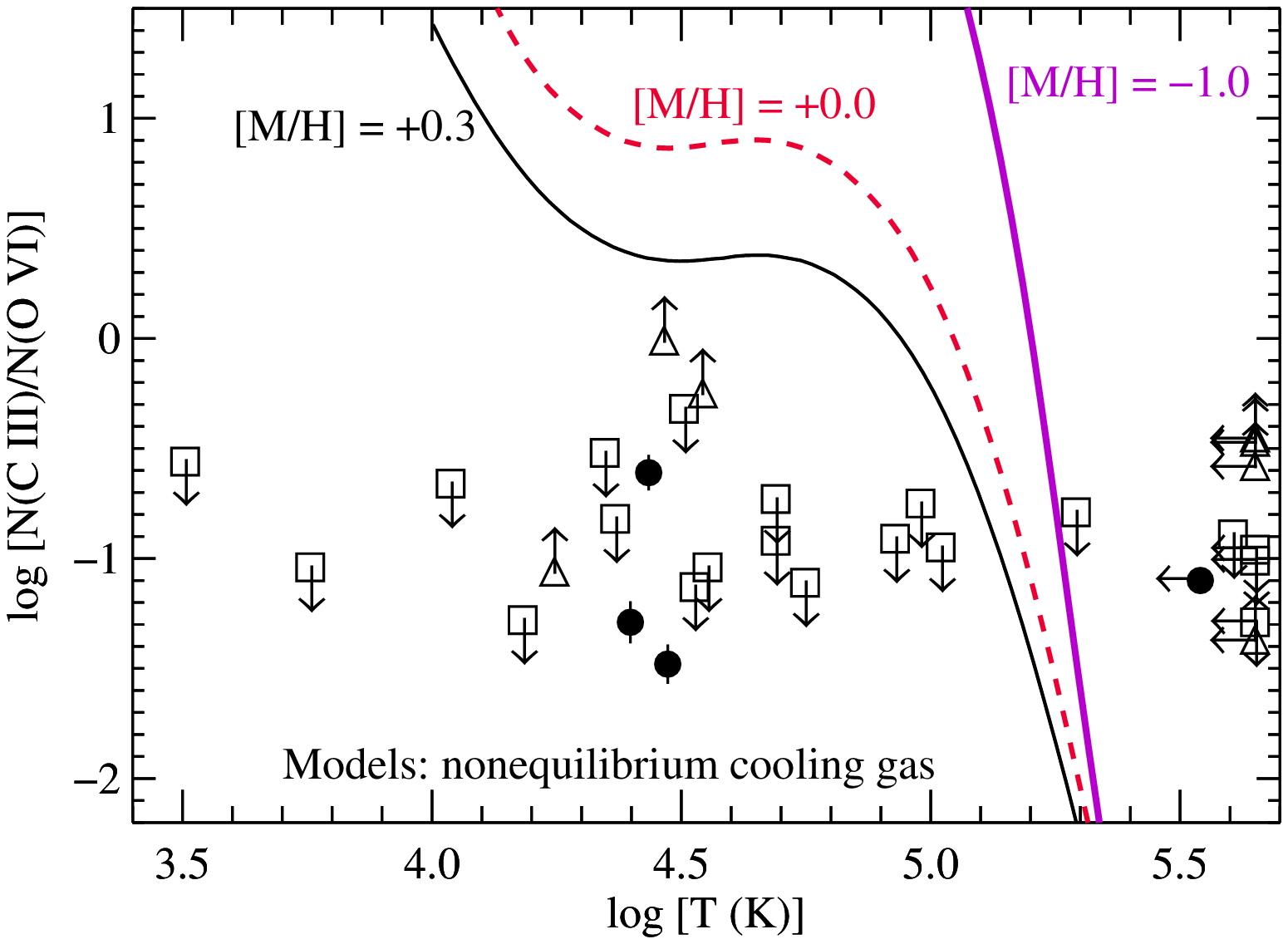}
\caption{Comparison of the observed $N$(C~III)/$N$(O~VI)
ratios for the well-aligned components of intervening systems (as in
Figure~\ref{plainh1o6}) to the predictions for this ratio from the
nonequilibrium ionization models of Gnat \& Sternberg (2007).  The
temperatures for the observed components are the best values derived
from the O~VI and H~I $b-$values as described in \S
\ref{tempsec}; absorbers without temperature constraints are shown at
far right. This figure only shows intervening systems: squares
indicate cases where C~III is not detected and the ratios are
derived from the $3\sigma$ upper limits on $N$(C~III), triangles
indicate lower limits in cases where C~III could be
underestimated due to line saturation, and filled circles indicate
cases with good C~III measurements. The solid curves show the
model predictions for isochoric nonequilibrium cooling gas for
logarithmic metallicities [M/H] = $-1.0$ (purple line), 0.0 (red
line), and 0.3 (black line) from Gnat \& Sternberg (2007).  The
isobaric Gnat \& Sternberg (2007) models fit the observations even
more poorly than the isochoric models.\label{modelc3o6}}
\end{figure}

The Gnat \& Sternberg (2007) models have even greater difficulty with
the measured \ion{C}{3}/\ion{O}{6} ratios and limits.
Figure~\ref{modelc3o6} compares the observed \ion{C}{3}/\ion{O}{6}
ratios and limits to the predictions from the isochoric models of Gnat
\& Sternberg (2007).  In this figure we have retained the temperatures
from the model and we have plotted the measured ratios vs. the
observed temperatures and limits from Table~\ref{alignedtab}.  For the
majority of the matched intervening components, the models predict
\ion{C}{3}/\ion{O}{6} ratios that are one to several orders of
magnitude too high.  The isobaric models are not shown in
Figure~\ref{modelc3o6}, but these models predict even higher
\ion{C}{3}/\ion{O}{6} ratios and thus fit the observed data even more
poorly.

We conclude that the Gnat \& Sternberg (2007) nonequilibrium models
usually fail the match the observed column-density ratios in the
well-aligned \ion{O}{6} absorbers.  One caveat, however, is that Gnat
\& Sternberg (2007) neglect photoionization in their calculations.
Given the low densities expected for IGM absorbers (see above),
photoionization is likely to be important.  The combination of
nonequilibrium collisional ionization plus additional photoionization
from impinging UV light could bring the models into better agreement
with the observations because photoionization could significantly
reduce the \ion{H}{1} and \ion{C}{3} columns without reducing
$N$(\ion{O}{6}) much, especially if the ionizing radiation is
predominantly from stars (which do not emit many photons with
sufficient energy to ionize \ion{O}{5} into \ion{O}{6}).  Of course,
how photoionization affects species like \ion{C}{3} depends on the
physical conditions of the gas. If the ionization parameter is low,
photoionization could boost $N$(\ion{C}{3}) thereby leading to greater
discrepancy with the observations. New nonequilibrium models that
include photoionization would be valuable. We will show in \S
\ref{hybridsec} that simple models that include both photoionization
and collisional ionization can satisfy the observational constraints,
but realistic modeling of time-dependent effects is beyond the scope
of this paper.

\subsubsection{Photoionization Models\label{photoionsec}}

Is collisional ionization required at all in the intervening
absorbers?  Are the absorber properties consistent with the expected
characteristics of purely photoionized gas?  We have evaluated the
viability of pure photoionization as the ionization mechanism using
the photoionization code CLOUDY (v96.01, Ferland et al. 1998).  We
treat the absorbers as plane-parallel slabs photoionized by a
background UV radiation field with a specified shape and intensity.
Apart from the shape and intensity of the ionizing flux, the absorber
model properties depend primarily on the ionization parameter $U$ of
the gas ($U \equiv n_{\gamma}/n_{\rm H}$ = ionizing photon
density/total particle density), the metallicity, and the total
hydrogen column density.  We assume that background QSOs and AGNs
provide the bulk of the ionizing flux, and we use the Haardt \& Madau
(1996) QSO background but with the modifications provided by Haardt
(2006) [for motivation of the UV background update, see Madau et
al. 1999; Haardt \& Madau 2001; Shull et al. 1999; Scott et al. 2004].

Usually, photoionization models are constrained by requiring the
models to fit the measured column densities of multiple detected
metals such as \ion{Si}{3}, \ion{C}{3}, and \ion{C}{4} (in addition to
\ion{O}{6} and \ion{H}{1}).  With measurements from multiple species,
both the metallicity and the ionization parameter can be constrained
by models (e.g., Tripp et al. 2002; Savage et al. 2002; Lehner et
al. 2006). Unfortunately, in the majority of our intervening
\ion{O}{6} systems, we only detect \ion{O}{6} and \ion{H}{1} (and
occasionally \ion{C}{3}).  Nevertheless, we can certainly check to see
if the photoionization model predictions are {\it consistent} with the
measurements.  In \S \ref{collionsec}, we showed that the Gnat \&
Sternberg (2007) nonequilibrium collisional ionization models are
clearly not consistent with the observations; will we find a similar
result from photoionization models?

\begin{figure}
\centering
    \includegraphics[width=9.0cm, angle=0]{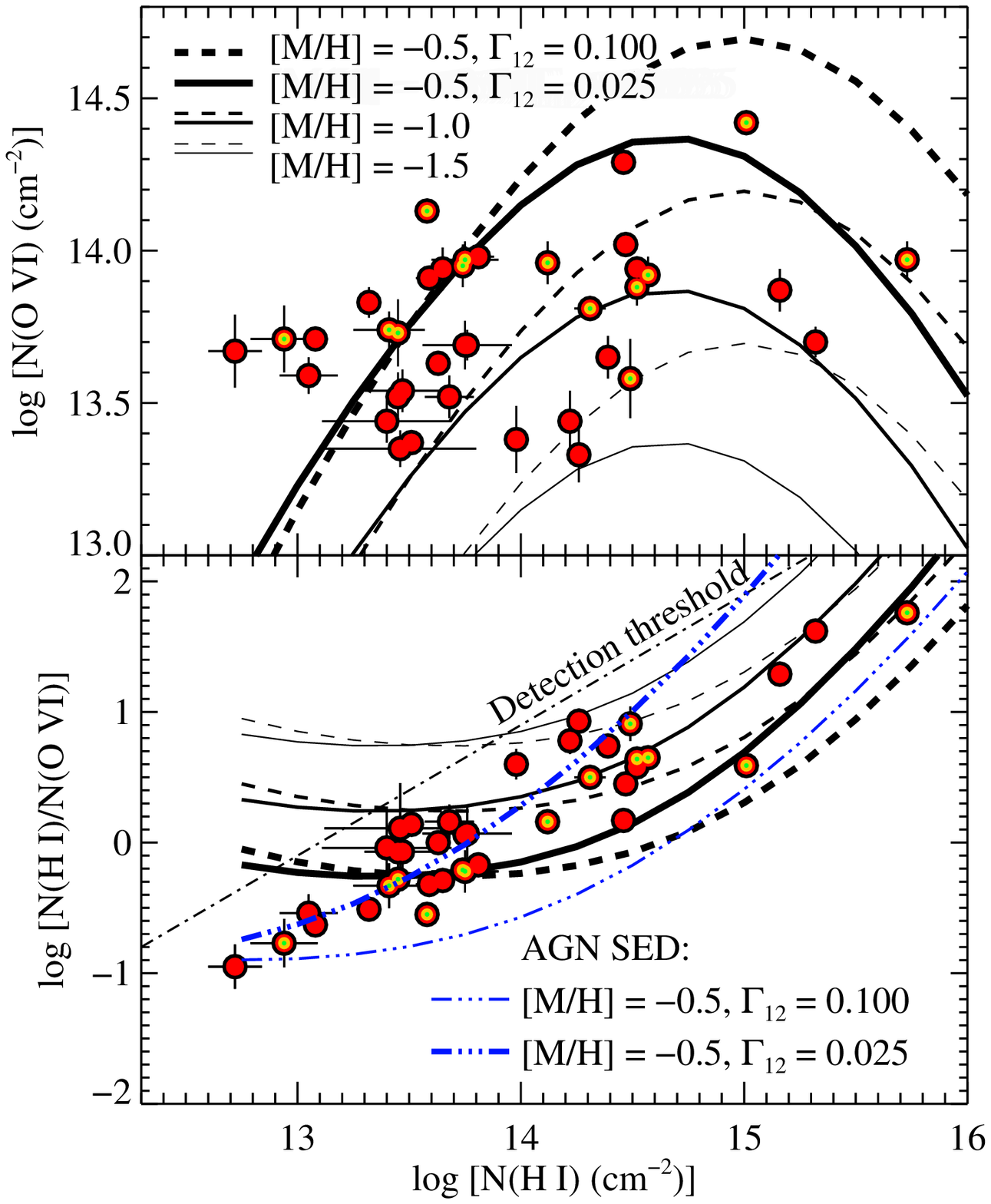}
\caption{Observed O~VI and H~I column densities for
matched components in intervening systems (filled circles); the
multicolor points represent components from complex/multiphase
absorbers. As in Figure~\ref{plainh1o6}, the upper panel shows log
$N$(O~VI) vs. log $N$(H~I), and the lower panel plots log
[$N$(H~I)/$N$(O~VI)] vs. log $N$(H~I).  These
observations are compared to the predicted column densities from six
photoionization models that assume that the gas density ($n_{\rm H}$)
and the H~I column density are related by equation
\ref{schayeeqn} (see Schaye 2001) and that the gas is photoionized by
the UV background light from QSOs/AGNs (Haardt \& Madau 1996; Haardt
2006).  The solid lines represent models that assume that the
H~I photoionization rate $\Gamma _{12} = 0.025$ (see text), and
dashed lines assume $\Gamma _{12} = 0.100$.  For each assumed value of
$\Gamma _{12}$, models are shown with overall metallicities of [M/H] =
$-1.5$ (thinnest lines), [M/H] = $-1.0$ (thicker lines), and [M/H] =
$-0.5$ (thickest lines).  In the lower panel an additional two
photoionization models that assume ionization by an AGN-dominated
spectral energy distribution (using the shape from Mathews \& Ferland
1987) are also compared to the data (triple-dot-dash lines) for two
values of $\Gamma _{12}$.
\label{photoschaye}}
\end{figure}

\begin{figure*}
   \includegraphics[width=17.0cm, angle=180]{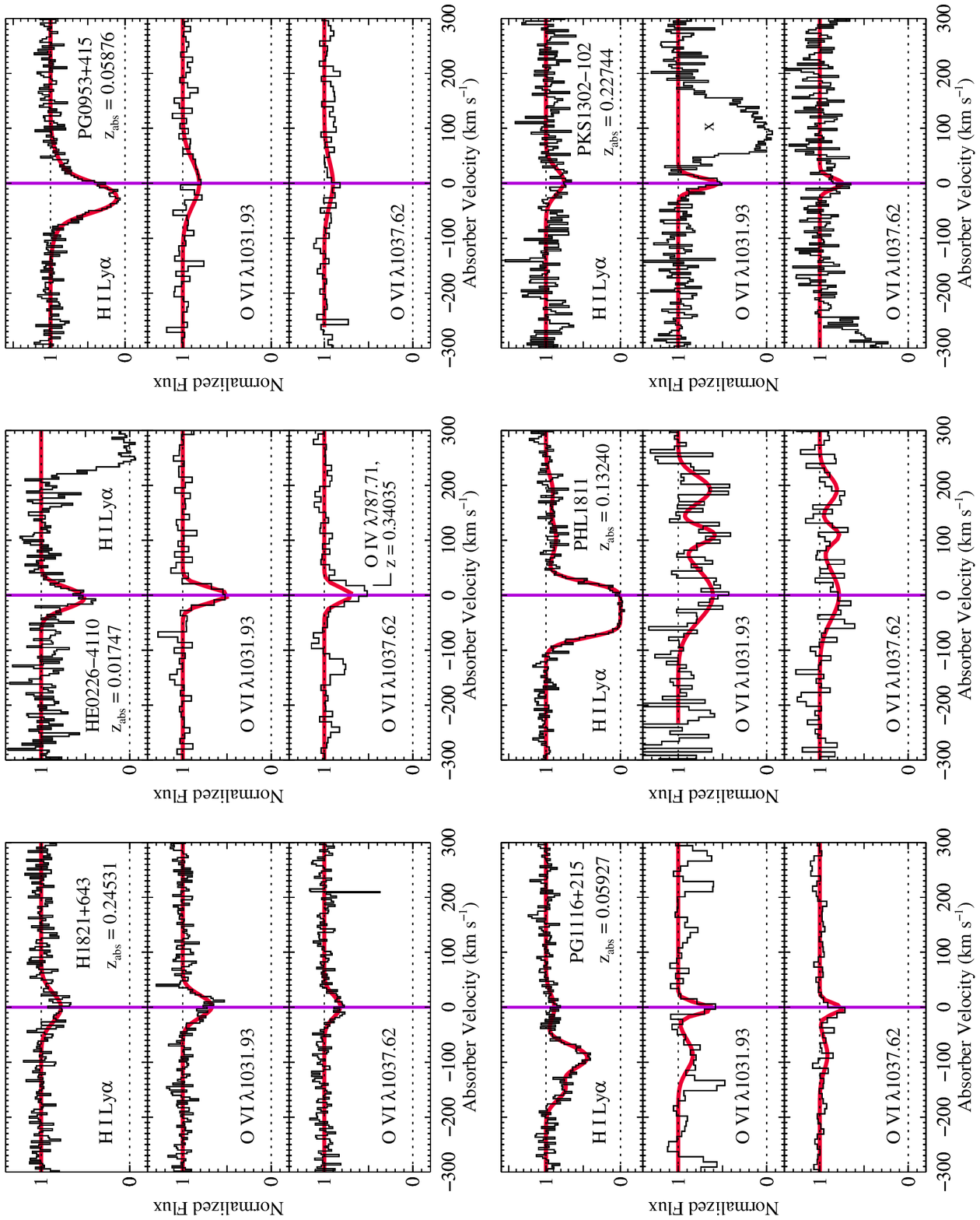}
\caption{Six of the seven O~VI absorbers with the lowest
H~I/O~VI ratios in individual components (the seventh
case with a low H~I/O~VI ratio is shown in
Figure~\ref{multi00334}). Each panel shows the H~I Ly$\alpha$
and O~VI $\lambda \lambda$1031.93,1037.62 absorption profiles
with sight line and absorber redshift indicated in the Ly$\alpha$
panel. At least one (but not necessarily all) of the components has a
low H~I/O~VI ratio (see
Table~\ref{compprop}).\label{lowratios}}
\end{figure*}

The ionization parameter depends on the gas density and the ionizing
radiation density.  To estimate the gas density, we can invoke the
arguments presented by Schaye (2001), who showed that if an
intergalactic \ion{H}{1} cloud is photoionized and in hydrostatic
equilibrium, then the characteristic size of the absorber will be the
Jeans length, and the total hydrogen number density $n_{\rm H}$ can be
directly estimated from the \ion{H}{1} column density:
\begin{equation}
n_{\rm H} \approx 10^{-5} \left[\frac{N({\rm H~I})}{2.3 \times
10^{13}}\right] ^{2/3} T_{4}^{0.17} \Gamma _{12}^{2/3}
\left(\frac{f_{g}}{0.16}\right)^{-1/3},  \label{schayeeqn}
\end{equation}
where we have rearranged Schaye's equation 8 and $T_{4} = T({\rm
K})/10^{4}$, $\Gamma$ is the hydrogen photoionization rate, $\Gamma
_{12} = \Gamma ({\rm s^{-1}})/1 \times 10^{-12}$, and $f_{g}$ is the
fraction of the total mass contributed by the baryonic gas.  Within a
reasonable range for gas temperatures of photoionized gas, the model
predictions are not overly sensitive to $T$; we assume $T$ = 20,000 K
for our calculations.  We also assume that $f_{g}$ is close to the
``universal'' value (i.e., $\Omega _{b}/\Omega _{m}$), and following
Schaye (2001) we use $f_{g}$ = 0.16.  While Schaye (2001) derived
equation~\ref{schayeeqn} analytically, very similar results for
photoionized gas have emerged from hydrodynamic simulations of the IGM
(see, e.g. equation 7 in Dav\'{e} et al. 1999), and the relationship
between overdensity and \ion{H}{1} column in the simulations has been
discussed in many papers.  We will use the results from Schaye (2001)
in the rest of this paper, but tests show that we would reach similar
conclusions if we used the relationship from the simulations instead.

The hydrogen photoionization rate depends on the shape and intensity
of the UV background.  The best observational constraints on the UV
background intensity at $z \approx$ 0 come from the observations of
Weymann et al. (2001). Based on deep upper limits on H$\alpha$
emission from the \ion{H}{1} 1225+01 intergalactic cloud (Giovanelli
\& Haynes 1989), Weymann et al. derive upper limits on the intensity
of the UV background at 1 Rydberg, $J_{\nu}$(1 Ryd).  Their upper
limit depends on the geometry of the cloud, and they derive limits
ranging from $J_{\nu}$(1 Ryd) $< 0.96 \times 10^{-23}$ to $< 3.84
\times 10^{-23}$ ergs cm$^{-2}$ s$^{-1}$ Hz$^{-1}$ str$^{-1}$.  For
the UV background from QSOs, these intensity limits imply
photoionization rates ranging from $\Gamma _{12} \approx 0.025$ to
$\Gamma _{12} \approx 0.10$.  These limits are in good agreement with
constraints obtained from other techniques (see \S 5 in Dav\'{e} \&
Tripp 2001).  We will use these two values for $\Gamma _{12}$ to
investigate the viability of photoionization models applied to low$-z$
\ion{O}{6} systems.  One appealing aspect of applying the Schaye
arguments to photoionization models is that this ensures that the
absorber size is a physically reasonable size -- the absorber size is
that which results from hydrostatic equilibrium.  Using Schaye's
equation 12, we find that the absorber sizes are $\lesssim$ 750 kpc
(450 kpc) for log $N$(\ion{H}{1}) $>$ 13.0 and $\Gamma _{12} = 0.025 \
(0.100)$.  The corresponding Hubble width due to Hubble broadening is
$b_{\rm H} \approx H(z)L/2 \lesssim 28 \ (17)$ km s$^{-1}$.  This
amount of Hubble broadening is quite reasonable compared to the
nonthermal broadening estimated for the aligned absorbers (see
Table~\ref{alignedtab}).

Using equation \ref{schayeeqn} to specify the gas density and using
the modified Haardt \& Madau (1996) UV background from QSOs,
Figure~\ref{photoschaye} compares the predicted \ion{O}{6} and
\ion{H}{1} column densities from several photoionization models to the
observed column densities from the matched intervening sample
(Table~\ref{alignedtab}).  In this figure, solid lines indicate models
that assume $\Gamma _{12}$ = 0.025, and dashed lines show models that
adopt $\Gamma _{12}$ = 0.100.  For each of these $\Gamma$ values, the
thin, thicker, and thickest lines represent models with logarithmic
metallicities [M/H] = $-1.5, -1.0,$ and $-0.5$, respectively.  In the
lower panel of Figure~\ref{photoschaye}, we also show the
\ion{H}{1}/\ion{O}{6} ratio from models that are photoionized by an
AGN spectral energy distribution (SED, triple-dot-dash lines), using
the Mathews \& Ferland (1987) characterization of the shape of an AGN
SED, with $\Gamma _{12}$ = 0.025 and 0.100.  The AGN SED model is
provided to illustrate the effect of changing the {\it shape} of the
ionizing flux field.

We have already noted that there is evidence of substantial
metallicity variations in the IGM.  It is also likely that the shape
as well as the intensity of the ionizing UV background varies
significantly in the IGM.  Regions that are closer to AGNs/QSOs could
naturally be exposed to a radiation field that is brighter but also
has a different shape.  The contribution from stellar flux escaping
from galaxies is uncertain but could also lead to ionizing radiation
shape and intensity variations.  Finally, the particle density can
vary from place to place, which changes the ionization parameter in
photoionization models.  Figure~\ref{photoschaye} shows that changes
in these variables can lead to significant variations in the
\ion{H}{1}/\ion{O}{6} ratio even in simple photoionized models.  We
see that when the \ion{O}{6} detection threshold is factored in,
photoionization models for \ion{H}{1} clouds in hydrostatic
equilibrium can easily explain the observed \ion{O}{6} and \ion{H}{1}
properties of many of the low$-z$ \ion{O}{6} absorbers.  The models
can produce the observed column densities, and moreover, the weak
correlation of $N$(\ion{O}{6}) with $N$(\ion{H}{1}) is a natural
result of these models (as long as the UV background and IGM
metallicity are variable in the low-z Universe).

The points with the lowest \ion{H}{1}/\ion{O}{6} ratios in
Figure~\ref{photoschaye} are not expected in the models photoionized
by the Haardt \& Madau (1996) UV background unless the metallicity is
quite high.  However, these points can be explained by models that
have a similar metallicity to the rest of the sample but are
photoionized by an AGN-like SED (such models are shown with the
triple-dot-dash lines in the lower panel of Figure~\ref{photoschaye}),
i.e., a harder radiation field.  One might wonder if these low
\ion{H}{1}/\ion{O}{6} ratios could be measurement artifacts, so we
show the absorption profiles for these cases in
Figure~\ref{lowratios}. From this figure, we 4 out of 7 of the
lowest-ratio cases are in systems with simple component structure, and
the lines are securely detected.  Therefore, most of the low ratios
are not simply due to measurement errors.  It would be interesting to
test whether the low-ratio absorbers are found in closer proximity to
a source of harder ionizing flux.  We do see that the intervening
systems with the lowest \ion{H}{1}/\ion{O}{6} ratios are close to the
ratios seen in the low-$N$(\ion{H}{1}) proximate absorbers (see
Figure~\ref{plainh1o6}).  The proximate absorbers are close (in
redshift) to the background QSO, so these cases are likely to be
ionized by a radiation field with an AGN SED shape.  The similarity of
the proximate absorbers to the low-ratio intervening systems is
consistent with the conclusion that a nearby AGN could be dominating
the radiation field in all of these cases.

\begin{figure}
\centering
    \includegraphics[width=9.0cm, angle=0]{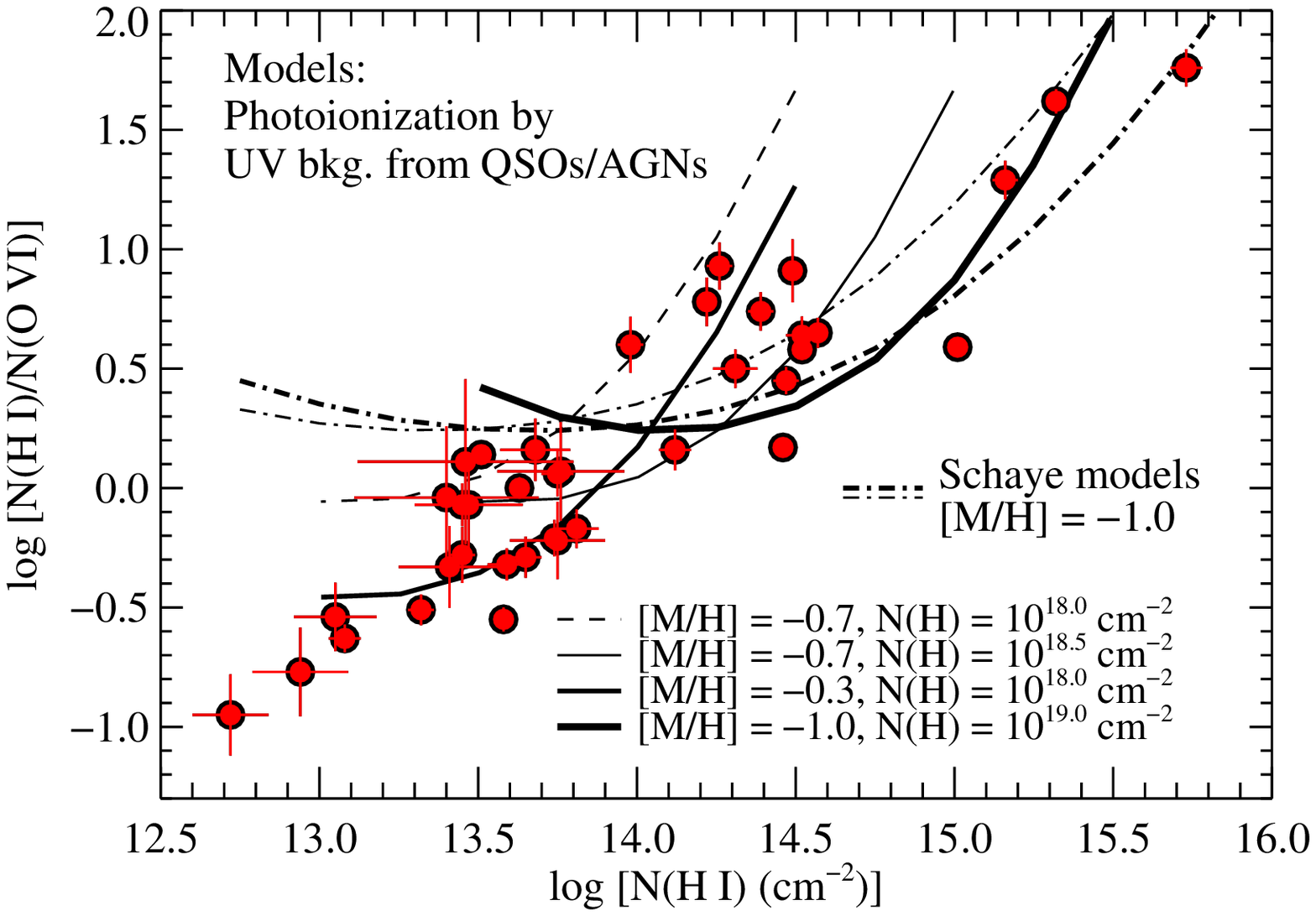}
\caption{Comparison of the observed H~I/O~VI ratios in
components of intervening systems (filled circles) to the
H~I/O~VI ratios predicted by generic photoionization
models (solid and dashed lines).  The models assume photoionization by
the UV background from quasars and AGNs according to the calculations
of Haardt \& Madau (1996). The variable parameters in the models shown
are the metallicity, [M/H], the total (H~I + H~II)
hydrogen column density $N$(H), and the ionization parameter. The
[M/H] and $N$(H) assumed for each curve are indicated in the
legend. Lower ionization parameters produce higher
H~I/O~VI ratios. For comparison, the photoionization
models that result from invoking Schaye's arguments (i.e., using
equation~\ref{schayeeqn}) with [M/H] = $-1.0$ are shown with dash-dot
lines; the thick dash-dot line assumes $\Gamma _{12} = 0.100$, and
the thin dash-dot line assumes $\Gamma _{12} =$
0.025. \label{cloudy_o6h1}}
\end{figure}

It is interesting to see that the physical framework of Schaye (2001)
naturally explains the observed \ion{O}{6} and \ion{H}{1} properties
of the well-aligned components, but we note that more generic
photoionization models that do not invoke equation~\ref{schayeeqn} are
also successful.  We show several more generic photoionization models
in Figure~\ref{cloudy_o6h1}.  In these models we have again assumed
the modified Haardt \& Madau (1996) UV background, but we have
selected various values of $N_{\rm total}$(H) and [M/H] without
requiring $n_{\rm H}$ and $N$(\ion{H}{1}) to be related by equation
\ref{schayeeqn}.  For a given model, the ionization parameter
decreases along the model curve from left to right in
Figure~\ref{cloudy_o6h1}. For purposes of illustration, in this figure
we show $N$(\ion{H}{1})/$N$(\ion{O}{6}) vs. $N$(\ion{H}{1}) and we
replot the two Schaye-based models from Figure~\ref{photoschaye} with
[M/H] = $-1.0$ and $\Gamma _{12}$ = 0.025,0.100.  The generic
photoionization models can also reproduce the observed \ion{O}{6} and
\ion{H}{1} column densities with reasonable values for $N_{\rm
total}$(H) and [M/H].  We do note that both the Schaye-based models
and the generic models require relatively high metallicities ($Z
\gtrsim 0.3 Z_{\odot}$) for some of the \ion{O}{6} systems, which
might be somewhat surprising.  However, similar metallicities have been
indicated by studies of comparable absorbers with detections of
multiple metals (e.g., Savage et al. 2002; Aracil et al. 2006; Lehner
et al. 2006).

\begin{figure}
\centering
    \includegraphics[width=9.0cm, angle=0]{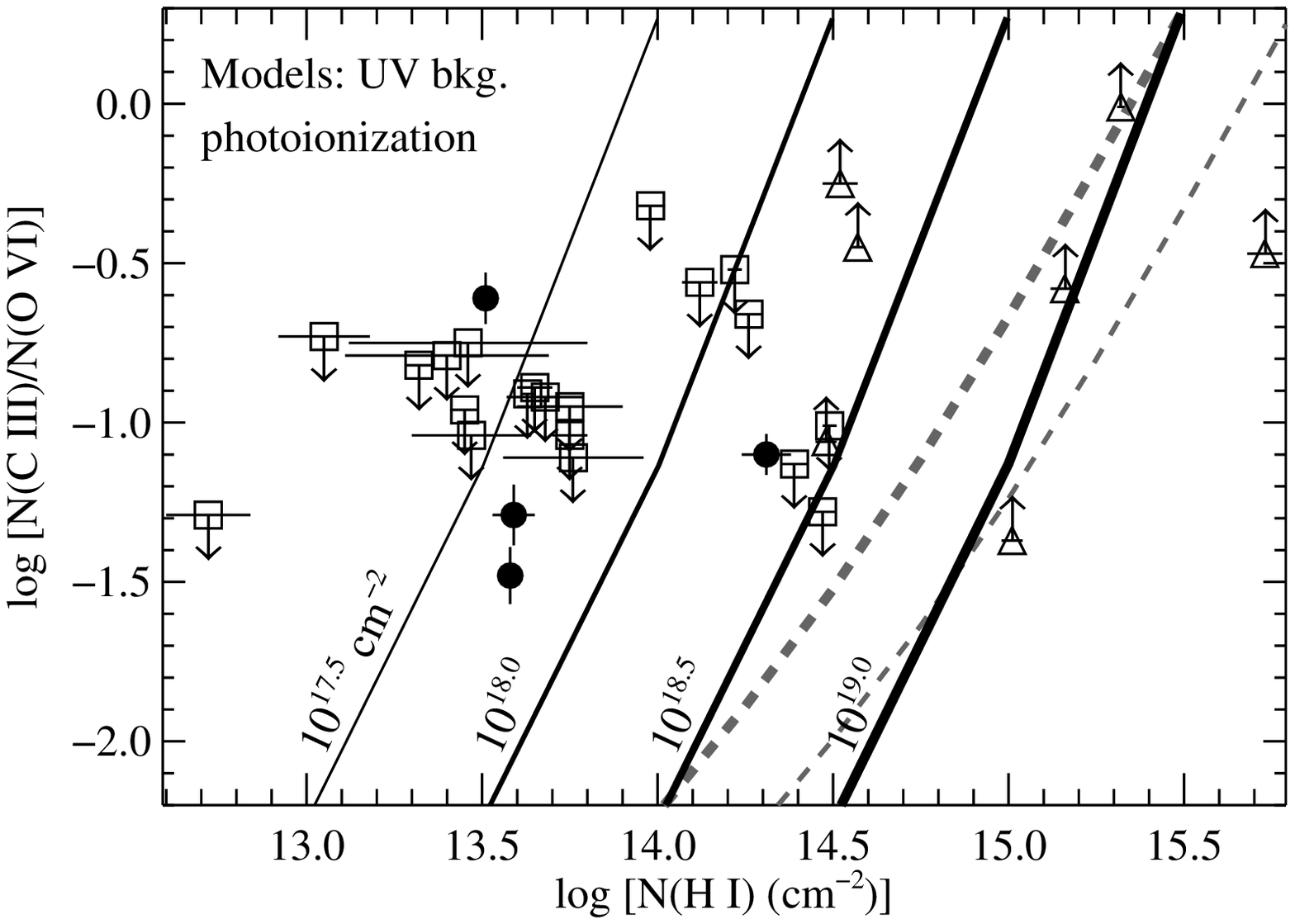}
\caption{Comparison of the observed C~III/O~VI ratios
vs. log $N$(H~I) in components of intervening systems (with
symbols coded as in Figure~\ref{modelc3o6}) to the
C~III/O~VI ratios as a function of log $N$(H~I)
predicted by generic photoionization models (solid lines). The
photoionization models assume the Haardt \& Madau (1996) UV background
from QSOs/AGNs. Four model curves are shown corresponding to total
(H~I + H~II) hydrogen columns of log $N$(H) = 17.5, 18.0,
18.5, and 19.0; each model is shown with a solid curve labeled with
its total H column [the line thickness increases with $N$(H)].  The
predicted C~III/O~VI ratios depend primarily on the
ionization parameter; increasing the ionization parameter decreases
the C~III/O~VI column density ratio.  Changing the overall
metallicity of the model does not move the curves in this plot because
$N$(O~VI) and $N$(C~III) increase (or decrease)
homologously with increasing (or decreasing) metallicity.  The gray
dashed lines represent the photoionization models that result from the
model of Schaye(2001) with $\Gamma _{12} = $ 0.025 (thick dashed line)
or $\Gamma _{12} = 0.100$ (thin dashed line).
\label{photmodelc3o6}}
\end{figure}

Evidently, purely photoionized gas can produce the observed \ion{O}{6}
and \ion{H}{1} column densities.  However, we can see from
Figures~\ref{photoschaye} and \ref{cloudy_o6h1} that there are
substantial degeneracies in the photoionization models when the only
observational constraints are from \ion{O}{6} and \ion{H}{1}, and it
would be very helpful to include constraints from other metals in the
ionization analysis.  Moreover, we found in \S \ref{collionsec} that
the nonequilibrium collisional ionization models were able to match
some of the \ion{H}{1}/\ion{O}{6} ratios but were much harder pressed
to fit the \ion{C}{3}/\ion{O}{6} constraints.  For both of these
reasons, we now examine whether our photoionization models are
consistent with the observed \ion{C}{3}/\ion{O}{6} constraints from
Table~\ref{alignedtab}.  In Figure~\ref{photmodelc3o6} we compare the
\ion{C}{3}/\ion{O}{6} constraints to the predictions from the
Schaye-based and generic photoionization models.  In this plot, the
model \ion{C}{3}/\ion{O}{6} ratios are not dependent on metallicity
because $N$(\ion{C}{3}) and $N$(\ion{O}{6}) change homologously with
metallicity.\footnote{We have assumed a solar C/O ratio for these
models, and moreover, we have assumed that the relative abundance of C
with respect to O does not depend on the overall metallicity.  Some
studies (e.g., Akerman et al. 2004) have indicated that C/O is not
constant as a function of metallicity and may even have a somewhat
complex dependence on metallicity.  Accounting for this complication
is beyond the scope of this paper. It would be useful to evaluate the
impact of departures from solar relative abundances in future
analyses.}  Figure~\ref{photmodelc3o6} shows that the photoionization
models are consistent with the constraints provided by the current
observations for a large fraction of the well-aligned components.  The
generic photoionization models are consistent with all of the
measurements provided that the total hydrogen column (\ion{H}{1} +
\ion{H}{2}) ranges from $10^{17}$ up to $\gtrsim 10^{19}$
cm$^{-2}$. We do notice that the models that employ equation
\ref{schayeeqn} with $\Gamma _{12} = 0.025$ or 0.100 appear to be most
consistent with the higher $N$(\ion{H}{1}) systems, and the
Schaye-based models do not fit the four cases that provide the best
\ion{C}{3} measurements.  Since most of the matched intervening
\ion{O}{6} systems only provide {\it upper limits} on $N$(\ion{C}{3}),
the data remain consistent with the Schaye-based models for most of
the measurements, but more sensitive measurements of (or upper limits
on) the \ion{C}{3} column densities would provide a useful test of the
Schaye-based photoionization predictions (as shown in \S
\ref{intfuturesec}, future \ion{C}{4} measurements would be similarly
useful for testing these models). The Schaye models move toward better
agreement with those points if $\Gamma _{12}$ is reduced.  Since the
best observational constraints still provide only upper limits on
$\Gamma _{12}$, this solution remains a possibility, but it appears
that $\Gamma _{12}$ must be reduced by a large amount.  The
Schaye-based models can also be moved closer to the well-measured
\ion{C}{3}/\ion{O}{6} ratios by using a UV background radiation field
that includes a greater contribution from {\it stellar} flux escaping
from galaxies, but this would upset the agreement with the
\ion{H}{1}/\ion{O}{6} models.  The discrepancy with the best
\ion{C}{3} measurements in Figure~\ref{photmodelc3o6} could be a
symptom of a serious problem with this model.  This could indicate
that the clouds are not in hydrostatic equilibrium, for example, or
that multiphase models are required even for the apparently simple
cases.  Future studies should revisit the Schaye-based models with a
larger sample of more-sensitive \ion{C}{3} and/or \ion{C}{4}
measurements.

\begin{figure}
\centering
    \includegraphics[width=9.0cm, angle=0]{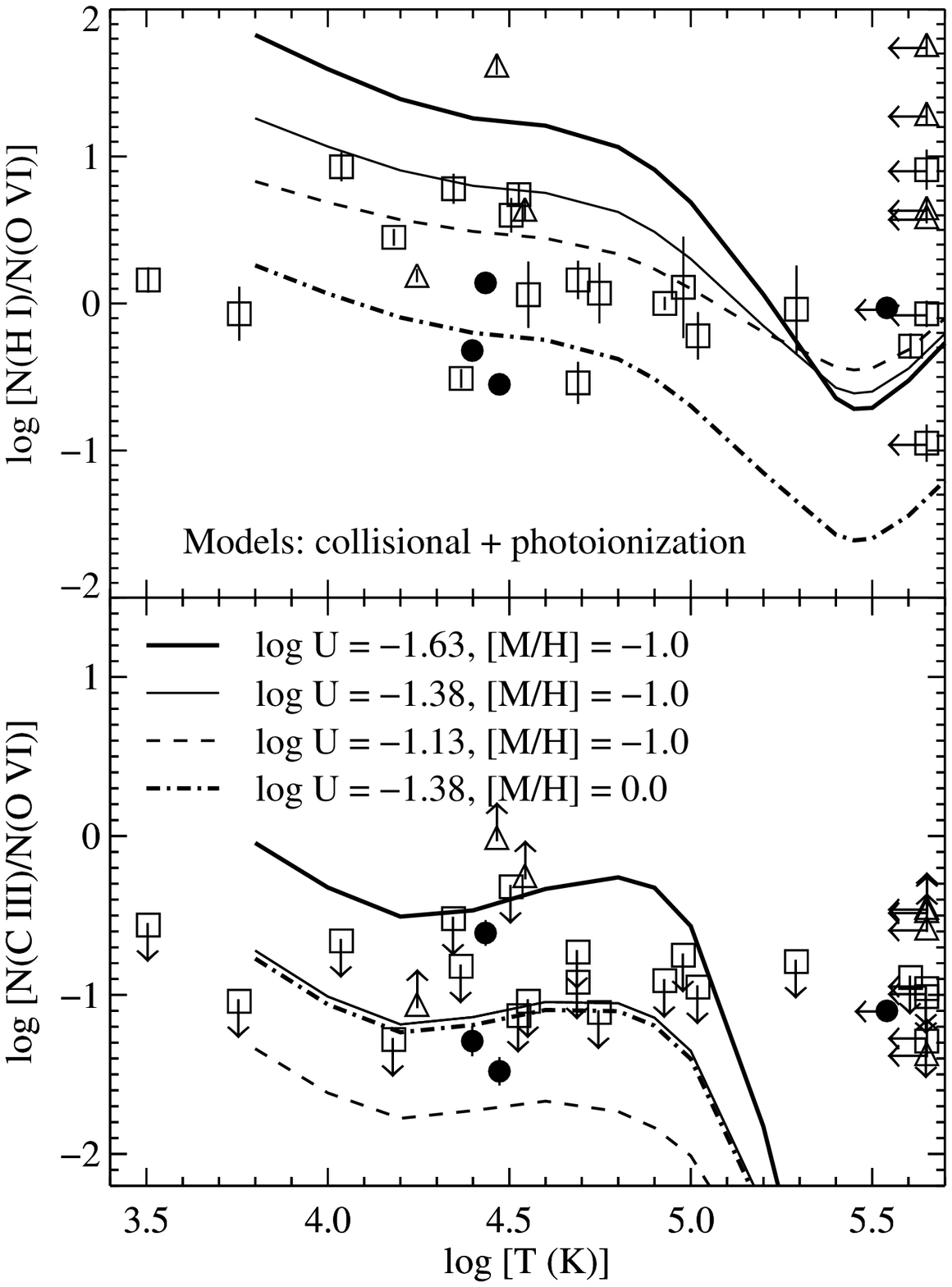}
\caption{Comparison of the observed H~I/O~VI ratios
(upper panel) and the observed C~III/O~VI ratios (lower
panel) to four ``hybrid'' ionization models that include both
collisional ionization and photoionization. Both panels show the same
ionization models, and the symbols have the same meaning as in
Figure~\ref{modelc3o6}. Three of the models assume $Z = 0.1 Z_{\odot}$
([M/H] = -1.0) with ionization parameters fixed at log $U = -1.63$
(thick solid line), log $U = -1.38$ (thin solid line), and log $U =
-1.13$ (dashed line).  The dash-dot line assumes $Z = 1.0 Z_{\odot}$
and log $U = -1.38$.  Changing the overall metallicity affects the
H~I/O~VI ratio but does not change the
C~III/O~VI ratio. The parameters of the models in this plot
were chosen for purposes of illustration; these are not fits to the
data.\label{hybridmod}}
\end{figure}

\subsubsection{Hybrid Models (Collisional and Photoionization)\label{hybridsec}}

We have shown that if the aligned \ion{O}{6} and \ion{H}{1} lines are
cospatial, then the available models in which the gas is predominantly
collisionally ionized to do not agree with the observed properties of
the \ion{O}{6} absorbers (\S \ref{collionsec}).  Predominantly
photoionized models can explain most of the measurements but often
require a relatively high metallicity (\S \ref{photoionsec}).
However, the best temperatures implied by the line widths (\S
\ref{tempsec}), while too cool to produce \ion{O}{6} in collisional
ionization equilibrium, are sometimes higher than the temperatures
expected in predominantly photoionized gas. This raises the question
of whether photoionization and collisional ionization are {\it both}
important in these absorbers.

We can provide a simplified evaluation of whether ``hybrid''
ionization models (i.e., including important contributions from
collisional and photoionization) are plausible using CLOUDY.  By
running CLOUDY with the plasma temperature locked at a specified input
value and including photoionization from the UV background, we can
assess the effects of the combined ionization mechanisms.  Such models
solve for ionization equilibrium but are {\it not} in thermal
equilbrium (heating does not balance cooling). In
Figure~\ref{hybridmod} we show CLOUDY models at various fixed
temperatures (shown on the x-axis) and photoionized by the Haardt \&
Madau (1996) QSO background flux. The models are compared to the
observed \ion{H}{1}/\ion{O}{6} and \ion{C}{3}/\ion{O}{6}
column-density ratios of the aligned components from
Table~\ref{alignedtab}.  These calculations have three adjustable
parameters: the temperature, metallicity, and ionization parameter.
The curves show models with [M/H] = $-1$ and [M/H] = 0 ($Z = 0.1
Z_{\odot}$ and $1.0 Z_{\odot}$) and three values for the ionization
parameter.  Our intention here is not to fit the ratios but to assess
whether or not the models are plausible.  We see that with reasonable
values for the temperature and ionization parameter, these hybrid
models can achieve good agreement with the observed ratios.  In some
cases, the models and measurements agree with [M/H] = $-1.0$.
However, in the small number of cases where good \ion{C}{3}
measurements are available, close inspection shows that high
metallicities are required in these hybrid models as well.  Changing
the overall metallicity does not change the \ion{C}{3}/\ion{O}{6}
ratio because both \ion{C}{3} and \ion{O}{6} change by the same
amount.  However, the \ion{H}{1}/\ion{O}{6} ratio is inversely
proportional to the metallicity.  We thus see from
Figure~\ref{hybridmod} that in some cases, a better overall fit to the
\ion{O}{6}, \ion{C}{3}, and \ion{H}{1} columns can be obtained with
higher metallicities.  As we discuss in \S 5, in the
future it will important to obtain good measurements of multiple metal
species in order to constrain the ionization and metallicity of as
many specific systems as possible.

\subsection{Physical Conditions in Complex Multiphase \ion{O}{6} Systems\label{complexmulti}}

Evidently, many of the well-aligned systems show compelling
indications of cool gas ($T < 10^{5}$ K), and among currently
available models, these cool absorbers are best explained as
photoionized gas.  Is there any observational evidence of the WHIM in
the nearby universe then?  X-ray absorption lines (e.g., the
\ion{O}{7} K$\alpha$ transition) are potentially useful for finding
WHIM gas because the X-ray absorbers persist in hotter plasmas. X-ray
spectroscopy has been used to search for the WHIM, and while some
redshifted X-ray absorption lines have been reported (e.g., Fang et
al. 2002; Mathur et al. 2003; Nicastro et al. 2005), the reliability
of the detections has been debated (Kaastra et al. 2006; Rasmussen et
al. 2002,2007; Bregman 2007; but see also Fang et al. 2007; Williams
et al. 2007).  X-ray absorption lines have been robustly detected at
$z_{\rm abs} \approx 0$ (e.g., Nicastro et al. 2002; Fang et
al. 2006), but there is evidence that the $z_{\rm abs} \approx 0$
X-ray absorbers arise in gas clouds that are relatively close to the
Milky Way (Futamoto et al. 2004; Wang et al. 2005; Fang et al. 2006),
perhaps even within a few kpc of the Milky Way disk (Yao \& Wang
2005), and thus are more likely to be hot Galactic ISM gas than WHIM
clouds.

In the ultraviolet bandpass, the strongest evidence of
WHIM-temperature plasma is the 3.9$\sigma$ detection of \ion{Ne}{8}
reported by Savage et al. (2005) in the strong \ion{O}{6} absorber at
$z_{\rm abs}$ = 0.20701 toward HE0226-4110.  In this case, the
\ion{Ne}{8} column cannot be produced in plausible photoionization
models, and the \ion{Ne}{8}/\ion{O}{6} ratio requires $T = 5.4 \times
10^{5}$ K in collisional ionization equilibrium. It is interesting to
note that this absorber is a complex case with multiple components in
the \ion{H}{1} profiles and low-ion metals that trace cooler (probably
photoionized) phases of the absorption system.  Other multiphase,
multicomponent \ion{O}{6} absorbers have similarly been argued to have
characteristics of warm-hot collisionally ionized gas (e.g., Chen \&
Prochaska 2000; Tripp et al. 2000; Prochaska et al. 2004; Shull et
al. 2003; Tumlinson et al. 2005). The fact that {\it all} of these
systems that present evidence of hot gas are multiphase absorbers
raises a question: is the warm-hot gas primarily located in the
complex multiphase absorbers? This is a crucial question for the
inventory of baryons in the low$-z$ IGM (see \S \ref{baryonicsec}).
We have classified roughly half of the intervening sample as
multiphase absorbers (\S \ref{absclass}).  Because they are
intrinsically complex, multiphase \ion{O}{6} systems often require
detailed analyses (e.g., Tripp et al. 2001; Richter et al. 2004;
Sembach et al. 2004; Savage et al. 2005; Tumlinson et al. 2005).
Complete analysis of the full sample of multiphase absorbers is beyond
the goals of this paper.  However, it is useful to present a few
examples of how the WHIM can be difficult to identify in multiphase
absorbers.

\begin{figure}
\centering
    \includegraphics[width=8.0cm, angle=0]{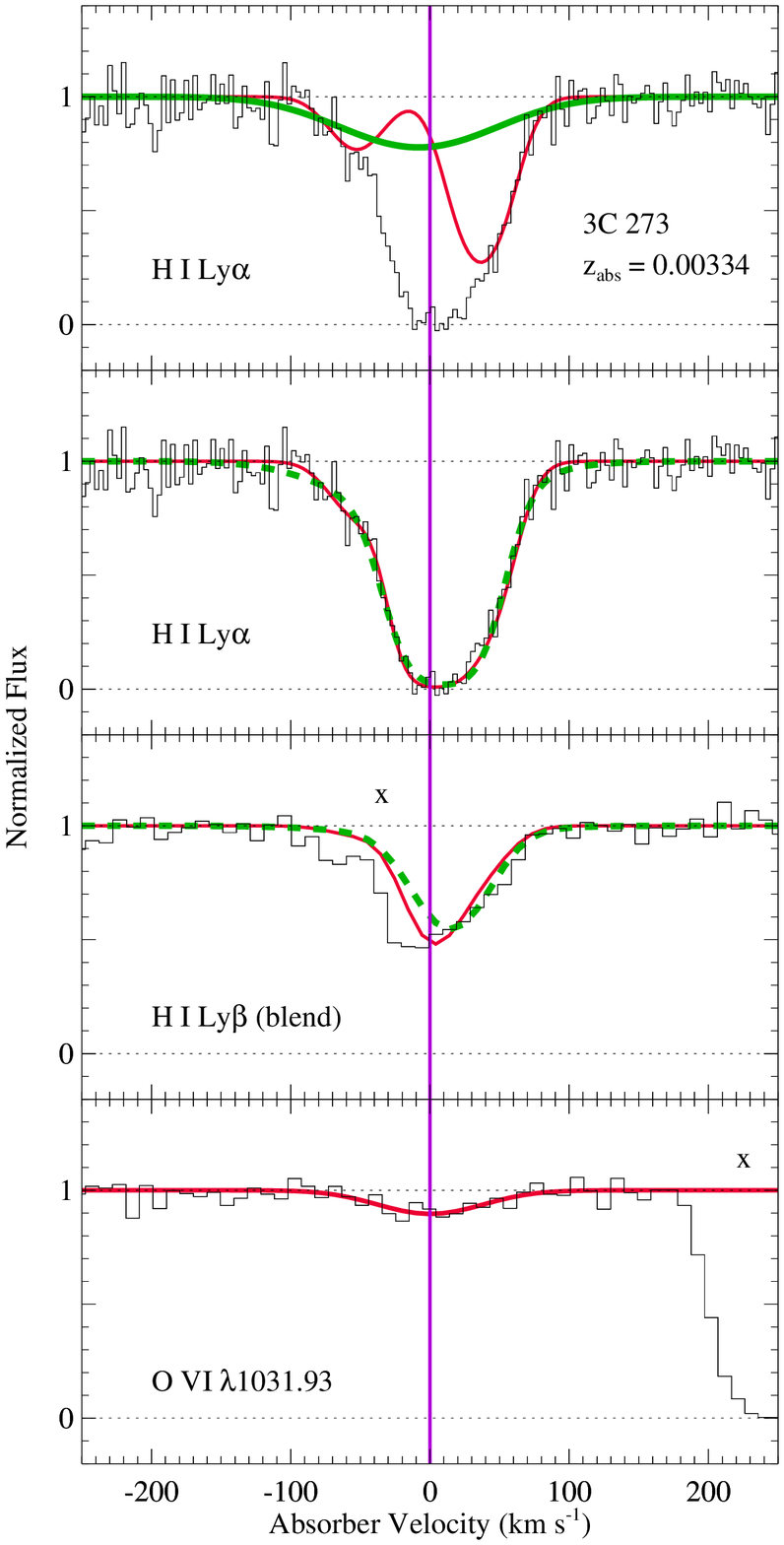}
\caption{Comparison of two comparably good fits to the H~I Ly$\alpha$
and Ly$\beta$ profiles of the O~VI absorber at $z_{\rm abs}$ = 0.00334
in the spectrum of 3C 273.0.  In the second and third panels, the thin
red line shows the three-component fit, and the thick dashed green
line shows an alternative fit that uses only two components (the
bottom panel shows the O~VI $\lambda$1031.93 line, which is fitted
with a single component).  The top panel shows the difference between
the alternative fits to the H~I profiles: the three-component fit uses
two narrow components (shown with a thin-red line) to fit the
inflections in the wings of the Ly$\alpha$ profile, but the two
component fit employs a single broad feature (shown with a thick green
line) to fit these features.  The H~I Ly$\beta$ profile is partially
affected by a blend with Milky Way H$_{2}$ absorption (see Sembach et
al. 2001). Xs mark unrelated absorption. \label{multi00334}}
\end{figure}

\begin{figure*}
\centering
    \includegraphics[width=8.0cm, angle=0]{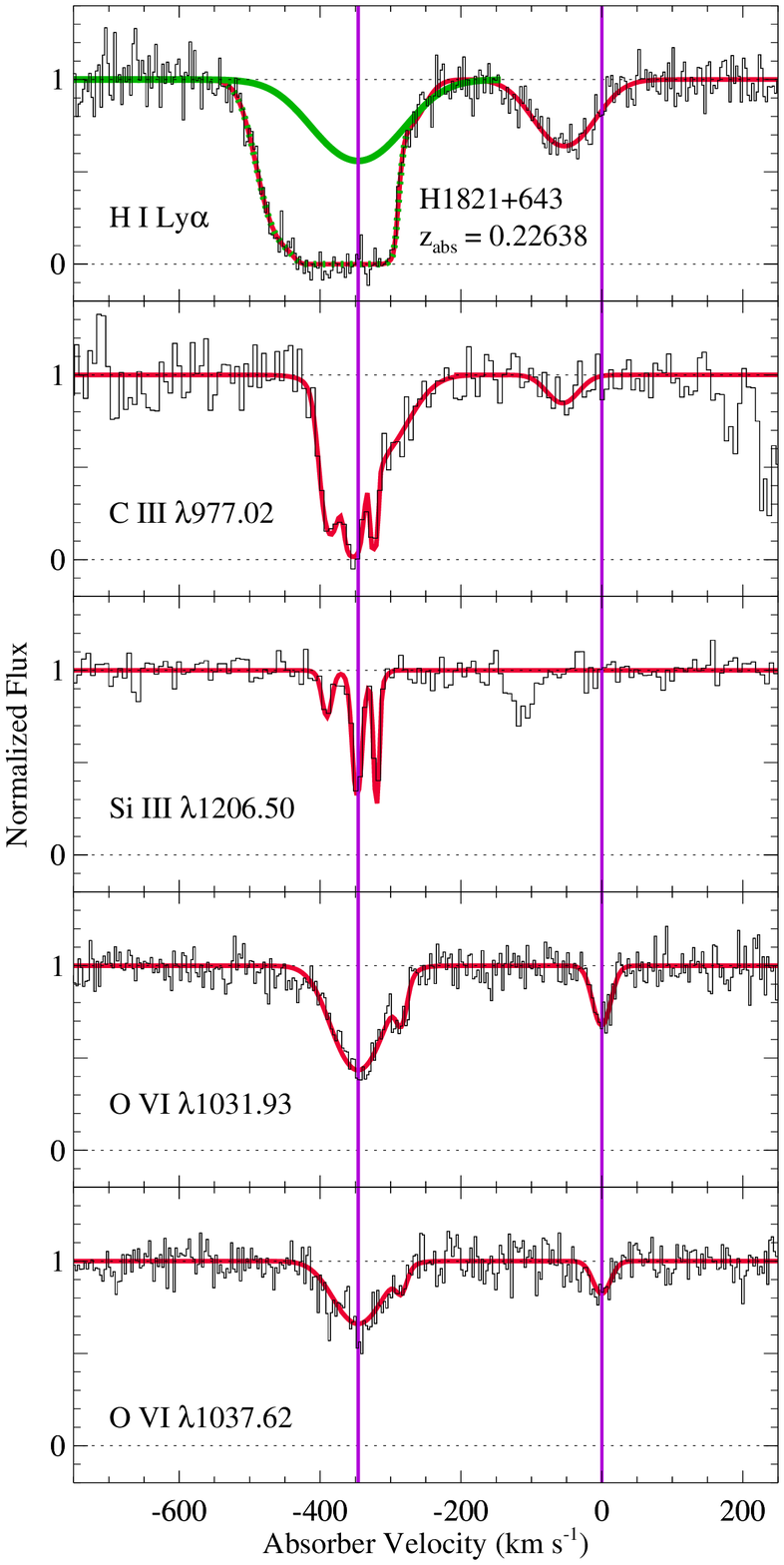}
    \includegraphics[width=8.0cm, angle=0]{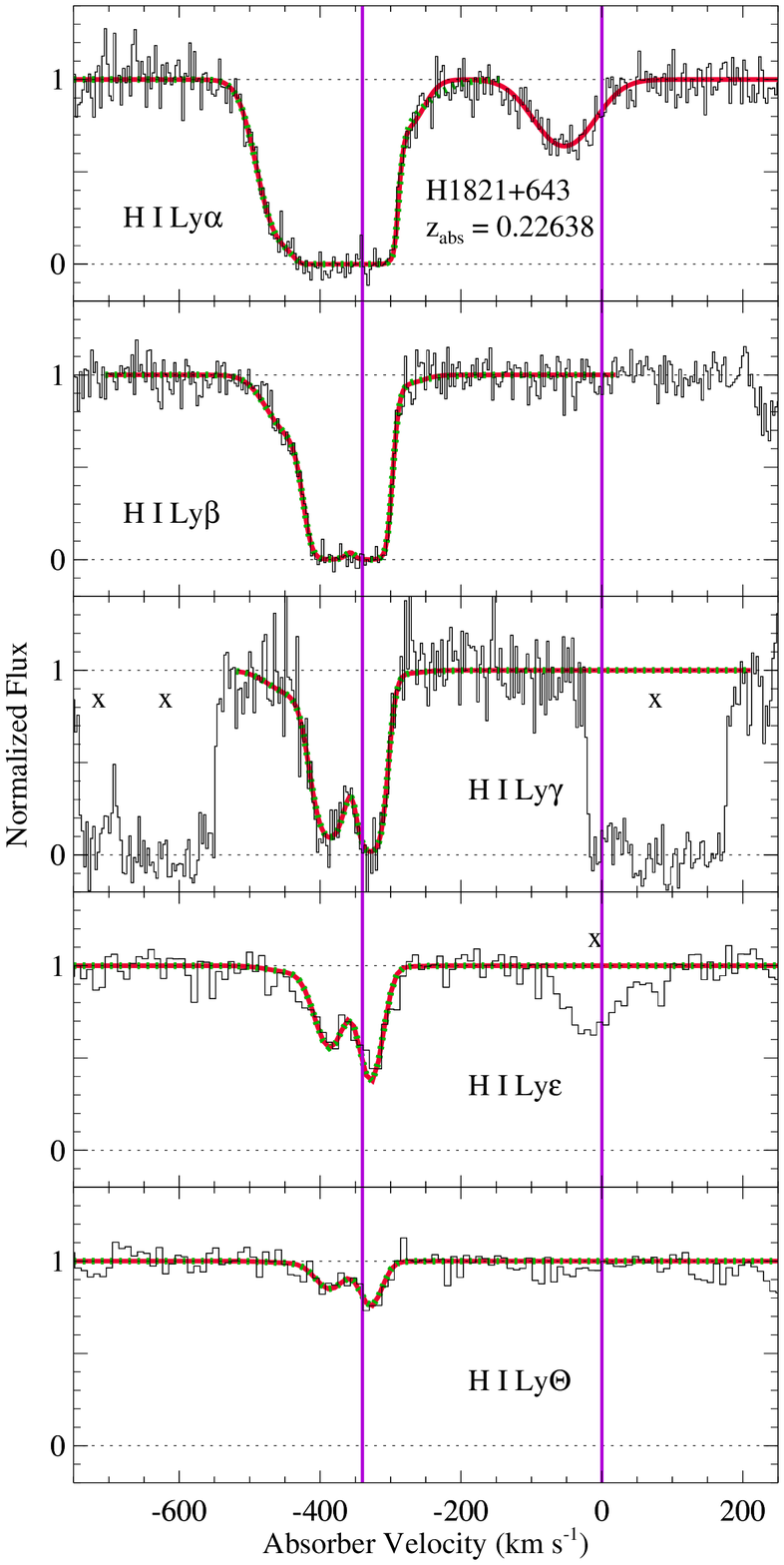}
\caption{Selected absorption profiles (histograms) of the complex,
multiphase O~VI absorbers at $z_{\rm abs}$ = 0.22496 and 0.22638
in the spectrum of H1821+643 (plotted in the $z_{\rm abs}$ = 0.22638
frame). Voigt-profile fits are overplotted with smooth lines in each
panel, and the vertical lines indicate the $v = 0$ km s$^{-1}$ points
for $z_{\rm abs}$ = 0.22496 and 0.22638 (i.e., the centroids of the
O~VI absorbers).  As discussed in the text, several aspects of
these absorbers suggest the presence of hot gas.  The H~I
panels show a comparison of two comparably good fits to the H~I
Lyman series lines in the complex, multiphase O~VI absorber at
$z_{\rm abs}$ = 0.22496. The red line shows the fit reported in
Table~\ref{compprop}, and the dotted green line shows an alternative
H~I fit with the velocity centroid of one H~I component
locked to the centroid of the O~VI lines.  The column density
and $b-$value of the H~I component with the locked velocity
centroid are freely varied in the alternative fit (only the centroid
is locked).  The locked component in the final fit is shown with a
thick green line in the top-left panel. Unrelated absorption lines are
denoted with Xs. \label{h1821complex}}
\end{figure*}

One problem with multiphase absorbers arises from limitations of
Voigt-profile fitting when dealing with complex component structure
(\S \ref{secmeas}).  In moderate S/N data, the appropriate number of
components to fit to a multicomponent profile can be ambiguous, and
this can have a substantial impact on the results for some of the
components. Figure~\ref{multi00334} shows the $z_{\rm abs}$ = 0.00334
absorber toward 3C 273.0, which provides an example of this
problem. In this case, the Ly$\alpha$ profile clearly shows a
significant but blended absorption feature at $v \approx -50$ km
s$^{-1}$. This feature cannot be caused by the main \ion{H}{1}
component at $v = 0$ km s$^{-1}$ and this motivates a fit with at
least two components.  However, the positive-velocity side of the
Ly$\alpha$ profile is also asymmetric with a ``ledge'' of pixels at $v
\approx 35$ km s$^{-1}$ (most easily seen in the upper panel of
Figure~\ref{multi00334}).  This asymmetry suggests the presence of a
third component.  But in these data, systematic noise could be the
origin of the ledge feature.  The second and third panels in
Figure~\ref{multi00334} compare the results from a two-component fit
(dashed green line) to the three-component fit (thin red line).  The
two-component fit provides a somewhat better reduced $\chi _{\nu}
^{2}$ = 1.15 (compared to $\chi _{\nu} ^{2}$ = 1.33 for the
three-component model) and thus was favored in Table~\ref{compprop}.
However, the improvement in $\chi _{\nu} ^{2}$ mainly results from a
better fit to the Ly$\beta$ profile, which is confused by blending
with Galactic H$_{2}$ absorption, and the two models fit the
Ly$\alpha$ profile comparably well.  The main \ion{H}{1} component
does not change much in the two- vs. three-component fit, but as shown
in the top panel of Figure~\ref{multi00334}, the properties of the
weaker component(s) are dramatically different in the two fits.  The
two-component fit requires a very broad \ion{H}{1} component that is
aligned with the \ion{O}{6} absorption within the $1\sigma$ centroid
uncertainties.  The three-component fit, on the other hand, results in
three narrow \ion{H}{1} lines (the main component plus the two weaker
components shown with a red line in the top panel of
Figure~\ref{multi00334}), and while the \ion{O}{6} is still aligned
with the main component, in this case $b$(\ion{O}{6}) $\gg
b$(\ion{H}{1}), which is unphysical.  The \ion{O}{6} $\lambda$1031.93
line is detected at the 4.5$\sigma$ level in the {\it FUSE} spectrum
of 3C 273.0, but the line is broad and shallow.  It is possible that
the \ion{O}{6} profile is composed of three components that align with
three \ion{H}{1} components, but we cannot discern the \ion{O}{6}
components at the available resolution and S/N ratio.  Thus, this
system is difficult to interpret.  To overcome these fitting problems
requires excellent data with high spectral resolution and high S/N.

This problem with identification of broad Ly$\alpha$ components in
multiphase systems is exacerbated by high \ion{H}{1} column densities.
Consider the strong \ion{O}{6} absorber at $z_{\rm abs}$ = 0.20266
toward PKS0312-770 (shown in Figures~\ref{speccontrast} and
\ref{pks0312multi}).  In this absorber, one or more of the components
have $N$(\ion{H}{1}) $\gtrsim 10^{17}$ cm$^{-2}$, which produces a
``boxcar'' Ly$\alpha$ absorption profile with a black core over a
large velocity range (see the upper left panel of
Figure~\ref{speccontrast}).  In this case, if there is a broad and
relatively shallow \ion{H}{1} component that arises in the
\ion{O}{6}-bearing gas, it would likely be impossible to detect.  Such
a component would have miniscule optical depth in higher Lyman series
lines and thus would be undetectable at available S/N levels, and in
the Ly$\alpha$ profile, it would be hidden in the black core of the
boxcar profile.

\begin{figure*}
\centering
   \includegraphics[width=17.0cm, angle=0]{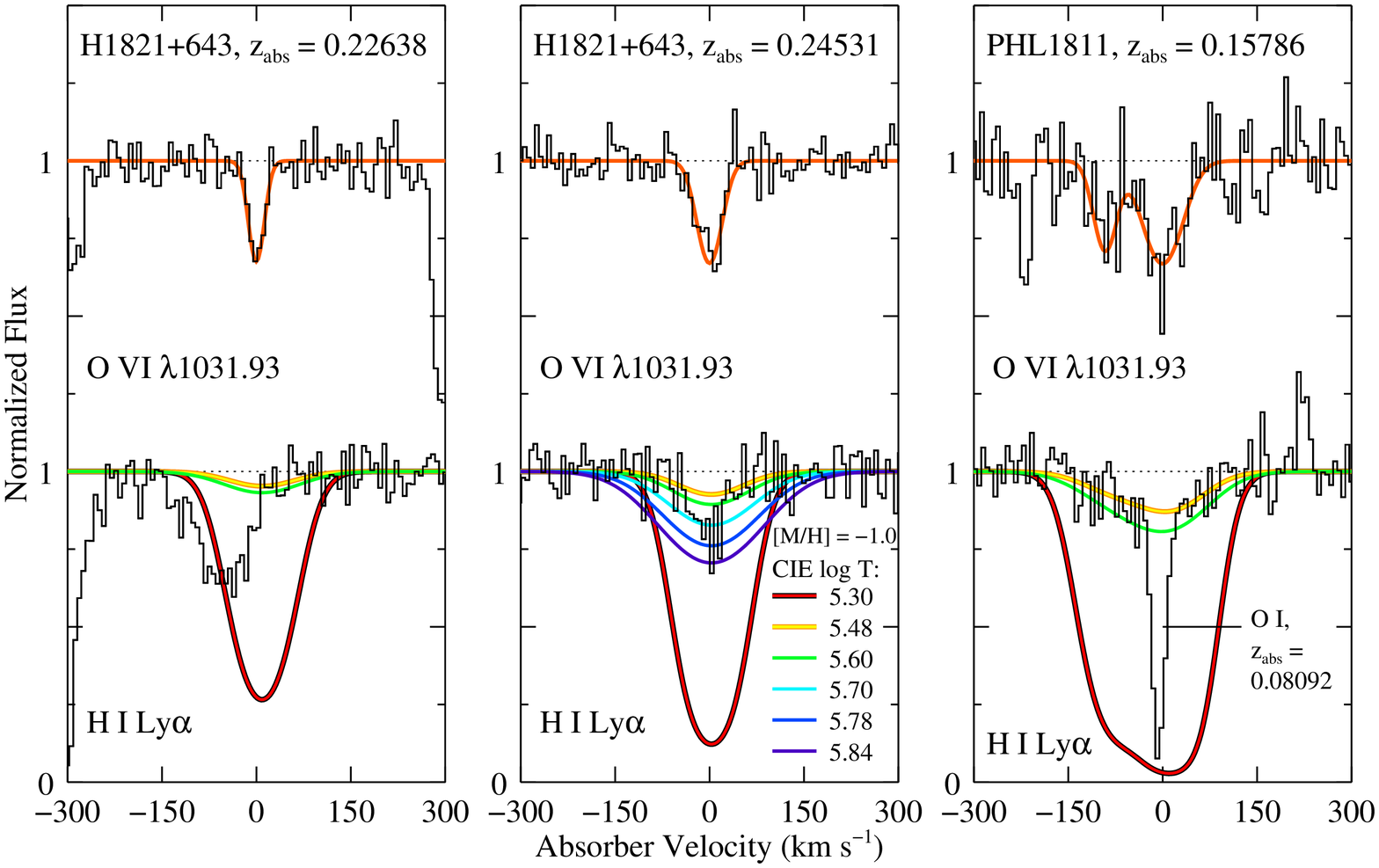}
\caption{Examples of the comparison of predicted H~I Ly$\alpha$
lines, assuming the O~VI arises in equilibrium collisionally
ionized gas with $Z = 0.1 Z_{\odot}$, to the observed Ly$\alpha$
absorption profiles.  The left and middle panels show the H1821+643
absorption systems at $z_{\rm abs}$ = 0.22638 and 0.24531,
respectively, and the right panel shows the PHL1811 system at $z_{\rm
abs}$ = 0.15786.  In each panel, the upper histogram indicates the
continuum-normalized O~VI $\lambda$1031.93 profile and the lower
histogram is the Ly$\alpha$ line.  Note that the PHL1811 Ly$\alpha$
line at $z_{\rm abs}$ = 0.15786 is blended with O~I
$\lambda$1302.17 absorption from the Lyman-limit absorber at $z_{\rm
abs}$ = 0.08092 discussed by Jenkins et al. (2005); the narrow core is
due to the O~I line. The smooth curve overplotted on the
O~VI line is the Voigt-profile fit from Table~\ref{compprop}.
Using the O~VI column density from the Voigt-profile fit, we
have determined the corresponding column density and $b-$value that
the H~I would have if it were in CIE with $Z = 0.1 Z_{\odot}$
and log $T$ = 5.30, 5.48, 5.60, 5.70, 5.78, and 5.84.  The predicted
Ly$\alpha$ profiles for the different assumed temperatures are
represented by different line shades as indicated in the key in the
middle panel. For purposes of discussion, we show the models for all
of these temperatures in the middle panel, but we only plot the models
with log $T$ = 5.30, 5.48, and 5.60 in the left and right panels.  In
this figure, the data have been binned to 7 km s$^{-1}$ pixels in
order to better reveal any weak structure in the wings of the
profiles.
\label{hiddenbla}}
\end{figure*}

And, there is some evidence that broad shallow Ly$\alpha$ components
could be present in the the high$-N$(\ion{H}{1}), multiphase
absorbers.  Figure~\ref{h1821complex} presents an example.  This
figure shows the H1821+643 \ion{O}{6} absorbers at $z_{\rm abs}$ =
0.22496 and 0.22638 (plotted in the $z_{\rm abs}$ = 0.22638
frame). Focusing first on the $z_{\rm abs}$ = 0.22497 system, we see
that this is a multiphase metal system that is similar to the
PKS0312-770 absorber shown in Figure~\ref{speccontrast}. Comparing the
\ion{C}{3}, \ion{Si}{3}, and \ion{O}{6} lines, we see that the
\ion{C}{3} and \ion{Si}{3} profiles show three narrow components that
are well-aligned.  In contrast, the \ion{O}{6} profiles are smoother
and clearly have different component structure.  These profile
differences are not due to thermal broadening and different species
masses because carbon and oxygen have similar masses.  Rather, these
differences are due to multiple phases in this absorber.  Now, turning
to the \ion{H}{1} Lyman series lines, we see that the two main,
high$-N$(\ion{H}{1}) components are well-constrained by the higher
series lines.  However, our original \ion{H}{1} fit (in
Table~\ref{compprop}) did not result in any broad Ly$\alpha$
components, nor did it indicate any components that are aligned with
the centroid of the \ion{O}{6} absorption.  Following our usual
procedure, the original fit allowed all component parameters ($v, b,$
and $N$) to freely vary during the fitting process.  Interestingly, if
we lock the velocity centroid of {\it only one} of the components to
equal the centroid of the strongest \ion{O}{6} component (and we allow
all the other parameters to vary freely), we obtain a comparably good
fit.  However, the locked component results in a {\it broad Ly$\alpha$
component} with $b$(\ion{H}{1}) = 81$^{+13}_{-11}$ km s$^{-1}$ and log
$N$(\ion{H}{1}) = 13.80$\pm$0.09 at the \ion{O}{6} velocity.  The two
fits to the \ion{H}{1} lines are shown in Figure~\ref{h1821complex}
with a red line and a dotted green line, and the broad Ly$\alpha$ line
resulting from the locked$-v$ fit is indicated with a thick green line
in the upper-left panel of Figure~\ref{h1821complex}.  We can see that
the two fits are nearly indistinguishable, and the fits have nearly
identical goodness of fit ($\chi _{\nu} ^{2}$ = 1.030 vs. $\chi _{\nu}
^{2}$ = 1.031). Between the two fits, the main two components (i.e.,
the two components evident in the higher Lyman series) change by tiny
amounts; the changes occur in the weaker components required to fit
the structure in the wings of the Ly$\alpha$ and $\beta$ profiles.
However, these changes in the weak outlying components have major
implications for the gas conditions. If we use the \ion{H}{1}
$b-$value from the locked fit together with the main-component
\ion{O}{6} $b-$value, these measurements imply that the temperature of
the \ion{O}{6} -bearing gas is log $T$ = 5.47.  Assuming the
\ion{H}{1} and \ion{O}{6} ion fractions from collisional ionization
equilibrium at this temperature, the measured column densities would
then imply an oxygen abundance of [O/H] = $-1.4$.

These examples show that broad \ion{H}{1} components could be present
in the complex profiles, and in some cases, the combination of broad
\ion{H}{1} and \ion{O}{6} is consistent with WHIM temperatures and
relatively low metallicities.  These results are intriguing, but we
must recognize that this only shows that hot gas is {\it allowed} by
the data; these analyses do not uniquely require the presence of WHIM
material.  Without additional information, it can be difficult to
discriminate between the fits shown in these examples.  As noted, high
S/N and spectral resolution would be helpful for constraining the
physical conditions in these absorbers.  Also, ancillary species such
as \ion{Ne}{8} can be immensely valuable for identification of hot gas
(e.g., Savage et al. 2005).  X-ray absorption lines would also be
useful, but the spectral resolution available in the X-ray bandpass is
still quite limited.

\subsubsection{Hidden Broad \ion{H}{1} Components?\label{hiddenblasec}}

Figure~\ref{h1821complex} demonstrates another notable characteristic
of multiphase absorbers.  We have remarked above that multiphase
systems have larger velocity offsets between the \ion{H}{1} and
\ion{O}{6} centroids (see Figure~\ref{veloffsets}).  The $z_{\rm abs}$
= 0.22638 absorber in Figure~\ref{h1821complex} shows an example of
one of the largest velocity offsets between \ion{H}{1} and \ion{O}{6}
in an intervening absorber: in this system, the \ion{O}{6} is detected
$v = 0$ km s$^{-1}$ while \ion{H}{1} and weak \ion{C}{3} are found at
$v = -54$ km s$^{-1}$.  Some \ion{H}{1} absorption is present in the
velocity range where \ion{O}{6} is detected, but this could be just
the wing of the \ion{H}{1} line at $-54$ km s$^{-1}$.  Why does this
well-detected \ion{O}{6} absorber have so little affiliated
\ion{H}{1}?  One possibility is that this absorber is similar to the
proximate absorbers and has a very high metallicity so that the
affiliated \ion{H}{1} column is too low to be detected.

However, another possibility is that this particular \ion{O}{6}
doublet arises in collisionally ionized hot plasma with a temperature
that makes the \ion{H}{1} difficult to detect.  The measured
\ion{O}{6} $b-$value is $16\pm 2$ km s$^{-1}$, which is consistent
with gas at $T \approx 10^{5.5}$.  The left panel of
Figure~\ref{hiddenbla} shows theoretical \ion{H}{1} Ly$\alpha$
profiles that would be expected to be affiliated with this \ion{O}{6}
absorber in collisionally ionized gas at log $T$ = 5.30, 5.48, and
5.60 with $Z = 0.1 Z_{\odot}$.  These predicted \ion{H}{1} profiles
use the Gnat \& Sternberg (2007) CIE ion fractions to predict the
\ion{H}{1} column density (based on the measured $N$(\ion{O}{6}) and
assumed metallicity), and the line widths are for the purely thermally
broadened case (nonthermal broadening can only make the \ion{H}{1}
lines broader).  We see that with log $T$ = 5.30, a strong (and easily
detected) \ion{H}{1} line is expected (red line in
Figure~\ref{hiddenbla}).  The \ion{H}{1} line is strong because even
though the \ion{H}{1} ion fraction is low at this $T$, the \ion{O}{6}
ion fraction drops very rapidly at log $T < 5.5$ (see
Figure~\ref{gnatfracs}), and consequently, $N$(\ion{H}{1}) has to be
high in order to produce the observed $N$(\ion{O}{6}).  Of course,
$N$(\ion{O}{6}) can be increased by increasing $Z$ and/or
$N$(\ion{H}{1}); an alternative solution is that $Z \gg 0.1
Z_{\odot}$.  However, we see from Figure~\ref{hiddenbla} that if log
$T$ = 5.48 or 5.60, close to the temperature of peak \ion{O}{6}
abundance in CIE, then the affiliated $N$(\ion{H}{1}) is low enough so
that the expected Ly$\alpha$ profile (yellow or green lines in
Figure~\ref{hiddenbla}) would be difficult or impossible to detect at
the current S/N levels.  Thus, this system could be in the warm-hot
range with a realistic metallicity, and the expected broad \ion{H}{1}
would be hidden in the noise.

This raises a question: could similar ``hidden'' broad components be
present in the aligned components discussed in \S \ref{tempsec}? The
middle panel of Figure~\ref{hiddenbla} explores this possibility.
This panel shows the well-aligned \ion{H}{1} and \ion{O}{6} absorption
lines at $z_{\rm abs}$ = 0.24531 in the spectrum of H1821+643 with
theoretical Ly$\alpha$ profiles, calculated as described in the
previous paragraph, for log $T$ = 5.30, 5.48, 5.60, 5.70, 5.78, and
5.84.  We see that none of these model Ly$\alpha$ profiles fit the
observed Ly$\alpha$ line very well.  However, if this is a multiphase
absorber with a narrow core from a lower-ionization phase that does
not contribute to the \ion{O}{6} absorption, then a broad component
would be allowed by the data as long as the temperature is in a narrow
range (or $Z \gg 0.1 Z_{\odot}$).  We see that as above, warm-hot gas
with log $T \approx 5.30$ or less is ruled out.  Likewise, models with
log $T \gtrsim$ 5.6 are ruled out; these models lead to Ly$\alpha$
profiles that are too strong and too broad.  We conclude that this
aligned \ion{O}{6} $-$ \ion{H}{1} absorber could harbor a hidden broad
Ly$\alpha$ component arising in the hot \ion{O}{6} phase, but only if
the absorber is multiphase and the temperature is fine-tuned to a
relatively narrow range near the \ion{O}{6} peak temperature and/or
the metallicity is relatively high.

We noted in \S \ref{tempsec} that some of the aligned absorbers are
outside of the physically allowed range of $b$(\ion{H}{1}) and
$b$(\ion{O}{6}).  Occasionally, these unphysical cases could be due to
coincidental alignment with unrelated absorption lines.  The right
panel of Figure~\ref{hiddenbla} shows an example of coincidental
alignment with unrelated absorption in the \ion{O}{6} absorber at
$z_{\rm abs}$ = 0.15786 toward PHL1811.  In this case, an apparently
very narrow \ion{H}{1} line seems to be aligned with the strongest
component of a multicomponent \ion{O}{6} profile, and the candidate
\ion{H}{1} line is too narrow (compared to \ion{O}{6}) to be in the
physically allowed regime.  This unphysical combination is due to
coincidental alignment of the \ion{O}{6} with an unrelated line: the
narrow component in this case is actually not \ion{H}{1} but rather is
the \ion{O}{1} $\lambda$1302.17 line from the Lyman limit absorber at
$z_{\rm abs}$ = 0.08092 (see Jenkins et al. 2005).  However, close
inpection of the profile reveals weak absorption in the wings of the
narrow \ion{O}{1} feature.  Those weak absorption wings do not appear
to be due to \ion{O}{1} because that component structure is not
corroborated by stronger metal-line profiles of other low ions at the
\ion{O}{1} redshift.  The weak components could be due to \ion{H}{1}
Ly$\alpha$ absorption at the \ion{O}{6} redshift. Interestingly, if
the two \ion{O}{6} components at $z_{\rm abs}$ = 0.15786 have log $T$
= 5.48 and $Z \approx 0.1 Z_{\odot}$, then the two expected broad
Ly$\alpha$ components associated with the \ion{O}{6} would explain the
weak components in the wings of the \ion{O}{1} profile (see the yellow
line in Figure~\ref{hiddenbla}).

Hidden broad Ly$\alpha$ lines are allowed for most of the intervening
\ion{O}{6} absorbers.  As shown in columns 4 and 5 of
Table~\ref{alignedtab}, the \ion{O}{6} line widths usually allow
temperatures near the \ion{O}{6} peak (log $T \approx$ 5.5), and given
the typical $N$(\ion{O}{6}), at this temperature an affiliated broad
Ly$\alpha$ component would usually be hidden in the noise.  This
explanation requires that essentially all of the intervening
\ion{O}{6} absorbers are multiphase because another (\ion{O}{6}-free)
phase is required to explain the rest of the Ly$\alpha$ absorption.
For many of the absorbers, the temperature or metallicity must be
tuned to avoid producing overly strong broad Ly$\alpha$.  However,
some of the strong multiphase systems have broad features that could
be due to \ion{O}{6} at somewhat lower temperatures (e.g., Tripp et
al. 2001).  This scenario is similar to the favored models for the
\ion{O}{6} HVCs seen in the vicinity of the Milky Way (Sembach et
al. 2003).

\subsubsection{\ion{O}{6} in Interface Layers?}

The correlation of log [$N$(\ion{H}{1})/$N$(\ion{O}{6})] with log
$N$(\ion{H}{1}) could be an indication that the \ion{O}{6} absorption
arises in some type of interface layer on the surface of a
lower-ionization \ion{H}{1} cloud.  In this argument, the \ion{O}{6}
column density is determined by the interface physics and is
independent of the \ion{H}{1} column density of the interior cloud.
This scenario is not required by the data: we have shown in
Figure~\ref{photoschaye} and \S \ref{photoionsec} that photoionization
models can explain the correlation if the IGM metallicity and UV
background shape and intensity are variable.  However, we have also
shown in the previous section that the current data cannot rule out
hidden broad Ly$\alpha$ lines that would indicate that the \ion{O}{6}
arises in a hot phase that is closely aligned in velocity with a
cooler lower-ionization phase that produces the bulk of the detected
\ion{H}{1} absorption.  

To evaluate this hypothesis, it is useful to consider {\it FUSE}
observations of \ion{O}{6} absorption detected in the very local
interstellar medium (LISM) within a few hundred parsecs of the Sun
(Oegerle et al. 2005; Savage \& Lehner 2006).  In these samples, the
short pathlength to the background star ($d \leq$ 230 pc) precludes
production of \ion{O}{6} by photoionization.  In addition, the Sun is
surrounded by a bubble of hot, X-ray emitting gas with $T \approx
10^{6}$ K (e.g., Smith et al. 2007), so this context provides
observational information on the properties of \ion{O}{6} absorption
arising in interface layers.

Interestingly, Savage \& Lehner (2006) find that find $\approx$40\% of
the \ion{O}{6} absorption line velocity centroids in the LISM are
aligned with \ion{C}{2} absorption centroids to within $\pm$ 10 km
s$^{-1}$ (see their Figure 11).  This supports the hypothesis that
\ion{O}{6} could originate in an interface without having different
kinematics (velocity centroids) compared to the low-ionization phase.
However, Savage \& Lehner also find that more than half of their local
\ion{O}{6} lines show clear and significant velocity offsets.
Overall, the velocity offsets of the LISM \ion{O}{6} absorbers appear
to be similar to those observed in the extragalactic \ion{O}{6}
systems of this paper.  However, there is a significant difference
between LISM \ion{O}{6} absorbers and the extragalactic systems: the
LISM \ion{O}{6} column densities are substantially lower.  Savage \&
Lehner find 12.38 $\leq $log $N$(\ion{O}{6}) $\leq$ 13.60 in the LISM
with a median of log $N$(\ion{O}{6}) = 13.10.  These observed LISM
columns are comparable to the $N$(\ion{O}{6}) values theoretically
predicted to be found in conductive interfaces between cool clouds and
hot gas (e.g., Slavin 1989; Borkowski et al. 1990), but the
extragalactic $N$(\ion{O}{6}) measurements are up to $\approx$50 times
higher.  Therefore, {\it many interfaces are required} to explain the
QSO \ion{O}{6} systems according to these interface models and LISM
observations.  This problem is exacerbated by the fact that the
extragalactic absorbers are likely to have significantly lower
metallicities than the LISM systems; lower metallicity will further
reduce the $N$(\ion{O}{6}) expected in an interface.  With the
requirement of many interfaces to produce the observed
$N$(\ion{O}{6}), we would expect to find complexity in the absorption
profiles (due to variations of the \ion{H}{1}/\ion{O}{6} ratios and
\ion{H}{1} vs. \ion{O}{6} kinematics from one cloud to the next within
the multilayer system).  Some of the QSO \ion{O}{6} profiles show
considerable complexity, but others appear to be relatively simple.
Given this problem, it is not clear if interface models can fit the
properties of the simple extragalactic systems.

However, we note that other models of interstellar \ion{O}{6}
production (see Table 1 in Indebetouw \& Shull 2004), such as
radiatively cooling clouds or bubbles blown by supernovae/stellar
winds, do predict \ion{O}{6} columns that are substantially higher.
We also note that in a survey of 148 Milky Way stars, Bowen et
al. (2007) find that the scatter in $N$(\ion{O}{6}) does not decrease
with increasing sight line pathlength, which indicates that the
\ion{O}{6} absorption lines do not arise in clouds with a fixed size
and density.  These factors suggest that collisionally ionized hot
\ion{O}{6} can originate in a variety of contexts with significant
variations in cloud characteristics.  The Milky Way \ion{O}{6} HVCs
(Sembach et al. 2003; Wakker et al. 2003) suffer a similar problem to
the extragalactic systems: the \ion{O}{6} in Galactic HVCs is thought
to arise in interfaces (e.g., Sembach et al. 2003; Ganguly et
al. 2005), but the observed HVC \ion{O}{6} columns tend to be
substantially greater than the expected $N$(\ion{O}{6}) in
interfaces. We conclude that the interface physics may not yet be
sufficiently well-understood or is oversimplified in available models.
While there are some significant discrepancies, the interface
hypothesis for the origin for extragalactic \ion{O}{6} absorbers is
viable and deserves further study.

\section{Discussion and Future Observations\label{discsec}}

\subsection{Intervening Absorbers}

\subsubsection{Baryonic Content and Missing Baryons \label{baryonicsec}}

What do these analyses imply about the baryonic content of the low$-z$
IGM and the missing baryons? Several previous papers have shown that
low-redshift \ion{O}{6} absorbers harbor roughly 5\% of the baryons in
the nearby Universe (Tripp et al. 2000,2006b; Savage et al. 2002;
Sembach et al. 2004; Danforth \& Shull 2005; Lehner et al. 2006).
Most of these papers have assumed that the mean metallicity of the
\ion{O}{6} systems is $Z_{\rm O~VI} = 0.1 Z_{\odot}$ and that the
\ion{O}{6} ion fraction ($f_{\rm O~VI} =$ \ion{O}{6}/O$_{\rm total}$)
is less than $\approx 0.2$.  The latter assumption is secure (see,
e.g., the Appendix in Tripp \& Savage 2000), but as noted by Lehner et
al. (2006) and as discussed above, many recent studies have found $Z >
0.1 Z_{\odot}$ in nearby IGM absorbers.  The absorber baryonic content
$\Omega _{\rm b}$(\ion{O}{6}) $\propto Z_{\rm O~VI} ^{-1} f_{\rm O~VI}
^{-1}$, so increased metallicity decreases the baryonic content by the
same factor.  The uncertain metallicity of the absorbers is currently
a large source of uncertainty in $\Omega _{\rm b}$(\ion{O}{6}).  New
measurements of $\Omega _{\rm b}$(\ion{O}{6}) should employ detailed
metallicity constraints for each absorber in the sample instead of
assuming a mean $Z_{\rm O~VI}$, or at least the assumed value for
$Z_{\rm O~VI}$ should be based on a sample of measured
metallicities. Unfortunately, many of the \ion{O}{6} systems are only
detected in \ion{O}{6} and \ion{H}{1}.  With only these two species,
the metallicity of the systems remains uncertain.  Better constraints
on $\Omega _{\rm b}$(\ion{O}{6}) will require secure measurements of
other metals in the \ion{O}{6} absorbers.

However, another problem with $\Omega _{\rm b}$(\ion{O}{6}) estimates
is that most previous studies have included {\it all} intervening
\ion{O}{6} absorption systems in the $\Omega _{\rm b}$(\ion{O}{6})
measurement regardless of the physical conditions of the
absorbers. The contribution to the baryon budget from low$-z$
photoionized gas has been separately estimated from studies of
Ly$\alpha$ lines (e.g., Penton et al. 2004; Lehner et al. 2007) and is
often included in baryon inventories (e.g., Fukugita et al. 1998), so
by including photoionized and collisionally ionized \ion{O}{6} systems
together in a single estimate of $\Omega _{\rm b}$(\ion{O}{6}), some
double-counting of baryons could occur.  To avoid this double
counting, $\Omega _{\rm b-Total}$(\ion{O}{6}) should be split into
$\Omega _{\rm b-WHIM}$(\ion{O}{6}) and $\Omega _{\rm
b-Photo}$(\ion{O}{6}), the baryonic content of WHIM and photoionized
\ion{O}{6} systems.  Our physical conditions analysis indicates that
at least 34\% of the intervening \ion{O}{6} components present
compelling evidence of cool temperatures (log $T <$ 5.0).  However, we
also found in \S \ref{tempsec} that at least 14\% of the \ion{O}{6}
components have log $T > 4.6$, which is hotter than the temperatures
typically expected in photoionized intergalactic gas.

The intermediate-temperature cases with log $T >$ 4.6 could arise in
plasma that was initially shock-heated and is now rapidly radiatively
cooling.  Shock-heated gas with $T < 10^{5}$ K is predicted by some
theoretical models, so these intermediate-temperature \ion{O}{6}
systems could be WHIM material.  Kang et al. (2005) report that in
their hydrodynamic models of cosmological structure growth, many
sheet-like structures are shock-heated and contain material that
conceptually should be classified as ``WHIM'' plasma, but in their
models, the shock velocities are lower, and they conclude that an
important fraction of the baryons are in the shocked IGM but with $T <
10^{5}$ K, which they refer to as the ``low-temperature'' WHIM.  Kang
et al. (2005) analyze the physical conditions and ionization of the
low-temperature WHIM, and they find that this IGM phase is
predominantly photoionized.  Moreover, they note that $\approx 60$\%
of the \ion{O}{6} absorbers originate in the low-temperature,
photoionized WHIM in their simulations.  The models of Kang et
al. (2005) are consistent with the observational constraints that we
have derived in this paper.

We have also noted that the \ion{O}{6} absorbers that provide the
strongest evidence of $T > 10^{5}$ K plasma tend to be complex,
multiphase absorbers with multiple \ion{H}{1} components (some of
which do not have associated \ion{O}{6}) and (in some cases) clear
absorption by low ionization stages.  In these multicomponent,
multiphase systems, it is easy to hide a broad \ion{H}{1} component
that would be associated with hot \ion{O}{6}.  We classify 53\% of the
intervening systems as complex/multiphase absorbers, and these complex
systems could be the primary harbor of the $T > 10^{5}$ K plasma.  In
fact, some of the simulations show that $T > 10^{5}$ K plasma is often
surrounded by cooler gas, and these simulations predict these
``embedded'' WHIM locations produce peaks in \ion{O}{6} density maps
(see, e.g., Figures 4-6 in Cen \& Ostriker 2006).  It seems likely
that random sight lines through the cosmological simulations would
often find that the WHIM is located in regions where cooler gas phases
are also present, and consequently \ion{O}{6} from these regions will
be affiliated with complex, multicomponent \ion{H}{1} and lower-ion
profiles at similar velocities, like some of the observed multiphase
\ion{O}{6} systems.  Theoretical predictions regarding the
characteristics of \ion{O}{6} and \ion{H}{1} profiles in various
environments would be helpful for checking the consistency of the
models and the observations.

It should also be borne in mind that the detectability of \ion{O}{6}
depends on the metallicity of the gas.  If the missing baryons are
predominantly located in gas that is shock-heated when it first
accretes into galaxy potentials, then the WHIM plasma could have a
relatively low metallicity, and \ion{O}{6} could be difficult to
detect in the WHIM phase in currently available data.  Instead of
\ion{O}{6}, broad \ion{H}{1} Ly$\alpha$ lines (BLAs) can be used to
search for the low$-z$ WHIM, and several studies have reported
detections of BLA candidates at low redshifts (Tripp et al. 2001;
Bowen et al. 2002; Richter et al. 2004,2006a,b; Sembach et al. 2004;
Lehner et al. 2007).  Only a small fraction of the reported BLAs are
also detected in \ion{O}{6} (the \ion{O}{6} doublet is covered but not
detected in most cases).  We can show that the lack of affiliated
\ion{O}{6} absorption is consistent with a WHIM origin for the
BLAs. The approximate carbon metallicity of the high-redshift IGM is
[C/H] $\lesssim -2.5$ (Schaye et al. 2003).  If the low-redshift WHIM
has a similar metallicity, then using the \ion{O}{6} and \ion{H}{1}
ion fractions from Gnat \& Sternberg (2007), we find that
$N$(\ion{O}{6})/$N$(\ion{H}{1}) $\lesssim 0.25$.  Since most low$-z$
BLAs have $N$(\ion{H}{1}) $\lesssim 10^{14}$ cm$^{-2}$ (Lehner et
al. 2007), in collisionally ionized plasma we predict that the
corresponding $N$(\ion{O}{6}) $\lesssim 10^{13.4}$ cm$^{-2}$, which is
comparable to the lowest \ion{O}{6} columns that we are currently able
to reliably detect.  Since the WHIM metallicity could be substantially
lower than $-2.5$, and because $N$(\ion{O}{6})/$N$(\ion{H}{1})
$\lesssim 0.25$ is derived at the peak of the \ion{O}{6}/\ion{H}{1}
vs. temperature curve, the \ion{O}{6}/\ion{H}{1} ratios could be
substantially lower in BLAs.  Thus, it may not be surprising that
\ion{O}{6} is not usually detected in the BLAs.  In a recent study of
the low$-z$ \ion{H}{1} lines detected in STIS echelle spectra, Lehner
et al. (2007) estimate that at least $\approx 20$\% of baryons could
be found in broad Ly$\alpha$ absorbers.

\subsubsection{Future Studies of Intervening Absorbers\label{intfuturesec}}

\begin{figure}
\centering
    \includegraphics[width=9.0cm, angle=0]{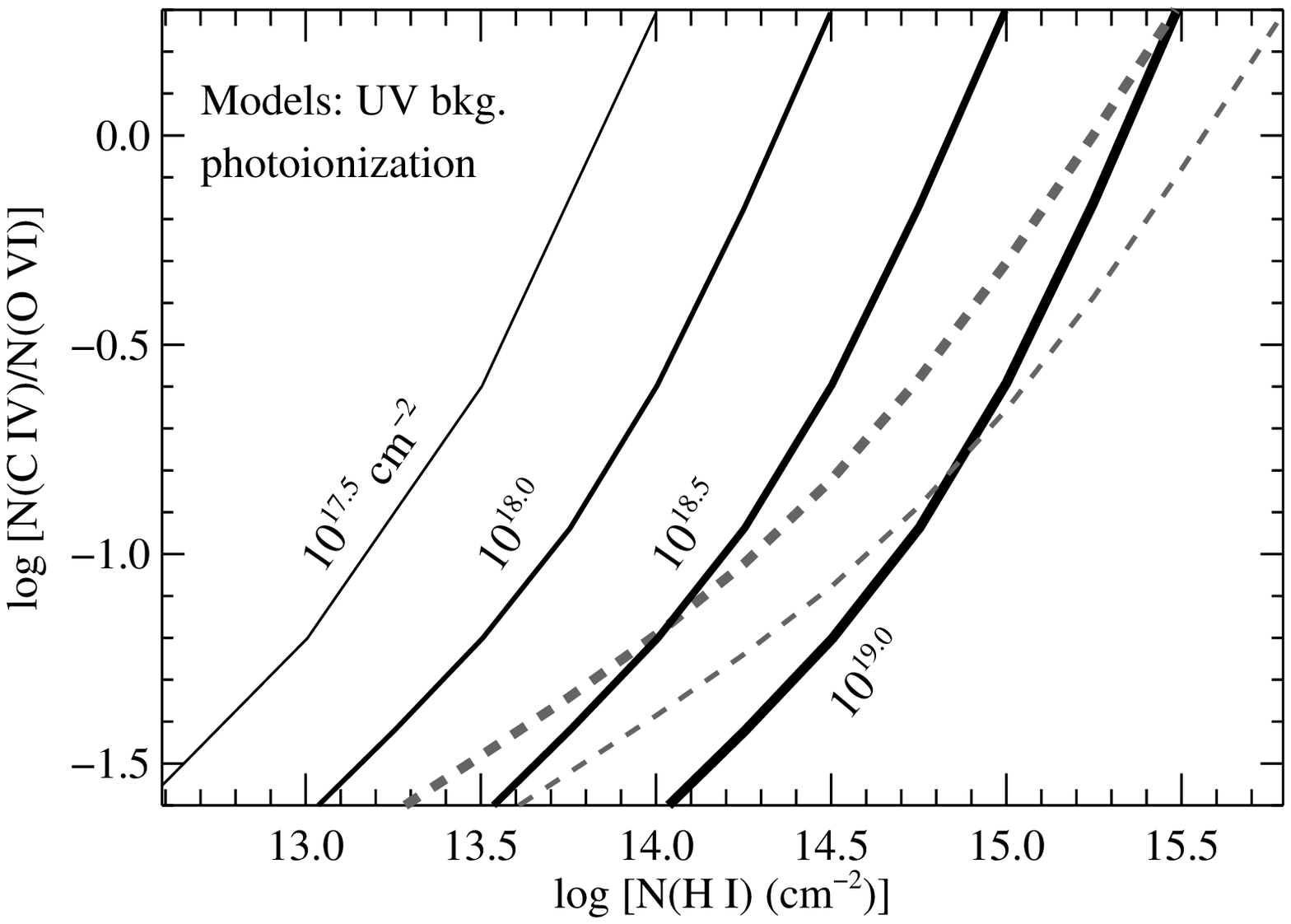}
\caption{Predicted C~IV/O~VI vs. $N$(H~I) from the
same photoionization models presented in Figure~\ref{photmodelc3o6}
with the same line codes. \label{photmodc4o6}}
\end{figure}

Several future observations would be valuable for gaining insight
about the low$-z$ IGM.  First, it will be important to obtain {\it
high signal-to-noise} spectra.  In addition to revealing weaker
\ion{O}{6} and ancillary metal lines, higher S/N ratios will help to
establish whether the BLAs are truly single-component, smooth and
broad Gaussian features.  Current data often {\it suggest} that
\ion{H}{1} lines are BLAs, but due to modest S/N, it remains possible
that these BLA candidates are actually unrecognized blends of several
adjacent components (see, e.g., Richter et al. 2006b).  With higher
S/N, it will often be possible to recognize profile asymmetries that
reflect multiple, blended components.  High S/N is also crucial for
probing the presence of hidden broad Ly$\alpha$ components that could
be associated with collisionally ionized \ion{O}{6} (\S
\ref{hiddenblasec}). Second, it will be useful to detect other
species, including \ion{O}{3}, \ion{O}{4}, and \ion{O}{5} as well as
ions of other elements, to enable more constrained modeling of the
ionization and metallicity of the absorbers.  We have discussed the
utility of \ion{C}{3} in this regard.  The \ion{Ne}{8} $\lambda
\lambda$770.41, 780.32 doublet is an important doublet for future
studies since it probes plasma with $T \approx 6 \times 10^{5}$ K
(Savage et al. 2005), but with the {\it HST} bandpass, this will only
be accessible in intermediate-redshift absorbers ($z_{\rm abs}
\gtrsim$ 0.50). The \ion{C}{4} $\lambda \lambda$1548.21, 1550.78
doublet will be an important metal to search for in many \ion{O}{6}
systems.  Figure~\ref{photmodc4o6} shows the \ion{C}{4}/\ion{O}{6}
ratios predicted by the various photoionization models considered in
\S \ref{photoionsec}.  From this figure, we see that the
\ion{C}{4}/\ion{O}{6} ratio is usually significantly greater than the
corresponding \ion{C}{3}/\ion{O}{6} ratio in photoionized \ion{O}{6}
systems.  The \ion{N}{5} $\lambda \lambda$1238.82, 1242.80 doublet is
also useful and has been occasionally detected (e.g., Savage et
al. 2002; Tumlinson et al. 2005). The \ion{N}{5} doublet is weaker and
more difficult to detect, but this can be overcome with high S/N
observations.  A greater difficulty with nitrogen is that it can be
underabundant due to nucleosynthesis effects in low-metallicity gas
(e.g., Henry et al. 2000), which can cause confusion in ionization
models.  Bearing such effects in mind, with measurement of two or more
metals as well as \ion{H}{1}, it will be possible to place specific,
precise constraints on the physical conditions and metal enrichment of
the low$-z$ IGM.

More detailed comparisons between the observations and theoretical
predictions from cosmological simulations would also be instructive.
The Kang et al. simulations predict that a large fraction of the
low$-z$ \ion{O}{6} absorbers should arise in relatively cool gas, but
it should be noted that other simulations have also indicated that
some fraction of the \ion{O}{6} absorption occurs in predominantly
photoionized gas (e.g., Cen et al. 2001; Fang \& Bryan 2001; Chen et
al. 2003). In the simulations, how do the \ion{O}{6} and \ion{H}{1}
profiles compare in photoionized regions?  Conversely, what are the
characteristics of the \ion{O}{6} and \ion{H}{1} profiles in the
warm-hot, shocked regions in the simulations?  For example, is the
WHIM expected to be typically found in close proximity to cooler gas
thereby resulting in complex, multiphase profiles?  Another question
is whether the similarity of many of the observed \ion{O}{6} and
\ion{H}{1} profiles could arise in multiphase gas in which the
\ion{O}{6} traces a hotter phase and the \ion{H}{1} arises in cooler
gas.  A wind-blown bubble-like feature might have hot \ion{O}{6} in an
interface just inside the bubble edge and cool \ion{H}{1} around the
outer periphery; could this configuration result in well-matched
\ion{O}{6} and \ion{H}{1} profiles?  Intuitively, differences in
factors such as temperature and turbulence in the hot vs. the cold gas
would be expected to impart different kinematic characteristics to the
\ion{O}{6} and \ion{H}{1} in this situation.  Detailed assessments of
the \ion{O}{6} and \ion{H}{1} characteristics in theoretical
simulations would provide valuable insight for interpreation of the
\ion{O}{6} absorbers.

\subsection{Proximate Absorbers}

\subsubsection{Evidence of AGN Feedback?}

One of the goals of our survey is to search for evidence of feedback
from outflows driven by QSOs/AGNs (see \S 1).  Some papers have argued
that a significant fraction of QSO absorption lines with $v_{\rm
displ} > 5000$ km s$^{-1}$ are actually intrinsic QSO absorption
systems that have been ejected by the QSO/AGN and accelerated to
dramatic outflow velocities (\S \ref{absclass}).  One of the QSOs in
our sample (3C 351.0) does reveal a complex, multicomponent \ion{O}{6}
system with 13 \ion{O}{6} components spread over $\approx 2800$ km
s$^{-1}$ at $z_{\rm abs} \approx z_{\rm QSO}$ (Yuan et al. 2002).
This complex proximate absorber, which does show evidence of partial
covering of the QSO continuum and broad emission line sources (see
Yuan et al.), is probably best described as a mini-BAL. Apart from
this particular system, we do not find any evidence of high-speed QSO
outflows in our sample.  We do see some clear differences between
\ion{O}{6} absorbers at $v_{\rm displ} < 5000$ km s$^{-1}$ and those
at $v_{\rm displ} \gg 5000$ km s$^{-1}$: (1) $dN/dz$ for \ion{O}{6}
increases substantially as $z_{\rm abs}$ approaches $z_{\rm QSO}$, (2)
we find a marginal indication, based on the proximate absorber
$b-$value distribution, that proximate \ion{O}{6} lines tend to be
narrower than the intervening \ion{O}{6} lines, and (3) $\approx$40 \%
of the proximate \ion{O}{6} systems have much lower
\ion{H}{1}/\ion{O}{6} ratios than the intervening systems.

\begin{figure}
\centering
    \includegraphics[width=9.0cm, angle=0]{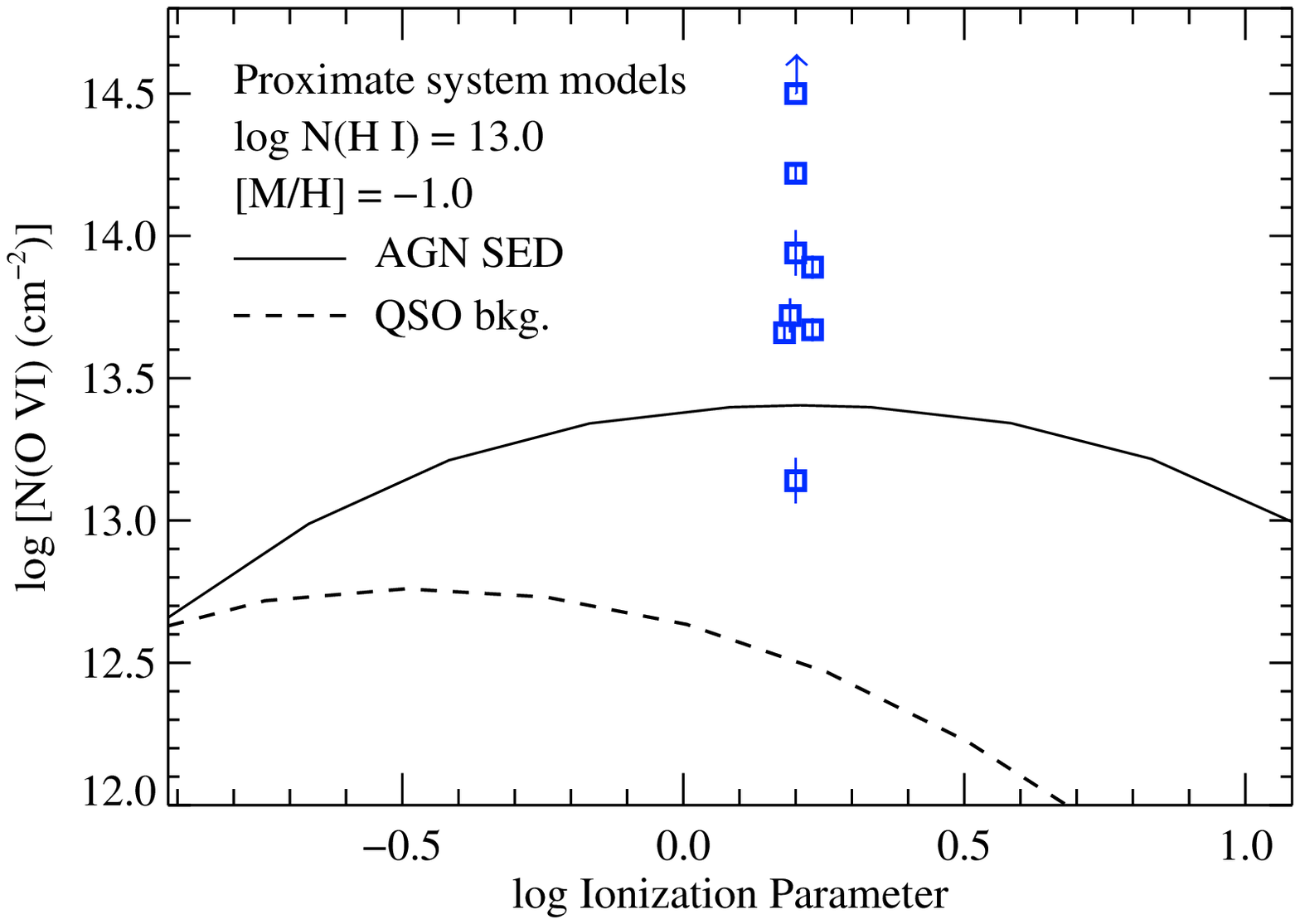}
\caption{Photoionization models with low H~I column densities
and moderately low metallicities for two ionizing radiation fields:
the approximate description of the spectral energy distribution of an
AGN from Mathews \& Ferland (1987) [solid line] and the QSO background
from Haardt \& Madau (2006) [dashed line].  These models have a fixed
H~I column density (log $N$(H~I) = 13.0) and metallicity
($Z = 0.1 Z_{\odot}$), and with these parameters, the predicted
O~VI column is shown as a function of the ionization parameter.
The open squares show the measured $N$(O~VI) values for the
eight proximate system components with log $N$(H~I) $\lesssim
13.1$ (one of the components is shown as a lower limit because the
O~VI column could be underestimated due to an uncertain amount
of line saturation). The open squares are arbitrarily plotted at the
ionization parameter where $N$(O~VI) peaks in the AGN SED model.
Detections of other metals (apart from O~VI) are required to
constrain the appropriate ionization parameter for each observed
absorber. However, moving away from the peak ionization parameter
increases the discrepancy between the observed and predicted
O~VI columns, which reinforces the conclusion that these
absorbers require [M/H] $\gg -1.0$.\label{highassc}}
\end{figure}

The lower \ion{H}{1}/\ion{O}{6} ratios in proximate systems could be
due to ionization effects (due to proximity to the bright backgound
QSO), higher metallicity, or both.  Figure~\ref{highassc} shows the
predicted $N$(\ion{O}{6}) from photoionized models with log
$N$(\ion{H}{1}) = 13.0, [M/H] = $-1.0$, and two different ionizing
radiation fields, the Mathews \& Ferland (1987) approximate description
of an AGN spectral energy distribution and the Haardt \& Madau (1996)
UV background that we have used previously.  We see that the Mathews
\& Ferland AGN SED does increase the amount of \ion{O}{6} predicted
from a low-$N$(\ion{H}{1}) cloud compared to the UV background
expected in the intervening IGM.  However, we also show the measured
$N$(\ion{O}{6}) that we find in eight proximate absorber components
that have log $N$(\ion{H}{1}) $\lesssim$ 13.0, and we see that even
the Mathews \& Ferland model falls short of the observed columns, in
some cases by more than an order of magnitude.  Realizing that most of
the eight cases in Figure~\ref{highassc} provide only upper limits on
$N$(\ion{H}{1}), and realizing that the ionization parameter does not
necessarily have the value that maximizes the predicted
$N$(\ion{O}{6}), we conclude that these proximate absorbers require
metallicities that are much greater than 0.1 $Z_{\odot}$.

Thus, for the subset of the proximate absorbers shown in
Figure~\ref{highassc}, we find possible evidence of feedback in the
form of metal enrichment.  However, the location of these absorbing
clouds is not clear.  If these clouds are located deep inside the QSO
host galaxy close to the QSO central engine, then they would not have
significant effects on galaxy evolution.  Also, the high metallicities
could simply reflect the high metallicity of a host galaxy that is old
enough to have significantly enriched its ISM.  The distance of the
proximate absorbers away from the QSO is a key uncertain parameter
that limits interpretation of these systems.  Some previous studies
have presented evidence that proximate absorption systems can be
several tens to several hundred kpc away from the QSO flux source
(e.g., Morris et al. 1986; Tripp et al. 1996; Hamann et al. 2001), but
other studies have revealed characteristics such as temporal
variability and partial covering that place proximate absorbers quite
close ($d \lesssim 1$ kpc) to the central engine (e.g., Hamann et
al. 1997b; Narayanan et al. 2004).

\subsubsection{Future Studies of Proximate Absorbers}

Several future observations would help to elucidate the nature and
significance of the proximate absorbers.  Monitoring the absorbers
for temporal variability would be valuable to constrain the plasma
densities and locations of the proximate systems.  Searches for
evidence of partial covering would be similarly useful, but this would
likely require high S/N spectra.  However, high S/N observations would
also be valuable for detecting weak \ion{H}{1} and other metal lines,
which would enable metallicity and physical conditions estimates for
specific systems.  Finally, thanks to the huge number of quasars
discovered by large-scale surveys such as SDSS, it has become possible
use background QSOs to probe absorption from gas affiliated with a
foreground, lower-redshift QSO (Bowen et al. 2006; Hennawi et
al. 2006).  These QSOs behind QSOs are found at various impact
parameters, and in principle, they can be used to test whether
proximate absorber properties depend on impact parameter and whether
sight lines directly to the QSOs have special characteristics.  For
example, are the low \ion{H}{1}/\ion{O}{6} ratios also observed in
sight lines to QSOs behind QSOs, or are these low ratios only found
when the foreground QSO itself is the background light source?
Observations of \ion{O}{6} in QSO-QSO pairs could provide insight,
especially at low redshifts where \ion{O}{6} is not badly
contaminated by Ly$\alpha$ forest lines.

\section{Summary\label{sumsec}}

To study the physical conditions, chemical enrichment, and baryonic
content of the low-redshift integalactic medium, we have surveyed the
\ion{O}{6} absorption lines of 16 low$-z$ QSOs using high-resolution
ultraviolet spectra.  We primarily use data obtained with the STIS
E140M echelle mode at 7 km s$^{-1}$ spectral resolution (FWHM), and we
supplement the STIS sample with {\it FUSE} data recorded at $20-25$ km
s$^{-1}$ resolution.  Signal-to-noise ratios (per resolution element)
range from 10 to 33 at $\lambda _{\rm obs}$ = 1300 \AA . We have examined
the properties of the intervening \ion{O}{6} absorbers with $z_{\rm
abs} \ll z_{\rm QSO}$ as well as the proximate \ion{O}{6} absorption
systems with $z_{\rm abs} \approx z_{\rm QSO}$.  From these data, we
obtain the following results:

\begin{enumerate}

\item We identify 51 intervening ($z_{\rm abs} \ll z_{\rm QSO}$)
\ion{O}{6} systems comprised of 77 individual components, and we find
14 proximate systems (within 5000 km s$^{-1}$ of $z_{\rm QSO}$)
containing 34 components. The intervening absorber redshifts range
from $z_{\rm abs}$ = 0.00210 to 0.49508, and the median redshift is
0.213.  For proximate systems, the system redshifts range from 0.15779
to 0.49246, and the median redshift is 0.267.  

\item We report redshifts, column densities, $b-$values, and component
velocity centroids for all \ion{O}{6} and \ion{H}{1} lines in the
identified \ion{O}{6} absorbers.  We also present comparisons of the
apparent column density profiles of \ion{O}{6} and \ion{H}{1} in these
systems.  Some of the absorbers are characterized by complex
kinematics, with comparison of the \ion{O}{6} and \ion{H}{1} profiles
revealing significantly different velocity centroids, line widths, and
different numbers of components within a system.  However, some of the
systems are characterized by relatively simple profile structures, and
moreover, in these cases the \ion{O}{6} and \ion{H}{1} profiles are
often well-aligned in velocity.

\item Based on the kinematics and ionization properties of the
absorption profiles, we classify the absorbers as simple (likely
single-phase) systems or complex/multiphase systems.  The multiphase
systems have a substantially greater spread in the distribution of
velocity offsets between the \ion{H}{1} and \ion{O}{6} centroids
(i.e., $v_{\rm H~I} - v_{\rm O~VI}$).  Transitions of low-ionization
metals are detected in some of the multiphase systems, and in some
cases, comparison of the low-ion and high-ion component structure
shows that the low ions and high ions arise in separate gas phases.

\item For intervening systems (components) with rest-frame equivalent
width $W_{\rm r} >$ 30 m\AA , the number of \ion{O}{6} absorbers per
unit redshift $dN/dz$ = 15.6$^{+2.9}_{-2.4}$ (21.0$^{+3.2}_{-2.8}$),
and this decreases to $dN/dz$ = 0.9$^{+1.0}_{-0.5}$
(0.3$^{+0.7}_{-0.3}$) for $W_{\rm r} >$ 300 m\AA . We also present the
differential $dN/dz$ distributions of the \ion{O}{6} components. We
compare the \ion{O}{6} $dN/dz$ measurements for intervening absorbers
to predictions from hydrodynamic cosmological simulations, and we find
reasonable agreement.

\item The number of \ion{O}{6} absorbers increases dramatically as
$z_{\rm abs}$ approaches $z_{\rm QSO}$; we find that $dN/dz$ is
$\approx 3 - 10$ times higher within 2500 km s$^{-1}$ of $z_{\rm
QSO}$.  While there are known low$-z$ proximate absorbers with $v_{\rm
displ} > 2500$ km s$^{-1}$ that originate in high-velocity QSO
outflows, we find little clear evidence of such AGN high-velocity
outflows in this sample.  The radio-loud QSO 3C 351.0 has a mini-BAL
outflow (Yuan et al. 2002), but apart from this sight line, most of
the proximate/intrinsic absorbers are found within 2500 km s$^{-1}$ of
$z_{\rm QSO}$.

\item The column density and differential $dN/dz$ distributions of the
intervening and proximate \ion{O}{6} absorbers are statistically
indistinguishable, but we find a weak indication that the proximate
absorbers are narrower (lower $b-$values).

\item A more distinguishing feature of the proximate absorbers is that
a subset of these systems has very low \ion{H}{1}/\ion{O}{6}
ratios. Many well-detected proximate \ion{O}{6} systems have little or
no affiliated \ion{H}{1} with log $N$(\ion{H}{1}) $<$ 13.0, which is
much lower than the typical $N$(\ion{H}{1}) detected in intervening
\ion{O}{6} absorbers.  We show that this could be partially due to
ionization effects, but the proximate absorbers must also have higher
metallicities than intervening systems.

\item In the intervening systems, we use the well-aligned \ion{O}{6}
and \ion{H}{1} components to derive constraints on the plasma
temperatures and the non-thermal broadening components of the
absorbers.  In many cases, the good alignment and similar shapes of
the \ion{O}{6} and \ion{H}{1} lines suggests that the \ion{O}{6} and
\ion{H}{1} absorption arises in the same gas.  Assuming that this is
the case (but see summary point 11 below), our analysis shows that the
well-aligned components are dominated by surprisingly cool gas clouds:
62\% of the aligned components indicate temperatures less than
$10^{5}$ K.  Considering the entire robust intervening sample
(including the \ion{O}{6} components that are {\it not} aligned with
\ion{H}{1} components), we place a lower limit on the cold fraction:
we find that $>$30\% of the intervening components have log $T
<$ 5.0.  These components are colder than expected in the canonical
WHIM models computed by hydrodynamic simulations but are consistent
with temperatures predicted for \ion{O}{6} absorbers in some of the
more recent simulations (e.g., Kang et al. 2005; Richter et
al. 2006). However, we also note that 26\% of the well-aligned
components (14\% of the entire robust sample) imply log $T >$
4.6, which is hotter than expected in generic photoionization models.

\item Motivated by the relatively cool temperatures implied by the
\ion{O}{6} and \ion{H}{1} line widths, we compare photoionization and
nonequilibrium (radiatively cooling) collisional ionization models to
the observations.  We find that most of these cool intervening
\ion{O}{6} absorbers are inconsistent with available equilibrium and
non-equilibrium collisional ionization models but are easily
understood if photoionized.  Photoionization can naturally explain the
observed correlation between log [$N$(\ion{H}{1})/$N$(\ion{O}{6})] and
log $N$(\ion{O}{6}) as long as there is variability in the intensity
and shape of the ionizing UV background and the IGM metallicity.  

\item Noting that roughly half of the intervening systems show
evidence of multiple physical phases, often with complex
multicomponent profile structure, we show that these multiphase
systems can accommodate the warm-hot gas.  However, while the data are
consistent with the presence of hot gas, hot gas is not uniquely
required by the data.  In these complex systems, compelling
identification of gas with $T > 10^{5}$ K requires high spectral
resolution and high signal-to-noise data.  Ancillary species such as
\ion{O}{3}, \ion{O}{4}, \ion{O}{5}, \ion{C}{4}, and \ion{Ne}{8} also
provide valuable insight on the physical conditions of these systems.

\item We show that the current data cannot rule out the possibility
that there are broad Ly$\alpha$ components hidden in the noise that
are associated with \ion{O}{6} lines, but this requires that the
\ion{O}{6} absorber temperatures are close to the temperature at which
\ion{O}{6} peaks in abundance in CIE or that the metallicities are high
(significantly lower or higher temperatures, or lower metallicities,
would lead to detectable broad Ly$\alpha$ components that are not
consistent with the data).  This hypothesis also requires that the hot
\ion{O}{6} phase is almost always affiliated with a cooler
low-ionization phase that is required to produce the bulk of the
narrow \ion{H}{1} absorption.  This situation could occur if the
\ion{O}{6} originates in interfaces on the low-ionization cloud
surfaces.  Current interface models have difficulting producing enough
\ion{O}{6} to match the QSO absorbers, but a variety of observations
suggest that the physical processes in interfaces are not yet
adequately understood.

\end{enumerate}

\acknowledgements

We appreciate helpful discussions with Bart Wakker, Jason
X. Prochaska, Sanchayeeta Borthakur, and Hsiao-Wen Chen, and we
especially thank J. Chris Howk and the anonymous referee for comments
that significantly improved this paper.  Several of the STIS
observations employed in this paper were obtained for {\it HST}
program 9184, with financial support through NASA grant HST
GO-9184.08-A.  We also appreciate and acknowledge extensive support
for this research from NASA LTSA grant NNG 04GG73G.  N.L. was also
supported by NASA grant FUSE-NNX07AK09G.  We thank the many people on
the {\it HST} and {\it FUSE} instrument and mission operations teams
for providing the high-quality instrumentation that made this research
possible. This research has made use of the NASA/IPAC Extragalactic
Database (NED), which is operated by the Jet Propulsion Laboratory,
California Institute of Technology, under contract with the National
Aeronautics and Space Administration.  This work is dedicated to the
memory of the first author's young brother Peter M. R. Tripp, who
passed away in an avalanche during the course of this project. His
interest in learning and discovery was inspiring to those who knew
him, and the enthusiasm and encouragement that he always shared
motivated much of the first author's career work.

\begin{figure}
\centering
   \includegraphics[width=8.5cm, angle=0]{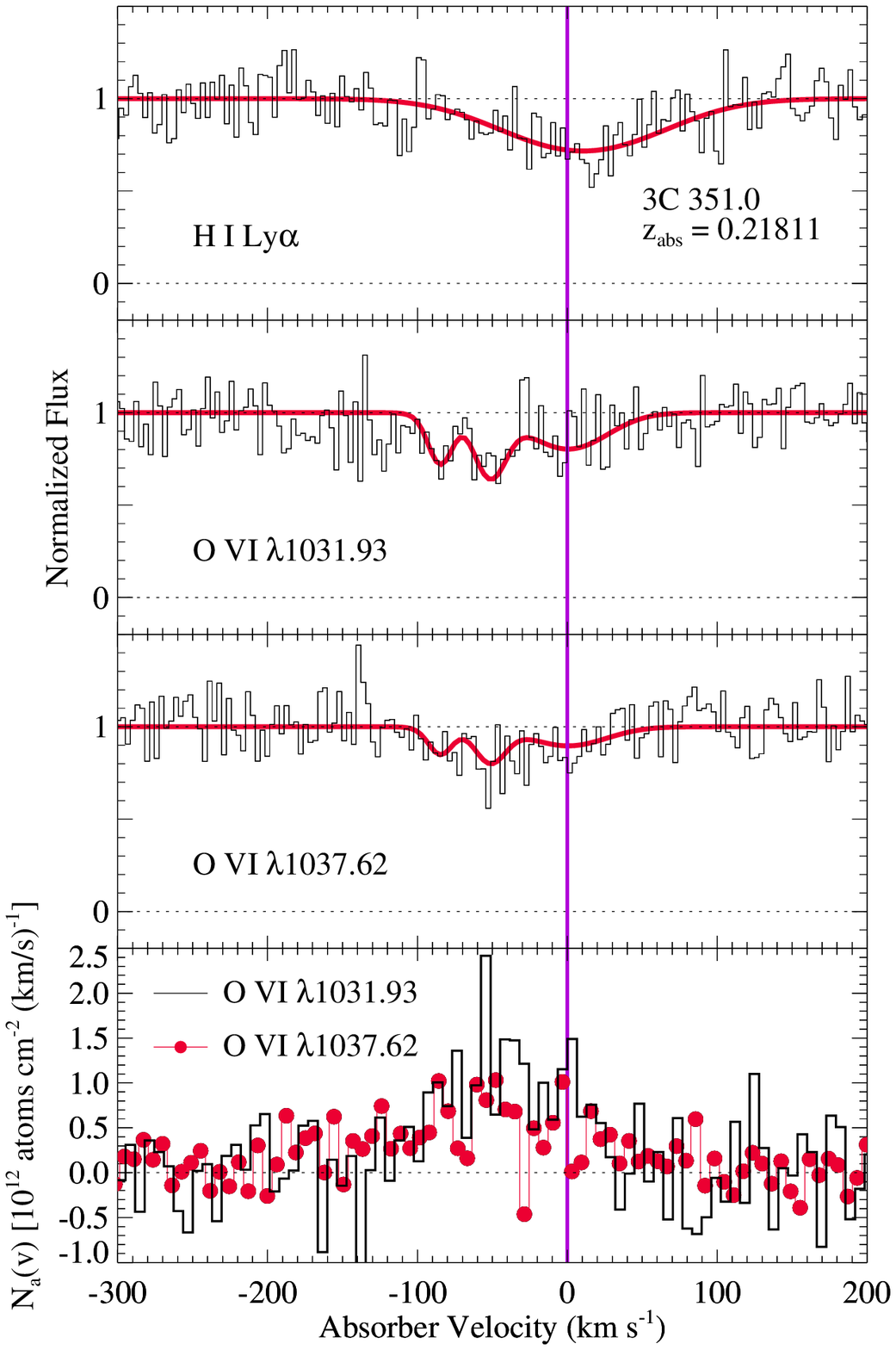}
\caption{Continuum-normalized absorption profiles of the H~I
Ly$\alpha$ and O~VI $\lambda \lambda$1031.93,1037.62 lines at
$z_{\rm abs}$ = 0.21811 in the spectrum of 3C 351.0 (upper three
panels), and comparison of the O~VI $N_{\rm a}(v)$ profiles
(lowest panel). Voigt-profile fits (see Table~\ref{compprop}) are
overplotted with smooth lines. In this figure, the $N_{\rm a}(v)$
profiles are binned to 7 km s$^{-1}$ pixels, but the absorption
profiles are shown at full resolution ($\approx 3.5$ km s$^{-1}$
pixels).\label{3c351_218}}
\end{figure}

\clearpage

\appendix

\section{Appendix: Comments on Line Identification, Blending, Hot Pixel Contamination, and Saturation in Invididual Systems}

(1.) {\it 3C 249.1, $z_{abs}$ = 0.24676.} In this system, \ion{H}{1}
   Ly$\beta$ is mildly blended with an unrelated line.  However, most
   of the Ly$\beta$ profile is free from the blend, and the unblended
   portion of Ly$\beta$ provides useful contraints and was included in
   the fit. Hot pixels are present in the STIS spectrum on both the
   blue and red sides of the \ion{O}{6} $\lambda$1037.62 line (see
   Figure~\ref{exsimstar}).  The \ion{O}{6} identification is secure
   because the $\lambda$1031.93 and $\lambda$1037.62 profiles agree
   well in the regions that are not affected by hot pixels.  However,
   the regions affected by hot pixels were excluded from the fit.

(2.) {\it 3C 249.1, $z_{abs}$ = 0.30811.} Both lines of the \ion{O}{6}
   doublet are detected at high significance at this redshift (see
   Figure~\ref{ascsample1}).  However, the \ion{O}{6} profiles are
   strongly saturated in the component at $v$ = 0 km s$^{-1}$. In
   addition, hot pixels are present in the core of the \ion{O}{6}
   $\lambda 1031.93$ line.  Consequently, the line parameters are
   highly uncertain for the $v$ = 0 km s$^{-1}$ component.

(3.) {\it 3C 249.1, $z_{abs}$ = 0.31364.} The \ion{O}{6}
   $\lambda$1037.62 line at this redshift is only detected at the
   1.9$\sigma$ level. However, the strength of the 2$\sigma$ feature
   is in good agreement with the expected strength implied by the
   well-detected \ion{O}{6} $\lambda$1031.93 line.  In addition, the
   \ion{O}{6} $\lambda$1031.93 line is well-aligned with the
   \ion{H}{1} Ly$\alpha$, Ly$\beta$, and Ly$\gamma$ lines detected at
   this redshift.  The metal profiles marginally suggest the presence
   of a second component, but better S/N is required to verify and
   measure the second component.

(4.) {\it 3C 273.0, $z_{abs}$ = 0.00334.} The blue side of the
   Ly$\beta$ profile is blended with a Galactic H$_{2}$ absorption
   line (see Sembach et al. 2001). Consequently, only the red side of
   the Ly$\beta$ line (which is free from blending) was used to
   constrain the fit.  The \ion{O}{6} $\lambda$1037.62 line is
   severely blended with Galactic H$_{2}$ absorption, so only the
   \ion{O}{6} $\lambda$1031.93 line can be measured.  The \ion{O}{6}
   $\lambda$1031.93 line is affiliated with a well-detected \ion{H}{1}
   absorber at the same redshift (Sembach et al. 2001).  This absorber
   clearly shows evidence of multiple components in the \ion{H}{1}
   Ly$\alpha$ profile (see Figure~\ref{multi00334}).  The \ion{O}{6}
   profile, on the other hand, only shows one clear component.
   However, the \ion{O}{6} line is broad and shallow, and the breadth
   of the \ion{O}{6} feature is consistent with the velocity range
   spanned by the \ion{H}{1} absorption.  The degree of
   velocity-centroid alignment of the \ion{O}{6} with \ion{H}{1} is
   ambiguous: the \ion{O}{6} is aligned, to within the 2$\sigma$
   uncertainty, with both \ion{H}{1} components derived from profile
   fitting in Table~\ref{compprop}.  Following the conventions
   outlined in \S \ref{aligndefsec}, we assign the \ion{O}{6} to the
   broader \ion{H}{1} component.  In addition, as discussed in \S
   \ref{complexmulti}, the breadth of the \ion{H}{1} component at $v =
   -8$ km s$^{-1}$ depends critically on the number of components
   assumed for the fit.

(5.) {\it 3C 273.0, $z_{abs}$ = 0.12003.} At this redshift, the
   \ion{O}{6} doublet is covered by both the STIS and the {\it FUSE}
   spectrum of 3C 273.0.  Both lines of the \ion{O}{6} doublet are
   clearly detected in the {\it FUSE} spectrum of 3C 273.0 (see Figure
   3 in Tripp et al. 2006b).  In the STIS spectrum, the \ion{O}{6}
   lines are located in a region of rapidly decreasing
   signal-to-noise.  The STIS spectrum shows the \ion{O}{6} 1037.62
   line at 4.0$\sigma$ significance, but the significance of the
   1031.93 line is $< 2\sigma$.  Consequently, we base our fit on the
   {\it FUSE} data, but we note that fitting the STIS data yeilds
   consistent (but noisier) results.

(6.) {\it 3C 273.0, $z_{abs}$ = 0.15779.} The \ion{O}{6} 1031.93 line
   is detected at 4.5$\sigma$ significance, but the weaker \ion{O}{6}
   $\lambda$1037.62 line is not detected. The \ion{O}{6}
   identification is favored based on the precise alignment of
   \ion{O}{6} $\lambda$1031.93 candidate with an \ion{H}{1} Ly$\alpha$
   line at the same redshift (see Figure~\ref{navplots}).

(7.) {\it 3C 351.0, $z_{abs}$ = 0.21811.} Figure~\ref{3c351_218} shows
   the \ion{H}{1} Ly$\alpha$ and \ion{O}{6} lines that we detect in
   this system; the upper panels show the absorption profiles and the
   lowest panel compares the \ion{O}{6} $N_{\rm a}(v)$ profiles.  The
   \ion{O}{6} $\lambda \lambda 1031.93,1037.62$ lines are detected at
   the 5.5$\sigma$ and 3.7$\sigma$ levels, respectively, and the
   $N_{\rm a}(v)$ profiles are in good agreement. The \ion{O}{6}
   profiles are broad and shallow and hence are sensitive to continuum
   placement.  Thom \& Chen (2008) do not agree with this system
   identification; the most likely source of this discrepancy is
   continuum placement, but differences in data reduction procedures
   could play a role.  Higher S/N observations with COS would be
   valuable for confirmation of broad and shallow lines such as these.
   Similar component structure is evident in the \ion{O}{6}
   $\lambda$1031.93 and \ion{O}{6} $\lambda$1037.62 profiles, and the
   similarity of the component structure favors a multicomponent
   fit. However, the \ion{O}{6} profiles are moderately noisy.  We
   flag these measurements with a colon because while three components
   are suggested by the \ion{O}{6} data, better S/N is needed to
   robustly establish that three components are present.  If we fit
   the \ion{O}{6} lines with a single component instead of the
   three-component fit listed in Table~\ref{compprop}, we obtain
   $b$(\ion{O}{6}) = 82$\pm$13 km s$^{-1}$ and log $N$(\ion{O}{6}) =
   14.06$\pm$0.05 for the single line.

(8.) {\it 3C 351.0, $z_{abs}$ = 0.22111.} An archival {\it FUSE}
   spectrum of 3C 351.0 shows that this is an optically-thick Lyman
   limit absorber with $N$(\ion{H}{1}) $> 10^{17}$ cm$^{-2}$.  The
   Ly$\alpha$ profile is strongly saturated but shows complex
   structure at the edges of the profile. This structure could be
   partly due to damping wings, but this profile structure cannot be
   unambigously attributed to damping wings.  Consequently, the
   \ion{H}{1} column density is highly uncertain.  The \ion{O}{6}
   $\lambda$1031.93 line is severely blended with the Galactic
   \ion{Si}{2} $\lambda$1260.42 line, and consequently
   $\lambda$1031.93 cannot be measured.  However, many metal lines are
   detected at the redshift of this strong Lyman limit system
   including transitions of \ion{C}{2}, \ion{N}{2}, \ion{Si}{2},
   \ion{Si}{3}, and \ion{Si}{4}.  The \ion{O}{6} $\lambda$1037.62 line
   is identified based on its alignment with the other metals at this
   redshift.  All available \ion{H}{1} lines are strongly saturated at
   the velocities of the metal lines, so the degree of alignment of
   the \ion{O}{6} and \ion{H}{1} lines cannot be evaluated.  However,
   analysis of the low- and high-ionization metals lines indicates the
   presence of multiple phases, so this system is classified as a
   complex absorber.

(9.) {\it H1821+643, $z_{abs}$ = 0.02438.} \ion{H}{1} Ly$\beta$ is
   mildly blended with a Galactic H$_{2}$ absorption line (see Sembach
   et al. 2008), but the Ly$\beta$ line is mostly free from the blend.
   The blended portion of Ly$\beta$ was excluded from the
   fit. \ion{O}{6} $\lambda$1037.62 is lost in a blend with Milky Way
   \ion{Fe}{2} and H$_{2}$ absorption.  The \ion{O}{6} identification
   is based on the precise alignment of \ion{O}{6} $\lambda$1031.93
   with Ly$\alpha$ and Ly$\beta$ lines at the same redshift.

(10.) {\it H1821+643, $z_{abs}$ = 0.12143.} Thom \& Chen (2008)
   challenge this \ion{O}{6} identification, noting that ``there is
   strong absorption at the \ion{O}{6} 1037 position, but no
   \ion{O}{6} 1031, which should be easily detected, given the
   strength of the weaker supposed \ion{O}{6} 1037 line.'' However, as
   discussed in detail by Tripp et al. (2001), the \ion{O}{6}
   $\lambda$1037.62 line is significantly blended with strong
   \ion{H}{1} Ly$\delta$ absorption from the absorber at $z_{\rm abs}$
   = 0.22496 (see Figure 2 in Tripp et al. 2001), and it appears that
   Thom \& Chen (2008) have not taken this serious blend into account.
   Moreover, Thom \& Chen base their conclusions on the STIS data
   only, which have S/N $\lesssim$ 3 per pixel in this wavelength
   range, whereas the {\it FUSE} observations we used have S/N
   $\gtrsim$ 13 per pixel here.  When the {\it FUSE} data are employed
   and the Ly$\delta$ blend is accounted for, we find compelling
   evidence supporting this system.  Because of the strong blend, our
   \ion{O}{6} measurements are based on the 1031.93 line alone.  The
   \ion{O}{6} $\lambda$1031.93 line is detected at the 6.9$\sigma$
   level in our data.  While the blend hampers confirmation based on
   the 1037.63 line, we note that there are no other clear
   identifications for the 6.9$\sigma$ line at the 1031.93 wavelength.
   This is not an \ion{H}{1} Ly$\alpha$ line because the redshift
   places the line blueward of the Ly$\alpha$ region, nor is it a
   higher Lyman series \ion{H}{1} line because corresponding strong
   \ion{H}{1} lines would be obvious in the STIS spectrum but are not
   evident.

(11.) {\it H1821+643, $z_{abs}$ = 0.21331.} The \ion{O}{6}
    $\lambda$1037.62 line is blended with weak, high-velocity
    \ion{S}{2} $\lambda$1259.52 absorption from Milky Way gas (see
    Savage et al. 1995 and Tripp et al. 2003 for information about the
    Galactic high-velocity gas toward H1821+643).  Comparison of the
    Galactic \ion{S}{2} $\lambda$1259.52 and \ion{S}{2}
    $\lambda$1253.81 lines shows that there is excess optical depth in
    the 1259.52 \AA\ line, and the excess is consistent with the
    expected contribution from the \ion{O}{6} $\lambda$1037.62 line at
    $z_{\rm abs}$ = 0.21331 (based on the strength of the unblended
    \ion{O}{6} $\lambda$1031.93 line), which supports the
    identification of \ion{O}{6} at this redshift.  In addition, the
    \ion{O}{6} $\lambda$1031.93 line is aligned with \ion{H}{1}
    Ly$\alpha, \beta, \ {\rm and} \ \gamma$ lines at the same $z_{\rm
    abs}$.

(12.) {\it HE0226-4110, $z_{abs}$ = 0.01747.} The \ion{O}{6}
    $\lambda$1037.62 line is blended with \ion{O}{4} $\lambda$787.71
    at $z_{\rm abs}$ = 0.34035 (see Lehner et al. 2006 and
    Figure~\ref{lowratios}). The \ion{O}{6} identification is favored
    based on the precise alignment of \ion{O}{6} $\lambda$1031.93
    candidate with an \ion{H}{1} Ly$\alpha$ line at the same redshift.
    We note that the comparison of the $N_{\rm a}$ profiles in
    Figure~\ref{navplots} does not show the absorption in the wings of
    the Ly$\alpha$ line very clearly; Figure~\ref{lowratios} more
    clearly shows how the \ion{H}{1} line is slightly broader than the
    \ion{O}{6} lines.

(13.) {\it HE0226-4110, $z_{abs}$ = 0.20701.} Detailed analysis of
    this system has been presented by Savage et al. (2005).  The
    \ion{H}{1} Ly$\gamma$ line is recorded in both the STIS spectrum
    and the {\it FUSE} spectrum of HE0226-4110.  The apparent
    component structure in the STIS recording of the Ly$\gamma$ line
    is incompatible with the {\it FUSE} recording of Ly$\gamma$ and
    with the other (higher) Lyman series lines (see Savage et
    al. 2005), and the STIS Ly$\gamma$ line was excluded from the fit.

(14.) {\it HE0226-4110, $z_{abs}$ = 0.35525.} A weaker line offset by
    +40 km s$^{-1}$ is present next to the main component that is
    clearly detected in the \ion{O}{6} $\lambda$1031.93 and \ion{O}{6}
    $\lambda$1037.62 profiles at this redshift.  The +40 km s$^{-1}$
    feature does not appear to be \ion{O}{6} because it is not
    confirmed by the $\lambda$1037.62 line.  However, the +40 km
    s$^{-1}$ feature would be relatively weak in the $\lambda$1037.62
    transition, and it could be hidden by noise.  Following Lehner et
    al. (2006), we do not include the 40 km s$^{-1}$ component in the
    \ion{O}{6} measurements; higher S/N data are needed to establish
    the identity of this feature.

(15.) {\it HE0226-4110, $z_{abs}$ = 0.42670.} At this redshift, the
    Ly$\alpha$ is redshifted beyond the long-wavelength cutoff of our
    STIS spectrum, and Ly$\beta$ line is not detected.  The \ion{O}{6}
    identification is based on the good agreement of the \ion{O}{6}
    $\lambda$1031.93 and $\lambda$1037.62 profiles (see Lehner et
    al. 2006).  The \ion{O}{6} $\lambda$1031.93 line is partially
    affected by hot pixels that were excluded from the fit.  We note
    that identification of this system depends critically on the STIS
    warm/hot pixel correction algorithm.  If we turn off the hot-pixel
    repair algorithm, we find that the \ion{O}{6} $\lambda$1037.62
    line is largely filled in by warm pixels.  It is important to
    obtain future observations of this system with COS in order to
    test the reliability of the identification and to expand the
    utility of this system with additional information (e.g., better
    \ion{H}{1} absorption constraints).

(16.) {\it HE0226-4110, $z_{abs}$ = 0.49246.} This complex,
    multispecies system has been analyzed in detail by Ganguly et
    al. (2006).  The \ion{O}{6} component at $v = 0$ km s$^{-1}$ is
    uncertain due to substantial saturation.  The \ion{O}{6}
    $\lambda$1037.62 line is partially blended with Galactic
    \ion{C}{4} (see Fox et al. 2005), but the distinctive component
    structure seen in the \ion{O}{6} $\lambda$1031.93 profile can be
    clearly recognized in the in the $\lambda$1037.62 profile as well
    (see Ganguly et al. 2006), so the identification is secure, and
    the weaker \ion{O}{6} components are well-constrained by the
    \ion{O}{6} $\lambda$1031.93 line.

\begin{figure}
\centering
    \includegraphics[width=8.5cm, angle=0]{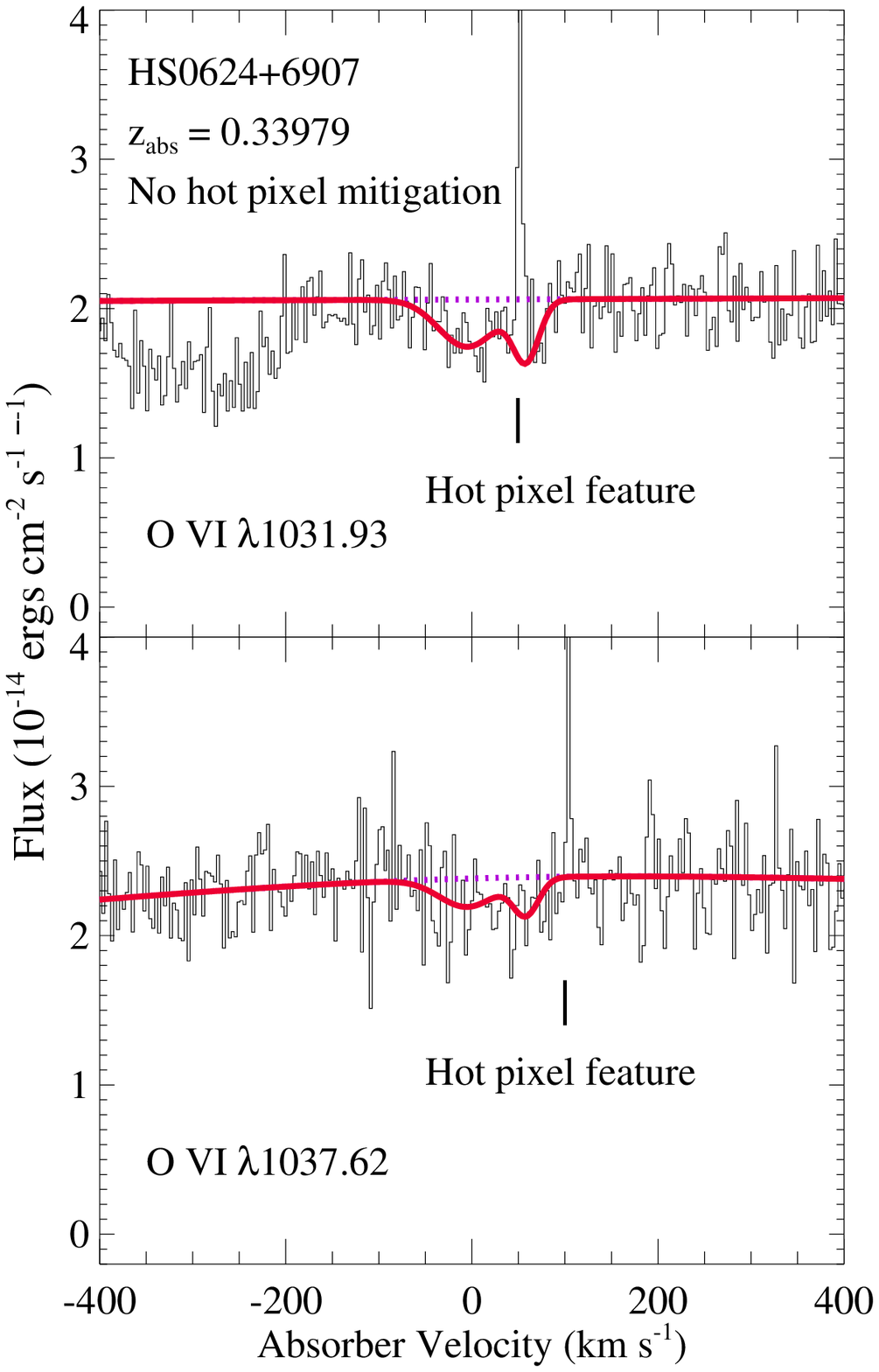}
    \includegraphics[width=8.5cm, angle=0]{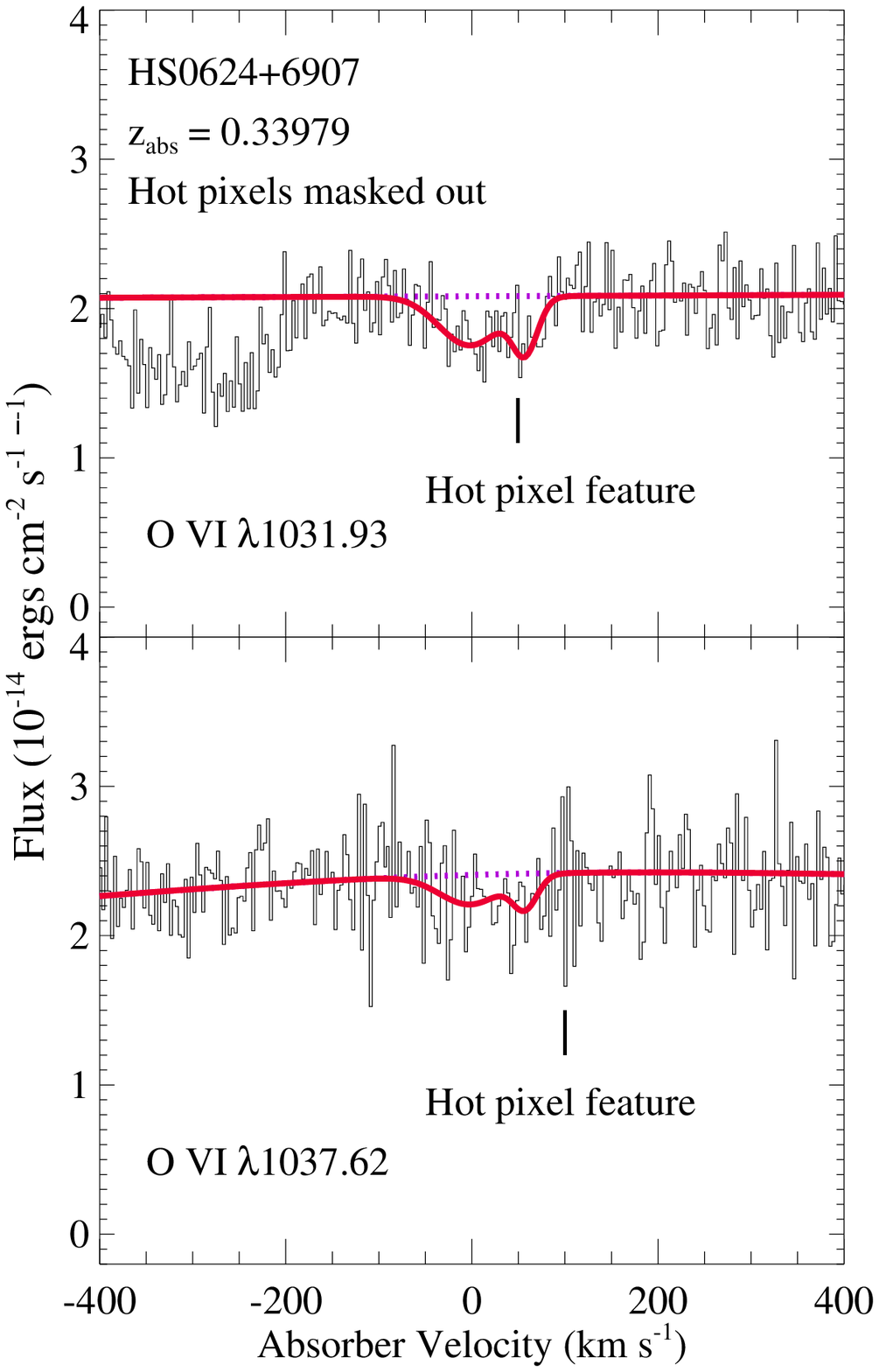}
\caption{Demonstration of hot pixel alleviation by masking (rejection)
of affected regions in individual exposures.  The panels show the
regions of the O~VI $\lambda$1031.93 line ({\it top panels}) and
the O~VI $\lambda$1037.62 line ({\it bottom panels}) at $z_{\rm
abs}$ = 0.33979 in the spectrum of HS0624+6907. The left panels show
the line profiles without any hot pixel alleviation; hot pixels are
clearly evident near both profiles.  The HS0624+6907 observations were
recorded in Jan. 2002 and Feb. 2002 (see Table~\ref{obslog}), and the
spectrum position on the detector was shifted between the two dates.
Consequently, the hot pixels are only present in the February data and
can be masked and rejected from the final coadded spectrum as shown at
right.\label{hotpixdemo}}
\end{figure}

(17.) {\it HS0624+6907, $z_{abs}$ = 0.33979.} \ion{H}{1} Ly$\beta$ is
    mildly blended with an unrelated line.  However, most of the
    Ly$\beta$ profile is free from the blend, and the unblended
    portion of Ly$\beta$ was included in the fit. As shown in the left
    panel of Figure~\ref{hotpixdemo}, hox pixel features are present
    within the \ion{O}{6} $\lambda$1031.93 line and at the red edge of
    the \ion{O}{6} $\lambda$1037.62 profile.  Fortunately, in this
    case the QSO was observed on two different dates (in January 2002
    and February 2002, see Table~\ref{obslog}), and the position of
    the spectrum on the detector was shifted between these two dates.
    Inspection of the data reveals that the hot pixel features are
    only present in the February 2002 data.  As shown in the right
    panel of Figure~\ref{hotpixdemo}, by masking and rejecting the
    affected hot pixels in the February data, we can suppress this
    problem with a minimal loss of the signal-to-noise.  

(18.) {\it HS0624+6907, $z_{abs}$ = 0.37053.} In this proximate
    absorber, \ion{H}{1} Ly$\alpha$ is not detected despite good
    signal-to-noise (see Figure~\ref{ascsample1}).  As shown in
    Figure~\ref{ascsample1}, the \ion{O}{6} identification is quite
    secure; both lines of the \ion{O}{6} doublet show multiple
    components and are in excellent agreement.

(19.) {\it PG0953+415, $z_{abs}$ = 0.06807.} We have carried out
    extensive investigations of this absorber in previous papers
    (Savage et al. 2002; Tripp et al. 2006). Comparison of the
    \ion{O}{6} $\lambda$1031.93 and $\lambda$1037.62 lines indicates
    moderate saturation, and application of the method of Jenkins
    (1996) indicates that $N$(\ion{O}{6}) could be 0.25 dex higher.

(20.) {\it PG0953+415, $z_{abs}$ = 0.14231.} The Ly$\beta$ profile is
    partially blended with an \ion{H}{1} Ly$\delta$ line from $z_{\rm
    abs}$ = 0.23351.  The blended part of the Ly$\beta$ line was not
    used in the fit. However, the Ly$\alpha$ line has a complex
    profile with many components (see Tripp \& Savage 2000), and the
    unblended portion of the Ly$\beta$ line provides useful
    constraints for the fit.

(21.) {\it PG0953+415, $z_{abs}$ = 0.22974.} In this proximate
    absorber of PG0953+415, \ion{H}{1} Ly$\alpha$ is not detected
    despite good signal-to-noise. The \ion{O}{6} identification is
    based on the good agreement of the \ion{O}{6} lines over a large
    portion of both profiles: the $\lambda$1031.93 and
    $\lambda$1037.62 profiles agree well over $\approx 20$ pixels
    between $v = -30$ km s$^{-1}$ and $v = 40$ km s$^{-1}$.  However,
    the \ion{O}{6} profiles are discrepant at $v < -30$ km
    s$^{-1}$. While this discrepancy has the appearance of a hot pixel
    feature, comparison of the data from 1998 December 4 and 1998
    December 11 shows the same profile structure.  The location of the
    spectrum on the detector was shifted between 1998 December 4 and
    1998 December 11, so this discrepancy cannot be due to hot pixels.
    We conclude that the \ion{O}{6} $\lambda$1031.93 line is blended
    with an unrelated Ly$\alpha$ line on the blue side of the profile.
    This part of the $\lambda$1031.93 profile was excluded from the
    Voigt profile fit.

(22.) {\it PG0953+415, $z_{abs}$ = 0.23351.} A Ly$\delta$ line with
    approximately correct strength is detected at this redshift but
    was not used in the fit due to blending with Ly$\beta$ from the
    complex, multicomponent absorber at $z_{\rm abs}$ = 0.14231.

(23.) {\it PG1116+215, $z_{abs}$ = 0.05927.} Both lines of the
    \ion{O}{6} doublet are detected and in excellent agreement at $v =
    0$ km s$^{-1}$ (see Figure~\ref{lowratios}). The \ion{O}{6}
    $\lambda$1031.93 line is also detected at $v = -84$ km s$^{-1}$,
    but the \ion{O}{6} $\lambda$1037.62 line is not significantly
    detected at that velocity.  A small portion of the detected
    \ion{O}{6} $\lambda$1031.93 component at $v = -84$ km s$^{-1}$ is
    blended with Galactic H$_{2}$ (see Sembach et al. 2004).  The $v =
    -84$ km s$^{-1}$ component is also identified as \ion{O}{6} based
    on the good agreement of the \ion{O}{6} and \ion{H}{1} Ly$\alpha$
    line shapes at $v = -84$ km s$^{-1}$ (not including the portion
    blended with H$_{2}$, which was also excluded from the fit), as
    shown in Figure~\ref{navplots}.

(24.) {\it PG1116+215, $z_{abs}$ = 0.13849.} The \ion{O}{6} doublet at
    this redshift is detected with both {\it FUSE} and STIS (see
    Sembach et al. 2004).  The fit reported here is based on the STIS
    data.  This system, which has a high \ion{H}{1} column and is
    detected in many Lyman series lines, has been analyzed in detail
    by Sembach et al. (2004). The \ion{O}{6} lines are aligned with
    the \ion{H}{1} lines, but analysis of the many low-ionization
    metal absorption lines detected in this system clearly establishes
    that this is a complex multiphase absorber (see Sembach et
    al. 2004).

(25.) {\it PG1116+215, $z_{abs}$ = 0.16553.} The \ion{H}{1} components
    at $v = -12, 143, 170,$ and 342 km s$^{-1}$ are well-constrained
    by the detected absorption lines. Additional absorption is clearly
    and significantly detected in other velocity ranges in the
    Ly$\alpha$ profile, e.g., at $v \approx$ 70 km s$^{-1}$, and
    components were added to the fit to account for this additional
    absorption.  However, these components are not well-constrained
    due to blending with the adjacent features.

(26.) {\it PG1116+215, $z_{abs}$ = 0.17340.} The \ion{O}{6}
    $\lambda$1031.93 line is detected at 5.8$\sigma$ significance and
    is well-aligned with an \ion{H}{1} line at the same redshift.  In
    addition, the \ion{O}{6} $\lambda$1037.62 line is detected with
    the expected wavelength and relative strength at 2.9$\sigma$.

(27.) {\it PG1216+069, $z_{abs}$ = 0.12360.} The Ly$\alpha$ profile
    has good signal-to-noise and shows clear inflections and
    asymmetries that reveal the complicated component structure
    including at least eight components; four of the components are
    clearly evident in the Ly$\beta$ profile as well (see
    Figure~\ref{intsample2}).  However, all of the components are
    either strong and significantly saturated (in both Ly$\alpha$ and
    Ly$\beta$) or are highly blended with adjacent strong components.
    Moreover, some of the saturated components show that there are
    errors in the flux zero level of the Ly$\alpha$ line, and the
    Ly$\beta$ line is relatively noisy.  Ly$\gamma$ lines are also
    evident (see Tripp et al. 2005), but the Ly$\gamma$ data are too
    noisy to usefully constrain the fits. The \ion{H}{1} component
    parameters are highly uncertain due to these combined problems.
    Both lines of the \ion{O}{6} doublet are strong and are clearly
    detected with component structure similar to the Ly$\beta$
    components. The apparent column density profiles of the \ion{O}{6}
    doublet are in good agreement (see Figure~\ref{intsample2}), which
    suggests that the \ion{O}{6} lines are not badly saturated.
    However, the \ion{O}{6} profiles are also relatively noisy.

\begin{figure}
\centering
   \includegraphics[width=8.5cm, angle=0]{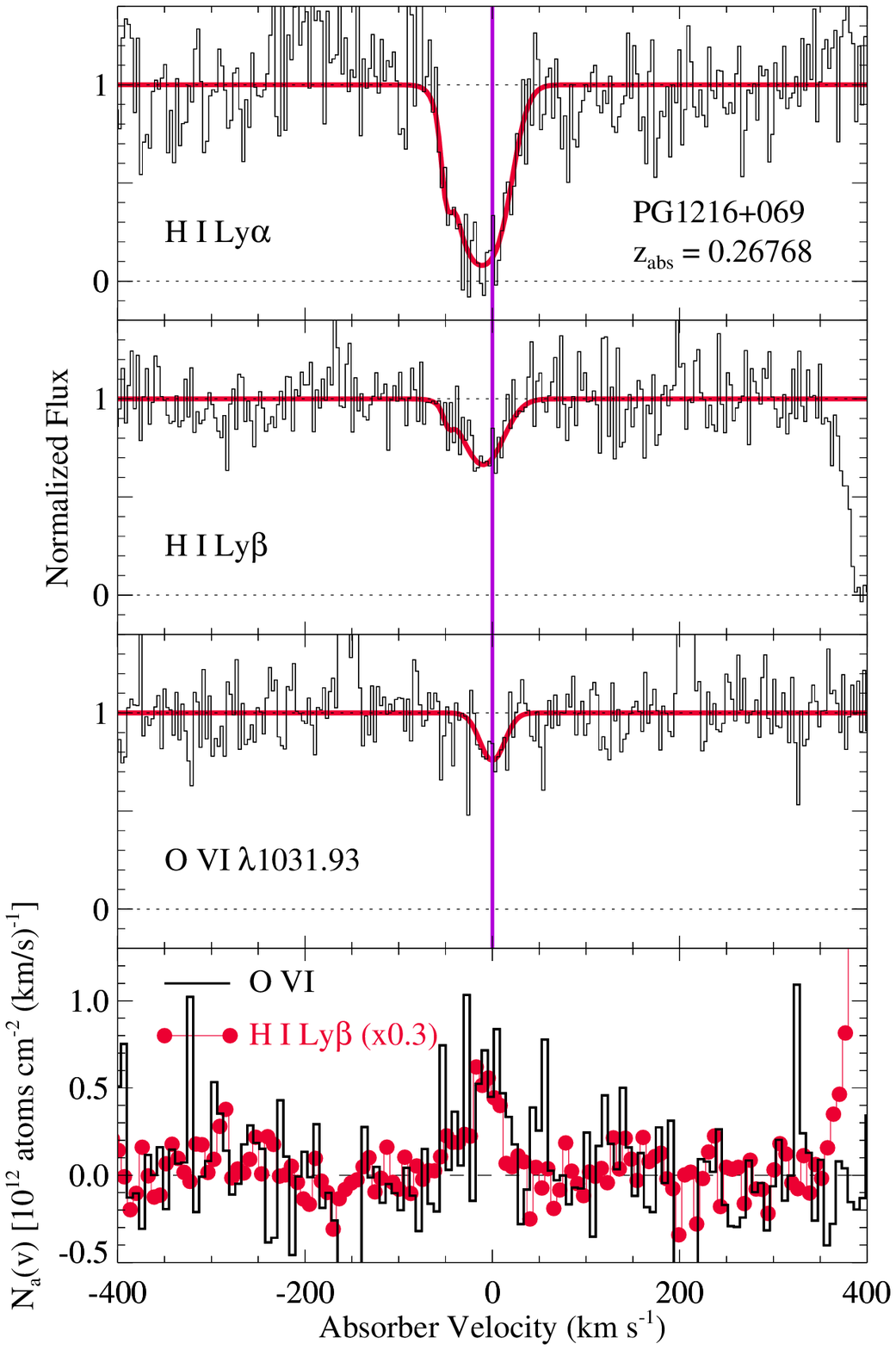}
   \includegraphics[width=8.5cm, angle=0]{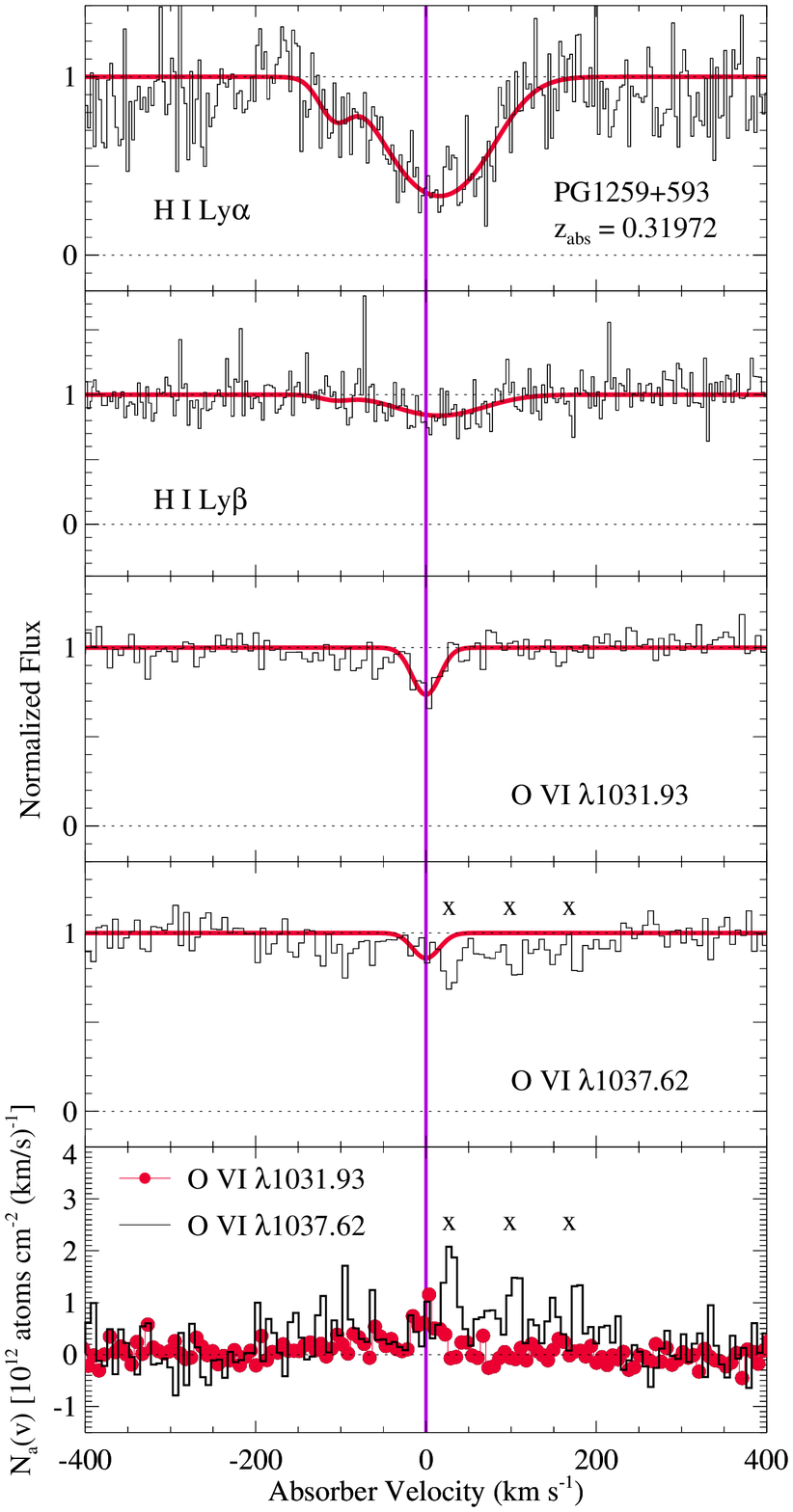}
\caption{{\it Left panels:} Continuum-normalized absorption profiles
of the H~I Ly$\alpha$, Ly$\beta$, and O~VI
$\lambda$1031.93 lines at $z_{\rm abs}$ = 0.26768 in the spectrum of
PG1216+069 (upper three panels), and comparison of the O~VI
$\lambda$1031.93 and Ly$\beta$ $N_{\rm a}(v)$ profiles (lowest
panel). Voigt-profile fits (see Table~\ref{compprop}) are overplotted
with smooth lines. As in other figures, unrelated lines are marked
with an `x'. {\it Right panels:} Absorption profiles of the H~I
and O~VI lines detected at $z_{\rm abs}$ = 0.31972 in the
spectrum of PG1259+593.  In this stack, the lowest panel compares the
$N_{\rm a}(v)$ profiles of the O~VI $\lambda$1031.93 and
$\lambda$1037.62 lines. In this figure, the $N_{\rm a}(v)$ profiles
are binned to 7 km s$^{-1}$ pixels. \label{pg1216_26768}}
\end{figure}

(28.) {\it PG1216+069, $z_{abs}$ = 0.26768.} The \ion{O}{6}
    $\lambda$1037.62 line cannot be measured because it is lost in a
    blend with a strongly saturated \ion{H}{1} Ly$\beta$ absorption
    line from $z_{\rm abs}$ = 0.28189.  As shown in
    Figure~\ref{pg1216_26768}, the \ion{O}{6} identification at
    $z_{\rm abs}$ = 0.26768 is based on the alignment of \ion{O}{6}
    $\lambda$1031.93 with Ly$\alpha$ and Ly$\beta$ at the same
    redshift.  At this redshift, initial inspection indentified a
    candidate \ion{C}{3} $\lambda$977.02 line that is somewhat blended
    with \ion{N}{5} $\lambda$1238.82 absorption from the Milky Way.
    The \ion{C}{3} candidate cannot be a second component of Galactic
    \ion{N}{5} absorption because it is not evident in the profile of
    the other line of the \ion{N}{5} doublet.  However, closer
    inspection reveals that this line is not \ion{C}{3} but rather is
    the \ion{H}{1} Ly$\gamma$ line from the strong \ion{H}{1} system
    at $z_{\rm abs}$ = 0.27353, so we can only place an upper limit on
    \ion{C}{3} absorption at this redshift.

(29.) {\it PG1259+593, $z_{abs}$ = 0.04637.} Only the \ion{O}{6}
    $\lambda$1031.93 line is detected at this redshift (\ion{O}{6}
    $\lambda$1037.62 is redshifted into a relatively noisy region of
    the {\it FUSE} spectrum that is only recorded by the SiC
    channels). Nevertheless, the identification is secure because the
    \ion{O}{6} $\lambda$1031.93 profile has a distinctive
    two-component structure that matches the component structure seen
    in the \ion{C}{3} and \ion{C}{4} lines detected at the same
    redshift.  An unrelated \ion{O}{4} $\lambda$787.71 line is located
    near the Ly$\gamma$ profile; the region affected by the \ion{O}{4}
    feature was excluded from the fit.

(30.) {\it PG1259+593, $z_{abs}$ = 0.21949.} Several components due to
    Galactic \ion{S}{2} $\lambda$1250.58 are located on the blue side
    of the Ly$\beta$ line (see Richter et al. 2004); the velocity
    range affected by the Milky Way \ion{S}{2} absorption was excluded
    from the fit.

\begin{figure}
\centering
   \includegraphics[width=8.5cm, angle=0]{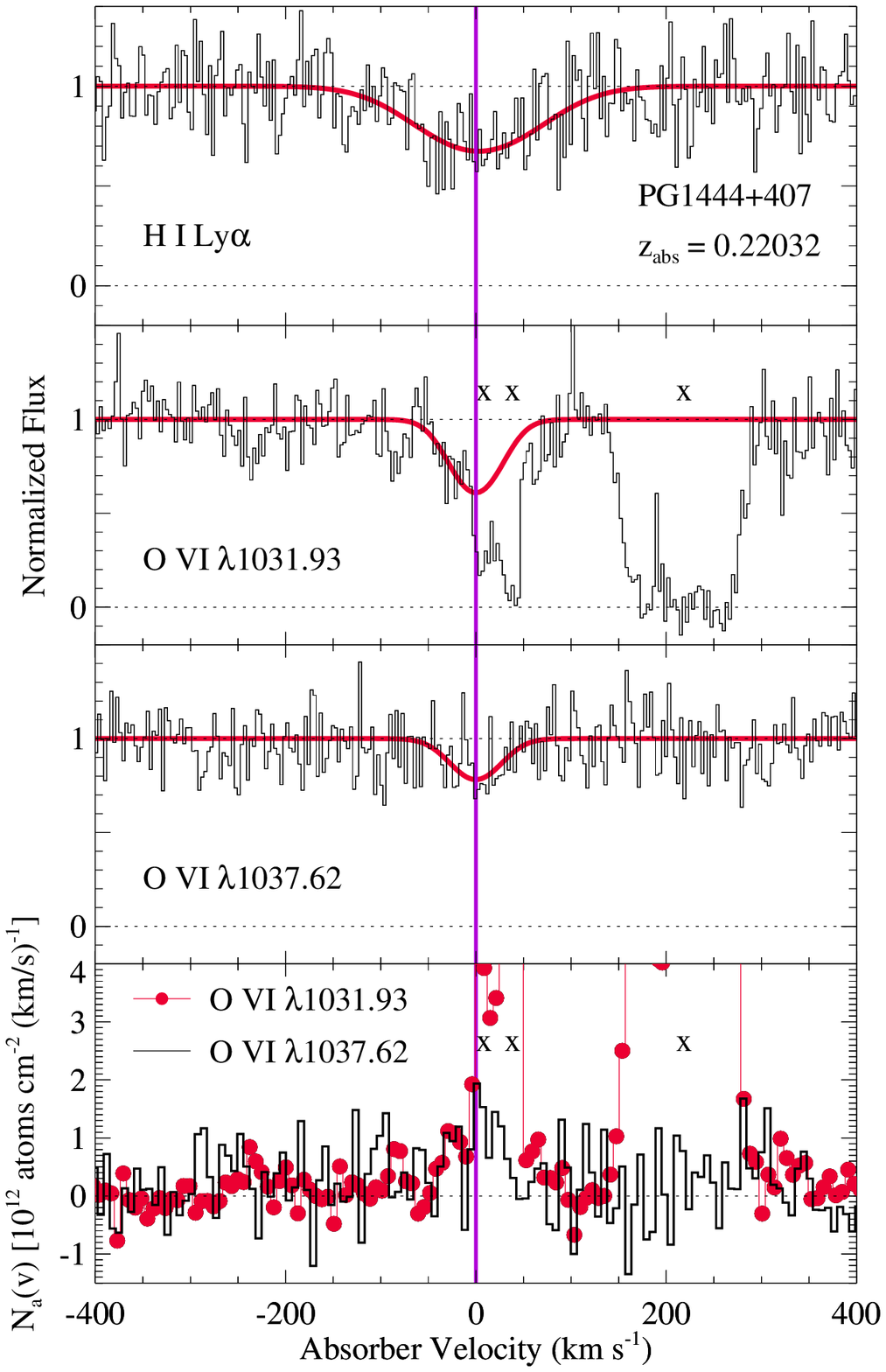}
   \includegraphics[width=8.5cm, angle=0]{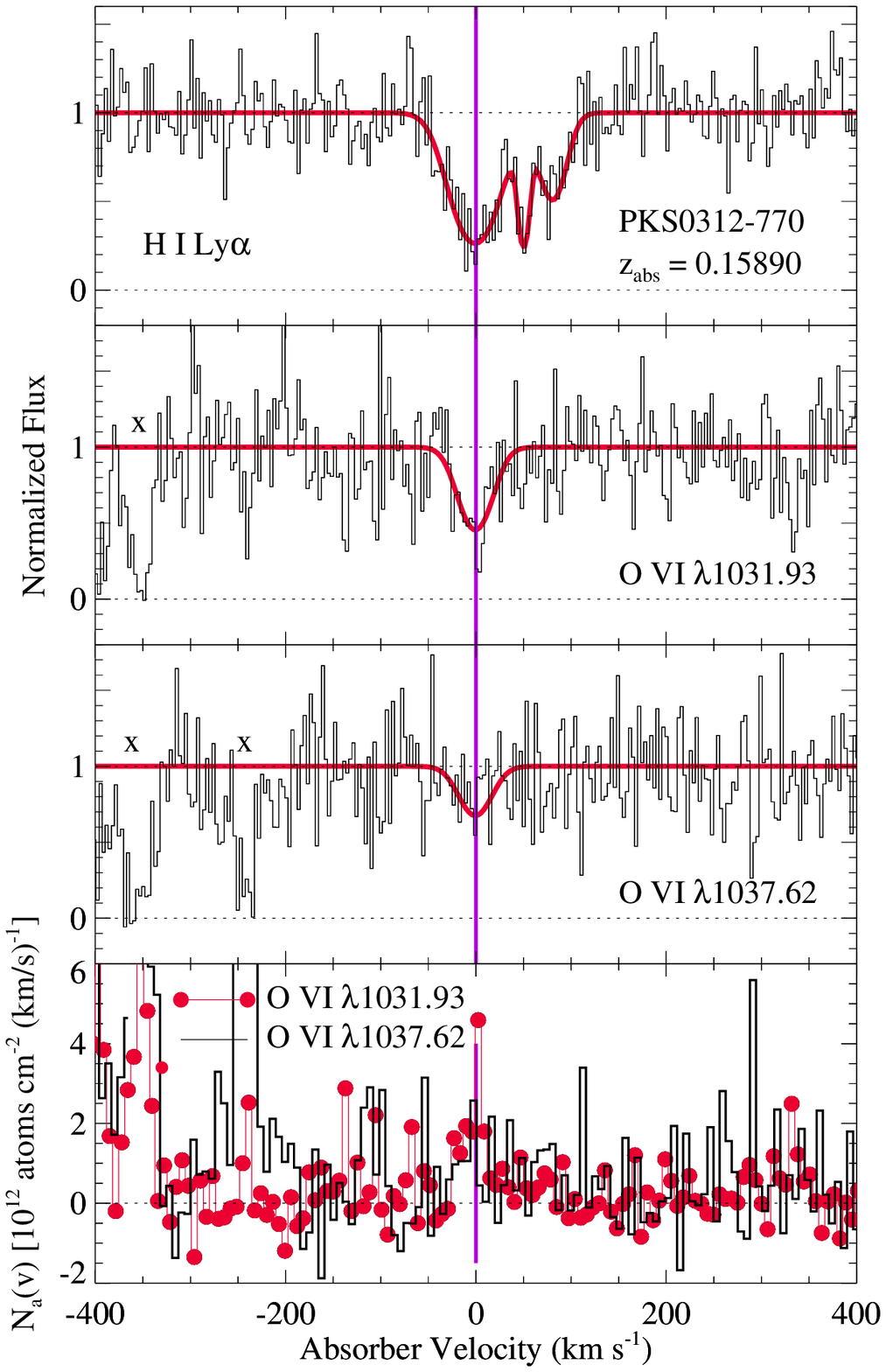}
\caption{{\it Left panels:} Continuum-normalized absorption profiles
of the H~I Ly$\alpha$ and O~VI $\lambda \lambda$1031.93,
1037.62 lines at $z_{\rm abs}$ = 0.22032 in the spectrum of PG1444+407
(upper three panels), and comparison of the O~VI
$\lambda$1031.93 and $\lambda$1037.62 $N_{\rm a}(v)$ profiles (lowest
panel). Voigt-profile fits (see Table~\ref{compprop}) are overplotted
with smooth lines. Unrelated lines are marked with an `x'. The
O~VI $\lambda$1031.93 line is blended with Milky Way S~II
$\lambda$1259.52 absorption, but excess absorption is present in the
blend that cannot be attributed to S~II.  As shown, the excess
absorption is consistent with the expected strength of the O~VI
$\lambda$1031.93 line. {\it Right panels:} The same data for the
O~VI absorber at $z_{\rm abs}$ = 0.15890 in the spectrum of
PKS0312-770. In this figure, the $N_{\rm a}(v)$ profiles are binned to
7 km s$^{-1}$ pixels. \label{pg1444_22032}}
\end{figure}

(31.) {\it PG1259+593, $z_{abs}$ = 0.31972.} The \ion{O}{6}
    $\lambda$1037.62 line is quite weak and mildly blended with
    high-velocity \ion{Ni}{2} absorption from the Milky Way (see
    Richter et al. 2004).  As shown in the right panels of
    Figure~\ref{pg1216_26768}, the \ion{O}{6} $\lambda 1031.93$ line
    is clearly detected, and an absorption feature with the right
    relative strength (compared to $\lambda$1031.93) is present at the
    expected wavelength of \ion{O}{6} $\lambda$1037.62, but because it
    is weak and blended with Galactic \ion{Ni}{2}, our fit is based on
    the \ion{O}{6} $\lambda$1031.93 line only.

(32.) {\it PG1444+407, $z_{abs}$ = 0.22032.} The \ion{H}{1} Ly$\alpha$
    and \ion{O}{6} $\lambda \lambda$1031.93, 1037.62 lines for this
    absorption system are shown in the left panels of
    Figure~\ref{pg1444_22032}.  The \ion{O}{6} $\lambda$1031.93 line
    at $z_{abs}$ = 0.22032 is blended with the Galactic \ion{S}{2}
    $\lambda$1259.52 line.  However, comparison of the Galactic
    \ion{S}{2} 1253.81 and \ion{S}{2} $\lambda$1259.52 profiles shows
    a clear and significant excess of absorption at the expected
    velocity of \ion{O}{6} $\lambda$1031.93 at $z_{\rm abs}$ =
    0.22032.  Moreover, the excess absorption has precisely the
    expected strength compared to the (unblended) \ion{O}{6} $\lambda$
    1037.62 line, as can be seen in the comparison of the \ion{O}{6}
    $N_{\rm a}(v)$ profiles shown in Figure~\ref{pg1444_22032}. Only
    the unblended portion of \ion{O}{6} $\lambda$1031.93 was included
    in the fit.

(33.) {\it PG1444+407, $z_{abs}$ = 0.26738.} The Ly$\beta$ line is
    slightly blended with an unrelated weak line; the region affected
    by the blend was excluded from the fit.  More importantly, the
    Ly$\alpha$ line is located at the peak of the broad Ly$\alpha$
    emission line, and this introduces significant continuum placement
    uncertainty.  We note that a broad and shallow feature is located
    just blueward of the Ly$\alpha$ line.  This feature could be due
    to additional weak \ion{H}{1} absorption, but its significance is
    highly dependent on the uncertain continuum placement.
    Consequently, we did not include the broad, shallow feature in the
    fit.

(34.) {\it PHL1811, $z_{abs}$ = 0.07765.} Only the stronger \ion{O}{6}
    $\lambda$1031.93 line is detected at $> 3\sigma$ significance.
    However, many metals are detected at this redshift including
    \ion{C}{2} $\lambda 1334.53$, \ion{Si}{2} $\lambda 1260.42$,
    \ion{C}{3} $\lambda 977.02$, \ion{C}{4} $\lambda 1548.20$,
    (marginal) \ion{Si}{4} $\lambda 1393.76$, and several \ion{H}{1}
    Lyman series lines.  The strongest \ion{C}{4} component shows a
    $\approx -25$ km s$^{-1}$ offset from the lower ionization metals,
    but the \ion{O}{6} $\lambda$1031.93 line is aligned with the
    \ion{C}{4} $\lambda$1548.20 transition.

(35.) {\it PHL1811, $z_{abs}$ = 0.15786.} The Ly$\alpha$ profile at
    $z_{\rm abs}$ = 0.15786 is blended with \ion{O}{1}
    $\lambda$1302.17 absorption from the Lyman-limit system at $z_{\rm
    abs}$ = 0.08092 (see Jenkins et al. 2005 and
    Figure~\ref{hiddenbla}). The narrow core in this blend is
    predominantly due to the Lyman-limit \ion{O}{1} line.  However,
    close inspection of this profile (see Figure~\ref{hiddenbla})
    reveals weak component absorption straddling the narrow core on
    the short- and long-wavelength sides.  We cannot corroborate that
    the weaker components are also \ion{O}{1}; similar component
    structure is not clearly evident in the other profiles of
    low-ionization metals in this Lyman-limit absorber, which suggests
    that these weaker components could be unrelated to the \ion{O}{1}
    and could be Ly$\alpha$ at $z_{\rm abs}$ = 0.15786.  However,
    better S/N data are needed to reliably determine the origin of the
    weak components and to accurately deblend and measure their
    parameters, so we flag this absorption with a colon in
    Table~\ref{compprop} to reflect the substantial uncertainty of
    this Ly$\alpha$ case.

(36.) {\it PHL1811, $z_{abs}$ = 0.17650.} At this redshift, \ion{O}{6}
    $\lambda 1031.93$ falls in the saturated core of the Milky Way
    damped Ly$\alpha$ line.  The \ion{O}{6} $\lambda 1037.62$ line is
    identified based on its alignment with multiple Lyman series
    lines, and this is supported by the detection of \ion{C}{3}
    $\lambda 977.02$ and \ion{Si}{3} $\lambda 1206.5$ in this
    absorber.  However, we note that the \ion{O}{6} is offset by $\sim
    -25$ km s$^{-1}$ compared to the \ion{C}{3} and \ion{Si}{3} lines.

\begin{figure}
\centering
   \includegraphics[width=8.5cm, angle=0]{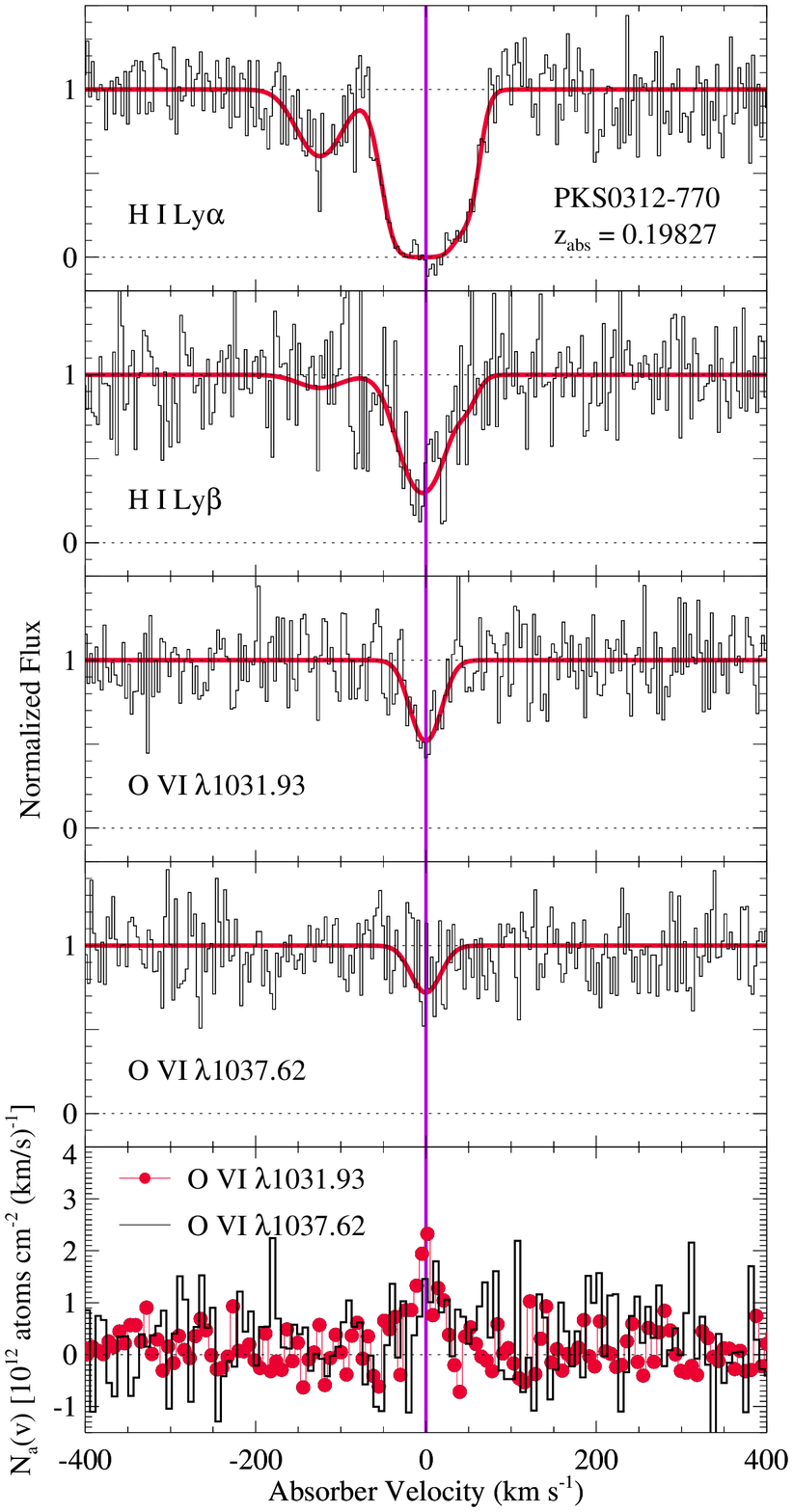}
   \includegraphics[width=8.5cm, angle=0]{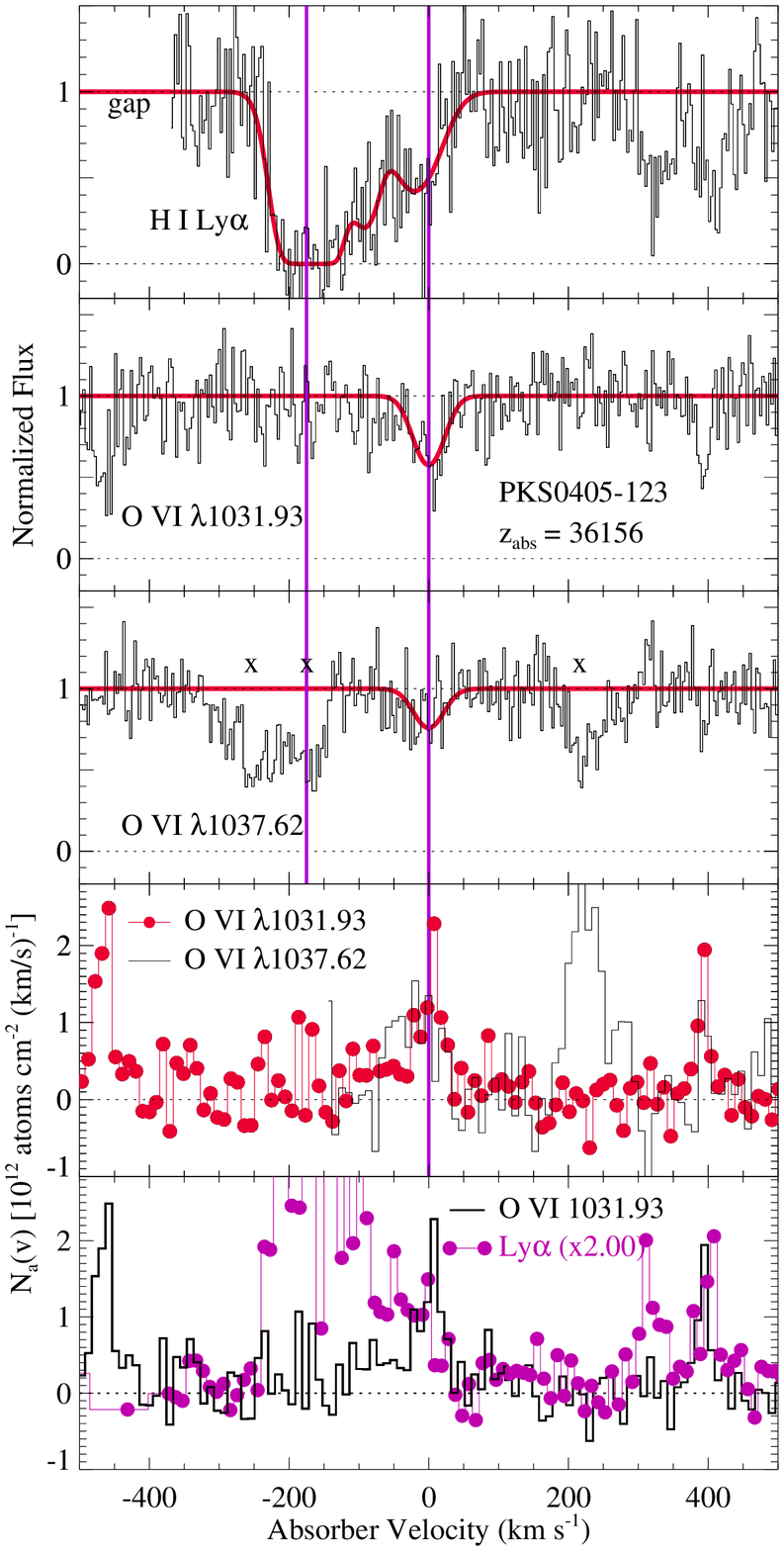}
\caption{{\it Left panels:} Continuum-normalized absorption profiles
of the H~I Ly$\alpha$, Ly$\beta$, and O~VI $\lambda
\lambda$1031.93, 1037.62 lines at $z_{\rm abs}$ = 0.19827 in the
spectrum of PKS0312-770 (upper four panels), and comparison of the
O~VI $\lambda$1031.93 and $\lambda$1037.62 $N_{\rm a}(v)$
profiles (lowest panel). Voigt-profile fits (see Table~\ref{compprop})
are overlaid with smooth lines. {\it Right panels:} Same data for the
absorber at $z_{\rm abs}$ = 0.36156 in the spectrum of PKS0405-123.
In this stack, the lowest two panels show the comparison of the
$N_{\rm a}(v)$ profiles of the two lines of the O~VI doublet
(second panel from bottom) and a comparison of the O~VI $\lambda
1031.93$ and H~I Ly$\alpha$ $N_{\rm a}(v)$ profiles (lowest
panel). Unrelated lines are marked with an `x'. In this figure, the
$N_{\rm a}(v)$ profiles are binned to 7 km s$^{-1}$
pixels. \label{pks0312_19827}}
\end{figure}

\begin{figure}
\centering
   \includegraphics[width=17.0cm, angle=0]{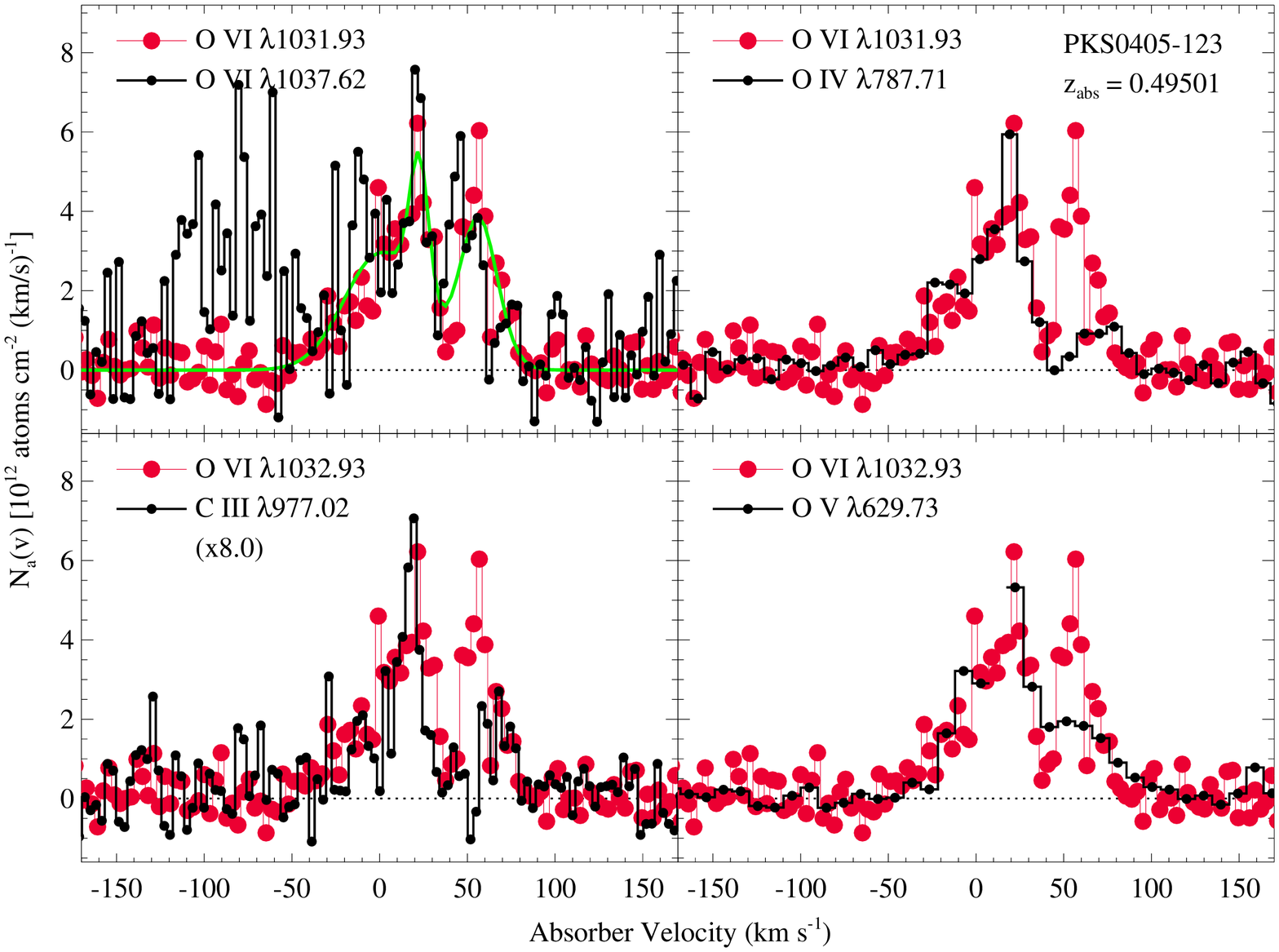}
\caption{Comparison of apparent column density profiles of several
metals in the O~VI absorption system at $z_{\rm abs}$ = 0.49501
in the spectrum of PKS0405-123.  {\it Upper left:} O~VI
$\lambda$1031.93 (red large circles) vs. O~VI $\lambda$1037.62
(small black circles).  The thick green line shows the O~VI
profile implied by the Voigt-profile fit listed in
Table~\ref{compprop}. {\it Upper right:} O~VI $\lambda$1031.93
(red large circles) vs. O~IV $\lambda$787.71. {\it Lower left:}
O~VI $\lambda$1031.93 (red large circles) vs. C~III
$\lambda$977.02. {\it Lower right:} O~VI $\lambda$1031.93 (red
large circles) vs. O~V $\lambda$629.73.  Several pixels are not
plotted in the core of the O~V profile because it is saturated
in that region. The O~IV and O~V profiles are from the
{\it FUSE} observations reported by Prochaska et al. (2004), and the
C~III and O~VI data are measured from the STIS E140M
spectrum.  Most of these profiles show two weaker components flanking
the deepest central component, but as discussed in the text, an offset
is apparent between the peak of the O~VI $\lambda$1031.93 and
$\lambda$1037.62 lines in the reddest component.\label{pks0405fig}}
\end{figure}

(37.) {\it PKS0312-770, $z_{abs}$ = 0.15890.} \ion{O}{6}
    $\lambda$1031.93 is detected at 5.6$\sigma$ significance, and the
    corresponding \ion{O}{6} $\lambda$1037.62 line is also detected,
    but only at 2.4$\sigma$ significance.  As shown in
    Figure~\ref{pg1444_22032}, the $N_{\rm a}(v)$ profiles of the two
    lines of the \ion{O}{6} doublet are in good agreement, and the
    marginal detection of the $\lambda$1037.62 line supports the
    \ion{O}{6} identification.  The \ion{O}{6} identification is
    also supported by the precise alignment of the \ion{O}{6}
    $\lambda$1031.93 line with \ion{H}{1} Ly$\alpha$ at the same
    redshift.

(38.) {\it PKS0312-770, $z_{abs}$ = 0.19827.} \ion{O}{6}
    $\lambda$1031.93 is detected at 4.7$\sigma$ significance, but the
    \ion{O}{6} $\lambda$1037.62 line is not significantly detected in
    the data at full resolution. However, as shown in the left panels
    of Figure~\ref{pks0312_19827}, if we mildly bin the data to 7 km
    s$^{-1}$ pixels to improve the S/N, we find a feature in the
    spectrum at the expected wavelength of the $\lambda$1037.62 line
    that is fully consistent with the detected \ion{O}{6}
    $\lambda$1031.93 line (compare the $N_{\rm a}(v)$ profiles shown
    in the lowest left-hand panel of Figure~\ref{pks0312_19827}). The
    \ion{O}{6} identification is bolstered by the close alignment of
    \ion{O}{6} $\lambda$1031.93 with Ly$\alpha$ and Ly$\beta$ at the
    same redshift (see the left panels in Figure~\ref{pks0312_19827}).

(39.) {\it PKS0312-770, $z_{abs}$ = 0.20266.} All accessible
    \ion{H}{1} lines in the STIS bandpass (Ly$\alpha$, Ly$\beta$,
    Ly$\gamma$) are strong and highly saturated.  Moreover, the
    Ly$\beta$ and Ly$\gamma$ profiles show complex component structure
    with at least five distinct components. Since most of the
    \ion{H}{1} components are black in the line cores, the \ion{H}{1}
    profile parameters are poorly constrained, and we have not
    attempted to fit the \ion{H}{1} lines.  Brief inspection of an
    archival {\it FUSE} spectrum reveals that this is an optically
    thick Lyman limit absorber.  While the \ion{H}{1} component
    properties are poorly constrained, comparison of the low- and
    high-ionization metal lines reveals that this is a complex
    multiphase system (see \S \ref{classmulti} and
    Figure~\ref{pks0312multi}).  In addition, the individual
    components are spread over a large velocity range with low- and
    high-ionization components detected at velocities ranging from
    $-204$ to +135 km s$^{-1}$.

(40.) {\it PKS0405-123, $z_{abs}$ = 0.16692.} Many Lyman series lines
    are available for constraining the \ion{H}{1} column density at
    this redshift, and the absence of strong Lyman limit absorption
    places a firm upper limit on the total \ion{H}{1} column density
    (Prochaska et al. 2004).  Nevertheless, the parameters of the
    individual \ion{H}{1} components are highly uncertain in this
    system.  The Lyman series lines clearly require a multicomponent
    fit, but the close spacing and blending of the components causes
    the component parameter uncertainties to be substantial.  Thus,
    the degree of alignment of the \ion{H}{1} and \ion{O}{6}
    components is highly uncertain.  Nevertheless, comparison on the
    low- and high-ionization metal lines shows that this is a complex
    multiphase case (see Chen \& Prochaska 2000).

(41.) {\it PKS0405-123, $z_{abs}$ = 0.36156.} The \ion{H}{1}
    Ly$\alpha$ and \ion{O}{6} $\lambda \lambda$1031.93, 1037.62
    absorption profiles and apparent column densities are compared in
    the right-hand stack of Figure~\ref{pks0312_19827}.  The
    \ion{O}{6} $N_{\rm a}(v)$ profiles are seen to be in reasonable
    agreement.  We note that the \ion{O}{6} $\lambda$1037.62 profile
    shows a small excess of absorption on the blue side compared to
    the $\lambda$1031.93 line; this could be due to blending with an
    unrelated line (several unrelated lines are readily apparent in
    the vicinity of $\lambda$1037.62), but this could also simply be a
    noise feature.  An inflection is also evident in the Ly$\alpha$
    profile at the velocity of the \ion{O}{6} lines.  It is
    interesting to note that this system has a relatively high
    $N$(\ion{H}{1}) and is detected in several Lyman series lines, but
    no \ion{O}{6} absorption is evident at the velocity of the main
    component where the strong \ion{H}{1} lines are found (see
    Figure~\ref{pks0312_19827}).  Instead, the \ion{O}{6} is in the
    wing of the profile near the weak \ion{H}{1} inflection component.
    Several other absorbers show similar offsets between the strong
    main \ion{H}{1} absorption component and the \ion{O}{6} lines (see
    Figure~\ref{navplots}).

(42.) {\it PKS0405-123, $z_{abs}$ = 0.36335.} The \ion{O}{6}
    $\lambda$1031.93 line is detected at the 4.5$\sigma$ level, but
    \ion{O}{6} $\lambda$1037.62 is measured at only 1.7$\sigma$
    significance.  However, the apparent column density profiles of
    the two \ion{O}{6} lines agree precisely; the wavelength
    separation and relative strength of the lines makes the \ion{O}{6}
    identification compelling. The corresponding \ion{H}{1} absorption
    is relatively weak, and moreover, the \ion{H}{1} Ly$\alpha$ line
    is strongly blended with Galactic \ion{C}{1} and \ion{C}{1}*
    absorption lines from the \ion{C}{1} $\lambda$1656 multiplet.  We
    can see that \ion{H}{1} Ly$\alpha$ absorption is present at this
    redshift because the Galactic \ion{C}{1}* $\lambda$1657.38 line is
    clearly too strong compared to other \ion{C}{1}* lines in the
    PKS0405-123 spectrum.  However, the \ion{H}{1} Ly$\alpha$ line is
    difficult to measure reliably due to this strong blending with
    Galactic \ion{C}{1}* $\lambda$1657.38.

\begin{figure}
\centering
   \includegraphics[width=8.5cm, angle=0]{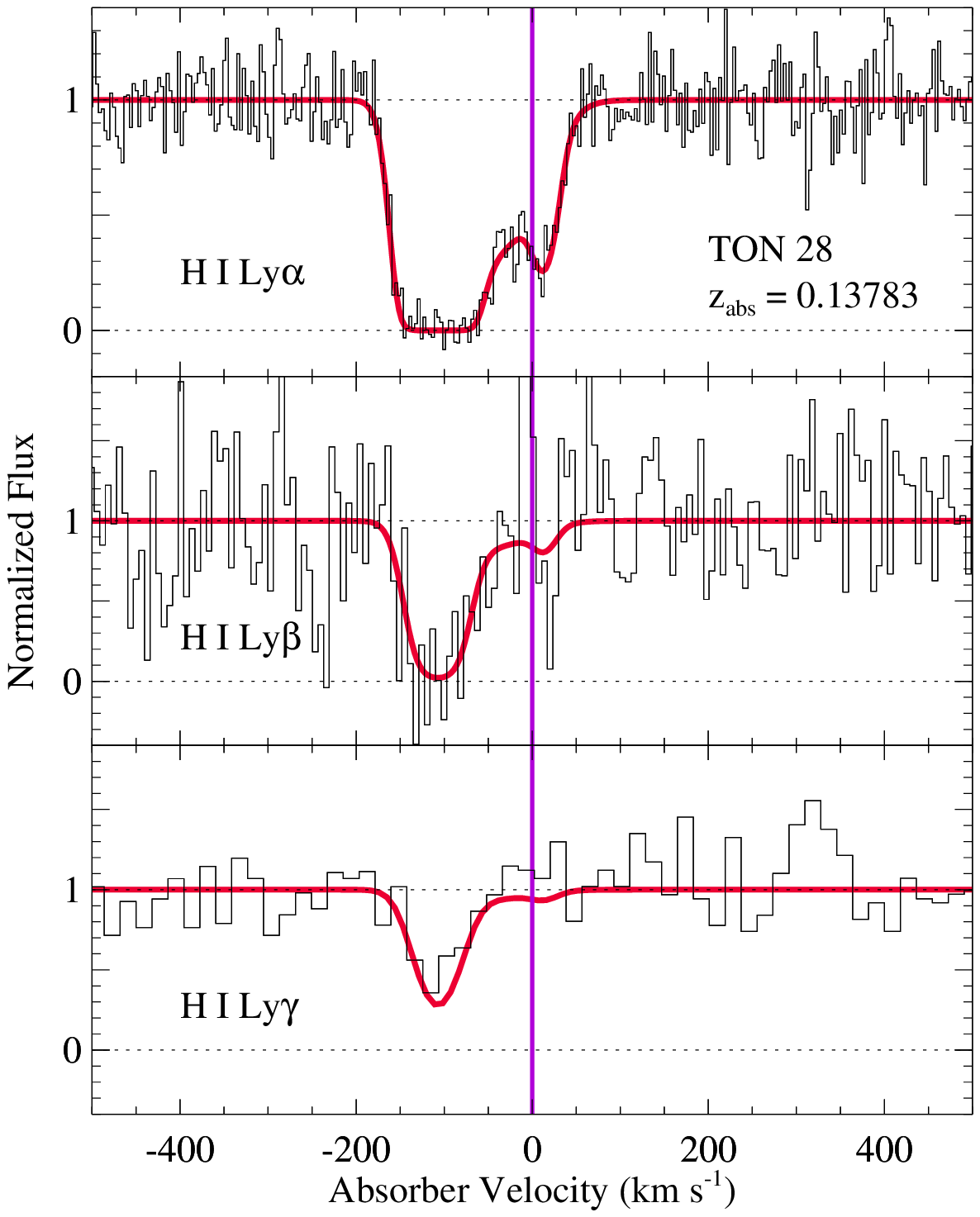}
   \includegraphics[width=8.5cm, angle=0]{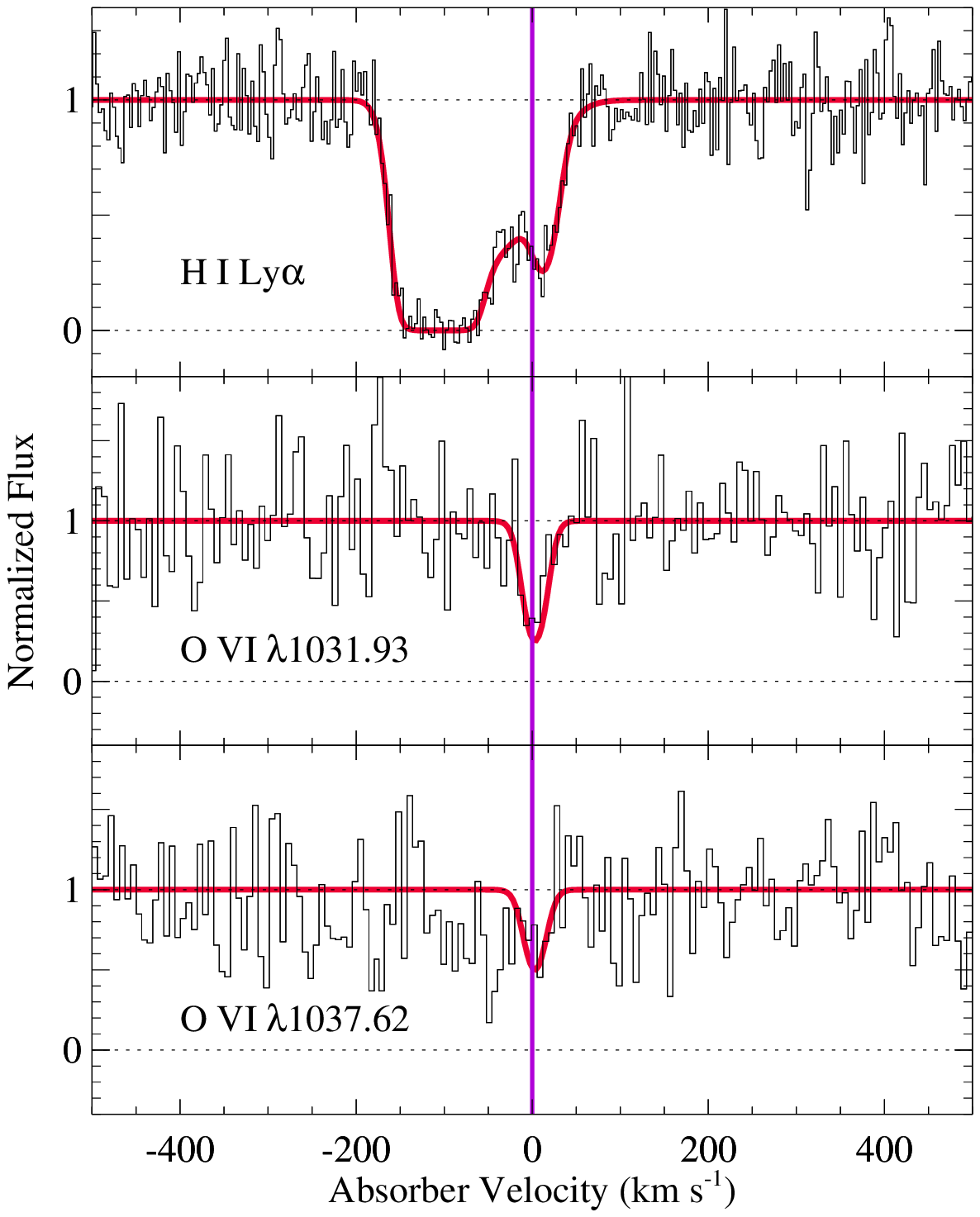}
\caption{{\it Left panels:} Continuum-normalized absorption profiles
of the H~I Ly$\alpha, \beta, \gamma$ and O~VI $\lambda
\lambda$1031.93,1037.62 lines in the absorption system at $z_{\rm
abs}$ = 0.13783 toward TON28.\label{ton28_13783}}
\end{figure}

(43.) {\it PKS0405-123, $z_{abs}$ = 0.49501.} In this case, the
    \ion{H}{1} Ly$\alpha$ line is redshifted beyond the long
    wavelength cutoff of the STIS spectrum, but Ly$\beta$ and
    Ly$\gamma$ are detected at the 4.1$\sigma$ and 2.0$\sigma$ levels,
    respectively.  This absorber is detected in a variety of metals
    (Prochaska et al. 2004), and many of the metal profiles, including
    the \ion{O}{6} lines, show evidence of multiple
    components. However, it should be noted that there is a
    discrepancy evident in one of the components of the \ion{O}{6}
    $\lambda \lambda$1031.93,1037.62 lines.  To show this, we compare
    the apparent column density profiles of the \ion{O}{6} $\lambda
    \lambda$1031.93,1037.62 lines in Figure~\ref{pks0405fig}. We also
    compare the \ion{O}{6} $\lambda$1031.93 $N_{\rm a}(v)$ profile to
    those of \ion{C}{3} $\lambda$977.02, \ion{O}{4} $\lambda$787.71,
    and \ion{O}{5} $\lambda$629.73 in Figure~\ref{pks0405fig}. The
    \ion{O}{6} The \ion{O}{6} $\lambda$1031.93 and $\lambda$1037.62
    profiles agree well in the stronger component at $v \approx$ 0 km
    s$^{-1}$. Looking closely at the strongest component, we can see
    that the profile is asymmetric with a sharp edge on the red side
    and a more gradually decreasing apparent column density on the
    blue side.  This asymmetry suggests that the main feature is a
    blend of two components, and this is corroborated by the
    \ion{C}{3} and \ion{O}{4} profiles, which also show an extra
    component on the blue side.  A third component is evident at $v
    \approx$ 70 km s$^{-1}$ in the \ion{C}{3} and \ion{O}{4}
    transitions.  This component appears to be present in the
    \ion{O}{5} and \ion{O}{6} profiles as well, but at a somewhat
    lower velocity ($v = 57$ km s$^{-1}$).  However, there is a $2-3$
    pixel offset between the peak of the \ion{O}{6} $\lambda$1031.93
    and \ion{O}{6} $\lambda$1037.62 profiles in the $v = 57$ km
    s$^{-1}$ component (see Figure~\ref{pks0405fig}), in contrast to
    the $v \approx$ 0 km s$^{-1}$ component in which the \ion{O}{6}
    profiles agree well.  Initially, this offset appeared to be due to
    a hot pixel feature falling in the middle of the \ion{O}{6}
    $\lambda$1031.93 line, but this cannot be the cause because the
    PKS0405-123 observations were obtained on two separate occasions
    (see Table~\ref{obslog}), and the spectrum detector position was
    shifted between the two visits.  The same component structure is
    evident in the \ion{O}{6} profiles extracted separately from the
    two visits, so this problem is not due to a hot-pixel
    feature. Apart from this small offset, the $N_{\rm a}(v)$ profiles
    of the \ion{O}{6} lines at $v = 57$ km s$^{-1}$ appear to be quite
    consistent: the relative strength and shape of the two lines are
    in agreement.  This suggests that the offset could be caused by an
    instrumental calibration problem.  For example, the STIS geometric
    distortion can cause offsets of this magnitude if not properly
    corrected (Walsh et al. 2001; Ma\'{i}z-Apell\'{a}niz \& \'{U}beda
    2004), and some problems with the distortion correction have been
    noted (Ma\'{i}z-Apell\'{a}niz \& \'{U}beda 2004).  Evidence of
    wavelength calibration problems have also been noted when
    comparing lines that should have identical component structure
    (e.g., Jenkins \& Tripp 2001; Tripp et al. 2005).  While we have
    found these problems to be relatively rare, the stability of the
    STIS gemometric distortion correction has not been studied
    systematically, and it remains possible that the discrepancy in
    the $v = 57$ km s$^{-1}$ component is caused by a calibration
    problem such as this.  Nevertheless, we flag the \ion{O}{6} lines
    in the $v = 57$ km s$^{-1}$ component with a colon in
    Table~\ref{compprop} because of this disagreement, and we treat
    this component as an insecure identification.  With the new Cosmic
    Origins Spectrograph, it will be possible to reobserve PKS0405-123
    to determine if the offset is due to a STIS instrumental problem.
    The \ion{O}{6} $\lambda$1037.62 line at this redshift falls close
    to Galactic \ion{C}{4} $\lambda$1550.78 (which is the source of
    the extra absorption evident at $v < -40$ km s$^{-1}$ in
    Figure~\ref{pks0405fig}), but from the corresponding Galactic
    \ion{C}{4} $\lambda$1548.20 line, we can see that the Milky Way
    \ion{C}{4} has little impact on the redshifted \ion{O}{6}
    $\lambda$1037.62 profile.  We give this system an uncertain
    classification due to the insecure identification of the
    \ion{O}{6} component at $v = 57$ km s$^{-1}$ and the fact that the
    Ly$\alpha$ profile, which is needed to detect low-$N$(\ion{H}{1})
    components, has not been observed at high resolution.

(44.) {\it PKS1302-102, $z_{abs}$ = 0.19159.} The profile fitting code
    obtains its best fit to the \ion{H}{1} lines with a narrow and
    deep component superimposed on a broad and shallow component at
    the same velocity.  The absorption that requires the broad
    component could be nearly as well-fit with multiple narrower
    components, and higher S/N is required to break this
    degeneracy. Only \ion{O}{6} $\lambda$1031.93 is detected, but the
    \ion{O}{6} line is precisely aligned with the \ion{H}{1} lines,
    and \ion{C}{3} $\lambda$977.02 and \ion{Si}{3} $\lambda$1206.50
    are also detected with good significance at this redshift.

(45.) {\it PKS1302-102, $z_{abs}$ = 0.22563.} Some parts of the
    Ly$\beta$ profile are affected by blends; those regions of
    Ly$\beta$ were excluded from the fit.  The \ion{O}{6}
    $\lambda$1037.62 line is also located next to an unrelated strong
    line.  However, the \ion{O}{6} 1031.93 and 1037.62 $N_{\rm a}(v)$
    profiles show excellent agreement over most of the velocity range
    where $\lambda$1031.93 is clearly detected, and the velocity range
    of the \ion{O}{6} $\lambda$1037.62 line that is affected by the
    adjacent interloper was excluded from the fit.

(46.) {\it TON28, $z_{abs}$ = 0.13783.} The \ion{O}{6}
    $\lambda$1037.62 line is only detected at the 2.2$\sigma$ level,
    but as shown in Figure~\ref{ton28_13783}, its wavelength and
    relative strength agree with the corresponding \ion{O}{6}
    $\lambda$1031.93 line. In addition, we can see from this figure
    that the \ion{O}{6} lines are aligned with a clearly detected
    component in the corresponding \ion{H}{1} Ly$\alpha$ line.  

(47.) {\it TON28, $z_{abs}$ = 0.27340.} The \ion{H}{1} Ly$\alpha$ line
    is strongly blended with Milky Way \ion{C}{4} $\lambda$1548.20.
    However, the \ion{H}{1} at this redshift is securely identified
    and measured based on the well-detected Ly$\beta$ and Ly$\gamma$
    lines, and comparison of the Galactic \ion{C}{4} $\lambda$1548.20
    and $\lambda$1550.78 apparent column density profiles verifies
    that substantial extra optical depth (due to the blended
    Ly$\alpha$ line) is present in the Galactic \ion{C}{4}
    $\lambda$1548.20 profile.

\begin{deluxetable}{llcccccclc}
\tablewidth{0pc}
\tabletypesize{\footnotesize}
\tablecaption{Sight Line Properties and Log of STIS E140M Echelle Observations\label{obslog}}
\tablehead{ \ & \ & \multicolumn{4}{c}{\underline{\ \ Galactic Coords. and Column Densities \ \ }} & \multicolumn{4}{c}{\underline{ \ \ \ \ \ \ \ \ \ \ STIS Observations \ \ \ \ \ \ \ \ \ \ }} \\
QSO & $z_{\rm QSO}$ & $l$ & $b$ & log $N$(\ion{H}{1}) & log $N$(H$_{2}$) & Dates & Integration & STIS & S/N \\
       Sight Line \ \ \ \ \ \ \ \ \        &      \        & (deg) & (deg) & \ & \ & (yr.mn.dy)  & Time (ksec) & Aperture & Ratio \\ 
(1) & (2)           & (3) & (4) & (5)                 & (6)              & (7)   & (8)         & (9) & (10) }
\startdata
3C 249.1\dotfill     & 0.3115 & 130.4 & 38.5 & 20.45 & 19.09 & 2002.04.16 & 8.888 & 0.06$\times$0.2 & 12 \\
        \            &    \   &  \    &   \  & \     & \ & 2002.10.03 & 12.168 & \ & \\
        \            &    \   &  \    &   \  & \     & \ & 2002.10.04 & 14.496 & \ & \\
        \            &    \   &  \    &   \  & \     & \ & 2002.10.05 & 12.168 & \ & \\
        \            &    \   &  \    &   \  & \     & \ & 2002.10.08 & 12.168 & \ & \\
        \            &    \   &  \    &   \  & \     & \ & 2002.10.09 & 8.888  & \ & \\ \hline
3C 273.0\dotfill     & 0.1583 & 290.0 & 64.4 & 20.22 & 15.92 & 2000.05.02 & 4.987 & 0.2$\times$0.2 & 33 \\
        \            &    \   &  \    &   \  & \     & \ & 2000.06.21 & 4.987 & \ & \\
\enddata

\tablecomments{(1) Target QSO, (2) - (4) QSO redshift and Galactic
coordinates from the NASA Extragalactic Database (NED), (5) total
Galactic H~I column density determined from 21 cm emission; all
measurements are from Wakker et al. (2003) except 3C 351.0 (from
Lockman \& Savage 1995) and PKS0312-770 (from Smoker et al. 2005), (6)
total Galactic H$_{2}$ column density from Wakker (2006); independent
measurements of $N$(H$_{2}$) from the {\it FUSE} data can be found in
Gillmon et al. (2006); in most cases, the Wakker (2006) and Gillmon
et al. (2006) measurements are in good agreement, but in a few
instances (e.g., H1821+643), there are large discrepancies, (7) STIS
observation date, (8) total integration time of exposures obtained on
the date in column 7, (9) the STIS aperture used for the observations
(aperture dimensions in arcsec), and (10) the mean signal-to-noise
ratio per resolution element in a 2 \AA\ continuum region at $\lambda
_{\rm ob} \approx$ 1300 \AA .}

\end{deluxetable}

\clearpage

{\LongTables
\begin{deluxetable}{lcllccc}
\tablecolumns{7}
\tablewidth{0pt}
\tabletypesize{\footnotesize}
\tablecaption{O~VI Absorbers: Integrated Total Equivalent Widths and Column Densities\label{intprop}}
\tablehead{\colhead{Sight line \ \ \ \ }  & \colhead{$z_{\rm abs}$} & \colhead{ \ \ \ \ Species} & \colhead{ \ $W_{\rm r}$ } & \colhead{log $N$} & \colhead{Component Velocities} & \colhead{Spectrograph} \\
 \colhead{} & \colhead{} & \colhead{} & \colhead{(m\AA )} & \colhead{} & \colhead{(km s$^{-1}$)} & \colhead{} \\ 
\colhead{ \ \ \ \ (1)} & \colhead{(2)} & \colhead{\ \ \ \ \ \ \ (3)} & \colhead{\ (4)} & \colhead{(5)} & \colhead{(6)} & \colhead{(7)}}
\startdata
3C 249.1\dotfill & 0.24676 & \ion{H}{1} Ly$\alpha$ & 468.3$\pm$24.5 & $>14.19$\tablenotemark{a} & $-79,0$ & STIS \\ 
   \             &    \    & \ion{O}{6} 1031.93 & 86.7$\pm$9.1 & 13.93$\pm$0.05 & 0 & STIS \\ 
   \             &    \    & \ion{O}{6} 1037.62 & 32.6$\pm$7.0 & 13.81$\pm$0.10 & 0 & STIS \\ \hline 
3C 249.1\dotfill & 0.30811 & \ion{H}{1} Ly$\alpha$ & 318.7$\pm$13.5 & 13.99$\pm$0.03 & $-14,12$ & STIS \\ 
   \             &    \    & \ion{O}{6} 1031.93 & 274.5$\pm$12.8 & $>14.61$\tablenotemark{a} & $-43,0$ & STIS \\ 
   \             &    \    & \ion{O}{6} 1037.62 & 203.1$\pm$8.0 & 14.76$\pm$0.04\tablenotemark{b} & $-43,0$ & STIS \\ \hline 
3C 249.1\dotfill & 0.31364 & \ion{H}{1} Ly$\alpha$ & 148.1$\pm$7.8 & $>13.78$\tablenotemark{a} & $-3$ & STIS \\ 
   \             &    \    & \ion{O}{6} 1031.93 & 58.8$\pm$10.6 & 13.75$\pm$0.08 & 0 & STIS \\
\enddata
\end{deluxetable}
}


{\LongTables
\begin{deluxetable}{lcccccclc}
\tablewidth{0pc}
\tabletypesize{\footnotesize}
\tablecaption{Profile-Fitting Parameters for Intervening and Proximate \ion{O}{6} Absorbers\label{compprop}}
\tablehead{Sight line \ \ \ \ \ \ \ \ \ & $z_{\rm abs}$\tablenotemark{a} & Species & Fitted & $v$\tablenotemark{b} & $b$\tablenotemark{b} & log [$N$(cm$^{-2}$)]\tablenotemark{b} & Component\tablenotemark{c} & Notes\tablenotemark{d} \\
              \       &  (Class)      &   \     & Lines  & (km s$^{-1}$) & (km s$^{-1}$) & \ & Number (type) & \\
(1) & (2) & (3) & (4) & (5) & (6) & (7) & \ \ \ \ \ (8) & (9) }
\startdata 
3C 249.1\dotfill & 0.24676 & \ion{H}{1} & Ly$\alpha, \beta$ & $-79\pm$3 & 8$^{+9}_{-4}$ & 12.57$\pm$0.18 & \ \ 1 (offset)  & 1 \\ 
    \     &  (Int-simple*) &   \        &    \            & $0\pm$1 & 36$\pm$2 & 14.52$\pm$0.02  & \ \ 2 (aligned) & \\ 
        \        &     \   & \ion{O}{6} & 1031.93,1037.62 & $0\pm$3 & 32$\pm$3 & 13.94$\pm$0.04 & \ \ 2 (aligned) & \\ \hline 
3C 249.1\dotfill & 0.30811 & \ion{H}{1} & Ly$\alpha , \beta$ & $-14\pm$2 & 47$\pm 2$ & 13.91$\pm$0.03 & \ \ 1 (offset) & 2 \\ 
    \     & (Prox-complex) &   \        &    \            & $12\pm$2 & 15$\pm$4 & 13.32$\pm$0.11  & \ \ 2 (offset) & \\ 
        \        &    \    & \ion{O}{6} & 1031.93,1037.62 & $-43\pm$1 & 16$\pm$2 & 13.84$\pm$0.04 & \ \ 3 (offset) & \\ 
    \     &    \    &   \        &    \            & 0$\pm$1: & 22$\pm$1: & $\gtrsim$14.68: & \ \ 4 (offset:) & \\ \hline 
3C 249.1\dotfill & 0.31364 & \ion{H}{1} & Ly$\alpha , \beta , \gamma$ & $-3\pm$1 & 12$\pm 1$ & 14.05$\pm$0.03 & \ \ 1 (aligned) & 3 \\
\enddata

\tablenotetext{a}{Absorber systemic redshift.  In this
paper, we use the centroid of the strongest O~VI component to
define the systemic redshift of the absorber. For reasons noted in the
text, in this paper absorption systems detected within 5000 km
s$^{-1}$ of the QSO redshift are classified as ``proximate'' (Prox.)
absorbers.  Systems that displaced by more than 5000 km s$^{-1}$ from
$z_{\rm QSO}$ are classified as ``intervening'' (Int.) systems.  We
also sort the absorbers into three categories based on the
correspondence of their O~VI and H~I profiles; systems in
which all H~I and O~VI components are aligned (see \S
\ref{aligndefsec}) are classified as ``simple'', while absorbers that
show significant velocity offsets between H~I and O~VI
components are classified as ``complex''. These absorber
classifications are listed in parentheses below the absorber redshift
in this column.}

\tablenotetext{b}{For each component, we list the velocity
centroid ($v$), Doppler parameter ($b$), and column density ($N$)
obtained from Voigt-profile fitting using the software of Fitzpatrick
\& Spitzer (1997).  The listed parameter uncertainties primarily
reflect the random noise in the measurements.  In some cases,
systematic uncertainties due to effects such as line saturation and
blending can be larger than the formal uncertainties listed in the
table.  For these more uncertain measurements, we list the formal
error bars indicated by the fitting, but we flag the measurements with
a colon. Comments on the source of the additional systematic
uncertainties are presented in the Appendix.}

\tablenotetext{c}{Comments on line identification, hot
pixels, blending, and saturation issues are provided for specfic
systems in the Appendix.  The comments in the Appendix are numbered,
and for cross-referencing convenience, the number in the notes column
indicates the comment number in the Appendix.}

\end{deluxetable}
}

\begin{deluxetable}{ccccc}
\tablewidth{0pc}
\tablecaption{Intervening O~VI Absorbers: Number per Redshift Interval\label{dndztab}}
\tablehead{Limiting $W_{\rm r}$ & Number of & Total $\Delta z$ & $dN/dz$ & $dN/dz$ \\
 (m\AA ) & O~VI Systems & \ & (Systems) & (Components) \\
(1) & (2) & (3) & (4) & (5)}
\startdata
30  & 41 & 2.623 & 15.6$^{+2.9}_{-2.4}$ & 21.0$^{+3.2}_{-2.8}$  \vspace{0.15cm} \\ 
70  & 27 & 3.063 & 8.8$^{+2.1}_{-1.7}$ & 9.5$^{+2.1}_{-1.8}$ \vspace{0.15cm} \\  
100 & 14 & 3.121 & 4.5$^{+1.5}_{-1.2}$ & 4.2$^{+1.5}_{-1.2}$ \vspace{0.15cm} \\  
200 &  6 & 3.177 & 2.2$^{+1.2}_{-0.8}$ & 1.9$^{+1.1}_{-0.8}$ \vspace{0.15cm} \\  
300 &  3 & 3.179 & 0.9$^{+1.0}_{-0.5}$ & 0.3$^{+0.7}_{-0.3}$     
\enddata
\tablecomments{(1) Limiting rest-frame equivalent width of the sample
(the corresponding quantities listed in columns 2-4 are for all
absorbers in the 16 QSO sight lines with rest-frame equivalent width
greater than the limiting $W_{\rm r}$); (2) number of absorption
systems in the sample (as discussed in the text, absorption
``systems'' can include multiple discrete components); (3) total
redshift path (summed over all 16 sight lines) over which the spectra
have sufficient S/N and resolution to detect lines with equivalent
width greater than the limiting $W_{\rm r}$; (4) number of O~VI
{\it systems} per unit redshift; (5) number of discrete O~VI
{\it components} per unit redshift.  Uncertainties in the $dN/dz$
measurements in columns (4) and (5) were calculated using the
small-number statistics tables of Gehrels (1986).}
\end{deluxetable}

\begin{deluxetable}{ccccc}
\tablewidth{0pc}
\tablecaption{Proximate O~VI Absorbers: Number per Redshift Interval\label{asscdndztab}}
\tablehead{Limiting $W_{\rm r}$ & Number of & Total $\Delta z$ & $dN/dz$ & $dN/dz$ \\
 (m\AA ) & O~VI Systems & \ & (Systems) & (Components) \\
(1) & (2) & (3) & (4) & (5)}
\startdata
\multicolumn{5}{c}{\underline{Absorbers within 5000 km s$^{-1}$ of $z_{\rm QSO}$}} \vspace{0.15cm} \\ 
30  & 12 & 0.357 & 33.6$^{+12.8}_{-9.6}$ & 72.8$^{+17.3}_{-14.2}$  \vspace{0.15cm} \\
70  &  9 & 0.387 & 23.3$^{+10.6}_{-7.6}$ & 25.8$^{+11.1}_{-8.0}$  \vspace{0.15cm} \\ 
100 &  8 & 0.389 & 20.6$^{+10.1}_{-7.2}$ & 18.0$^{+9.7}_{-6.6}$ \vspace{0.15cm}  \\ 
200 &  4 & 0.390 & 10.3$^{+8.1}_{-5.0}$ & 7.7$^{+7.5}_{-4.2}$ \vspace{0.15cm}  \\ 
300 &  2 & 0.391 & 5.1$^{+6.8}_{-3.3}$ & $< 4.7 (1\sigma)$  \vspace{0.15cm} \\ 
 \  &  \ &   \   &  \                   &   \\
\multicolumn{5}{c}{\underline{Absorbers within 2500 km s$^{-1}$ of $z_{\rm QSO}$}} \vspace{0.15cm} \\ 
30  & 11 & 0.214 & 51.4$^{+20.7}_{-15.3}$ & 107.5$^{+27.4}_{-22.3}$  \vspace{0.15cm} \\ 
70  &  7 & 0.228 & 30.7$^{+16.2}_{-11.3}$ & 39.5$^{+18.0}_{-12.9}$  \vspace{0.15cm} \\
100 &  6 & 0.230 & 26.1$^{+15.6}_{-10.4}$ & 26.1$^{+15.6}_{-10.4}$  \vspace{0.15cm} \\
200 &  3 & 0.230 & 13.0$^{+12.7}_{-7.1}$ & 8.7$^{+11.5}_{-5.6}$  \vspace{0.15cm} \\
300 &  2 & 0.232 & 8.6$^{+11.4}_{-5.5}$ & $< 7.9 (1\sigma )$ \vspace{0.15cm}  \\
\enddata
\tablecomments{The quantities in columns 1 - 5 are described in the notes of Table~\ref{dndztab}. The 3C 351.0 sight line was not included in the sample used to derive the proximate absorber quantities listed in this table because the 3C 351.0 spectrum has a prominent mini-BAL type proximate absorption system (see Yuan et al. 2002 and discussion in text).  Inclusion of this sight line would dramatically increase the proximate system $dN/dz$ measurements listed here.}
\end{deluxetable}

\begin{deluxetable}{lccccccccc}
\tablewidth{0pc}
\tabletypesize{\scriptsize}
\tablecaption{Statistics of O~VI Column Density and Doppler Parameter Distributions\label{mediantab}}
\tablehead{ \ & \   & \multicolumn{2}{c}{$N$(O~VI)} & \ & \ &
\multicolumn{2}{c}{$b$(O~VI)} & \ & \\
 \ & No. & \multicolumn{2}{c}{\underline{ \ \ \ ($10^{13}$ cm$^{-2}$) \ \ \ }} & \multicolumn{2}{c}{\underline{ \ \ KS Test \ \ }} & \multicolumn{2}{c}{\underline{ \ \ \ \ \ (km s$^{-1}$) \ \ \ \ \ }} & \multicolumn{2}{c}{\underline{ \ \ KS Test \ \ }} \\
 \ \ \ \ \ \ \ \ \ \ \ \ \ \ \ \ \ \ \ \ Sample  & Components & Median & Mean$\pm \sigma$ & D & Prob & Median & Mean$\pm \sigma$ & D & Prob \\ 
 \ \ \ \ \ \ \ \ \ \ \ \ \ \ \ \ \ \ \ \ (1) & (2) & (3) & (4) & (5) & (6) & (7) & (8) & (9) & (10) } 
\startdata
Int (Robust)\dotfill                                     & 70 & 6.5 & 9.2$\pm$8.8   & \nodata & \nodata & 24 & 27$\pm$14 & \nodata & \nodata \\
Int (Full)\dotfill                                       & 77 & 5.6 & 8.7$\pm$8.6   & \nodata & \nodata & 24 & 26$\pm$14 & \nodata & \nodata \\
Asc (Robust,$v_{\rm displ} < 5000$ km s$^{-1}$)\dotfill  & 28 & 5.2 & 6.7$\pm$4.6   & 0.171   & 0.558   & 17 & 20$\pm$11 & 0.307   & 0.036 \\
Asc (Robust,$v_{\rm displ} < 2500$ km s$^{-1}$)\dotfill  & 24 & 4.7 & 6.3$\pm$4.1   & 0.185   & 0.533   & 16 & 19$\pm$11 & 0.355   & 0.016 \\
Asc (Full,$v_{\rm displ} < 5000$ km s$^{-1}$)\dotfill    & 34 & 5.8 & 10.0$\pm$11.4 & 0.079   & 0.998   & 17 & 19$\pm$11 & 0.270   & 0.053 \\
Asc (Full,$v_{\rm displ} < 2500$ km s$^{-1}$)\dotfill    & 30 & 5.2 & 10.1$\pm$11.9 & 0.097   & 0.981   & 16 & 18$\pm$11 & 0.306   & 0.027 \\
\enddata
\tablecomments{For six samples of intervening and proximate
absorption lines, the columns list the following information: (1) The
name of the sample (see \S \ref{basicstats} for the sample
definitions). (2) The number of individual components in the
sample. (3) The median O~VI column density and (4) the mean
$N$(O~VI) $\pm$ the standard deviation. (5) The
Kolmogorov-Smirnov D statistic and (6) the probability of the null
hypothesis (i.e., the probability that the two distributions are drawn
from the same parent distribution) found from comparing the
intervening absorber column density distribution to the corresponding
proximate absorber distribution (e.g., the robust intervening sample
is always compared to the robust proximate sample). (7) The median
O~VI Doppler parameter and (8) the mean Doppler parameter $\pm$
the standard deviation. (9) The KS test D and (10) KS test probability
of the null hypothesis from the comparison of the intervening and
proximate $b-$value distributions.}
\end{deluxetable}

\clearpage
\begin{landscape}
\begin{deluxetable}{lcccccccrrc}
\tabletypesize{\footnotesize}
\tablewidth{0pc}
\tablecaption{Column Density Ratios and Temperature Constraints of Well-Aligned Components in Intervening O~VI Absorbers\label{alignedtab}}
\tablehead{ \  & \ & \ & \multicolumn{2}{c}{\underline{Temperature: O~VI only}} & \multicolumn{2}{c}{\underline{Temperature: H~I+O~VI}} & \ & \ & \ & \\
\ \ \ \ \ Sightline \ \ \ \ \ & $z_{\rm abs}$ & $v$ (km s$^{-1}$) & log $T_{\rm best}$ & log $T_{3 \sigma}$ & log $T_{\rm best}$ & log $T_{3\sigma}$ & $b_{\rm nt}$ (km s$^{-1}$) & log $\frac{N({\rm H \ I})}{N({\rm O \ VI})}$ & log $\frac{N({\rm C \ III})}{N({\rm O \ VI})}$ & Class \\
\ \ \ \ \ \ \ \ \ (1) & (2) & (3) & (4) & (5) & (6) & (7) \ \ \ \ & (8) \ & (9) \ \ \ \ & (10) \ \ \ \ & (11) }
\startdata
3C 249.1\dotfill    & 0.24676 & 0      & $<$6.02 & $<$6.26 & 4.26    & $<$4.89 & 33 & 0.58 \ \ \ \    & \nodata \ \ \ & Simple* \\
3C 273.0\dotfill    & 0.00334 & 0      & $<$6.40 & $<$6.76 & 5.29    & $<$6.33 & 49 & 0.75 \ \ \ \    & $< -0.79$ \ \ \ & Complex \\
3C 273.0\dotfill    & 0.12003 & 0      & $<$4.79 & $<$5.30 & 4.43    & $<$4.60 & 6  & 0.14 \ \ \ \    & $-0.61$ \ \ \ & Simple \\
3C 351.0\dotfill    & 0.31658 & $-51$  & $<$5.63 & $<$5.96 & 4.53    & $<$4.92 & 20 & 0.74 \ \ \ \    & $< -1.13$ \ \ \ & Simple \\
              \     & \       & 0      & $<$5.63 & $<$5.82 & 4.18    & $<$4.70 & 21 & 0.45 \ \ \ \    & $< -1.28$ \ \ \ & \\
              \     & \       & 70     & $<$6.07 & $<$6.30 & \nodata & \nodata & \nodata & $-0.17$ \ \ \ \ & \nodata \ \ \ & \\
H1821+643\dotfill   & 0.02438 & 0      & $<$5.63 & $<$6.18 & 4.35    & $<$4.86 & 20 & 0.78 \ \ \ \    & $< -0.52$ \ \ \ & Simple \\
H1821+643\dotfill   & 0.12143 & 0      & $<$6.74 & $<$7.05 & 5.02    & $<$5.79 & 75 & $-0.22$ \ \ \ \ & $< -0.95$  \ \ \ & Complex \\
\enddata

\tablecomments{Columns in this table list the following data for
components with O~VI and H~I components that have aligned
centroids (see \S \ref{aligndefsec}): (1) sightline, (2) absorption
system redshift, (3) velocity of the particular component from which
the temperature and column density ratios were measured, (4)
temperature upper limit implied by the best-fit O~VI $b-$value
(equation~\ref{tvsb}), (5) $3\sigma$ upper limit on $T$ indicating by
propagating the uncertainty in $b$(O~VI) through
equation~\ref{tvsb}, (6) the best measurement of the gas temperature
found by solving equation~\ref{fulltvsb} using the O~VI and
H~I $b-$values (log $T$, with $T$ measured in Kelvins), (7) the
$3\sigma$ upper limit on the temperature found by propagating the
$b-$value uncertainties through equation~\ref{fulltvsb}, (8) the
nonthermal broadening component corresponding to the best temperature
in column 6, (9) log of the H~I/O~VI column-density
ratio, (10) log of the C~III/O~VI column-density ratio, and
(11) the absorber classification (\S \ref{classmulti}).}

\end{deluxetable}
\clearpage
\end{landscape}

\end{document}